\pdfoutput=1 
\documentclass[a4paper,11pt,openany]{scrbook}
\usepackage{graphicx}
\usepackage{amssymb}
\usepackage{epstopdf}
\usepackage[final]{pdfpages}
\usepackage{amsmath}  
\usepackage{amsthm}   
\usepackage{graphicx}
\usepackage{mathrsfs}
\usepackage{floatflt}
\usepackage[mathscr]{eucal}
\usepackage{units}
\usepackage{color}
\usepackage{fancyhdr}
\usepackage{caption}

\usepackage[utf8]{inputenc} 
\usepackage{textcomp} 
\usepackage{bm}  

\usepackage{memhfixc}  

\usepackage[pdftex,bookmarks,colorlinks,breaklinks]{hyperref}  
\definecolor{dullmagenta}{rgb}{0.4,0,0.4}   
\definecolor{darkblue}{rgb}{0,0,0.4}
\hypersetup{linkcolor=black,citecolor=black,filecolor=black,urlcolor=black} 

\newcommand{\sss}[1]{{\scriptscriptstyle{#1}}}
\newcommand{\mean}[1]{\left\langle #1 \right\rangle}
\newcommand{\abs}[1]{\left| #1 \right|}
\newcommand{\order}[1]{\mathscr{O}\!\left(#1\right)}
\newcommand{\diag}{\mathrm{diag}}

\newcommand{\Mpc}{~\mathrm{Mpc}}
\newcommand{\eV}{~\mathrm{eV}} 
\newcommand{\MeV}{~\mathrm{MeV}} 
\newcommand{\TeV}{~\mathrm{TeV}}
\newcommand{\GeV}{~\mathrm{GeV}}
\newcommand{\km}{~\mathrm{km}}
\newcommand{\scnd}{~\mathrm{s}}

\newcommand{\efolds}{$e$-folds\,}
\newcommand{\efold}{$e$-fold\,}

\newcommand{\dd}{\mathrm{d}}
\newcommand{\uPl}{\mathrm{Pl}}
\newcommand{\uin}{\mathrm{in}}
\newcommand{\udm}{\mathrm{dm}}
\newcommand{\ude}{\mathrm{de}}

\newcommand{\uend}{\mathrm{end}}
\newcommand{\ueff}{\mathrm{eff}}
\newcommand{\ueq}{\mathrm{eq}}

\newcommand{\urad}{\mathrm{rad}}

\newcommand{\uphys}{\mathrm{phys}}
\newcommand{\uPgiP}{\mathrm{(gi)}}
\newcommand{\uPVP}{{\tiny \mathrm{(V)}}}
\newcommand{\uleft}{\mathrm{left}}
\newcommand{\uright}{\mathrm{right}}
\newcommand{\uclosed}{\mathrm{closed}}
\newcommand{\uopen}{\mathrm{open}}
\newcommand{\ue}{\mathrm{e}}
\newcommand{\uc}{\mathrm{c}}
\newcommand{\ud}{\mathrm{d}}
\newcommand{\ub}{\mathrm{b}}
\newcommand{\us}{\mathrm{s}}

\newcommand{\um}{\mathrm{m}}
\newcommand{\uhc}{\mathrm{hc}}
\newcommand{\udec}{\mathrm{dec}}
\newcommand{\umod}{\mathrm{mod}}
\newcommand{\uC}{\mathrm{C}}
\newcommand{\uK}{\mathrm{K}}

\newcommand{\uCS}{\mathrm{CS}}
\newcommand{\uNS}{\mathrm{NS}}
\newcommand{\uR}{\mathrm{R}}
\newcommand{\uS}{{_\mathrm{S}}}

\newcommand{\uPSP}{{\tiny \mathrm{(S)}}}

\newcommand{\uT}{{_\mathrm{T}}}
\newcommand{\uPTP}{{\tiny \mathrm{(T)}}}

\newcommand{\uDBI}{\sss{\mathrm{DBI}}}
\newcommand{\uUV}{\sss{\mathrm{UV}}}

\newcommand{\utotal}{\mathrm{tot}}

\newcommand{\usssS}{\sss{\uS}}
\newcommand{\usssT}{\sss{\uT}}
\newcommand{\usssPl}{\sss{\uPl}}

\newcommand{\calH}{\mathcal{H}}
\newcommand{\calP}{\mathcal{P}}
\newcommand{\calN}{\mathcal{N}}
\newcommand{\calL}{\mathcal{L}}

\newcommand{\cS}{c_\usssS}
\newcommand{\nS}{n_\usssS}

\newcommand{\QQoverTT}{\frac{Q^{2}_{\mathrm{rms-PS}}}{T^{2}}}

\newcommand{\ini}{\uin}

\newcommand{\ie}{\emph{i.e.}\,}
\newcommand{\etc}{\emph{etc.}\,}
\newcommand{\eg}{\emph{e.g.}\,}

\newcommand{\apriori}{\emph{a priori}\,}

\newcommand{\gs}{g_\us}
\newcommand{\ells}{\ell_\us}
\newcommand{\alphas}{\alpha'}

\newcommand{\mpl}{m_\usssPl}

\newcommand{\Mpl}{M_\usssPl}
\newcommand{\ms}{m_{\sss{s}}}

\newcommand{\Dbar}{$\overline{\mathrm{D}3}$\ }

\newcommand{\epsone}{{\epsilon_1}}
\newcommand{\epstwo}{{\epsilon_2}}

\newcommand{\phiend}{\phi_{\mathrm{end}}}

\newcommand{\phiin}{\phi_\uin}

\newcommand{\calM}{\mathcal{M}}

\newcommand{\scrT}{\mathscr{T}}
\newcommand{\scrM}{\mathscr{M}}

\newcommand{\kstar}{k_*}

\newcommand{\OmegaCDM}{\Omega_\udm}
\newcommand{\OmegaT}{{\Omega_\utotal}}
\newcommand{\OmegaL}{{\Omega_{\Lambda}}}

\newcommand{\OmegaB}{\Omega_\ub}
\newcommand{\OmegaM}{\Omega_\um}

\newcommand{\Ricci}{R}
\newcommand{\cl}{\mathrm{cl}}

\newcommand{\beq}{\begin{equation}}
\newcommand{\eeq}{\end{equation}}
\newcommand{\bea}{\begin{eqnarray}}
\newcommand{\eea}{\end{eqnarray}}

\newcommand{\Lcdm}{$\Lambda$CDM}
\newcommand{\rhocrit}{\rho_{\sss{\mathrm{crit}}}}
\newcommand{\rhotot}{\rho_\utotal}
\newcommand{\const}{\mathrm{const.}}

\newcommand{\Sgrav}{\mathcal{S}_{\mathrm{grav}}}
\newcommand{\Smatter}{\mathcal{S}_{\mathrm{matter}}}
\newcommand{\Sphi}{\mathcal{S}_{\phi}}

\newcommand{\action}{\mathcal{S}}

\newcommand{\Lphi}{\mathcal{L}_{\phi}}

\newcommand{\phicl}{\phi_\cl}

\newcommand{\drm}{{\mathrm d}}

\newcommand{\bra}[1]{\langle #1|}
\newcommand{\ket}[1]{|#1\rangle}

\newcommand{\beqa}{\begin{eqnarray}}
\newcommand{\eeqa}{\end{eqnarray}}

\newcommand{\alphap}{\alpha'}

\newcommand{\bear}{\begin{array}}  \newcommand{\eear}{\end{array}}
\newcommand{\bef}{\begin{figure}}  \newcommand{\eef}{\end{figure}}
\newcommand{\bec}{\begin{center}}  \newcommand{\eec}{\end{center}}

\title{Primordial Fluctuations in String Cosmology}
\author{Larissa C. Lorenz}
\date{\today}

\begin{document}

\frontmatter
\begin{titlepage}
\begin{minipage}{0.5\textwidth}
\includegraphics[width=0.5\textwidth]{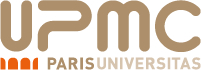}
\end{minipage}
\hfill
\begin{minipage}{0.5\textwidth}
\flushright\includegraphics[width=0.5\textwidth]{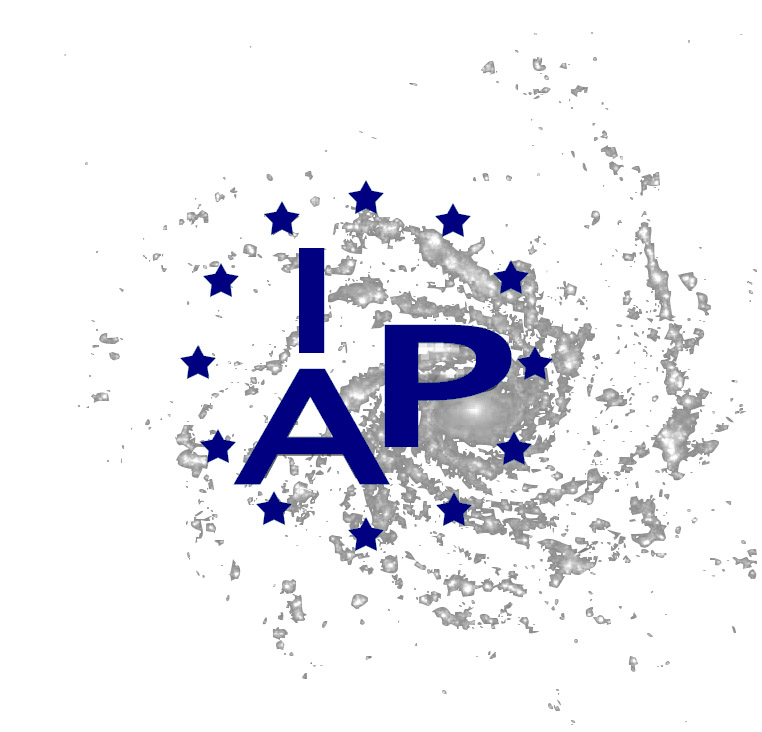}
\end{minipage}
\bigskip
\begin{center}
{\Large \textsc{Th\`ese de Doctorat de l'Universit\'e Pierre et Marie Curie}}
\bigskip

Sp\'ecialit\'e: Physique Th\'eorique

\bigskip
\bigskip

{\large Ecole Doctorale 107 ``Physique de la r\'egion parisienne''}\\
{\large Institut d'Astrophysique de Paris}

\bigskip
\bigskip

Present\'ee par

\textsc{Larissa C. Lorenz}

\bigskip
\bigskip

Pour obtenir le grade de

\textsc{Docteur de l'Universit\'e Pierre et Marie Curie}
\end{center}

\bigskip
\bigskip
{\bf Sujet de la th\`ese:}
\begin{center}
{\Large \textsc{Fluctuations primordiales en cosmologie des cordes}}

\bigskip
\bigskip
soutenue le 26 juin 2009 devant le jury compos\'e de

\vfill
{\large
\begin{tabular}{lp{8cm}r}
M.& J\'er\^ome \textsc{Martin}&directeur de th\`ese\\
M.&Robert \textsc{Brandenberger}&co-directeur de th\`ese\\
M.&Edmund \textsc{Copeland}&rapporteur\\
M.&William H. \textsc{Kinney}&rapporteur\\
M.&Michael \textsc{Joyce}&pr\'esident\\
M.&Philippe \textsc{Brax}&examinateur\\
M.&Fernando \textsc{Quevedo}&examinateur
\end{tabular}}
\end{center}

\newpage
\thispagestyle{empty}
\mbox{}
\newpage
\thispagestyle{empty}
\begin{center}
\emph{\LARGE Acknowledgements}
\end{center}

\parindent = 0.4in

I would like to express my heartfelt gratitude to, and my profound respect for, my \emph{directeur de th\`ese} J\'er\^ome Martin who has inspired, encouraged and supported me throughout this thesis. On the map of modern cosmology, he guided me to the main sights without telling me where to go.\\
No less have I benefited from the ideas and invaluable advice of my \emph{co-directeur} Robert Brandenberger. Under his kind hospitality in Montr\'eal I found the key to doors that I had long been trying to open.

Many thanks go to my \emph{rapporteurs} Edmund Copeland and William H. Kinney for their patience in light of the size of the manuscript, as well as to the members of my \emph{jury de th\`ese}, Philippe Brax, Michael Joyce and Fernando Quevedo. I deeply appreciated their time and effort, and their scienfic interest in my work is a source of continuous encouragement to me.

Finally, I would like to thank everyone who has been part of my Paris adventure, from its rocky start to the soft landing. Whether you were with me \emph{sur place} or present through your caring counsel in calls, emails and letters, whether we spent \emph{un petit instant} together or have shared the better part of our lives, you hold the most special of places in my heart because: \emph{We'll always have Paris.}

\parindent = 0.0in
\newpage
\thispagestyle{empty}
\mbox{}
\newpage
\thispagestyle{empty}
\begin{flushright}
{\LARGE \emph{F\"ur Stefan}}
\end{flushright}
\newpage
\thispagestyle{empty}
\mbox{}
\newpage

\markboth{}{Abstract}
\begin{center}
\textsc{Primordial Fluctuations in String Cosmology}
\end{center}

This thesis is dedicated to the study of inflationary scenarios based on string theory. Inflation is a brief period of accelerated expansion in the very early Universe which gives natural explanations for problems of the Standard Big Bang Model (SBBM) of cosmology. The phenomenological success of inflation provides the motivation for numerous efforts to establish its connection with particle physics. As a promising contender for a unified theory at very high energy scales, string theory is an obvious framework to look for the theoretical foundations of inflation; we therefore try to determine the characteristic features of string-inspired inflationary scenarios, along with their observational consequences. In particular, we shall be interested in models of brane inflation, in which the inflaton field has non-canonical dymanics and where the era of reheating is interpreted as the mutual annihilation of (anti-)branes.

We present first a detailed comparison of the ``KKLMMT brane inflation'' scenario to data from the Wilkinson Microwave Anisotropy Probe (WMAP) satellite. We then turn our attention to the consequences of modified dynamics for the inflaton: outside the string theory context, these scenarios are known as $k$-inflation, and we establish their observational predictions with particular regard for the differences to the standard case (where the inflaton kinetic term is canonical). We study in detail the case of ``Dirac Born Infeld (DBI) inflation'', which is the string-inspired subclass of $k$-inflationary models. In a second step, we then again compare the obtained predictions to WMAP. Two other publications consider the end of brane inflation: firstly, we calculate the behaviour of entropy perturbations at the onset of brane--anti-brane annihilation in the KKLMMT model and, secondly, we study reheating of the Universe following multiple brane collisions in a novel scenario called ``monodromy inflation''.
\bigskip
{\bf Keywords:} cosmology, inflation, perturbations, cosmic microwave background radiation, string theory, branes
\end{titlepage}

\thispagestyle{empty}
\mbox{}
\newpage
\markboth{Introduction}{Introduction}
\section*{Introduction}{\markboth{Introduction}{}}

The scenario of inflation has shaped the subject of cosmology ever since its invention over 25 years ago. Originally intended to dispose of some of the Standard Big Bang Model's (SBBM) shortcomings, it was soon realized that inflation may be responsible for a powerful manifestation of our Universe's quantum nature: assuming the very early Universe was dominated by a scalar field (the so-called inflaton), the origin of today's cosmic structures such as galaxies, clusters, and filaments can be traced back to this field's quantum fluctuations. Stretched by the quasi-exponential expansion of spacetime, these fluctuations served as the primordial seeds for inhomogeneities that later grew under the influence of gravitational instability.

As a discipline of physics, cosmology is as much interested in explaining \emph{how} a phenomenon occurs (at the technical level, \ie through its formulation in mathematical terms) as it tries to understand \emph{why} it takes place. Technically, the inflationary scenario is appealing because of its simple and economic nature. However, its origin from an underlying physical theory has remained elusive because the (scalar) inflaton is difficult to explain from the field content of the Standard Model (SM) of particle physics.\\
On the other hand, given that inflation possibly took place while the Universe's energy density was beyond the SM's predictive regime, we may have a better chance of finding the inflaton among the degrees of freedom of a Grand Unified Theory (GUT). In terms of both the effort invested and the progress achieved, string theory clearly occupies a privileged position among GUT candidates. Moreover, it is particularly rich in scalar fields, whose stringy interpretation ranges from coupling constants to the geometrical detail of extra dimensions. From a cosmological viewpoint, it hence seems natural to closely examine these string-inspired scalar fields with respect to their ``inflationary'' use. In recent years, this has been met with a warm welcome on the side of string theory.\\
While one might not appraise string theory for its mathematical simplicity, 
the attempt to construct a theory that has but one parameter (namely the length of fundamental strings $\ells$) has proven to be very fruitful. However, the lack of string theoretic predictions verifiable in accelerator experiments has been felt more and more clearly over the years. Therefore cosmology, as a different arena (besides particle physics) for string theory's concepts to surface, is of utmost interest. It is the purpose of this thesis to walk the line between these two subjects, the confrontation of which has already produced a wealth of literature.

In Part \ref{part:cosmo}, we lay the cosmological ground work for our analysis, introducing the concepts of modern cosmology including the scenario of inflation. We discuss the treatment of perturbations in inflation as well as their relation to today's observable quantities, explaining how to pass from primordial perturbation spectra through the phases of reheating and SBBM evolution to measurements of cosmic microwave background (CMB) anisotropies. The global picture (\ie beyond our own Hubble horizon) of the inflationary Universe is considered in the context of ``stochastic inflation''.

Open issues with inflation and the need for proper theoretical anchorage in high energy physics lead us to an in-depth discussion of string cosmology in Part \ref{part:string}: following a self-contained introduction to string theory, we discuss the progress and pitfalls on the way to viable string-inspired models of the early Universe. We give a short overview of the different types of string inflationary models constructed, with a clear focus on scenarios of brane inflation, which are behind the original scientific results presented in this thesis. In particular, we derive the archetype of these models, the so-called KKLMMT scenario, from its foundation in type IIB superstring theory.

Part \ref{part:results} assembles the new scientific results obtained over the course of this thesis. Firstly, the KKLMMT scenario of brane--anti-brane inflation was compared with the WMAP3 data using the Monte Carlo Markov Chain approach. A thorough derivation of the model's cosmological parameters from the type IIB string theory background was applied when restricting their \emph{a priori} exploration range for consistency, and as well when interpreting the \emph{a posteriori} probability distributions. These results are presented in the article included in Chapter \ref{chapter:WMAP3-paper}.\\
In the following Chapter, we take a closer look at the modified dynamics of the inflaton field if it is an open string mode, as is the case for brane inflation. Traditionally called $k$-inflation, scenarios with a non-canonical term in string theory are of the Dirac Born Infeld (DBI) type. The calculation of perturbation spectra must be adapted to this new situation because the $k$-inflationary perturbations propagate at a non-trivial sound speed $\cS$. We derive the scalar and tensor spectra in the $k$-inflation analogue of the slow roll approximation (where changes in $\cS$, on top of those in the Hubble parameter $H$, must be small), using the so-called uniform approximation. Also presented in Chapter \ref{chapter:kinf-WMAP5-paper} is a comparison of these $k$-inflationary power spectra (and their DBI subclass) to the WMAP5 data.\\
When the brane and the anti-brane start their annihilation at the end of inflation in the KKLMMT scenario, a tachyon appears and, for a short period of time, the evolution can be described in terms of a two-field scenario with a new, adjusted potential. This two-field phase (where both inflaton and tachyon are dynamic) is the subject of the publication in Chapter \ref{chapter:tachyon-paper}. It is found that entropy perturbations between the two fields can grow exponentially during a brief time interval, and they may accumulate enough to induce a sizable contribution to the comoving curvature perturbation (which is due to purely adiabatic perturbations in single field scenarios).\\
Reheating is also considered in Chapter \ref{chapter:monodromy-paper}, but this time in the context of a different brane inflation model: in type IIA string theory, the inflaton field can be associated with the ``wrapping'' of a D4-brane along a direction of monodromy in the extra-dimensional geometry. (A simple example of monodromy is compactification on two twisted tori.) Seeking to minimize its world volume (and hence its energy), the D4 will unwrap, traversing the monodromic dimension multiple times. If a D6-brane, on which the Standard Model of particle physics is located, sits at a fixed position in the monodromy direction, it will be hit multiple times by the unwrapping D4. However, as shown in the article of Chapter \ref{chapter:monodromy-paper}, the energy transfer from the inflaton towards the Standard Model is negligible during these collisions, with reheating taking place instantaneously at the last brane encounter.

Finally, in Part \ref{part:conclusions} we conclude with general remarks on the lessons learnt from the present work. We recall their broader scientific context and comment on the prospects in store over the upcoming years. Keeping unbridled optimism at arm's length, we are nonetheless confident that such prospects are bright.

\newpage
\thispagestyle{empty}
\mbox{}
\newpage
\thispagestyle{empty}
\mbox{}
\newpage
\markboth{Contents}{Contents}
\tableofcontents


\vfill
\section*{Remarks on Notation}

\textsc{Indices}\\
Small greek indices \(\mu,\,\nu\dots\) run from 0 to 3, where \(x^{0}=t\) is cosmic time. Small Latin indices \(m,\, n\dots\) run from 0 to 9, i.e. over \emph{all} dimensions, compactified or not. An exception is \(i\), which from time to time will be taken to run from 1 to 3 only, such that \((x^{0},\,x^{i}),\,i=1,2,3\) is ordinary four-dimensional spacetime, and the pair $a\,,b$, which we will sometimes use for the $(p+1)$-spacetime directions aligned with the world volume of a D$p$-brane. Capital Latin indices \(A,\,B\dots\) run from 4 to 9 (or from 5 to 9 occasionally), that is, over extra (compactified) dimensions only.

\textsc{Units}\\
Our units are such that $\hbar=c=1$, \ie in particular space and time are measured in the same units, which are inverse to the unit of mass.

\mainmatter
\part{Introducing Modern Cosmology}\label{part:cosmo}
\chapter{Standard Cosmology and Inflation}\label{chapter:cosmo-infl}
\begin{quotation}
\emph{In this Chapter, the cornerstones of modern cosmology are presented in broad terms. Based on four key observational facts, the Standard Big Bang Model (SBBM) offers a natural explanation for the sequence of events in the Universe ever since the hot Big Bang 13.7 billion years ago. Questions left unanswered by the SBBM can, to a large extent, be resolved by the scenario of inflation, whose mechanism we discuss at the background level.}
\end{quotation}

\section{The Standard Big Bang Model}\label{sec:SBBM}

\begin{figure}[t]
\begin{center}
\includegraphics[width=\textwidth,clip=true]{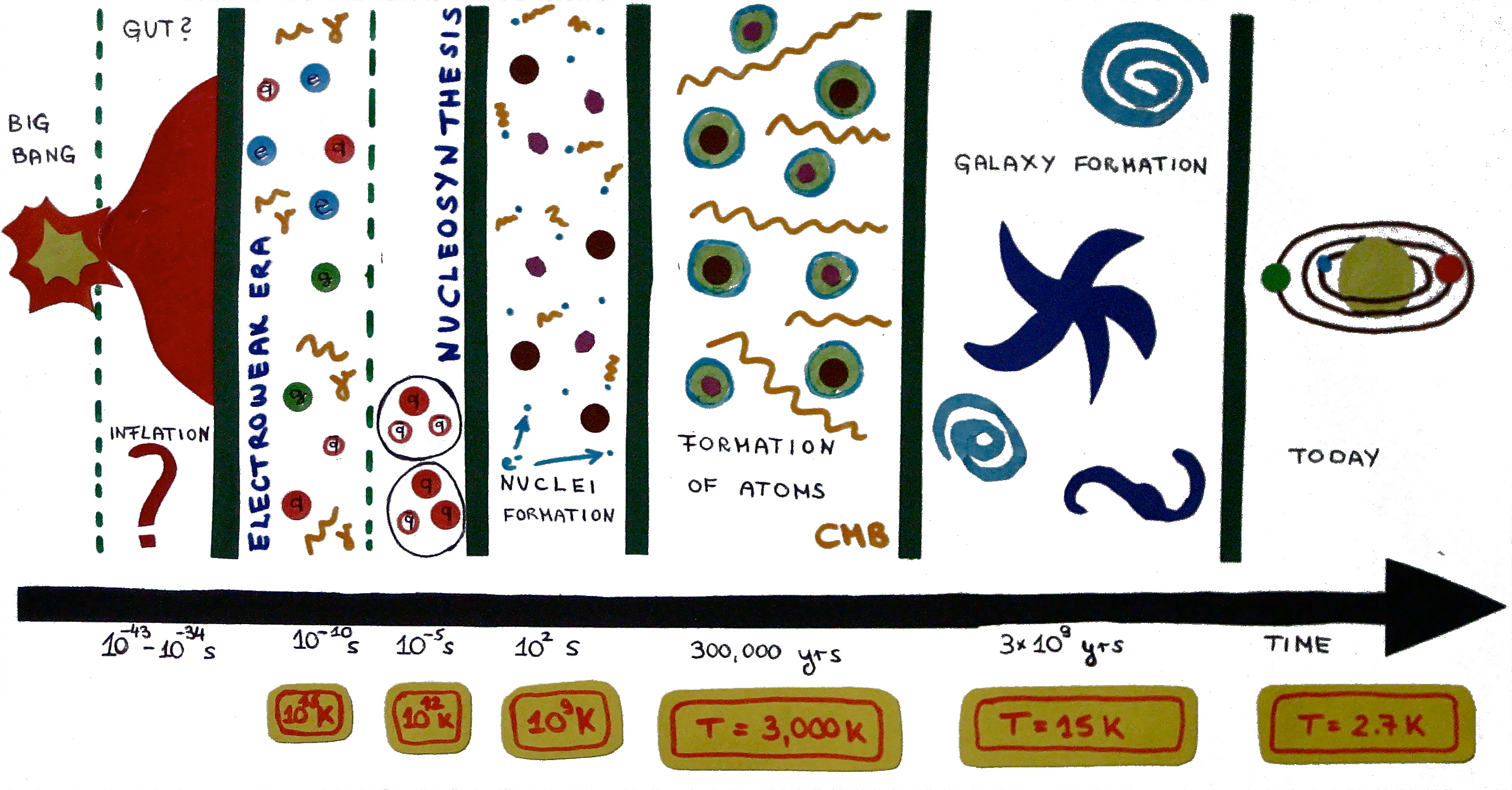}
\caption[Brief history of the Universe]{\small Overview of the Universe's history.}
\label{fig:history-univ}
\end{center}
\end{figure}

Our understanding of the present-day Universe and its history is summarized in the elaborate framework of the Standard Big Bang Model (SBBM) of cosmology. The interplay between Nature's interactions, \ie gravity, electromagnetism, the weak and the strong forces, allows an explanation of the evolution of the Universe considerably far back into the past. Using our knowledge of electroweak unification around an energy density of $\rho^{1/4}\simeq\order{100\MeV}$, one can extend this description and obtain a consistent picture from the hot ``primordial soup'' state shortly after the Big Bang until today's large scale structures of filaments, clusters of galaxies and their substructures. We now briefly recall some elements of the SBBM picture crucial for the purposes of this thesis, along with their experimental evidence. A wide-angle sketch of events can be found in Fig.~\ref{fig:history-univ}, the details being the subject of a broad range of textbooks \cite{Mukhanov:2005sc,peter:cosmo,Liddle:1998ew,Peebles:1994xt,Kolb:1990aa}.

On cosmological scales, gravity is the largely dominant interaction. It is described by the theory of General Relativity (GR), whose fundamental quantity is the metric tensor $g_{\mu\nu}$ of spacetime, from which the line elemend $\dd s^{2}$ is calculated as
\beq\label{eq:linelement}
\dd s^{2}=g_{\mu\nu}\,\dd x^{\mu}\,\dd x^{\nu}\,.
\eeq
The \emph{Einstein Hilbert action}, describing the dynamics of this metric, reads
\beq\label{eq:Sgrav}
\Sgrav=-\frac{1}{2\kappa}\int\dd^{4} x\sqrt{-g}\,R\,,
\eeq
with the coupling constant 
$\kappa\equiv8\pi\,G=8\pi/\mpl^{2}=1/\Mpl^{2}$, $g$ the determinant of $g_{\mu\nu}$, and $R$ the Ricci curvature scalar 
$R\equiv g^{\mu\nu}R_{\mu\nu}$, constructed from the Ricci tensor $R_{\mu\nu}$. Assuming that matter in the Universe is described by an action $\Smatter$, leading to an energy-momentum tensor $T_{\mu\nu}\propto\delta\Smatter/\delta g^{\mu\nu}$, one obtains for the system $\Sgrav+\Smatter$ the \emph{Einstein equations of motion}\footnote{We have not explicitly included a cosmological constant $\Lambda$ in Eq.~(\ref{eq:Einstein}). Such a term can always be considered as the potential energy of a scalar field (see below) at its equilibrium position $\bar{\phi}$, $\Lambda=8\pi \,G\,V(\bar{\phi})$, and hence as a part of the energy momentum tensor $T_{\mu\nu}$.}
\beq\label{eq:Einstein}
R_{\mu\nu}-\frac{1}{2}\,g_{\mu\nu}R=\kappa\,T_{\mu\nu}\,.
\eeq
To put these equations to use, we have to \emph{i)} choose an ansatz for the metric $g_{\mu\nu}$ in Eq.~(\ref{eq:linelement}) and \emph{ii)} specify the description of matter in the Universe, \ie the form of $T_{\mu\nu}$.

\subsection{A Homogeneous and Isotropic Universe}\label{subsec:homogeneous}
When formalizing $g_{\mu\nu}$ for the Universe as a whole, the observation of homogeneity and isotropy on large scales $\order{\sim100\Mpc}$ determines the metric up to an arbitrary function of time $a(t)$ (known as the scale factor), and a discrete parameter $\mathcal{K}=-1,0,+1$ encoding the spatial curvature (open, flat or closed). The resulting ansatz is called the \emph{Friedmann Lema\^{i}tre Robertson Walker (FLRW) metric},
\beq\label{eq:FLRWmetric}
\dd s^{2}=-\dd t^{2}+a^{2}(t)\left[\frac{\dd r^{2}}{1-\mathcal{K} r^{2}}+r^{2}\left(\dd \theta^{2}+\sin^{2}\theta\,\dd\phi^{2}\right)\right]\,,
\eeq
which, when plugged into the Einstein equations (\ref{eq:Einstein}), leads to time derivatives of $a(t)$ on the left hand side. When it comes to the description of matter, the homogeneity and isotropy requirements are met by an ideal fluid, for which 
\beq\label{eq:Tmunu}
T_{\mu\nu}=\frac{2}{\sqrt{-g}}\,\frac{\delta\Smatter}{\delta g^{\mu\nu}}\,,
\eeq
so that ${T^{\mu}_{}}_{\nu}=\diag(-\rho,p,p,p)$. The pressure $p$ and energy density $\rho$ are related by an equation of state, $p=w\rho$, where $w$ can often be chosen constant. Thus, remarkably, the homogeneous and isotropic Einstein equations have been reduced to
\bea
H^{2}&=&\frac{\kappa}{3}\,\rho-\frac{\mathcal{K}}{a^{2}}\,,\label{eq:Friedmann}\\
\frac{\ddot{a}}{a}&=&-\frac{\kappa}{6}\,\left(\rho+3 p\right)\,\label{eq:Raychaudhuri},
\eea
known as the \emph{Friedmann} and the \emph{Raychaudhuri equation}, respectively. The quantity $H(t)\equiv\dot{a}/a$ is called the \emph{Hubble parameter} and measures the change in the scale factor of the Universe, where $H>0$ corresponds to expansion. In the following, we assume universal expansion as the default case; contracting universes with $H<0$ are briefly mentioned in Section \ref{sec:alternatives}.\\
In Eq.~(\ref{eq:FLRWmetric}), the four-dimensional spacetime coordinates read $(t,r,\theta,\phi)$. It is often useful to trade cosmic time $t$ for the conformal time coordinate $\eta$ defined by $\dd t=a(\eta)\,\dd\eta$, so that the scale factor can be factored out and the FLRW metric reads
\beq\label{eq:FLRWconformal}
\dd s^{2}=a^{2}(\eta)\left\{-\dd\eta^{2}+\left[\frac{\dd r^{2}}{1-\mathcal{K} r^{2}}+r^{2}\left(\dd \theta^{2}+\sin^{2}\theta\,\dd\phi^{2}\right)\right]\right\}\,.
\eeq 
One can then define the quantity $\calH(\eta)\equiv a'/a=a(t)\,H(t)$, and Eqs.~(\ref{eq:Friedmann}) and (\ref{eq:Raychaudhuri}) then read
\bea
\calH^{2}&=&\frac{\kappa}{3}\,a^{2}\rho-\mathcal{K}\,,\label{eq:Friedmann-conformal}\\
\frac{a''}{a}-\calH^{2}&=&-\frac{\kappa}{6}\,a^{2}\left(\rho+3 p\right)\,\label{eq:Raychaudhuri-conformal}.
\eea
Evidently, as seen from Eq.~(\ref{eq:Friedmann}), $H$ has dimensions of $\mpl$ and therefore provides a default mass scale, while its inverse $H^{-1}$ is known as one Hubble time (or Hubble length, with our choice of units), \ie the standard time scale for processes occurring in the Universe. Finally, length scales can be compared with $H^{-1}$, but also with the ``comoving Hubble length'' $1/(aH)=1/\calH$. 
We now drop the curvature term in Eq.~(\ref{eq:Friedmann}), setting $\mathcal{K}=0$ for a spatially flat universe, which is well-motivated by observations \cite{Spergel:2003cb}.

\subsection{Present Composition of the Universe}\label{subsec:composition}
We shall assume that globally, the FLRW metric (\ref{eq:FLRWmetric}) [or (\ref{eq:FLRWconformal})] describes the Universe throughout its history, \ie also today. We now focus on the matter side of the Einstein equations and discuss the present composition of the Universe. To this end, we define the so-called ``critical density'' $\rhocrit$ in a flat universe with respect to the Hubble parameter at the present epoch\footnote{All quantities referring to their current (present-day) value are designated by a subscript ``0'' [or occasionally a superscript ``(0)''].} $H_{0}$,
\beq
\rhocrit=\frac{3}{\kappa}\,H_{0}^{2}\,.
\eeq
The total energy density $\rhotot$ is a sum of different contributions, $\rhotot=\sum_{i}\rho_{i}^{(0)}$, where each part $\rho_{i}^{(0)}$ stands for an ideal fluid $i$ with its own equation of state parameter $w_{i}$. The dimensionless quantities $\Omega_{i}^{(0)}=\rho_{i}^{(0)}/\rhocrit$ then allow to re-write the Friedmann equation (\ref{eq:Friedmann}) today simply as $\sum_{i}\Omega_{i}^{(0)}=1$. [If one wants to keep the term containing $\mathcal{K}$ in Eq.~(\ref{eq:Friedmann}), one may define $\Omega_{\mathcal{K}}^{(0)}=-\mathcal{K}/(a_{0}^{2}H_{0}^{2})$ for the contribution of the curvature, and hence $\sum_{i}\Omega_{i}^{(0)}+\Omega_{\mathcal{K}}^{(0)}=1$.] In the present Universe, the four components contributing to this sum, their equations of state and their relative importance are \cite{Dunkley:2008ie}:

\textsc{Radiation}\\
All the photons in the Universe [most of which belong to the Cosmic Microwave Background (CMB), see below], with their equation of state $p_{\mathrm{rad}}=(1/3)\,\rho_{\mathrm{rad}}$ represent a tiny fraction of $\rhotot$, with $\Omega_{\mathrm{rad}}^{(0)}\approx 10^{-5}$. This equation of state also holds for any gas of relativistic particles such as weakly interacting cosmic neutrinos, which are therefore summarily counted with the photons into\footnote{Strictly speaking, massive neutrinos fell out of equilibrium recently when the temperature dropped below $T\leq0.01\eV$, and only photons remain relativistic today \cite{Durrer:CMB}.} ${\Omega}_{\mathrm{rad}}^{(0)}$.

\textsc{Baryonic Matter}\\
The constituents of ordinary matter (atoms, nuclei \etc) behave non-relativistically (\ie without pressure) at present, hence their equation of state is $p_\ub=0$ and the baryonic contribution to $\OmegaT$ (given that baryons are \emph{much} heavier than leptons) amounts to $\OmegaB^{(0)}\approx 0.04$.

\textsc{Nonbaryonic (or ``Dark'') Matter}\\
For a consistent explanation of many observational facts, ranging from galaxy formation to the CMB, it is necessary to postulate the existence of another non-relativistic matter component in the Universe, commonly referred to as ``Dark Matter'', equally with an equation of state $p_\udm=0$ and contributing a percentage of $\OmegaCDM^{(0)}\approx0.26$. The nature of Dark Matter is the subject of active study and related to theories beyond the Standard Model of particle physics, see Chapter \ref{chapter:infl-guts}.

\textsc{Dark Energy}\\
Evidently, after summing over radiation, baryonic and non-baryonic matter, the bulk part of the Universe's energy density is still missing, which (together with evidence from supernovae and other observations) motivates the introduction of ``Dark Energy''. Its equation of state should be very close to $p_\ude\approx-\rho_\ude$, which makes a cosmological constant $\Lambda$ (for which $w_{\Lambda}=-1$ exactly) the frontrunner candidate for Dark Energy. Here, we shall treat both terms as synonyms. This component accounts for the major contribution to $\rhotot$, \ie $\Omega_\ude\approx0.7$\,.

\subsection{The History of the Universe}\label{subsec:history}
Combination of Eqs.~(\ref{eq:Friedmann}) and (\ref{eq:Raychaudhuri}), as well as conservation of the energy-momentum tensor, $\nabla_{\mu}T^{\mu\nu}=0$, yields the continuity equation
\beq\label{eq:continuity}
\dot{\rho}=-3H(\rho+p)\,,
\eeq
which holds for $\rhotot$, but also for each individual component $\rho_{i}$ if the interaction between the different fluids $i$ is purely gravitational. Using $H=\dot{a}/a$ and $p_{i}=w_{i}\,\rho_{i}$, it is easy to solve for $\rho_{i}$ as a function of the scale factor $a$:
\beq
\rho_{i}=\rho_{i}^{(\uin)}\,\left(\frac{a_{\uin}}{a}\right)^{3(1+w_{i})}
\eeq
The index ``in'' refers to the energy density and the scale factor at some initial time; note in particular that in an expanding universe like ours one has $a>a_{\uin}$. Defining an initial fraction $f_{i}=\rho_{i}^{(\uin)}/\rhotot^{(\uin)}$ \cite{Burgess:2007pz}, and limiting ourselves to the four components enumerated above, \ie radiation ($w_{\mathrm{rad}}=1/3$), baryonic and Dark Matter ($w_{\mathrm{b}}=w_\udm=0$) and Dark Energy ($w_\ude=-1$), we see that their contributions to the total energy density vary as
\beq\label{eq:rhototvariation}
\rhotot=\rhotot^{(\uin)}\left[f_\ude+\left(f_{\mathrm{b}}+f_\udm\right)\left(\frac{a_{\uin}}{a}\right)^{3}+f_{\mathrm{rad}}\left(\frac{a_{\uin}}{a}\right)^{4}\right]\,.
\eeq
Since the scale factor grows with time, and the terms on the right hand side in Eq.~(\ref{eq:rhototvariation}) scale with different negative powers of $a$, we see that each of them dominates $\rhotot$ during a separate era of the Universe's history.

\textsc{Present/Future: Dark Energy domination}\\
In the far future, only the term $\rho_\ude^{(\uin)}=\const$ will be left on the right hand side of Eq.~(\ref{eq:rhototvariation}), and hence solely feeding into the Hubble parameter $H$ through Eq.~(\ref{eq:Friedmann}). One can then integrate the Friedmann equation to obtain the evolution of the scale factor with time, which gives
\beq\label{eq:aoft-deSitter}
a(t)=a_\ude\,e^{H_\ude(t-t_\ude)}\,,
\eeq
where $a_\ude=a(t_\ude)$ is a normalization constant calculated at a $t=t_\ude$, and $H_\ude$ is the (constant) Hubble parameter during this (final) phase of evolution.  A period of exponential expansion with constant Hubble parameter is known as a \emph{de Sitter phase}. According to the best observational evidence today, Dark Energy ``recently'' became the biggest contributor to our Universe's energy density, \ie around redshift 
$z_\ude\approx 1$.

\textsc{Recent past/Present: matter dominated era}\\
In the recent past, \ie until about $\order{10^9}$ years ago, our Universe was dominated by matter\footnote{As we discussed in Section \ref{subsec:composition}, the total matter contribution of $\OmegaB+\OmegaCDM\approx0.3$ today is more or less of the same order as the inferred Dark Energy $\OmegaL\approx0.7$\,, giving rise to the so-called ``coincidence problem''.}. Hence, keeping only the second term in Eq.~(\ref{eq:rhototvariation}) and integrating Eq.~(\ref{eq:Friedmann}), we find that during this era the scale factor changes as
\beq
a(t)=a_{\mathrm{m}}\left(\frac{t}{t_{\mathrm{m}}}\right)^{2/3}\,,
\eeq
where again $a_{\mathrm{m}}=a(t_{\mathrm{m}})$. This era was proceeded by a period of radiation domination; the point in time when matter and radiation contributed in equal parts to $\rhotot$, it is also referred to as the time of radiation matter equality $t_\ueq$, which measured in redshift occurs at $z_\ueq\approx3600$. 

\textsc{Past: Radiation dominated era}\\
Finally, going back far enough into the Universe's past, it is the last term and hence radiation that should have dominated the energy density of the Universe $\rhotot$. Again, an integration of the Friedmann equation yields a power law for the growth of the scale factor:
\beq
a(t)=a_{\mathrm{rad}}\left(\frac{t}{t_{\mathrm{rad}}}\right)^{1/2}
\eeq
The fact that the radiation contribution to the total energy density in Eq.~(\ref{eq:rhototvariation}) dies away more quickly than that of non-relativistic matter is easily understood from the additional decrease of the individual photon energies caused by the universal expansion. Photons only are subject to this effect on top of the number density dilution that both photons and massive particles experience.

Evidently, between these three limiting cases, one may still use Eq.~(\ref{eq:Friedmann}) in combination with Eq.~(\ref{eq:rhototvariation}) to find $t(a)$ [and possibly $a(t)$ by inversion] from $t-t_{\uin}=\int^{a}_{a_{\uin}}(\dd\tilde{a}/\tilde{a})\,\left[(3/\kappa)\,\rhotot^{-1}\right]^{1/2}$. In general, a universe filled predominantly with an ideal fluid whose equation of state is $p=w\,\rho$ (with constant $w$) will expand as
\beq\label{eq:scalefactor-general}
a(t)=a_{0}\,t^{2/(3+3w)},\qquad a(\eta)=a_{0}\,\eta^{2/(1+3w)}
\eeq
in cosmic time $t$ or conformal time $\eta$, respectively. During a de Sitter phase\footnote{Note that in a de Sitter spacetime, the conformal time $\eta$ is negative and runs from $\eta=]-\infty,0]$.} it follows that
\beq\label{eq:aofeta-deSitter}
a(\eta)=-\frac{1}{H\eta}\,,
\eeq
while $a\propto\eta$ for a radiation dominated universe, and $a\propto\eta^{2}$ for pressureless matter.

For a comoving observer, there is a fundamental difference between the eras of radiation and matter domination and a Dark Energy universe: remember that the quantity $1/(aH)$ provides a measure for the observer's accessible part of the Universe. As is easily seen from Eq.~(\ref{eq:scalefactor-general}), while a material with $w>-1/3$ provides the bulk energy density of the Universe, it holds that $\dd/\dd t\,\left(aH\right)^{-1}>0$, and hence the comoving Hubble radius grows. For a Dark Energy (or cosmological constant) dominated Universe, on the other hand, $\dd/\dd t\,\left(aH\right)^{-1}<0$, and so the comoving Hubble radius shrinks, gradually hiding ever-growing parts of the Universe from the observer's view. This property of exponentially expanding spacetimes becomes crucial when we discuss inflation in the following Section.

\subsection{The SBBM's Observational Pillars}\label{subsec:pillars}
We established that, according to the SBBM, the Universe evolved from an initial state of very high (radiation dominated) energy density into today's composition commonly called ``\Lcdm'' for a cold Dark Matter Universe with cosmological constant $\Lambda$. The sketch of events provided in Fig.~\ref{fig:history-univ} contains some milestones of this evolution -- but how can we know this much about events so far in the past? The following observational facts, successfully explained by the SBBM, are usually quoted as its ``pillars''.

\begin{figure}[t]
\begin{center}
\begin{minipage}[c]{0.5\textwidth}
\includegraphics[width=\textwidth]{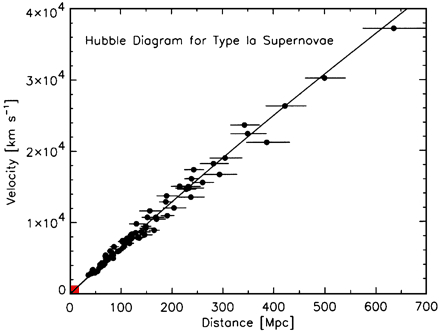}
\end{minipage}
\hfill
\begin{minipage}{0.4\textwidth}
\includegraphics[width=\textwidth]{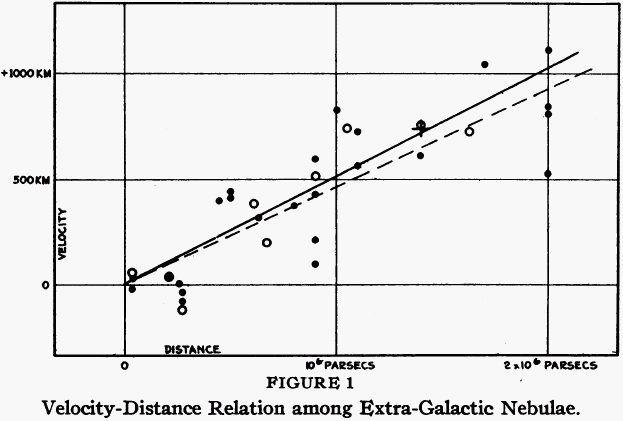}
\end{minipage}
\caption[Modern and original Hubble diagrams showing galaxy velocities as a function of their distance. (Sources: \cite{KirshnerPNAS}, \cite{Hubble:1929ig})]{\small \emph{Left:} The Hubble diagram for type Ia supernovae. The scatter about the line corresponds to statistical distance errors of $<10$~\% per object. The small red region in the lower left marks the span of Hubble's original Hubble diagram from 1929. (Figure from \cite{KirshnerPNAS}) \emph{Right:} Hubble's diagram, leading to the hypothesis of an expanding universe with a linear expansion law $v=H\, r$.  (Figure from \cite{Hubble:1929ig})}
\label{fig:sbbm-expansion}
\end{center}
\end{figure}

\textsc{Expansion of the Universe}\\
First observed by Hubble in 1929 \cite{Hubble:1929ig}, many experiments have since confirmed that galaxies (with the exception of the closest ones) are receding from us, with a velocity proportional to their distance, leading to redshift in their observed spectra (see Fig.~\ref{fig:sbbm-expansion}). The present expansion rate is written as
\beq\label{eq:Hubble-today}
H_{0}=100\,h\, \frac{\km}{\scnd\Mpc}\,,
\eeq
where the current best measurement of $h$ is $0.719^{+0.026}_{-0.027}$ \cite{Dunkley:2008ie}. Over the last decade, evidence has accumulated \cite{Perlmutter:1998np,Riess:1998cb} that galaxies at \emph{very} high redshift are moving away from us with even higher speed than predicted by the Hubble law $v=H_{0}\, r$. This provides independent evidence for the existence of Dark Energy, the component held responsible for the missing amount of energy density to render the Universe flat.

\textsc{Abundance of light elements}\\
Observations of stars, galaxies and the interstellar medium allow an estimate of the relative abundance of light elements [\ie of Hydrogen (which is largely dominant at 75\%), Helium (providing most of the remaining 25\%), Lithium, and Beryllium (of which, like for the heavier elements, there are only trace amounts)]. From our understanding of nuclear physics we can infer the conditions under which these elements were first produced during nucleosynthesis around redshift $z\approx 10^{10}$. Reproducing the observed abundances places tight constraints on this primordial environment \cite{Steigman:2003gc,Durrer:CMB} (see Fig.~1.3).

\begin{figure}[t]
\begin{center}
\begin{minipage}{0.6\textwidth}
\includegraphics{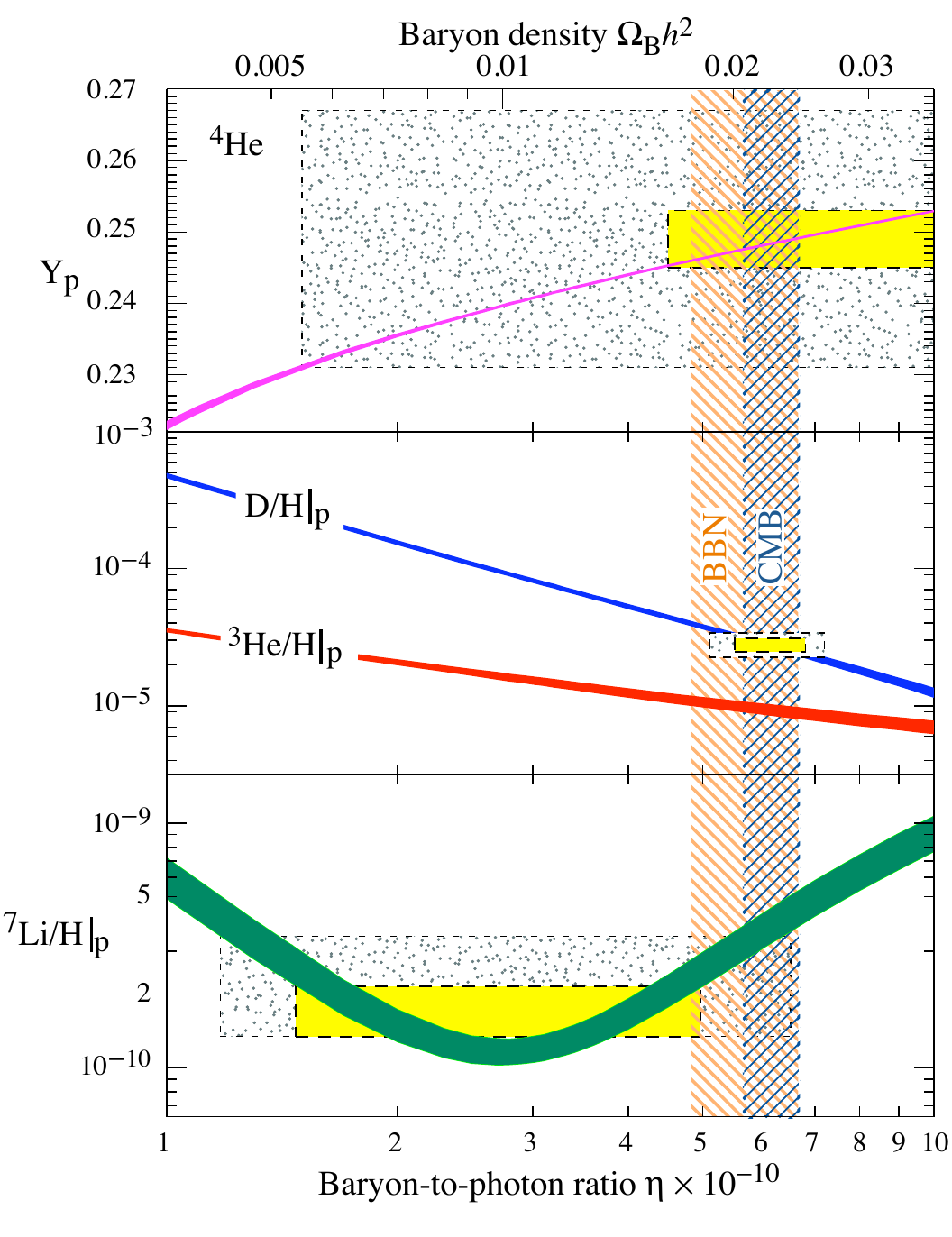}
\end{minipage}
\hfill
\begin{minipage}[b]{0.3\textwidth}
\caption[Primordial element abundance. (Source: \cite{Yao:2006px})]{\small The abundance of light elements as a function of the number ratio of baryons to photons $n_\ub/n_\gamma$. The bands show the regions in agreement with the observed quantities of the respective nuclei. Compatibility with CMB anisotropies is a stronger constraint than nucleosynthesis alone \cite{Durrer:CMB}. (Figure from \cite{Yao:2006px})}
\end{minipage}
\end{center}
\label{fig:sbbm-nucleo}
\end{figure}

\textsc{CMB as the echo of the Big Bang}\\
The cosmic microwave background radiation, first detected by Penzias and Wilson in 1965 \cite{Penzias:1965wn}, has become the prime object of study for cosmologists, with the most recent measurements performed by the COBE \cite{Smoot:1992td} and WMAP \cite{Spergel:2003cb} satellites, and more data to come from the Planck mission \cite{Toffolatti:1997dk}. In the SBBM, the CMB photons can naturally be explained as the ``echo'' of the Big Bang released when the Universe was about $t_{\mathrm{dec}}\approx300,000$ years old (\ie at $z_{\mathrm{dec}}\approx1100$, after the onset of the matter dominated epoch) and first became transparent to photons: for $t<t_{\mathrm{dec}}$, photons continuously scattered with free electrons, their mean path length being very short. Once the electrons were caught by the first nuclei to form neutral atoms, the photons decoupled and began to free-stream. (This is also called ``recombination'', though nuclei and electrons had never been combined before.) More precisely, recombination occurs as a two-step process with Helium becoming neutral earlier. However, since Helium only accounts for a quarter of the Universe's nuclei, it is only somewhat later at Hydrogen recombination that most of the photons are released \cite{Mukhanov:2005sc,Durrer:CMB}. They reach us today with the same shape of their temperature distribution as at $t_{\mathrm{dec}}$, \ie an almost perfect blackbody spectrum, with its central temperature redshifted by the amount of expansion $\order{10^{3}}$ that has occurred since. To high precision, their average temperature is $T_{\mathrm{CMB}}=2.725$~K, which tells us that at recombination, the Universe was homogeneous and isotropic on all scales up to the present horizon to at least one part in $10,000$.\\
The utmost importance of the microwave background radiation in modern cosmology warrants two more comments: first, the high homogeneity of the CMB is a good justification of the FLRW metric ansatz for the metric made in Eq.~(\ref{eq:FLRWmetric}). Second, the experiments cited above are  precisely dedicated to measuring the \emph{deviation} of the CMB from homogeneity, \ie to determining its tiny inhomogeneities of $\order{10^{-5}}$. Their spectrum is one of the key predictions of the theory of inflation we discuss below.

\begin{figure}[b]
\begin{center}
\includegraphics[width=\textwidth]{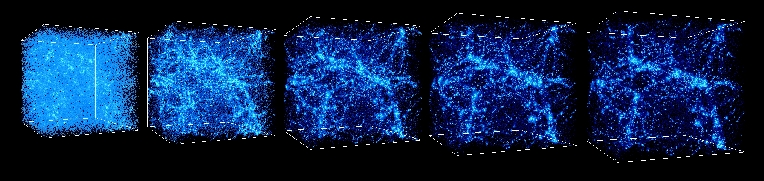}
\caption[Illustration of SBBM structure formation. (Source: simulations performed at the National Center for Supercomputer Applications
by A. Kravtsov (The University of Chicago) and A. Klypin (New Mexico State University), visualizations by A. Kravtsov)]{\small Movie stills from a simulation of structure formation in the \Lcdm\, model. The side of the cube corresponds to 140 million light years, and the first picture is taken at redshift $z=30$, down to today on the right. (Figure by A. Kravtsov, from \texttt{http://cosmicweb.uchicago.edu/filaments.html})}
\label{fig:sbbm-structure}
\end{center}
\end{figure}

\textsc{Formation of structure via gravitational instability}\\
Under the influence of gravitational instability, initial matter inhomogeneities in an expanding universe grow larger with time, at a rate determined by the composition of this universe. Initially, one may describe such perturbations using a linear treatment, while the formation of the first stars and galaxies occurs in the nonlinear regime \cite{Blumenthal:1984bp}. Very sophisticated simulations (see Fig.~\ref{fig:sbbm-structure}) are available for the process of structure formation, which notably allow the r\^{o}le of Dark Matter to be understood \cite{Springel:2005nw}: without it, matter radiation equality $t_\ueq$ (which crucially influences the speed of perturbation growth) would occur too late. Also, Dark Matter ensured that the period of matter domination lasted sufficiently long for successful structure formation before Dark Energy becomes too dominant.

The last two points are closely related in the SBBM: CMB temperature fluctuations serve as an input for the theory of structure formation, leading to predictions for the galaxy distribution, which 
are well confirmed by observations. The initial fluctuation spectrum for successful structure formation must be nearly scale invariant, and recent measurements (see Section \ref{subsec:WMAP5}) confirm that the CMB spectrum indeed has this property. However, both the high homogeneity of the CMB radiation and the spectrum of its tiny fluctuations have no natural explanation in the SBBM framework, and it is a phenomenological requirement that the initial perturbation spectrum should be nearly flat. We now discuss how this situation may be remedied by an early period of ``inflation''.

\section{The Scenario of Inflation}\label{sec:inflation}
In the previous Section, the successes of the SBBM were sketched in very broad brushstrokes. There are, however, some striking observational facts which remain unexplained within this framework, of which we now cite three prominent examples.

\textsc{Horizon problem}\\
We saw that the SBBM traces the \emph{origin} of the CMB back to the time of decoupling, but how can one understand the (almost perfect) \emph{homogeneity} in temperature observed for CMB photons from opposite directions? Turning the clock of expansion backwards within the SBBM framework, one finds that only photons from very close regions (within $\sim1\deg$ of angular separation) in the sky had had time to establish thermal equilibrium by the time of decoupling.

\textsc{Flatness problem}\\
Why is the Universe on its spatial sections so close to flatness? SBBM evolution pushes the Universe away from flatness with time, and to be this close to flatness today, the Universe would have had to be flat with incredibly high precision initially.

\textsc{The problem of unwanted relics}\\
The Grand Unified Theories (GUTs) valid at high energy densities shortly after the Big Bang predict magnetic monopoles (and other relics) produced during phase transitions, but these are not observed today. Where have they gone?

Detailed descriptions of how each of these problems (and several others) arise within the SBBM are readily found in the literature, see \eg \cite{Mukhanov:2005sc,Kinney:2009vz,Martin:2003bt}.   Here, we content ourselves with stating that a remarkably effective and elegant solution to many of these problems at once has been found\footnote{An early inflationary era is now often considered an integral part of standard cosmology. Here, we still distinguish between the ``traditional'' SBBM and its ``add-on'' element inflation.}: the scenario of inflation \cite{Guth:1980zm}. In this Section, we discuss the mechanism of inflation at the background level, along with its consequences for the evolution of a FLRW universe.

Technically, the term ``inflation'' describes a period of accelerated spacetime expansion (\ie the second time derivative of the scale factor $\ddot{a}>0$) in the very early Universe, which should have occurred between $10^{-43}$ and $10^{-34}\scnd$ after the Big Bang. 
From Eq.~(\ref{eq:Raychaudhuri}) it is straightforward to see that for $\ddot{a}>0$ in a FLRW universe, we need an ideal fluid with
\beq\label{eq:inflcond}
\rho+3\,p<0\,,
\eeq
which violates the strong energy dominance condition $\rho+3\,p\geq0$ and is tantamount to negative pressure. Recalling that $p$ and $\rho$ are related by the equation of state, $p=w\,\rho$, we see that Eq.~(\ref{eq:inflcond}) means $w<-1/3$. While strange in ``every day'' matter, the condition (\ref{eq:inflcond}) can be met if the ideal fluid filling the Universe is a scalar field $\phi$, whose action may be written as
\beq\label{eq:Sphi}
\Sphi=-\int\dd^{4}x\sqrt{-g}\left[\frac{1}{2}\,g^{\mu\nu}\partial_{\mu}\phi\,\partial_{\nu}\phi+V(\phi)\right].
\eeq
The function $V(\phi)$ is called the scalar field's potential and contains its mass term as well as interactions. The energy-momentum tensor for $\phi$ [compare Eq.~(\ref{eq:Tmunu})],
\beq
T_{\mu\nu}=\partial_{\mu}\phi\,\partial_{\nu}\phi-g_{\mu\nu}\left[\frac{1}{2}\,g^{\mu\nu}\partial_{\mu}\phi\,\partial_{\nu}\phi+V(\phi)\right]\,,
\eeq
allows us to identify energy density and pressure [note that in a FLRW universe, the field is spatially homogeneous, \ie $\phi=\phi(t)$ only] as
\beq\label{eq:rho-and-p}
\rho=\frac{1}{2}\,\dot{\phi}^{2}+V(\phi),\qquad p=\frac{1}{2}\,\dot{\phi}^{2}-V(\phi)\,.
\eeq
Hence, for a scalar field $p=-\rho+\dot{\phi}^{2}$, and while $\dot{\phi}^{2}\ll V(\phi)$, we can obtain\footnote{For a positive potential, $w$ is bounded from below by the cosmological constant value $w=-1$, and consequently the weak energy condition $\rho+p\geq0$ still holds \cite{Mukhanov:2005sc}.} $w< -1/3$ in agreement with Eq.~(\ref{eq:inflcond}). However, $w$ for a scalar field changes with time since $\phi$ evolves according to its equation of motion obtained from varying Eq.~(\ref{eq:Sphi}),
\beq\label{eq:KleinGordon}
\ddot{\phi}+3H\dot{\phi}+\frac{\dd V}{\dd\phi}=0\,,
\eeq
where the Hubble parameter $H$ in the friction term is obtained from Eq.~(\ref{eq:Friedmann}) with $\rho$ from Eq.~(\ref{eq:rho-and-p}). Note that while $\dot{\phi}^{2}\ll V(\phi)$, $H^{2}$ is essentially given by the potential\footnote{We quantify this statement in the next Chapter.} $V(\phi)$.\\
For later convenience, we also write down the equation of motion for a time and space dependent field $\phi=\phi(x^{\mu})$, obtained from variation of the action (\ref{eq:Sphi}):
\beq\label{eq:KG-inhomogeneous}
\frac{1}{\sqrt{-g}}\,\partial_{\mu}\left(\sqrt{-g}\,g^{\mu\nu}\partial_{\nu}\phi\right)+\frac{\dd V}{\dd\phi}=0
\eeq
For a thorough explanation of how an inflationary phase solves the SBBM problems, we again refer the reader to the literature. With respect to the problem of unwanted relics, it is clear that inflation simply dilutes their densities to less than one relic per Hubble volume. The flatness problem is solved because the short period of inflation drives the Universe so close to spatial flatness that the entire ensuing SBBM evolution cannot push it significantly away again. Inflation also predicts such drastic, quasi-exponential expansion of spacetime that CMB photons from the entire sky were in causal contact (\ie thermal equilibrium) before.\\
The Universe has not been inflating throughout its history (though, as discussed above, its expansion recently started to speed up again under the influence of Dark Energy), nor has it always been dominated by scalar field matter. Therefore, the realization that a scalar ``inflaton'' field can drive a period of early accelerated expansion (and hence solve SBBM problems) immediately raises two questions: How long did this inflationary era last, and how did it end? How can one ensure a smooth transition towards the (initially decelerated) SBBM evolution, more precisely to the radiation dominated epoch?

\subsection{Quantifying Inflation}\label{subsec:quantifying}
In Section \ref{subsec:homogeneous}, we already used two different time coordinates, cosmic time $t$ and conformal time $\eta$. For measuring the amount of expansion during inflation, it is useful to define yet another measure of time called the number of ``\efolds'' $N$, related to cosmic time by $\dd N=H\dd t$. We choose to define $N$ counting from the beginning of inflation \emph{onwards}, \ie $N(t_\uin)=0$. [Alternatively, the number of \efolds can be specified \emph{backwards} from the end of inflation, \ie $N(t_\uend)=0$, so that $N(t)$ measures the amount of inflation still to occur.] Using $H=\dot{a}/a$, we see that
\beq
a(N_\uin)=\exp\left({\int_{t_\uin}^{t_\uend} \dd t\,H}\right)=\exp\left({\int_{N_\uin}^{N_\uend}\dd N}\right)=e^{N_\uend-N_\uin}\,.
\eeq
Commonly, between 40 and 60 \efolds of expansion are considered necessary to solve the SBBM problems cited above \cite{Mukhanov:2005sc,peter:cosmo}. Note, however, that this number depends on the energy scale of inflation, and that the cited range $N=40-60$ is for a potential with $V^{1/4}\simeq 10^{10}-10^{16}\GeV$.  The number N can be smaller if the inflationary energy is lower, the only firm lower bound being that $N$ should cover the entire range of observed scales (from today's Hubble radius to \eg the scale enclosing $10^{4}m_\odot$), which imposes $N>14$ \cite{Lyth:2007qh}. (Put a different way, the ``observational window'' of scales is $\Delta N=14$ \efolds wide.)\\
The only \apriori upper bound on the total amount of inflation $N_\utotal$ comes from the restriction that the initial field value $\phiin$ must not lie in the regime of super-Planckian energy densities, \ie $V^{1/4}(\phiin)\leq1$. With this constraint, however, the expansion of the Universe during inflation can still be \emph{much} larger than the minimum requirement, of $\order{10^{{10}^{10}}}$ \cite{Linde:2007fr}. Additional restrictions may, on the other hand, be well-motivated for an inflaton field originating from a Grand Unified Theory, as we shall see in later Chapters.

\begin{figure}[t]
\begin{center}
\includegraphics[width=\textwidth]{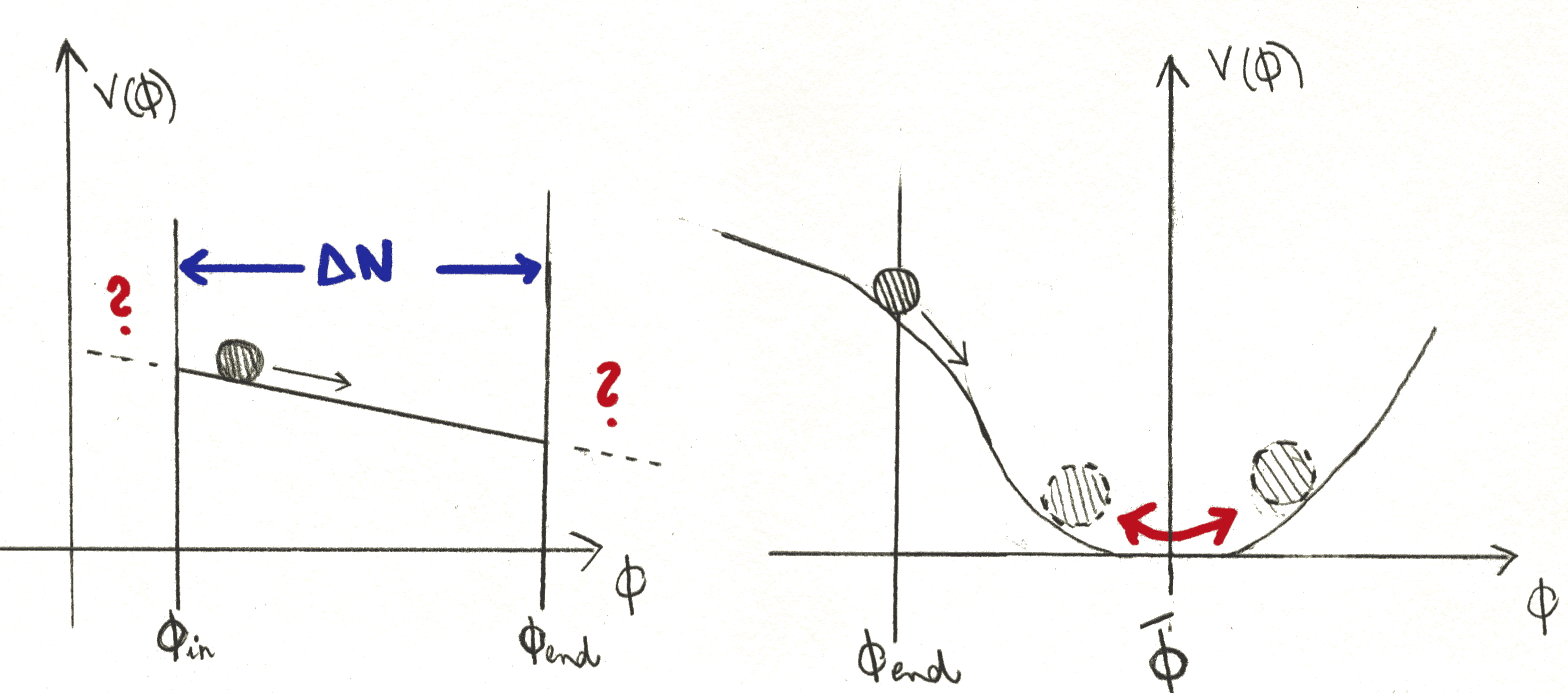}
\caption[Inflation at the background level: slow roll on a flat potential, and reheating at the end of inflation.]{\small \emph{Left:} Sketch of a flat stretch of a single field inflaton potential $V(\phi)$. Inflation occurs while the field $\phi$ rolls from an initial to a final field value, expanding the Universe by a given number of \efolds. The amount of \efolds required depends on the scale of inflation, see the discussion in the text. The minimum requirement to produce all observed scales today is $\Delta N=14$ \cite{Lyth:2007qh}. Note that only the part of the potential corresponding to these scales can be constrained from observations of the CMB power spectrum (see following Chapter). \emph{Right:} At the end of inflation, the inflaton field $\phi$ drops down into a minimum and starts oscillating. It decays and transfers its energy to the particles of the Standard Model, triggering a phase of reheating from which the Big Bang model can start out.}
\label{fig:inflation-bg}
\end{center}
\end{figure}

Assume the condition $\dot{\phi}^{2}\ll V(\phi)$ (and hence $\ddot{a}>0$) is satisfied intially 
at some field value $\phi_\uin$ -- how can we make sure inflation lasts? One may think of the inflaton $\phi$ as a ball up on an inclined surface, starting out with negligible initial velocity compared to its potential energy at some position $\phi_\uin$ (see Fig.~\ref{fig:inflation-bg}). If the surface is sufficiently flat, the ball does not pick up speed too quickly because of the friction term in Eq.~(\ref{eq:KleinGordon}), and $\dot{\phi}^{2}\ll V(\phi)$ continues to hold for some time. Indeed, returning to the field $\phi$, we see from Eq.~(\ref{eq:KleinGordon}) that, if the ``driving force'' $\dd V/\dd\phi$ is small, the change in velocity $\dd\dot{\phi}/\dd t=\ddot{\phi}$ will be negligible, and the Klein-Gordon equation may be approximated as\footnote{Again, we formalize this quantitatively below.} $3H\dot{\phi}\simeq-\dd V/\dd\phi$.\\
Recall also from Eq.~(\ref{eq:Friedmann}) that $\dot{\phi}^{2}\ll V(\phi)$ translates into $H^{2}\simeq (\kappa/3)\,V(\phi)$. Hence, if the potential $V(\phi)$ is almost flat, the Hubble scale changes little with time and may be considered constant during one Hubble time $H^{-1}$ of expansion. Why not set $H=\const$, corresponding to a perfectly flat potential? Indeed, we already encountered a constant Hubble parameter in Section \ref{subsec:history} when we studied the case of Dark Energy domination in the far future, giving rise to a de Sitter universe. Note, however, that at present we try to describe a phase of inflationary expansion that we know must end after a certain amount $N_\utotal$ of expansion. Considered over the entire duration of inflation, the Hubble parameter $H$ does change, but during each individual Hubble time interval $H^{-1}$, its variation is small. We cannot allow $H$ to stay exactly constant as inflation would then never come to a halt. In this sense, inflation may be called a period of \emph{quasi-de Sitter expansion}.

\subsection{Ending Inflation}\label{subsec:endinf}
After the final \efold of inflation, a smooth transition to a radiation dominated universe must take place, in which the SBBM can start out. (For this transition, the term ``graceful exit'' has been coined, and it was indeed a major problem in the first models of inflation.) Hence we need a mechanism for transferring the energy density stored in the inflaton field to the SBBM degrees of freedom such as photons and matter fields. This phase is called ``reheating'', referring to the fact that by the end of inflation the blown-up Universe is empty and cold, and one must convert the inflaton into relativistic matter to start the Big Bang evolution \cite{Kofman:1994rk}.\\
In the picture of the inflaton rolling down a potential, the phase of reheating sets in when $\phi$ reaches some field value $\phi_\uend$ where the field velocity becomes too fast and the potential energy $V(\phi_\uend)$ no longer dominates. $\phi$ then starts oscillating around a minimum of its potential 
(see Fig.~\ref{fig:inflation-bg}). In this phase, couplings of the inflaton to other matter fields become important and are commonly summarized in an additional friction term $\propto\dot{\phi}$ in the Klein Gordon equation (\ref{eq:KleinGordon}), parametrized with a coefficient $\Gamma_\ueff$. (Coupling to bosonic degrees of freedom is more effective than to fermionic fields, therefore only the former are typically considered.) Though difficult to treat analytically, the phase of reheating is nevertheless crucial for predictions of inflationary observables:  it is during reheating that today's observed scales are tied to those during inflation (see the next Chapter), therefore the number of additional \efolds produced in the reheating phase (which can be important in some models) has to be taken into account.\\
If the oscillating inflaton were to decay only in the usual, perturbative way (\ie each of the particles of the homogeneous background field decays independently), a large number of oscillations would be required. However, parametric resonance decay can also occur if coherence\footnote{In this sense, the Universe filled with a coherently oscillating inflaton behaves like a Bose Einstein condensate and hence is anything but hot \cite{Kinney:2009vz}.} of the inflaton is taken into account (leading to the terminology of ``preheating'', for parametric reheating) \cite{Traschen:1990sw,Kofman:1994rk,Kofman:2008zz}. This nonperturbative, resonant conversion of the inflaton energy into other scalar particles is much more efficient than the regular couplings (typically a few oscillations suffice). The effective mass and momenta of the particles produced through resonance can exceed the inflaton mass, which brings GUT energy scales back into play, even if inflation terminates at lower energy density. In this case, however, one must be careful not to re-produce unwanted heavy relics (whose density, given that inflation is essentially over, cannot be diluted away anymore). In a region of the inflaton potential where $V''<0$, also ``tachyonic preheating'' can occur, which is even more efficient than preheating and may require only a single inflaton oscillation \cite{Felder:2000hj}.

\subsection{Building Inflationary Models}\label{subsec:modelbuilding}
In conclusion, when building a model of inflation, one faces two choices, \emph{i)} the form of the potential $V(\phi)$ and \emph{ii)} the way of ending inflation. Hence, these serve as criteria for the classification of inflationary models, which may be subdivided into the categories of \cite{Linde:2007fr,liddle:inflation} \emph{large field}, \emph{small field} and \emph{hybrid inflation} models. An extensive review of inflaton potential shapes (as well as their motivation from high energy physics, which we treat in Chapter \ref{chapter:infl-guts}) can be found in \cite{Lyth:1998xn}.

\begin{figure}[t]
\begin{center}
\includegraphics[width=0.65\textwidth]{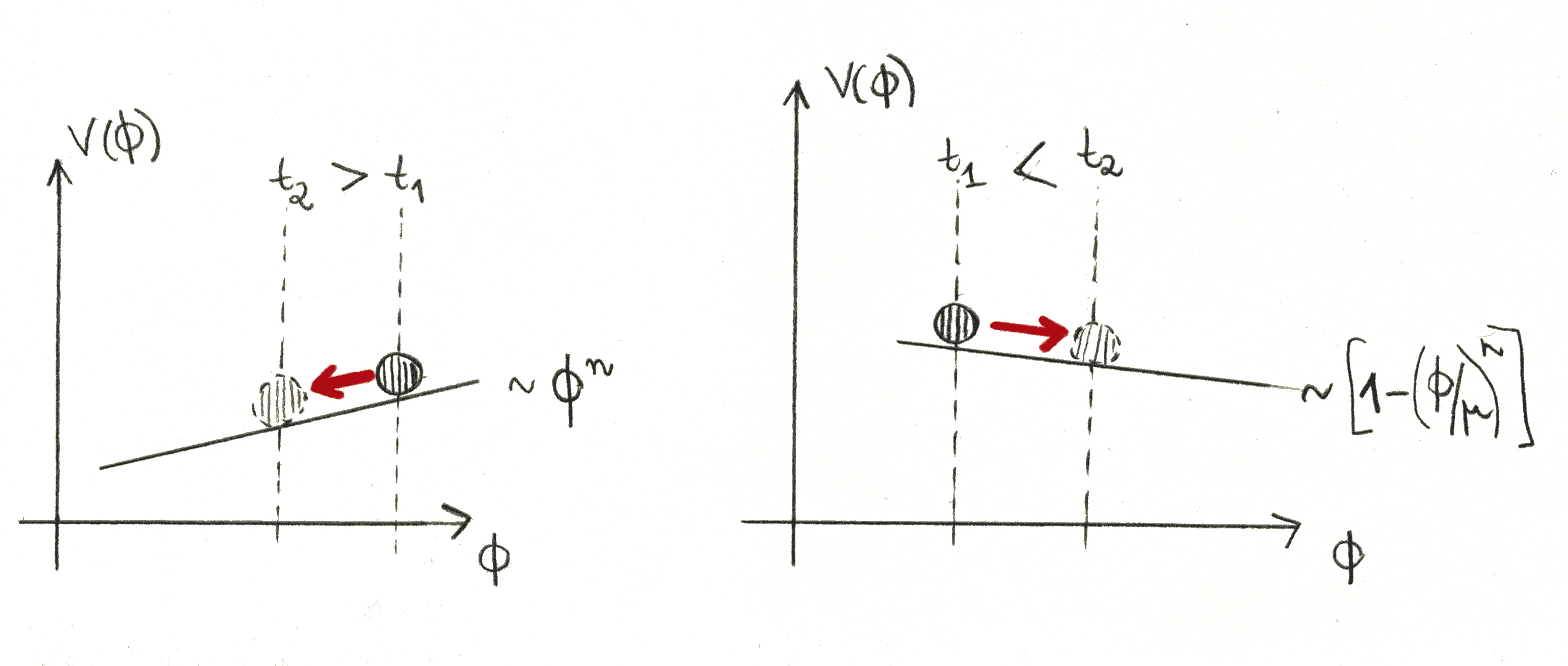}
\hfill
\includegraphics[width=0.25\textwidth]{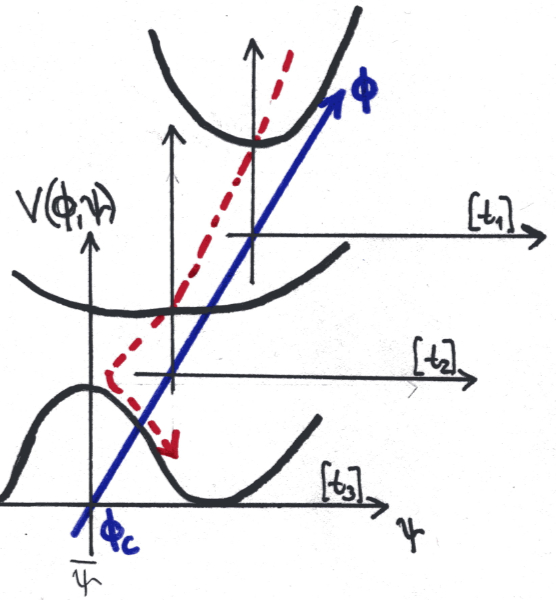}
\caption[Types of inflaton potentials: large field, small field and hybrid inflation scenarios.]{\small \emph{Left:} The inflaton $\phi$ rolling ``from right to left'' on a potential of the large field type (\ref{eq:largefield}). The field is at its largest value initially, rolling towards smaller values. \emph{Center:} An example of a small field potential (\ref{eq:smallfield}), where the direction of motion is inversed. Inflation occurs for small values of $\phi$. \emph{Right:} In hybrid models of inflation with a two-field potential (\ref{eq:hybrid}), one field ($\psi$) is trapped at a minimum during inflation (which is non-zero and provides most of the energy density). Along the inflationary direction $\phi$, the potential is flat but has a slope towards the waterfall point $\phi_\uc$, where the minimum of $\psi$ turns into a maximum: the steepest direction in field space is now $\psi$, and the system moves towards its true minimum. Note that at the critical point $\phi_\uc$, the energy density is still non-zero.}
\label{fig:potentials}
\end{center}
\end{figure}

Let us write a simple general inflaton potential as $V(\phi)=\Lambda^{4}\,f\left(\frac{\phi}{\mu}\right)$  \cite{Kinney:2009vz}, where the parameters $\Lambda$ (the ``height'' of the potential) and $\mu$ (the ``width'' of the potential) have dimensions of mass. By $f$ we denote some general function of different shape for each class of models \cite{Kinney:2009vz}. For large field models, typically one has
\beq\label{eq:largefield}
V(\phi)=\Lambda_{\mathrm{lf}}^{4}\left(\frac{\phi}{\mpl}\right)^{n}\,,\qquad n\geq2\,.
\eeq
The potential can also be a sum of such terms, \ie a polynomial, and the $n$ must be such that the potential is positive-definite. (Fractional values of $n$ are also possible, and for example potential in the model treated in Chapter \ref{chapter:monodromy-paper} is of this form.) These potentials are also referred to as ``chaotic inflation'' \cite{Linde:1981mu, Linde:1984st}. This nomenclature \cite{Vilenkin:2004vx} is due to the arbitrary choice of the initial field value $\phi_\uin$ in different parts of the Universe, \ie the background field may be distributed chaotically. (We discuss this in more detail in the next Chapter.) The case $n=2$, \ie a simple mass term $V(\phi)=(m^{2}/2)\,\phi^{2}$, is among the oldest models of inflation proposed \cite{Linde:1981mu} (and fits the data impressively well, see Section \ref{subsec:WMAP5}), as is the quartic potential with $n=4$ and $V(\phi)=(\lambda/4)\,\phi^{4}$ (which seems excluded by recent observations). Note that for monomial potentials like Eq.~(\ref{eq:largefield}), there is but one free parameter. The exponential potential $V(\phi)\propto \exp\left[-\phi/\left(p^{1/2}\,\mpl\right)\right]$, leading to ``power law inflation'' \cite{Lucchin:1984yf} with the scale factor growing as a power law of time, $a\propto t^{p}$, also belongs in the large field category. The initial field value $\phi_\ini$ measured in Planck units is large in this class of models, typically 
$\phi_\uin/\mpl\sim\order{1-10}$, but avoiding the region where $V\simeq\mpl^{4}$. During inflation, the field moves towards smaller $\phi$ until it encounters $\phi_\uend<\phi_\uin$, where inflation (but not the inflaton) comes to a halt (see Fig.~\ref{fig:potentials}). Then the oscillatory phase of reheating with particle production takes over.

In the class of small field models, the values of $\phi$ during inflation are small compared to a given scale $\mu$ (often the Planck mass), and the inflaton's direction of motion is typically inversed (the field rolls from small towards larger values). One may write the shape of the potential as (see Fig.~\ref{fig:potentials})
\beq\label{eq:smallfield}
V(\phi)=\Lambda_{\mathrm{sf}}^{4}\left[1-\left(\frac{\phi}{\mu}\right)^{n}\right]\,,\qquad n>0\,,
\eeq
as it can, for example, arise from spontaneous symmetry breaking. Again it is assumed that after reaching $\phi_\uend>\phi_\uin$, the inflaton drops into a minimum where it oscillates. Note that there are two parameters ($\Lambda_{\mathrm{sf}}$ and $\mu$) in Eq.~(\ref{eq:smallfield}), which should eventually be fixed by observations.

Hybrid inflationary models \cite{Linde:1993cn} involve two scalar fields $\phi$ and $\psi$ with a typical potential
\beq\label{eq:hybrid}
V(\phi,\psi)=\frac{\lambda}{4}\,\left(\psi^{2}-\bar{\psi}^{2}\right)^{2}+\frac{1}{2}\,g^{2}\phi^{2}\psi^{2}+\tilde{V}(\phi)\,,
\eeq
where the potential for $\phi$ is \eg of the form $\tilde{V}(\phi)=\Lambda^{4}\left[1+(\phi/\mu)^{n}\right]$. During inflation, only the field $\phi$ is dynamic, while $\psi$ is ``trapped'' at a minimum $\bar{\psi}$ (see Fig.~\ref{fig:potentials}) and provides a non-vanishing potential energy. The effective mass of $\psi$ evolves while $\phi$ is slow rolling, $m^{2}_{\psi}=g^{2}\left(\phi^{2}-\phi_\uc^{2}\right)$, and becomes negative at $\phi_\uc^{2}=\lambda\bar{\psi}^{2}/g^{2}$, where $\bar{\psi}$ changes from a minimum to a maximum. This usually marks the end of inflation, and $\psi$ rolls off in a different direction in field space towards the global minimum, around which it starts oscillating. Therefore, the tasks of driving and ending inflation are split among the fields $\phi$ and $\psi$ in hybrid inflation, and notably the potential energy is not yet minimized when $\phi$ reaches the ``waterfall point'' $\phi_\uc$.

\subsection{Inflation with Modified Kinetics}\label{subsec:k-inflation}
In the Lagrangian of Eq.~(\ref{eq:Sphi}), the scalar field $\phi$ has a canonic kinetic term, which notably inspired our analogy of a ball rolling down an inclined surface, picking up speed. However, \apriori we can extend our notion of scalar field dynamics and allow the Lagrangian to be a general function of inflaton derivatives, while keeping the GR part of the theory unchanged. It can be shown that in this case accelerated expansion may be obtained on potentials too steep for conventional slow roll: this mechanism is known as $k$-inflation \cite{ArmendarizPicon:1999rj,Garriga:1999vw,ArmendarizPicon:2000yi}, where the prefix $k$ stands for ``kinetically driven''. The matter Lagrangian of such a theory may be written as\footnote{We do not change the notation of $\calL_{\phi}$, $p$, $\rho$ \etc when referring to a non-canonic theory. In each case, it will be clear from the context whether the dynamics of the scalar field under scrutiny are standard or not.}
\beq\label{eq:kinflation}
\calL_{\phi}=p(X, \phi)\,,\qquad X=\frac{1}{2}\,g_{\mu\nu}\,\partial^{\mu}\phi\,\partial^{\nu}\phi\,.
\eeq
Note that $p(X,\phi)$ cannot be a function of $X$ only to successfully drive inflation, as it is notably the $\phi$ dependence that will ensure a ``graceful exit'' in this scenario \cite{Mukhanov:2005sc,ArmendarizPicon:1999rj}. It is usually assumed that $p(X, \phi)$ admits an expansion $p(X, \phi)=K(\phi)\,X+L(\phi)\,X^{2}+\dots$. We therefore frequently choose the notation $\calL_{\phi}=F\left(\phi,g_{\mu\nu}\,\partial^{\mu}\phi\,\partial^{\nu}\phi\right)-V(\phi)$, keeping the potential term explicit.\\
For the theory to be well-defined, one must have $\partial p/\partial X>0$ and $2X\,\partial^{2} p/\partial X^{2}+\partial p/\partial X>0$ \cite{Bruneton:2007si}. The function $p(X,\phi)$ can be interpreted as the effective pressure, with the corresponding energy density $\rho(X,\phi)$ given by
\beq\label{eq:rho-kinf}
\rho=2X\,\frac{\partial p}{\partial X}-p\,.
\eeq
The continuity equation (\ref{eq:continuity}) remains formally unchanged with these redefinitions of $p$ and $\rho$. Recall that inflation can take place while the effective equation of state $p=w\,\rho$ obeys $w<-1/3$, which for the Lagrangian of Eq.~(\ref{eq:kinflation}) corresponds to $X\,\partial p/\partial X\ll -p$. While this condition holds, the Universe can inflate because the action depends non-linearly on the kinetic energy though there is no slow roll for the field $\phi$. Note that the resulting Hubbe parameter is still given by $H^{2}=(\kappa/3)\,\rho$ with the energy density defined in Eq.~(\ref{eq:rho-kinf}).

\bigskip
For completeness, let us state that it is not inevitable to invoke a scalar field to achieve inflationary expansion \cite{Mukhanov:2005sc,liddle:inflation}. The dynamics discussed above may also be obtained from adding terms to the Einstein Hilbert action (\ref{eq:Sgrav}), \ie higher powers of the Ricci curvature scalar $R$ and of $R_{\mu\nu}R^{\mu\nu}$. The resulting equations of motion contain higher order derivatives, which means that the gravitational field has degrees of freedom on top of gravitational waves, which usually include a scalar field. When the additional terms in the Einstein Hilbert action are restricted to functions $f(R)$, these are called  ``scalar tensor theories'' \cite{Jordan:1949,Brans:1961sx}. There exists a one-to-one correspondence between these terms and certain shapes of the potential $V(\phi)$.\\
In this Section, we discussed inflation in the early Universe, its duration and its end. We have, however, so far ignored a most fundamental question: what is the nature of the scalar field $\phi$? We postpone this question for one more Chapter and first turn to the theory of inflation at the perturbative level.

\chapter{The Inflationary Toolkit}\label{chapter:toolkit}
\begin{quotation}
\emph{We explain why the inflationary scenario is considered a good explanation for the origin of structure in the Universe: when perturbed around its bulk, a scalar field in a FLRW universe predicts the spectra of primordial scalar perturbations to be nearly scale invariant, \ie ideally suited as a starting point for SBBM structure formation. The most important tool for observational verification is to map out temperature fluctuations of the photons released at last scattering, which are obtained from measurements of the Cosmic Microwave Background (CMB) radiation. We show how CMB observations are related to the inflationary calculations, and quote recent experimental results along with their consequences for inflationary parameters. The Chapter concludes with a brief discussion of the physics at very small (Planckian) and very large scales, at which inflation must be conceptually consistent.}
\end{quotation}

\section{Perturbations in Inflation}\label{sec:perturbations}
It is evident from the observation of galaxies, clusters \etc that matter is distributed inhomogeneously at scales below $\order{\sim100\Mpc}$, which makes it necessary to study the Einstein equations (\ref{eq:Einstein}) beyond the FLRW approximation. In this Section, we therefore push both the scalar field $\phi$ and the metric tensor $g_{\mu\nu}$ to linear order in inhomogeneous perturbations. There are good physical arguments for this since in a universe governed by quantum mechanics, tiny fluctuations $\delta\phi$ and $\delta g_{\mu\nu}$ of quantum origin are unavoidably present.\\
The linear perturbative treatment can be reduced to the study of three dynamical degrees of freedom, corresponding to the inflaton itself and the two polarization modes of the graviton. We now show how the primordial spectra of these fluctuations may be obtained as a generic prediction of inflation, and expressed in terms of background quantities. When used as an input for the SBBM plasma and particle physics machinery, the perturbations of the inflaton field eventually produce an imprint in the CMB in terms of temperature fluctuations. The tensor perturbations of the metric lead to a gravitational wave background (\ie ripples in spacetime itself), which, if they were detected by future experiments, would provide us with the earliest snapshot possible of the young Universe.

\subsection{Matter and Metric Perturbations at Linear Order}\label{subsec:linearpert}

If the Universe is filled with scalar field matter, perturbing the right hand side of the Einstein equations (\ref{eq:Einstein}) amounts to setting\footnote{A subscript ``0'', or superscript ``(0)'', in this Section denotes background quantities, and not present-day values.}
\beq\label{eq:phipert-conformal}
\phi(\eta,\vec{x})=\phi_{0}(\eta)+\delta\phi(\eta,\vec{x})\,.
\eeq
We work in conformal time $\eta$ in the following. The left hand (gravity) side of Eq.~(\ref{eq:Einstein}) may be perturbed by writing the line element as
\begin{equation}\label{eq:perturbedline}
ds^{2}= g_{\mu\nu}^{(0)}\,\dd x^{\mu}\,\dd x^{\nu}+\delta g_{\mu\nu}\,\dd x^{\mu}\,\dd x^{\nu}\,,
\end{equation}
where the background part \( g_{\mu\nu}^{(0)}\,\dd x^{\mu}\,\dd x^{\nu}\) is given by Eq.~(\ref{eq:FLRWconformal}). In a straightforward generalization of vector decomposition into a gradient and a curl part, the metric perturbations \(\delta g_{\mu\nu}\) in Eq.~(\ref{eq:perturbedline}) may be decomposed into a scalar, vector and tensor part with respect to their transformation properties under the group of spatial rotations and translations:
\begin{equation}
\delta g_{\mu\nu}=\delta g_{\mu\nu}^\uPSP+\delta g_{\mu\nu}^\uPVP+\delta g_{\mu\nu}^\uPTP
\end{equation}
The three perturbations decouple at linear order and can therefore be treated independently. As stated above, the tensor part $\delta g_{\mu\nu}^\uPTP$ leads to purely gravitational waves\footnote{which are automatically gauge-invariant}, while $\delta g_{\mu\nu}^\uPSP$ couples to the matter perturbation $\delta\phi$. We drop the vector contribution $\delta g_{\mu\nu}^\uPVP$ since it decays with time in the absence of vector stress energy perturbations.

\subsubsection{Scalar perturbations}
Starting our perturbative analysis with $\delta g_{\mu\nu}^\uPSP$, we note that in their most general form, scalar metric perturbations can be expressed in terms of four scalar functions 
\(\varphi\), \(\psi\), \(B\) and \(E\) \cite{Mukhanov:1990me},
\begin{equation}\label{eq:scalarpertmatrix}
\delta g_{\mu\nu}^\uPSP=a^{2}\left(\begin{array}{cc} 2\varphi & B_{,i} \\ B_{,i} & 2\psi\, \gamma_{ij}+2E_{,ij} \end{array} \right)\,,
\end{equation}
where $\gamma_{ij}$ is the metric on the spatial sections, hence in a flat universe $\gamma_{ij}=\delta_{ij}$. The full line element (\ref{eq:perturbedline}) for the background and (only) scalar metric perturbations reads
\begin{equation}\label{eq:linewithfunctions}
\dd s^{2}=a^{2}\left\{(1+2\varphi)\dd \eta^{2}+2B_{,i}\dd x^{i}\,\dd\eta-\left[(1-2\psi)\gamma_{ij}-2E_{,ij}\right]\dd x^{i}\,\dd x^{j}\right\}\,.
\end{equation}
However, the functions \(\varphi\), \(\psi\), \(B\) and \(E\) are related to $\delta\phi$ via the Einstein equations (which are by construction invariant under general coordinate transformations), therefore it is crucial to separate physical degrees of freedom from gauge modes. The simplest linear combinations of metric scalar functions that are gauge-invariant at linear order are
\begin{equation}\label{eq:bardeenpots}
\Phi=\varphi-\frac{1}{a}\,[a(B-E')]'\qquad\textnormal{and}\qquad\Psi=\psi+\frac{a'}{a}\,(B-E')\,,
\end{equation}
the so-called Bardeen potentials \cite{Bardeen:1980kt}. Likewise, it is possible to construct a gauge-invariant quantity \(\delta\phi^\uPgiP\) from the perturbation \(\delta\phi\) in Eq.~(\ref{eq:phipert-conformal}) by
\begin{equation}\label{eq:giscalarperturbation}
\delta\phi^\uPgiP=\delta\phi + \phi'_{0}(B-E')\,.
\end{equation}
From a physical point of view, there are only two degrees of freedom in Eq.~(\ref{eq:linewithfunctions}), which can be represented by the potentials $\Phi$ and $\Psi$. As a consequence, two superfluous gauge degrees of freedom may be removed \eg by choosing the so-called longitudinal gauge 
in which \(B=E=0\). This then leads to the identifications \(\Phi=\varphi\) and \(\Psi=\psi\).\\
The evolution of $\Phi$ and $\Psi$ is obtained from a perturbative expansion of the Einstein equations (\ref{eq:Einstein}) and the general Klein Gordon equation (\ref{eq:KG-inhomogeneous}) using Eqs.~(\ref{eq:phipert-conformal}) and (\ref{eq:linewithfunctions}). 
In particular, for the ideal fluid $\phi$ there is no anisotropic stress, and it is found that\footnote{By \(\delta T^{\mu}{}_{\nu}^\uPgiP\), we denote the gauge-invariant perturbed energy momentum tensor that can be constructed using Eqs.~(\ref{eq:bardeenpots}) and (\ref{eq:giscalarperturbation}), see \cite{Mukhanov:1990me}.} \(\delta T^{i}{}_{j}^\uPgiP\propto \delta^{i}{}_{j}\) and the two Bardeen potentials \(\Phi=\Psi\). The remaining free metric perturbation variable \(\Phi\) then is a generalization of the Newtonian gravitational potential. [Therefore, the longitudinal gauge where one sets \(B=E=0\) is also referred to as ``conformal Newtonian gauge'' since $\Phi=\varphi$, where $\varphi$ appears in the $g_{00}$ component in Eq.~(\ref{eq:scalarpertmatrix}).] The perturbed Einstein equations then lead to the following system of equations for $\Phi$ \cite{Mukhanov:1990me}:
\begin{eqnarray}
\nabla^{2}\Phi-3\calH\Phi'-\left(\calH'+2\calH^{2}\right)\Phi&=& \frac{\kappa}{2}\left[\phi_{0}'\,{\delta\phi^\uPgiP}'+V_{,\phi}\,a^{2}\,\delta\phi^\uPgiP\right]\label{eq:Phi1}\\
\Phi'-\calH\Phi&=& \frac{\kappa}{2}\,\phi_{0}'\,\delta\phi^\uPgiP\\
\Phi''+3\calH\Phi'+\left(\calH'+2\calH^{2}\right)\Phi&=&\frac{\kappa}{2}\left[\phi_{0}'\,{\delta\phi^\uPgiP}'+V_{,\phi}\,a^{2}\,\delta\phi^\uPgiP\right]\label{eq:Phi3}
\end{eqnarray}
[We use the shorthand notation $V_{,\phi}=\left(\dd V/\dd\phi\right)_{\phi=\phi_{0}}$.] From Eq.~(\ref{eq:KG-inhomogeneous}), we also find that the background field $\phi_{0}$ in conformal time obeys
\beq\label{eq:phibg_conformal}
\phi''_{0}+2\calH\phi'_{0}+a^{2}\,V_{,\phi}=0\,,
\eeq
while at the perturbed order we obtain for $\delta\phi^\uPgiP$ as defined in Eq.~(\ref{eq:phipert-conformal}) the equation
\beq\label{eq:deltaphigi}
{\delta\phi^\uPgiP}''+2\calH\,{\delta\phi^\uPgiP}'-\nabla^{2}\delta\phi^\uPgiP+a^{2}V_{,\phi\phi}\,\delta\phi^\uPgiP=4\phi_{0}'\,\Phi'-2a^{2}V_{,\phi}\,\Phi\,,
\eeq
where evidently $V_{,\phi\phi}=\left(\dd^{2} V/\dd\phi^{2}\right)_{\phi=\phi_{0}}$. The perturbed Einstein equations together with Eq.~(\ref{eq:deltaphigi}) can, with the help of the background relation Eq.~(\ref{eq:phibg_conformal}), be combined into a single equation of motion,
\begin{equation}\label{eq:eofmPhi}
\Phi''+2\frac{(a/\phi'_{0})'}{(a/\phi'_{0})}\Phi'-\nabla^{2}\Phi+2\phi'_{0}\left(\frac{\calH}{\phi'_{0}}\right)'\Phi=0\,.
\end{equation}
Remarkably, we have therefore reduced the study of scalar perturbations around a homogeneous and isotropic cosmological background to the evolution of one single (classical) variable $\Phi$, which is the relativistic generalization of the Newtonian gravitational potential. Indeed, one may even use a variable redefintion\footnote{Note that $\Phi$ is dimensionless, and that the factor $1/\kappa$ in the definition of $u$ ensures that this property is preserved after the variable redefinition.} \(u=(2/\kappa)\,(a\Phi/\phi'_{0})\) to remove the friction term in Eq.~(\ref{eq:eofmPhi}), so that $u$ obeys
\begin{equation}\label{eq:eofmu}
u''-\nabla^{2}u-\frac{\theta''}{\theta}\,u=0,\qquad \theta\equiv\frac{\calH}{a\,\phi'_{0}}\,.
\end{equation}
Note that this equation has the form of a parametric harmonic oscillator, \ie an oscillator with time-dependent frequency.

\subsubsection{Quantization of scalar perturbations}
For consistency, a quantum treatment of perturbations has to start from the action of a system, in this case from $\action=\Sgrav+\Sphi$. Introducing as a new perturbation variable the ``comoving curvature perturbation'' $\mathscr{R}$,
\begin{equation}\label{eq:intrinsic}
\mathscr{R}=-\frac{a'}{a}\frac{\delta\phi}{\phi'_{0}}-\Phi\,,
\end{equation}
the second order perturbed action $\delta_{2}S=\delta_{2}\Sgrav+\delta_{2}\Sphi$ may be written as \cite{Mukhanov:1990me}
\begin{equation}\label{eq:actionintrinsic}
\delta_{2}S=\frac{1}{2}\int \dd^{4}x\,z^{2}\left({\mathscr{R}'}^{2}-\delta^{ij}\partial_{i}\mathscr{R}\,\partial_{j}\mathscr{R}\right)\,,\qquad
z\equiv\frac{a\,\phi'_{0}}{\calH}\,.
\end{equation}
[In Eq.~(\ref{eq:intrinsic}), $\mathscr{R}$ is written in longitudinal gauge. It can, however, be expressed in a gauge-invariant way, which we shall use in later Chapters.]  
Physically, $\mathscr{R}$ describes the curvature perturbation on hypersurfaces orthogonal to comoving worldlines. Much like the pair of variables $(\Phi,\,u)$ above with their respective equations of motion (\ref{eq:eofmPhi}) and (\ref{eq:eofmu}), let us define the \emph{Mukhanov Sasaki variable} \(v\) from\footnote{Note that the comoving curvature perturbation $\mathscr{R}$ is dimensionless, while $[v]=\mpl$, \ie the Mukhanov Sasaki variable has the dimensions of a scalar field.}
\begin{equation}\label{eq:Mukhanov-v}
v=a\left[\delta\phi^\uPgiP+\frac{\phi'_{0}}{\calH}\,\Phi\right]=-z\,\mathscr{R}\,,
\end{equation}
for which the perturbed action $\delta_{2}S$ reads
\begin{equation}\label{eq:startaction}
\delta_{2}S=\frac{1}{2}\int\dd^{4}x\,\left(v'^{2}-v_{,i}v_{,i}+\frac{z''}{z}\,v^{2}\right)\,.
\end{equation}
In changing from $\mathscr{R}$ to $v$, the overall time-dependent factor of \(z^{2}\) has been removed from Eq.~(\ref{eq:actionintrinsic}). Formally, it therefore looks like the action of a scalar field in flat Minkowski spacetime but with a time-dependent ``mass'' term \(z''/z\). As shown below, this similarity can be exploited to find a well-motivated vacuum definition for the perturbations. Following \cite{Mukhanov:1990me}, we now perform the quantization for Eq.~(\ref{eq:startaction}), translating the result\footnote{This is possible because Eq.~(\ref{eq:actionintrinsic}) and Eq.~(\ref{eq:startaction}) are equivalent up to boundary terms.} back into $\mathscr{R}$ using Eq.~(\ref{eq:Mukhanov-v}).

Defining the canonically conjugate momentum from Eq.~(\ref{eq:startaction}), \(\pi=\partial\calL/\partial v'=v'\), we promote \(v\) and \(\pi\) to operators \(\hat{v}\) and \(\hat{\pi}\) with equal-time commutation relations
\begin{eqnarray}
\left[\hat{v}(\eta, \vec{x}),\hat{v}(\eta, \vec{x}')\right]&=&0\label{secondcomm1}\,,\\
\left[\hat{\pi}(\eta, \vec{x}),\hat{\pi}(\eta, \vec{x}')\right]&=&0\,,\\
\left[\hat{v}(\eta, \vec{x}),\hat{\pi}(\eta, \vec{x}')\right]&=&i\delta(\vec{x}-\vec{x}')\label{secondcomm3}\,.
\end{eqnarray}
Variation of Eq.~(\ref{eq:startaction}) with respect to \(v\) gives for the evolution\footnote{We are hence working in the Heisenberg picture.} of \(\hat{v}\) that
\begin{equation}\label{eq:fieldequ}
\hat{v}''-\nabla^{2}\hat{v}-\frac{z''}{z}\hat{v}=0\,.
\end{equation}
In flat FLRW spacetime, we can use plane waves $\propto \exp(i\vec{k}\cdot\vec{x})$ for a Fourier expansion of \(\hat{v}\),
\begin{equation}\label{eq:basisexpansion}
\hat{v}=\frac{1}{(2\pi)^{3/2}}\int \dd^{3}\vec{k}\left[v_{k}(\eta)\,\hat{a}_{\vec{k}}\,e^{i\vec{k}\cdot\vec{x}}+v_{k}^{*}(\eta)\,\hat{a}_{\vec{k}}^{\dagger}\,e^{-i\vec{k}\cdot\vec{x}}\right]\,,
\end{equation}
where the creation and annihilation operators $\hat{a}_{\vec{k}},\hat{a}^{\dagger}_{\vec{k}}$ obey the standard commutation relations,
\begin{equation}
\left[\hat{a}_{\vec{k}},\hat{a}_{\vec{k}'}\right]=0\,,\qquad\left[\hat{a}_{\vec{k}}^{\dagger},\hat{a}_{\vec{k}'}^{\dagger}\right]=0\,,\qquad\left[\hat{a}_{\vec{k}},\hat{a}_{\vec{k}'}^{\dagger}\right]=i\delta^{3}(\vec{k}-\vec{k}').
\end{equation}
As a consequence, the number-valued mode functions \(v_{k}(\eta)\) in Eq.~(\ref{eq:basisexpansion}) each obey the Fourier transform of Eq.~(\ref{eq:fieldequ})\,,
\begin{equation}\label{eq:modeequation}
v''_{k}+\left(k^{2}-\frac{z''}{z}\right)v_{k}=0\,,\qquad
v'_{k}(\eta)v^{*}_{k}(\eta)-v'^{*}_{k}(\eta)v_{k}(\eta)=-i\,,
\end{equation}
where the constraint equation on the right (to be imposed on the solutions of the equation of motion) is called the \emph{Wronskian condition} and ensures that the commutation relations of Eqs.~(\ref{secondcomm1})-(\ref{secondcomm3}) still hold. Translating Eq.~(\ref{eq:modeequation}) back into the comoving curvature perturbation's Fourier modes \(\mathscr{R}_{k}\), one finds the equation of motion
\begin{equation}\label{eq:modesintrinsic}
\mathscr{R}''_{k}+2\,\frac{z'}{z}\,\mathscr{R}'_{k}+k^{2}\mathscr{R}_{k}=0.
\end{equation}

\subsubsection{Tensor perturbations}
Perturbations of tensor type in the metric are described by
\begin{equation}
\delta g_{\mu\nu}^\uPTP=a^{2}\left(\begin{array}{cc} 0&0 \\ 0 & h_{ij} \end{array} \right)\,,\qquad {h^{i}_{}}_{i}=0,\,\nabla_{i}\,{h^{i}_{}}_{j}=0\,,
\end{equation}
where the two conditions on the right express that $h_{ij}$ is traceless and transverse (as well as symmetric). 
The tensor analogue of Eq.~(\ref{eq:deltaphigi}) reads \cite{Mukhanov:1990me}
\beq\label{eq:tensorperturbation}
h''_{ij}+2\calH\,h'_{ij}-\nabla^{2}\,h_{ij}=0\,,
\eeq
Physically, two degrees of freedom arise from these perturbations, corresponding to the two polarization states of the graviton. We may therefore write $h_{ij}=\sum_{\diamond}\varphi_\uT^{\diamond}\,e_{ij}^{\diamond}$, where where $\diamond=+,\times$ and the $e_{ij}^{+,\times}$ are vectors of longitudinal and transverse polarization, respectively. The scalar functions $\varphi_\uT^{\diamond}$ evidently both obey Eq.~(\ref{eq:tensorperturbation}), and we drop the index $\diamond$ in the following.\\
For quantization, the variable $\varphi_\uT$ is again effectively redefined as\footnote{The resulting $v_\uT$ has the same dimension ($\mpl$) as a scalar field.} $v_\uT=a\,\varphi_\uT/\sqrt{2\kappa}$, which, much like the Mukhanov Sasaki variable $v$ in Eq.~(\ref{eq:modeequation}), after Fourier decomposition obeys an equation of motion [an equivalent Wronskian condition as in Eq.~(\ref{eq:modeequation}) also applies to its two solutions]:
\beq\label{eq:vtensor}
{v_{\uT}}_{k}''+\left(k^{2}-\frac{a''}{a}\right){v_{\uT}}_{k}=0
\eeq
Note again the parameteric oscillator structure of this equation.

\subsection{Solutions in the Sub- and Super-Hubble Limits}\label{subsec:subsuperlimits}
\begin{figure}[t]
\begin{center}
\includegraphics[width=\textwidth]{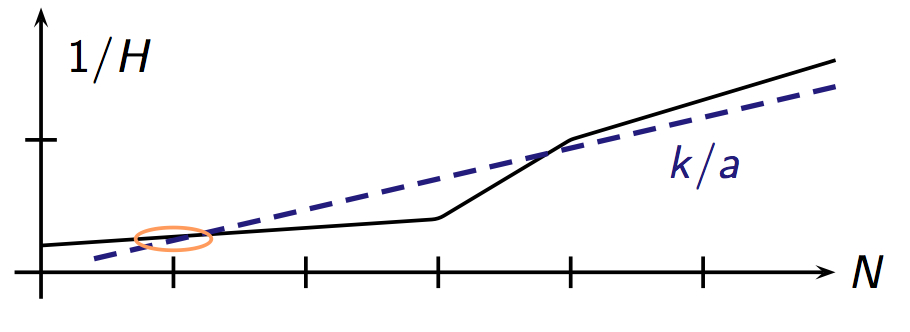}
\caption[Hubble radius crossing of a scale $k$ during and after inflation.]{\small Sketch of the Hubble crossing of a scale $k$ during and after inflation. The horizontal axis measures time (\efolds), the vertical axis denotes length scales (no units are given). The black solid line is the Hubble radius $H^{-1}$, which during inflation evolves slowly and is practically constant during one Hubble time. After inflation, it grows much faster first during the radiation and then during the matter dominated epoch. (The slope of $H^{-1}$ in these epochs is not to scale.) The blue dashed line is an exemplary scale $k$ that leaves the Hubble radius during inflation and crosses back into it (\ie becomes observable and starts evolving again) shortly before radiation matter equality. Such as scale therefore probes the physics of inflation [\ie the potential $V(\phi)$] at the time $N_{k}$ when it was ``frozen in''.}
\label{fig:horizoncrossing}
\end{center}
\end{figure}

Via Fourier transformation, we have formally introduced the ``labels'' $k$, \ie comoving wavenumbers for the modes, which correspond to a certain length scale given by the corresponding comoving wavelength $\lambda\propto 1/k$. This wavelength $\lambda$ can be compared with the comoving Hubble radius $(aH)^{-1}$ we defined earlier, which divides the modes into those which at a given time $\eta$ are
\begin{itemize}
\item {\bf sub-Hubble:} $k/aH>1$,
\item or {\bf super-Hubble:}  $k/aH<1$.
\end{itemize}
Another way to see this is that the product $a(\eta)\,\lambda$ corresponds to the \emph{physical} length scale $\lambda_\uphys(\eta)$ that grows with time in an expanding universe. The sub- and super-Hubble separation than is according to $\lambda_\uphys<H^{-1}$ or $\lambda_\uphys>H^{-1}$, respectively. However, the Hubble scale $H^{-1}$ changes slowly in an inflationary universe. Hence, a given mode can therefore start inside the Hubble radius, 
grow and cross out of it at a certain time $\eta_\uhc$ to continue as a super-Hubble mode. This is illustrated in Fig.~\ref{fig:horizoncrossing}. The instant $\eta_\uhc$, where ``hc'' stands for ``Hubble crossing'' is determined by $k=a(\eta_\uhc)\,H(\eta_\uhc)$. The perturbation evolution equations (\ref{eq:vtensor}) and (\ref{eq:modeequation}) may therefore be studied in these two distinct limits, depending whether a given $k$ is in- or outside the Hubble radius.

We now discuss tensor perturbations, whose equation of motion (\ref{eq:vtensor}) shows that they behave like a spectator field in an fixed cosmological background. \emph{Deep inside} the Hubble radius, the ``mass'' term $a''/a$ in Eq.~(\ref{eq:vtensor}) is much smaller than $k^{2}$, hence
\bea
{v_{\uT}}_{k}''+k^{2}\,{v_{\uT}}_{k}&\approx&0\,,\nonumber\\
{v_{\uT}}_{k}&\propto&A_{k}\,e^{-ik\eta}+B_{k}\,e^{ik\eta}\,.\label{eq:subhorizon-tensor}
\eea
The choice of constants $A_{k}, B_{k}$ corresponds to the choice of vacuum state for the quantized perturbations (see below), and they have to obey the Wronskian condition. \emph{Far outside} the Hubble radius, we drop the $k^{2}$ term in Eq.~(\ref{eq:vtensor}) and therefore find
\bea
{v_{\uT}}_{k}''-\frac{a''}{a}\,{v_{\uT}}_{k}&\approx&0\,,\nonumber\\
{{v_{\uT}}_{k}}&\propto&a\,.\label{eq:superhorizon-tensor}
\eea
How can we interpret this result? Recall that $v_\uT$ was a redefinition of $\varphi_\uT$, where $\varphi_\uT\propto{v_\uT}/a$. Therefore for the solution (\ref{eq:superhorizon-tensor}), each mode ${\varphi_{\uT}}_{k}$ simply approaches a ($k$-dependent) constant once outside the Hubble radius. This observation is at the heart of the notion of ``mode freezing'': in terms of the damped harmonic oscillator equation Eq.~(\ref{eq:tensorperturbation}), the ${\varphi_{\uT}}_{k}$ are dynamic while they are inside the Hubble radius for each $k$, but after they have passed the point where $k=a(\eta_\uhc)\, H(\eta_\uhc)$, ${\varphi_{\uT}}_{k}$ freezes to a constant value.

For the scalar mode functions, it is easy to see from Eq.~(\ref{eq:modeequation}) that they oscillate analogously to the solution (\ref{eq:subhorizon-tensor}) while inside the Hubble radius. Since Eqs.~(\ref{eq:tensorperturbation}) and (\ref{eq:modeequation}) are related by replacing $a$ with the function $z$, we conclude that outside the Hubble radius, scalar perturbations obey $v_{k}\propto z$, and therefore the modes of the comoving curvature perturbation $\mathscr{R}_{k}=-v_{k}/z$, like the ${\varphi_{\uT}}_{k}$ in the tensor case, approach a constant for $-k\eta\rightarrow0$.

\subsection{Hubble Slow Roll Parameters}
The evolution equations (\ref{eq:modeequation}) and (\ref{eq:vtensor}) have the same structure apart from the replacement of the function $z$ with the scale factor $a$. Let us rewrite this difference  in an astute way using $z=a\,\phi'_{0}/\calH=a\,\dot{\phi}_{0}/H$. From a time derivation of Eq.~(\ref{eq:Friedmann}) it follows, with the use of Eq.~(\ref{eq:KleinGordon}), that $\dot{H}=-(\kappa/2)\,\dot{\phi}_{0}^{2}$. Hence, we can write
\beq\label{eq:zexplained}
z=a\,\left(\frac{\dot{\phi}_{0}^{2}}{H}\right)^{1/2}=a\,\left(-\frac{2}{\kappa}\,\frac{\dot{H}}{H^{2}}\right)^{1/2}=a\,\sqrt{\frac{2}{\kappa}}\,\epsilon_{1}^{1/2}\,,\qquad\epsilon_{1}=-\frac{\dot{H}}{H^{2}}\,.
\eeq
The dimensionless quantity $\epsilon_{1}$ measures the change in the Hubbe parameter\footnote{Using $\calH=a\,H$, one finds $\epsilon_{1}=1-\calH'/\calH^{2}$.}, and it is a straightforward consequence of the definition $H\equiv\dot{a}/a$ that
\beq
\frac{\ddot{a}}{a}=H^{2}\left(1-\epsilon_{1}\right)
\eeq
and hence we need $\epsilon_{1}<1$ to ensure $\ddot{a}>0$ and continued cosmological inflation. Inflation ends when the field reaches the value $\phi_\uend$ such that $\epsilon_{1}(\phi_\uend)=1$.\\
So far, all expressions were exact. Recall that the limit of constant $H$ (and hence $\epsilon_{1}=0$) corresponds to de Sitter expansion [where $a=-1/(H\eta)$, see Eq.~(\ref{eq:aofeta-deSitter})]. It can then be shown \cite{Martin:2003bt} that, neglecting higher derivatives of the Hubble parameter, during the inflationary quasi de Sitter phase one may write
\beq\label{eq:a-quasidS}
a\approx-\frac{1}{H\eta}\,\frac{1}{1-\epsilon_{1}}\,,\qquad\textnormal{or}\quad \eta\approx-\frac{1}{aH}\,\frac{1}{1-\epsilon_{1}}\,.
\eeq
These expressions are the first step in a perturbative expansion of the inflationary background around de Sitter space, which may be formalized using the so-called Hubble slow roll parameters. 

The natural unit of time during inflation are \efolds, with $\dd N=H\dd t$, so that the definition of $\epsilon_{1}$ in Eq.~(\ref{eq:zexplained}) in terms of \efolds reads $\epsilon_{1}=-(\dd H/\dd N)/H=-\,\dd\ln H/\dd N$. This inspires the definition of a logarithmic derivative hierarchy of $\epsilon_{i}$ parameters defined by
\bea\label{eq:epsiparameters}
\epsilon_{i+1}&=&\frac{\dd\left|\ln\epsilon_{i}\right|}{\dd N}\,,\,i=0,1,\dots,\qquad\epsilon_{0}\equiv\frac{H_\ini}{H}\,.
\eea
$H_\ini$ is the value of the Hubbe parameter at the onset of inflation. All $\epsilon_{i}$ are of the same order, and expressed in terms of (cosmic) time derivatives of $H$ the first two members of the hierarchy\footnote{The third Hubble flow parameter is
\beq\label{eq:eps3}
\epsilon_{3}=7\epsilon_{1}-\epsilon_{2}-6\,\frac{\epsilon_{1}^{2}}{\epsilon_{2}}-\frac{1}{\epsilon_{1}\epsilon_{2}}\,\frac{\dddot{H}}{H^{4}}\,.
\eeq} read
\beq\label{eq:eps1eps2}
\epsilon_{1}=-\frac{\dot{H}}{H^{2}}\,,\quad\epsilon_{2}=2\epsilon_{1}-\frac{1}{\epsilon_{1}}\,\frac{\ddot{H}}{H^{3}}\,.
\eeq
From successive derivations of Eq.~(\ref{eq:Friedmann}) we find that
\beq\label{eq:slowrollexpressions}
\dot{\phi}_{0}^{2}=\frac{2}{\kappa}\,H^{2}\epsilon_{1}\,,\qquad\ddot{\phi}_{0}=\frac{H\dot{\phi}_{0}}{2}\,\left(\epsilon_{2}-2\epsilon_{1}\right)\,,
\eeq
which allows us to rewrite the background Friedmann and Klein Gordon equations (\ref{eq:Friedmann}) and (\ref{eq:KleinGordon}) as
\beq\label{eq:Friedmann-approx}
H^{2}=\frac{\kappa}{3}\,\frac{V}{1-\epsilon_{1}/3}\,,\qquad\dot{\phi}_{0}\left(6-2\epsilon_{1}+\epsilon_{2}\right)=-\frac{2V'}{H}\,.
\eeq

\subsubsection{Slow roll approximation and trajectory}
Using the $\epsilon_{i}$, we can employ Eqs.~(\ref{eq:Friedmann-approx}) to quantify the statements made in Section \ref{subsec:quantifying}:
\bea
\epsilon_{1}\ll1:&&H^{2}\approx\frac{\kappa}{3}\,V\label{eq:H-approx}\\
\epsilon_{1},\epsilon_{2}\ll1:&&\dot{\phi}_{0}\approx-\frac{V'}{3H}\label{eq:dphi-approx}
\eea
From Eq.~(\ref{eq:slowrollexpressions}), we see that $\epsilon_{1},\epsilon_{2}\ll1$ ensures potential energy domination \emph{and} negligible acceleration of the field. The situation where $\epsilon_{i}\ll 1$ for all $i$ is called the ``slow roll approximation''. It usually suffices to consider the first few members of the hierarchy, \ie $\epsilon_{1},\epsilon_{2},\epsilon_{3}$ (which, as we see below, are the ones appearing in the dominant terms of the perturbation spectrum, its spectral index and the running). Note that, while $\epsilon_{1}=1$ marks the end of inflation (and we therefore can write $\phi_\uend=\phi_{\epsilon_{1}}$), the slow roll approximation breaks down as soon as one of the $\epsilon_{i}=1$. Let us assume this happens when the field reaches $\phi_{\epsilon_{2}}$ determined by $\epsilon_{2}(\phi_{\epsilon_{2}})=1$: usually, the end of the slow roll regime and of inflation occur around the same time and $\phi_{\epsilon_{1}}\approx\phi_{\epsilon_{2}}$, but for precise calculations one must take into account whether for a particular model the slow roll approximation holds until the end of inflation or breaks down earlier.\\
While $\epsilon_{1},\epsilon_{2}\ll1$, we can use the definition of $N$ together with Eqs.~(\ref{eq:H-approx}) and (\ref{eq:dphi-approx}) to calculate
\beq\label{eq:Nofphi}
N=\int_{t_\uin}^{t}\dd t\,H=-\int_{\phi_\uin}^{\phi}\dd\phi\,\frac{3H^{2}}{V'}=-\int_{\phi_\uin}^{\phi}\dd\phi\,\frac{\kappa\,V}{V'}\,,
\eeq
which gives a function $N(\phi)$ measuring how many \efolds have already occurred when the field reaches the value $\phi$. [In Eq.~(\ref{eq:Nofphi}), we have written $V'\equiv\dd V/\dd\phi$, and we shall continue to use this notation for potential derivatives.] Consequently, the total number of \efolds $N_{\mathrm{tot}}$ is given by integrating from $\phi_\uin$ to $\phi_\uend$ in Eq.~(\ref{eq:Nofphi}). For certain choices of the potential $V(\phi)$, one can invert $N(\phi)$ to obtain the slow roll trajectory $\phi(N)$, which describes the time evolution of $\phi$ while $\epsilon_{1},\epsilon_{2}\ll1$.

In passing, let us note that Eq.~(\ref{eq:Nofphi}) suggests another interpretation of inflaton field perturbations $\delta\phi$: they correspond to small fluctuations in the total number of e-folds $\delta N$ for a given point in the Universe because inflation will end slightly earlier (or later). It is possible to formulate the perturbation theory on large scales (including the calculation of spectra) in terms of these fluctuations in the duration of inflation. This is the ``$\delta N$ formalism'' \cite{Sasaki:1995aw,Lyth:2004gb}, which is well suited to calculate perturbations in multifield scenarios (see Chapter \ref{chapter:ext-alt}).

\begin{figure}[h]
\begin{center}
\includegraphics[width=0.3\textwidth,clip=true]{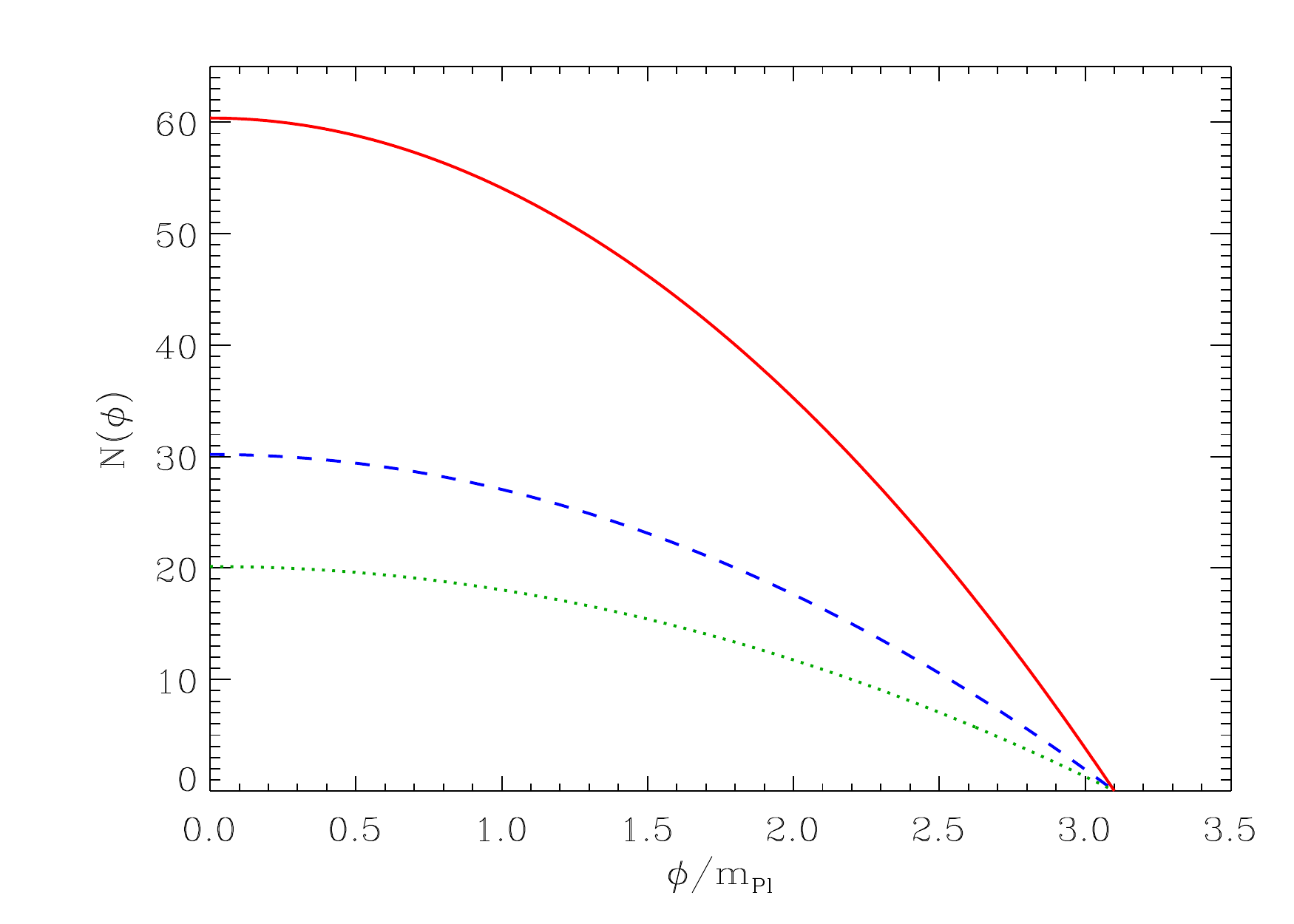}
\hfill
\includegraphics[width=0.3\textwidth,clip=true]{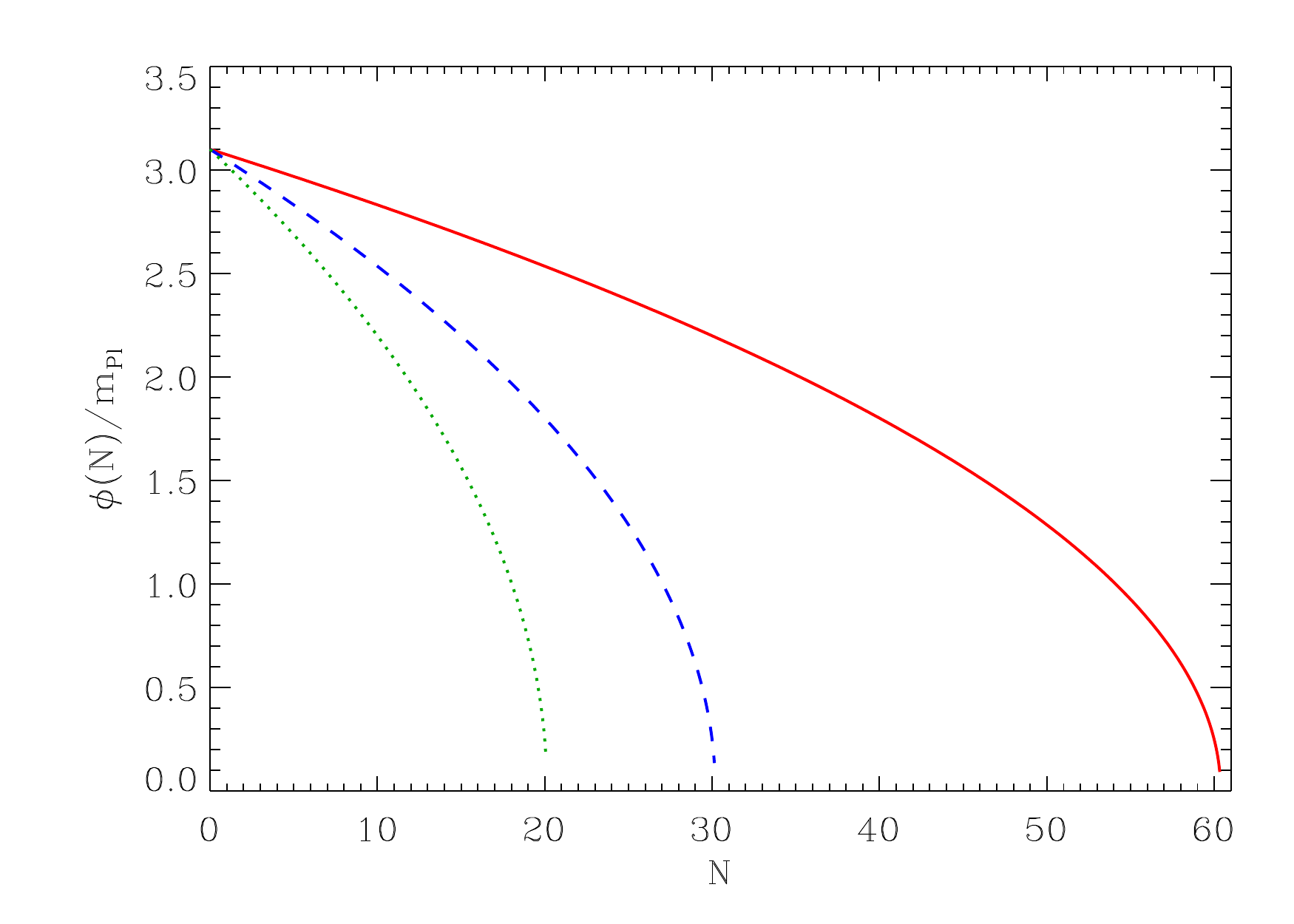}
\hfill
\includegraphics[width=0.3\textwidth,clip=true]{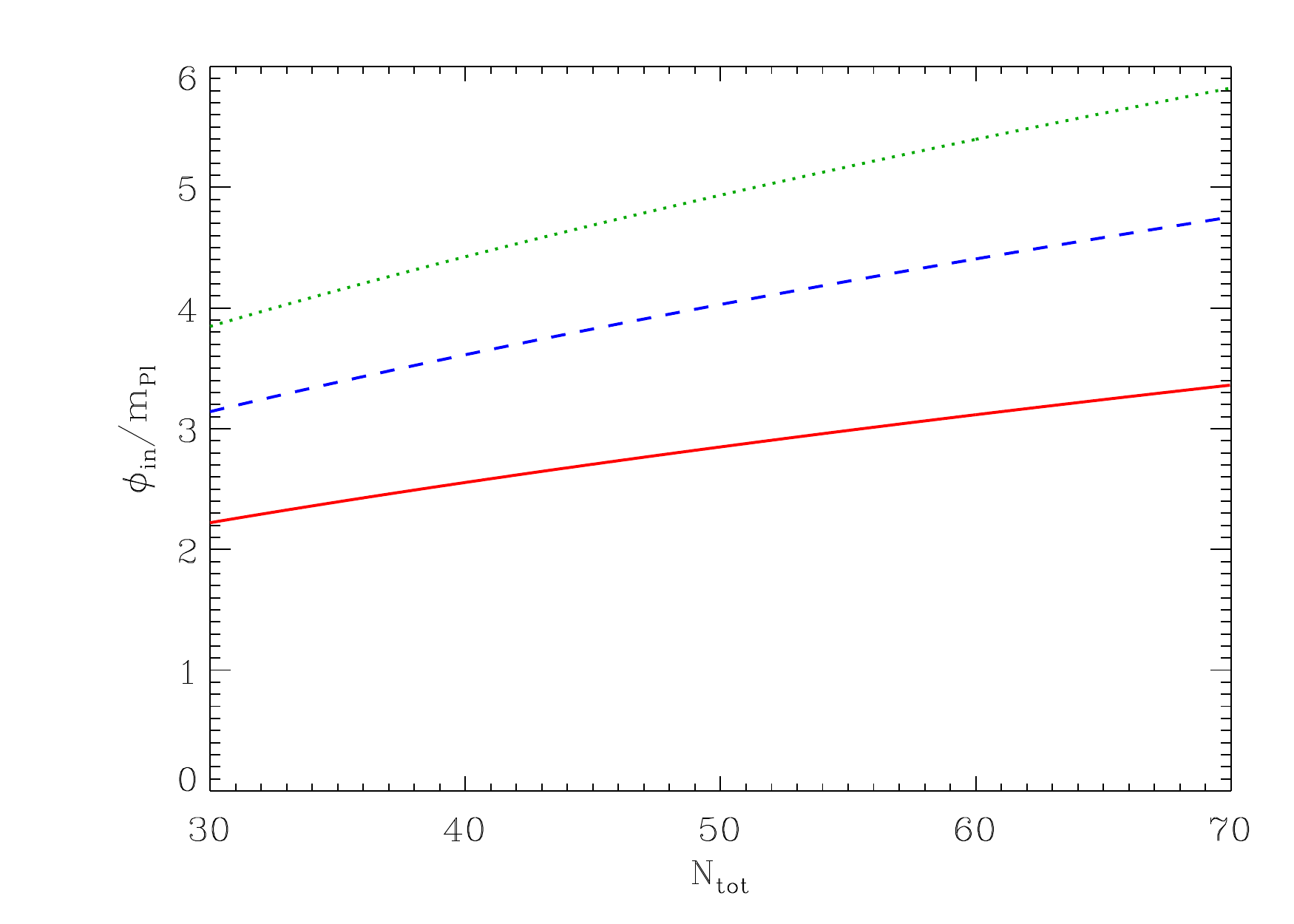}
\caption[\efolds, slow roll trajectory and initial field value for exemplary large field models.]{\small The number of \efolds $N(\phi)$ \emph{(left)}, the slow roll trajectory $\phi(N)/\mpl$ \emph{(center)} from Eq.~(\ref{eq:largefieldtrajectory}) and the initial field value $\phiin(N_{\mathrm{tot}})/\mpl$ \emph{(right)} from Eq.~(\ref{eq:Ntot-largefield}) for the large field models with $n=2,4,6$ (corresponding to the red solid, blue dashed, and green dotted curves, respectively) in Eq.~(\ref{eq:largefield}). Note that, as is seen from the left and center plots, the same initial field value $\phiin=3.1\mpl$ in the $n=2$ case produces the required $N\simeq60$ \efolds, while $N\simeq30$ and $N\simeq20$ in the cases $n=4,6$ with this starting value, respectively. The right plot shows that one would have to choose $\phiin\approx4\mpl$ for $n=4$ and $\phiin\approx5\mpl$ for $n=6$ to achieve $N\approx60$ for these potentials.}
\label{fig:largefieldexample}
\end{center}
\end{figure}

\begin{quotation}
{\bf Example: slow roll trajectory in large field inflation}\\
Inserting the large field potential of Eq.~(\ref{eq:largefield}) into Eq.~(\ref{eq:Nofphi}), one obtains that during the slow roll phase
\beq\label{eq:largefieldtrajectory}
N(\phi)=\frac{4\pi}{n}\left(\frac{\phi_\uin^{2}}{\mpl^{2}}-\frac{\phi^{2}}{\mpl^{2}}\right),\qquad\phi(N)=\left(\phi_\uin^{2}-\frac{n\,\mpl^{2}}{4\pi}\,N\right)^{1/2}\,.
\eeq
This is one example of a potential for which the exact trajectory during slow roll can be found by inversion. The two functions $N(\phi)$ and $\phi(N)$ for different $n$ are plotted in Fig.~\ref{fig:largefieldexample}.
\end{quotation}

\subsubsection{Characterizing the potential}
The $\epsilon_{i}$ parameters describe a perturbative expansion around a universe with a perfectly constant Hubble parameter $H$. Since, during slow roll, $H\propto V^{1/2}$, this can also be interpreted as an expansion around a ``perfectly flat potential''. While we may rely on Eqs.~(\ref{eq:H-approx}) and (\ref{eq:dphi-approx}), $\epsilon_{1}$ and $\epsilon_{2}$ are given by the slope and curvature of $V(\phi)$, respectively,
\beq\label{eq:sr-eps}
\epsilon_{1}\approx\frac{1}{2\kappa}\left(\frac{V'}{V}\right)^{2},\qquad\epsilon_{2}\approx\frac{2}{\kappa}\left[\left(\frac{V'}{V}\right)^{2}-\frac{V''}{V}\right]\,.
\eeq
As we show below, it is the essentially the parameters $\epsilon_{1}$ and $\epsilon_{2}$ that are constrained by observations. Using Eq.~(\ref{eq:sr-eps}), one may therefore learn something about the shape of the inflationary potential directly from CMB measurements.

\begin{figure}[h]
\begin{center}
\includegraphics[width=0.3\textwidth,clip]{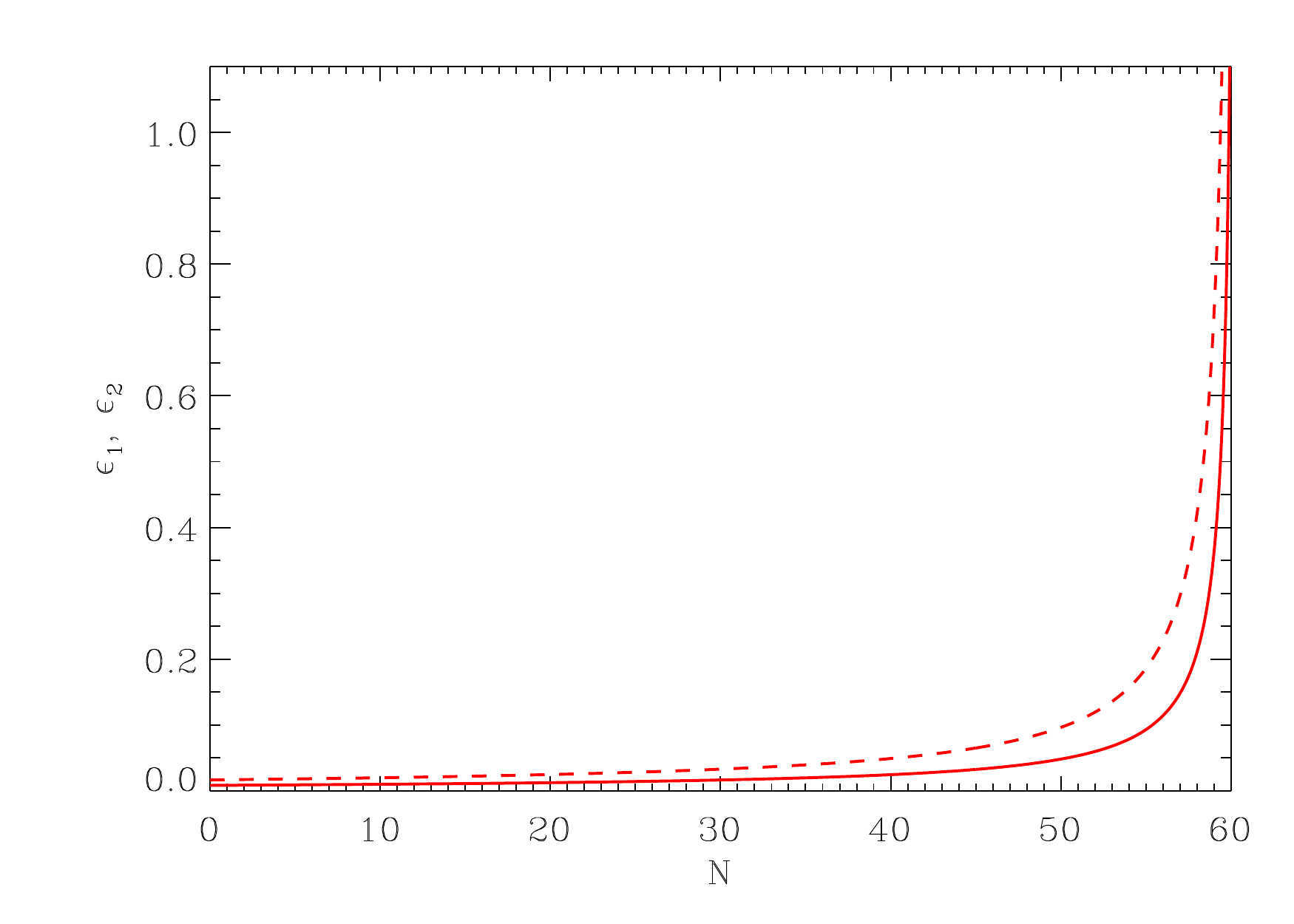}
\hfill
\includegraphics[width=0.3\textwidth,clip]{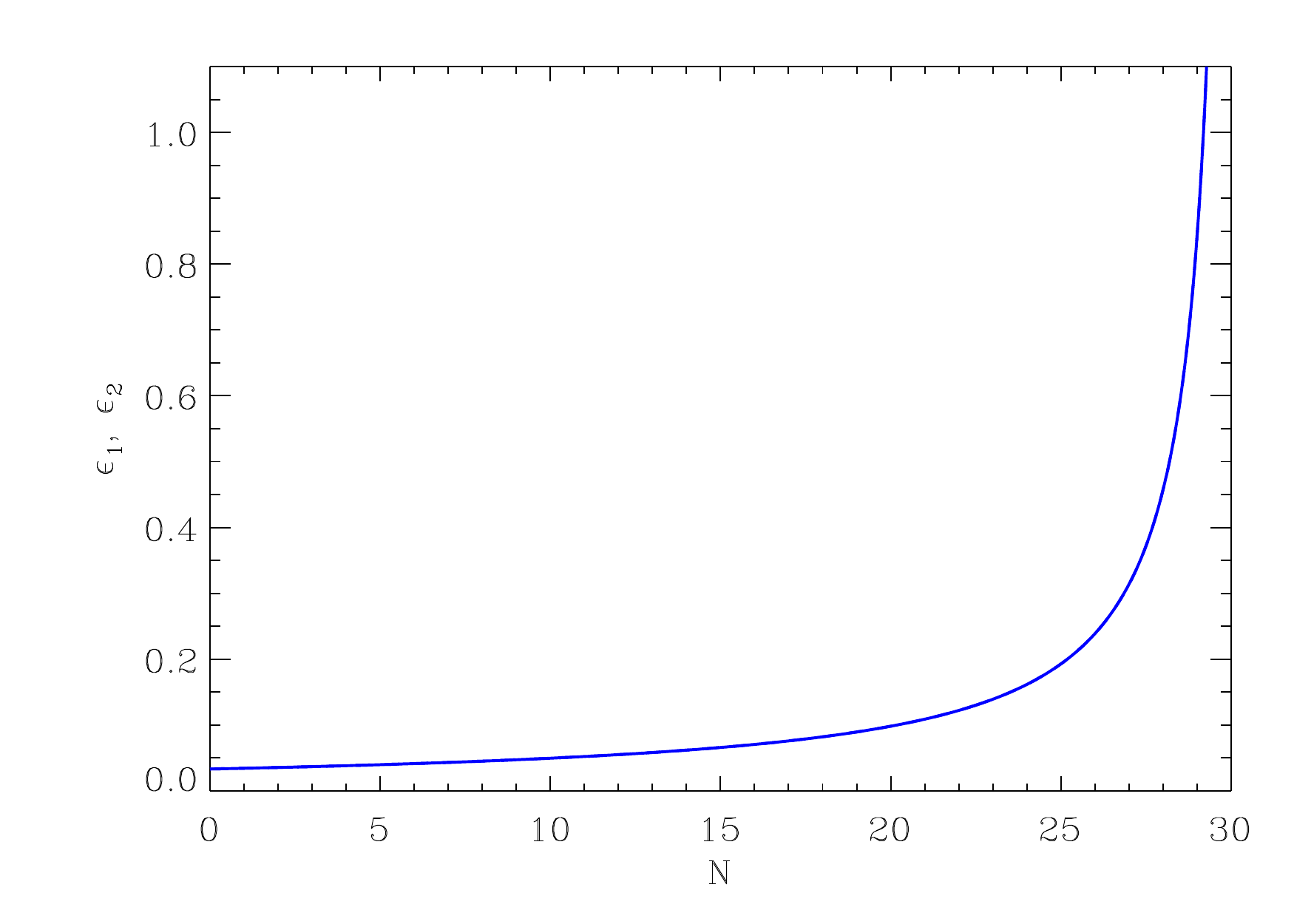}
\hfill
\includegraphics[width=0.3\textwidth,clip]{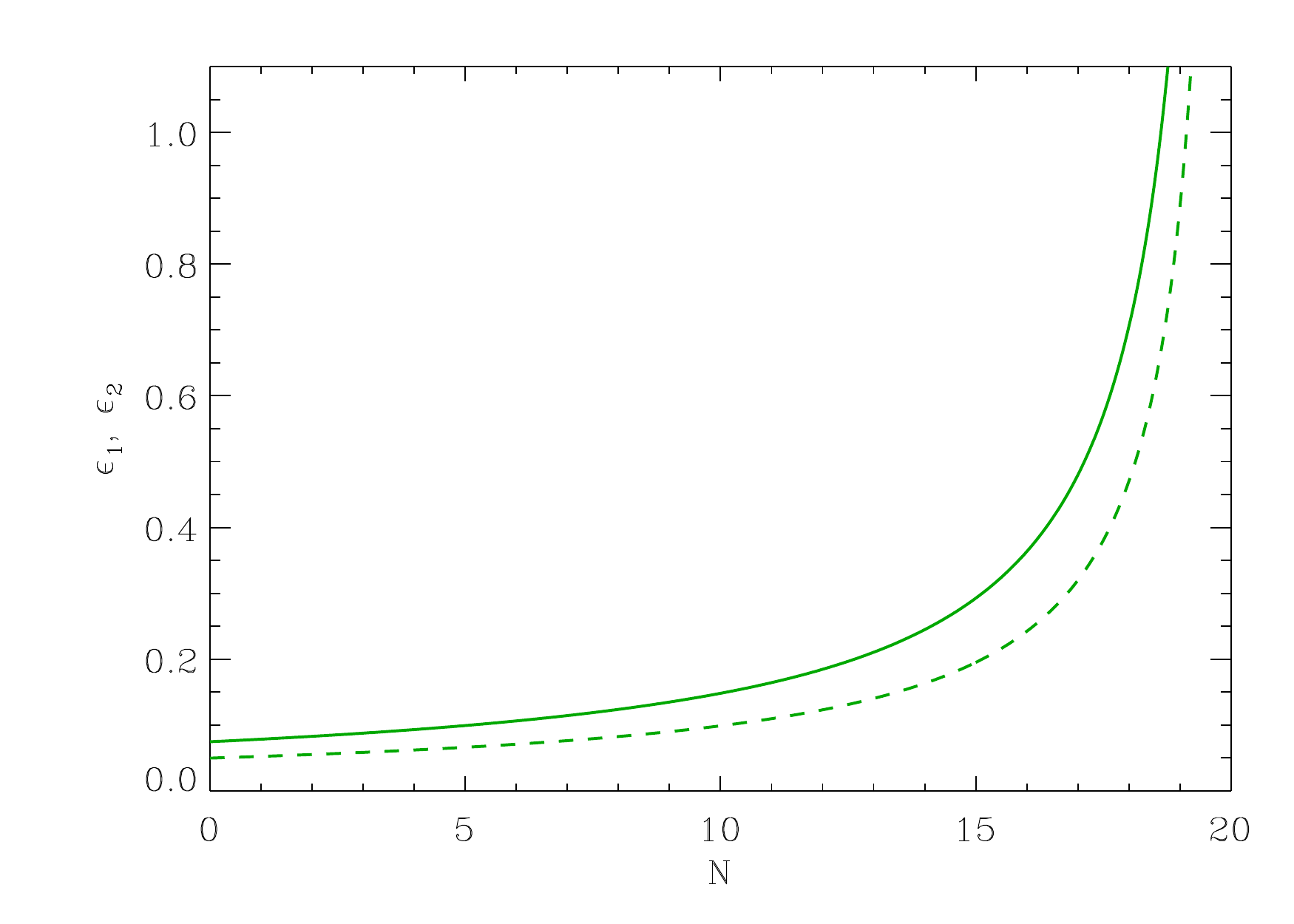}
\caption[Slow roll parameters $\epsone, \epstwo$ for large field inflation models of different powers $n$.]{\small Slow roll parameters $\epsone$ (solid lines) and $\epstwo$ (dotted lines) for the large field models of Eq.~(\ref{eq:largefield}), using Eq.~(\ref{eq:eps1ofN}) and $\epstwo=(4/n)\epsone$. On the \emph{left}, the case $n=2$, in the \emph{center} $n=4$ and on the \emph{right} $n=6$. Note that $\epsone$ and $\epstwo$ coincide for $n=4$, and that for $n<4$ the second slow roll parameter is always larger than $\epsone$, \ie the slow roll approximation breaks down before inflation ends. The opposite is the case for $n>4$: slow roll holds in this case until $\epsone=1$.}
\label{fig:largefieldparameters}
\end{center}
\end{figure}

\begin{quotation}
{\bf Example: $\epsilon_{1},\epsilon_{2}$ for the large field potential}\\
Using the equations (\ref{eq:sr-eps}), we find for the large field potential Eq.~(\ref{eq:largefield}) that
\bea
\epsilon_{1}=\frac{n^{2}}{16\pi}\left(\frac{\mpl}{\phi}\right)^{2}\,,&&\frac{\phi_{\epsilon_{1}}}{\mpl}=\frac{n}{4\sqrt{\pi}}\,,\\
\epsilon_{2}=\frac{n}{4\pi}\left(\frac{\mpl}{\phi}\right)^{2}=\frac{4}{n}\,\epsilon_{1}\,,&&\frac{\phi_{\epsilon_{2}}}{\mpl}=\frac{1}{2}\,\sqrt{\frac{n}{\pi}}=\frac{2}{\sqrt{n}}\,\frac{\phi_{\epsilon_{1}}}{\mpl}\,.
\eea
Note that for $n<4$, $\epsilon_{2}>\epsilon_{1}$, while $\epsilon_{2}$ is the smaller parameter of the two if $n$ exceeds 4. Consequently, we find $\phi_{\epsilon_{2}}>\phi_{\epsilon_{1}}$ (but of the same order) for $n<4$, with $\phi_{\epsilon_{2}}=\phi_{\epsilon_{1}}$ for $n=4$. Since the field moves ``from right to left'' in this model, it therefore encounters $\phi_{\epsilon_{2}}$ first and inflation continues very briefly after slow roll is no longer valid. Using Eq.~(\ref{eq:largefieldtrajectory}), the slow roll parameters are functions of the number of \efolds and the initial field value $\phiin$ only, with
\beq\label{eq:eps1ofN}
\epsilon_{1}=\frac{n^{2}}{16\pi}\left(\frac{\phiin^{2}}{\mpl^{2}}-\frac{n}{4\pi}\,N\right)^{-1}\,.
\eeq
The parameters $\epsilon_{1},\epsilon_{2}$ are plotted in Fig.~\ref{fig:largefieldparameters}.
Note also that from combining Eq.~(\ref{eq:largefieldtrajectory}) with $\phi_{\epsilon_{2}}$, the total number of \efolds depends on the initial field value only,
\beq\label{eq:Ntot-largefield}
N_{\mathrm{tot}}=\frac{4\pi}{n}\left[\frac{\phi_\uin^{2}}{\mpl^{2}}-\frac{n}{4\pi}\right]\,,\qquad\frac{\phi_\uin}{\mpl}=\sqrt{\frac{n}{4\pi}\,(N_{\mathrm{tot}}+1)}\,,
\eeq
and the required number of \efolds can be obtained if (for $n=2$) we have $\phi_\uin/\mpl\approx3$.
\end{quotation}

There exists a variety of alternative slow roll parameter sets to the one defined in Eq.~(\ref{eq:epsiparameters}), see, for example, \cite{Schwarz:2001vv,Schwarz:2004tz}. Very often, instead of the couple $(\epsilon_{1},\epsilon_{2})$, the two parameters $(\epsilon,\eta)$ are used, where $\epsilon=\epsilon_{1}$ and $\eta=\epsilon_{1}-\epsilon_{2}/2$. We use the parameter hierarchy (\ref{eq:epsiparameters}) because its members may be easily derived one after the other, and are all of the same order, which notably is important whenever higher order terms in the slow roll approximation are neglected.

\subsection{Exact Solutions and Primordial Perturbation Spectra}
We now use the slow roll parameters $\epsilon_{i}$ to express the ``mass'' terms $a''/a$ and $z''/z$ in the mode equations Eqs.~(\ref{eq:modeequation}) and (\ref{eq:vtensor}), respectively, as
\bea
\frac{a''}{a}&=&a^{2}H^{2}\left(2-\epsilon_{1}\right)\label{eq:tensorpotential}\,,\\
\frac{z''}{z}&=&a^{2}H^{2}\left(2-\epsilon_{1}+\frac{3}{2}\,\epsilon_{2}-\frac{1}{2}\,\epsilon_{1}\epsilon_{2}+\frac{1}{4}\,\epsilon_{2}^{2}+\frac{1}{2}\,\epsilon_{2}\epsilon_{3}\right)\nonumber\\
&\approx&a^{2}H^{2}\left(2-\epsilon_{1}+\frac{3}{2}\,\epsilon_{2}\right)\label{eq:scalarpotential}\,,
\eea
where the last line contains the first order part of $z''/z$ in the Hubble slow roll parameters only. Again, let us treat the simpler case of tensor perturbations first. Using the approximation (\ref{eq:a-quasidS}) to express $a$ in Eq.~(\ref{eq:tensorpotential}) (which is exact), the tensor mode equation at first order in the $\epsilon_{i}$ [using $1/(1-x)\approx 1+x$ for small $x$] takes the form of a Bessel equation,
\beq
{v_{\uT}}_{k}''+\left[k^{2}-\frac{1}{\eta^{2}}\left(\nu_\uT^{2}-\frac{1}{4}\right)\right]{v_{\uT}}_{k}=0\,,\qquad \nu_\uT^{2}\equiv\frac{9}{4}+3\epsilon_{1}\,.
\eeq
Its two solutions can be written in terms of the Bessel functions $J_{\nu_\uT}$ and $Y_{\nu_\uT}$ of first and second kind and of order $\nu_\uT$,
\beq\label{eq:vtensor-exact}
{v_{\uT}}_{k}^{1,2}\sim\sqrt{-k\eta}\,\left[J_{\nu_\uT}(-k\eta)\pm i Y_{\nu_\uT}(-k\eta)\right]\,,\qquad \nu_\uT\approx\frac{3}{2}+\epsilon_{1}\,.
\eeq
The complex combinations appearing in Eq.~(\ref{eq:vtensor-exact}) are also referred to as the Hankel functions of the first and second kind or order $\nu_\uT$,
\beq\label{eq:Hankel}
H^{(1,2)}_{\nu_\uT}(-k\eta)=J_{\nu_\uT}(-k\eta)\pm i Y_{\nu_\uT}(-k\eta)\,.
\eeq
We immediately see that for de Sitter space 
$\nu_\uT=3/2$, in which case the Bessel functions in Eq.~(\ref{eq:vtensor-exact}) become
\beq\label{eq:dSsolutions}
{v_{\uT}}_{k}\propto\left(\frac{k\eta\pm i}{k\eta}\right)e^{\pm ik\eta}.
\eeq
Note that this solution is valid for all $k$. It is evident that for $-k\eta\rightarrow-\infty$, the prefactor in Eq.~(\ref{eq:dSsolutions}) approaches one, and we recover the sub-Hubble solution (\ref{eq:subhorizon-tensor}). Eventually, however, we are interested in those modes which crossed out of the Hubble radius during inflation and hence reached their long wavelength limit $-k\eta\rightarrow 0$. We already know from Eq.~(\ref{eq:superhorizon-tensor}) that ${v_{\uT}}_{k}\propto a$ in this case, but have yet to determine the normalization constant.

The case of scalar perturbations is similar, but since their potential (\ref{eq:scalarpotential}) is different from the tensor case, the order $\nu_\uS$ in the Bessel equation is modified and one finds at first order
\beq
v_{k}''+\left[k^{2}-\frac{1}{\eta^{2}}\left(\nu_\uS^{2}-\frac{1}{4}\right)\right]v_{k}=0\,,\qquad \nu_\uS^{2}\equiv\frac{9}{4}+3\epsilon_{1}+\frac{3}{2}\,\epsilon_{2}
\eeq
with the solutions
\beq\label{eq:vscalar-exact}
v_{k}^{1,2}\sim\sqrt{-k\eta}\,\left[J_{\nu_\uS}(-k\eta)\pm i Y_{\nu_\uS}(-k\eta)\right]\,,\qquad \nu_\uS\approx\frac{3}{2}+\epsilon_{1}+\frac{1}{2}\,\epsilon_{2}\,.
\eeq
We now turn to the question of the right vacuum state for perturbations: once the normalization for Eq.~(\ref{eq:vscalar-exact}) fixed, we may calculate the exact behavior for all $k$.


\subsubsection{Choice of vacuum}
We showed earlier that the quantum treatment of perturbations has to start from the second-order perturbed action, which for tensor perturbations [compare Eq.~(\ref{eq:startaction})] reads
\begin{equation}\label{eq:tensoraction}
\delta_{2}S^\uPTP=\frac{1}{8}\sum_{\diamond}\int\dd^{4}x\,\left[\left({v_\uT^{\diamond}}'\right)^{2}-\left(v_\uT^{\diamond}\right)_{,i}\left(v_\uT^{\diamond}\right)_{,i}+\frac{a''}{a}\,{v_\uT^{\diamond}}^{2}\right]\,,
\end{equation}
where we have temporarily reintroduced the index $\diamond$ to account for the polarization states\footnote{Recall that we have defined $v_\uT$ in a dimensionful way, $v_\uT=a\varphi_\uT/\sqrt{2\kappa}$. Written in terms of the dimensionless original tensor $h_{ij}$, the action (\ref{eq:tensoraction}) reads $\delta_{2}S^\uPTP=[\mpl^{2}/(64\pi)]\int\dd^{4}x\,a^{2}\left[(h^{i}_{j})'(h^{j}_{i})'-\partial_{l}(h^{i}_{j})\,\partial^{l}(h^{j}_{i})\right]$, compare with Eq.~(\ref{eq:actionintrinsic}).}. Up to the (time-dependent) ``mass'' term \(a''/a\), this is the action of a scalar field in flat Minkowski spacetime with (time-independent) metric $g_{\mu\nu}=\eta_{\mu\nu}$. Its special symmetry properties allow to define a unique vacuum state $\ket{\Omega}$ for Minkowski space, which minimizes the energy at all times and is annihilated by all mode creation operators, $\hat{a}_{\vec{k}}\ket{\Omega}=0$. Moreover, the operators $\hat{a}_{\vec{k}}$ and their adjoints (and hence the notion of ``particles'') are uniquely defined in this case because there exists a privileged set of mode functions ${v_{\uT}}_{k}=\exp(-ik\eta)/\sqrt{2k}$ in the notation of the expansion for tensor perturbations analogous to Eq.~(\ref{eq:basisexpansion}).\\
One may still perform an expansion of the type (\ref{eq:basisexpansion}) in a time-dependent background (as we implicitly did above), however, the notion of $\hat{a}_{\vec{k}},\,\hat{a}^{\dagger}_{\vec{k}}$ is generally no longer unique in this case. Notably, different choices of operators and mode functions ${v_{\uT}}_{k}$ are related by \emph{Bogolyubov transformations} \cite{birrell-davies}, and therefore the definition of the vacuum state $\ket{\Omega}$ becomes ambiguous: there no longer exists a state minimizing the energy at all times. A common strategy consists of defining a vacuum at a fixed time $\eta_{0}$, which then only changes adiabatically.\\
The de Sitter background is special among time-dependent spacetimes in the sense that one may define the \emph{Bunch Davies vacuum}, corresponding to the vacuum state minimizing the energy for each mode in its infinite de Sitter past \(\eta\rightarrow-\infty\). Therefore the corresponding mode functions are equal to the Minkowski mode functions for \(\eta\rightarrow-\infty\), \ie when they are deep inside the Hubble radius in their short wavelength limit. Comparing the Minkowski mode functions with our approximate sub-Hubble solutions (\ref{eq:subhorizon-tensor}), and at the same time taking into account the Wronskian condition, we see that we must set $A_{k}=1,\,B_{k}=0$ in Eq.~(\ref{eq:subhorizon-tensor}) to recover Minkowski space in the infinite past. An intuitive illustration is that spacetime looks essentially flat to modes $k$ much smaller than the characteristic curvature scale (\ie the Hubbe radius) of the time-dependent background.

For the exact solution (\ref{eq:vtensor-exact}) it then follows that
\beq\label{eq:vacuumchoice}
{v_{\uT}}_{k}=\frac{\sqrt{-\pi\eta}}{2}\,i^{\nu_\uT+1/2}\,H^{(1)}_{\nu_\uT}(-k\eta)\,,
\eeq
with the Hankel function of the first kind defined in Eq.~(\ref{eq:Hankel}). Using its expansion in the super-Hubble limit $-k\eta\rightarrow0$, we finally find
\beq\label{eq:phit-superhorizon}
\frac{\left|{\varphi_{\uT}}_{k}\right|^{2}}{2\kappa}=2^{2\nu_\uT-3}\,\frac{H^{2}}{2k^{3}}\,\left[\frac{\Gamma(\nu_\uT)}{\Gamma(3/2)}\right]^{2}\left(\frac{k}{aH}\right)^{3-2\nu_\uT}\propto\frac{H^{2}}{2k^{3}}\,\left(\frac{k}{aH}\right)^{-2\epsilon_{1}}\qquad\textnormal{for}\,-k\eta\rightarrow 0\,,
\eeq
where we have used $\nu_\uT$ from Eq.~(\ref{eq:vtensor-exact}). To see that Eq.~(\ref{eq:phit-superhorizon}) indeed approaches a constant (as we know it must from the approximate solution in the super-Hubble limit), note that the combination $H^{2}(k/aH)^{3-2\nu_\uT}$ indeed obeys $\dd \left[H^{2}(k/aH)^{3-2\nu_\uT}\right]/\dd t=0$ at first order in $\epsilon_{i}$ \cite{Kinney:2009vz}.

A similar expression is obtained for the scalar Mukhanov Sasaki variable $v$ and the comoving curvature perturbation $\mathscr{R}=-v/z$, respectively. Note that $z\propto\sqrt{\epsilon_{1}}$ [see Eq.~(\ref{eq:zexplained})], and that this time the order of the Hankel function is $\nu_\uS$ from Eq.~(\ref{eq:vscalar-exact}), therefore we have
\beq\label{eq:intrinsic-superhorizon}
\left|\mathscr{R}_{k}\right|^{2}\propto\frac{H^{2}}{2\epsilon_{1}k^{3}}\,\left(\frac{k}{aH}\right)^{-2\epsilon_{1}-\epsilon_{2}}\qquad\textnormal{for}\,-k\eta\rightarrow 0\,.
\eeq
[Again, the combination $(H^{2}/\epsilon_{1})(k/aH)^{3-2\nu_\uS}$ is easily shown to be time-independent.] Since $\epsilon_{1}<1$ during inflation, the amplitude of scalar perturbations with wavenumbers $k$ that became super-Hubble before the end of inflation is enhanced with respect to tensors. Note also that for tensor perturbations, the original variable was $h_{ij}=\sum_{\diamond}\varphi_\uT^{\diamond}\,e_{ij}^{\diamond}$, \ie one must be careful to count both polarization states.

\subsubsection{Primordial spectra}\label{subsubsec:spectra}
The two-point equal time correlation function for tensor fluctuations in the vacuum,
\beq\label{eq:h-spectrum}
\bra{\Omega}h_{ij}(\eta,\vec{x})h^{ij}(\eta,\vec{x}')\ket{\Omega}=\int_{0}^{\infty}\frac{\dd k}{k}\,\frac{\sin\left(k\left|\vec{x}-\vec{x}'\right|\right)}{k\left|\vec{x}-\vec{x}'\right|}\,k^{3}\calP_\uT(k)\,,
\eeq
defines the tensor spectrum (recall that ${v_{\uT}}_{k}$ has dimensions of a scalar field)
\beq\label{eq:defPt}
k^{3}\calP_\uT(k)=\frac{16\,k^{3}}{\pi\mpl^{2}}\left|\frac{{v_{\uT}}_{k}}{a}\right|^{2}\,.
\eeq
An exactly analogous definition to Eq.~(\ref{eq:h-spectrum}) for the comoving curvature perturbation $\mathscr{R}$ allows to write the scalar spectrum as
\beq\label{eq:defPs}
k^{3}\calP_\uS(k)=\frac{k^{3}}{8\pi^{2}}\left|\frac{v_{k}}{z}\right|^{2}\,.
\eeq
After expanding the prefactors $2^{2\epsilon_{1}}$, $\Gamma(3/2+\epsilon_{1})/\Gamma(3/2)$ and $(k/aH)^{-2\epsilon_{1}}$ in Eq.~(\ref{eq:phit-superhorizon}) [and analogously in Eq.~(\ref{eq:intrinsic-superhorizon})] for small $\epsilon_{1},\epsilon_{2}$  \cite{peter:cosmo}, one obtains the spectra at first order in the slow roll parameters:
\bea
k^{3}\calP_\uS(k)&=&\frac{{H_{*}}^{2}}{\pi\epsilon\mpl^{2}}\left[1-2(C+1)\epsilon_{1}-C\epsilon_{2}-(2\epsilon_{1}+\epsilon_{2})\,\ln\left(\frac{k}{k_{*}}\right)\right]\label{eq:scalarspectrum}\\
k^{3}\calP_\uT(k)&=&\frac{16{H_{*}}^{2}}{\pi\mpl^{2}}\left[1-2(C+1)\epsilon_{1}-2\epsilon_{1}\,\ln\left(\frac{k}{k_{*}}\right)\right]\label{eq:tensorspectrum}
\eea
By $C$ we denote here a constant that arises from the prefactor expansion, $C=\gamma_{\mathrm{E}}+\ln 2-2$, with $\gamma_{\mathrm{E}}\simeq0.5772$ the Euler constant and hence $C\simeq -0.73$. The scale $k_{*}$ is called the pivot scale: it is picked \eg in the middle of the observational window and used to compare the above expressions to observations. The choice of a pivot point singles out the moment in time when $k_{*}$ left the Hubble horizon, $k_{*}=a(\eta_{*})\,H(\eta_{*})$, at which the zeroth order (in slow roll parameters) amplitude is calculated, hence the subscripts ``*'' in Eqs.~(\ref{eq:scalarspectrum}) and (\ref{eq:tensorspectrum}). 
A suitable pivot scale is, for example, $k_{*}=0.01\,h\Mpc^{-1}$ with $h$ the reduced Hubble parameter in Eq.~(\ref{eq:Hubble-today}). Note that the amplitude ratio $\calP_\uT/\calP_\uS$ at first order in slow roll is $\calP_\uT/\calP_\uS=16\epsilon_{1}$, \ie as stated above, tensor modes are suppressed by a factor $\epsilon_{1}$.

Spectra like Eqs.~(\ref{eq:scalarspectrum}) and (\ref{eq:tensorspectrum}) are characterized by their indices defined as
\beq\label{eq:indices}
n_\uS-1=\left[\frac{\dd\left(k^{3}\calP_\uS\right)}{\dd\ln k}\right]_{k=k_{*}}=-2\epsilon_{1}-\epsilon_{2},\qquad n_\uT=\left[\frac{\dd\left(k^{3}\calP_\uT\right)}{\dd\ln k}\right]_{k=k_{*}}=-2\epsilon_{1}\,,
\eeq
where we have suppressed the index ``*'' on the right hand side indicating that the slow roll parameters should be evaluated when $k_{*}$ left the Hubble radius. Note that these can also simply be read off Eqs.~(\ref{eq:scalarspectrum}) and (\ref{eq:tensorspectrum}) if we suppose that the spectra (\ref{eq:defPt}) and (\ref{eq:defPs}) are of power law form, \ie $\calP_\uT(k)\propto \mathscr{A}_\uT\,k^{n_\uT}$ and $\calP_\uS(k)\propto \mathscr{A}_\uS\,k^{n_\uS-1}$. The cases where $n_\uS=1$ and $n_\uT=0$, respectively, correspond to a scale invariant (flat) \emph{Harrison Zel\textquoteright{}dovich spectrum}. It follows from Eq.~(\ref{eq:indices}) that during slow roll inflation, \ie $\epsilon_{1},\epsilon_{2}\ll1$, both tensor and scalar perturbation spectra are almost flat. Note that the tensor index $n_\uT$ and the amplitude ratio $r=\calP_\uT/\calP_\uS$ are related by the ``consistency relation'' $r=-8n_\uT$.\\
Hence, at first order in the Hubble slow roll parameters, there are four observables, the amplitudes $\mathscr{A}_\uS$ and $\mathscr{A}_\uT$ and their indices $n_\uS, n_\uT$. Of these, at present $\mathscr{A}_\uS$ and $n_\uS$ have been measured (see Section \ref{subsec:WMAP5}), and there are upper bounds on $\mathscr{A}_\uT$ and $n_\uT$. We discuss these results and how they are obtained from CMB data below. It is hoped that future experiments like Planck \cite{Toffolatti:1997dk} will detect gravitational waves, which notably allows a test of the consistency relation.

Pushing even further, one may also define the running $\alpha_\uS,\alpha_\uT$ of the spectral indices $n_\uS$ and $n_\uT$ with wavenumber $k$ from
\beq
\alpha_\uS=\left[\frac{\dd n_\uS}{\dd \ln k}\right]_{k=k_{*}}=-2\epsilon_{1}\epsilon_{2}-\epsilon_{2}\epsilon_{3}\,,\qquad \alpha_\uT=\left[\frac{\dd n_\uT}{\dd\ln k}\right]_{k=k_{*}}=-2\epsilon_{1}\epsilon_{2}\,.
\eeq
The $\alpha_\uS,\alpha_\uT$ are therefore of second order in the slow roll parameters, \ie \emph{very} small, which emphasizes how close to scale invariance the inflationary power spectra of Eqs.~(\ref{eq:scalarspectrum}) and (\ref{eq:tensorspectrum}) are. (There also exists a second order consistency relation relating $r$, $\alpha_\uT$ and $n_\uS$ \cite{peter:cosmo}.) When measured at the scale $k_{*}$, one may develop the spectral amplitude of scalar perturbations (and analogously for tensors) around $k_{*}$ according to
\beq
\ln \mathscr{A}_\uS^{2}(k)=\ln\mathscr{A}_\uS^{2}(k_{*})+(n_\uS-1)\,\ln\frac{k}{k_{*}}+\frac{\alpha_\uS}{2}\,\ln^{2}\frac{k}{k_{*}}+\dots\,.
\eeq
Finally note from Eqs.~(\ref{eq:scalarspectrum}) and (\ref{eq:tensorspectrum}) that $\mathscr{A}_\uT\propto H^{2}$, while $\mathscr{A}_\uS\propto H^{2}/\epsilon_{1}$. It is therefore from the tensor amplitude that one may hope to uniquely fix the energy scale $H/\mpl$ during inflation, while scalar perturbations only determine the ratio between that scale and $\sqrt{\epsilon_{1}}$.

\subsubsection{Observable window of scales}
The choice of a pivot scale $k_{*}$ defines a new measure of time by singling out the Hubble crossing $k_{*}/[a(N_{*})\,H(N_{*})]=1$ of a given scale. In the same way, each scale $k$ is uniquely related to the \efold $N_{k}$ when it becomes super-Hubble, and we know that for observable scales today $N_{\mathrm{tot}}-N_{k}\approx 40-60$ \cite{Liddle:2003as} (recall that we count \efolds from the beginning of inflation onwards), where $N_{\mathrm{tot}}$ is the total expansion achieved while the inflaton field $\phi$ moves from $\phiin$ to $\phiend$. To obtain predictions for the primordial spectra (\ref{eq:scalarspectrum}) and (\ref{eq:tensorspectrum}) from a given inflationary model [\ie a choice of $V(\phi)$] at today's observed $k$, one must complete the following steps.

Firstly, integrate the (background) system of coupled differential equations for a FLRW universe filled by a scalar field with potential $V(\phi)$, given by Eqs.~(\ref{eq:Friedmann}) and (\ref{eq:KleinGordon}). It is usually most convenient to integrate the Klein Gordon equation written in terms of \efolds,
\beq\label{eq:KG-efolds}
\frac{\dd^{2}}{\dd N^{2}}\,\phi+\left(3H^{2}+\frac{\dd H/\dd N}{H}\right)\frac{\dd}{\dd N}\,\phi+V_{,\phi}=0\,.
\eeq
Note that the friction term may be written as $3H^{2}+(\dd H/\dd N)/H=H^{2}(3-\epsilon_{1})$. Moreover, numerical integration codes often use a vector of two variables $(\phi,\dot{\phi})$, whose evolution is coupled as $\dd\phi/\dd N=\dot{\phi}/H$ and $\dd\dot{\phi}/\dd N$ from Eq.~(\ref{eq:KG-efolds}). The choice of the initial field value at $t=0$ is made such that (at least) enough inflation is obtained to solve the SBBM problems. A good estimate for this is obtained using slow roll calculations [such as Eq.~(\ref{eq:Ntot-largefield}) for the large field case] because (in standard scenarios) most of the inflationary expansion takes place in this regime. After integrating $\phi$, the time evolution of $H$ and its derivatives is known.\\
Secondly, since the background integration provides us with exact knowledge of the $\epsilon_{i}$, we can hence determine when inflation ends, $\epsilon_{1}(t_{\epsilon_{1}})=1$, and when the slow roll approximation breaks down [as soon as one $\epsilon_{i}(t_{\epsilon_{i}})=1$]. 
While $\epsilon_{i}\ll1$, the slow roll trajectory $N(\phi)$ [which, again for the case of a large field potential, was calculated in Eq.~(\ref{eq:largefieldtrajectory})] should coincide with $\phi$ obtained from the exact numerical integration.\\
Thirdly, one can now integrate the perturbations' equations (\ref{eq:modeequation}) and (\ref{eq:vtensor}), where the ``mass'' terms (\ref{eq:tensorpotential}) and (\ref{eq:scalarpotential}) are functions of the $\epsilon_{i}$ as well as of $a$ and $H$ already determined from the background integration. [The range of $k$ to integrate is determined from their Hubble exit, $k=a(N_{k})\,H(N_{k})$, occurring around the ``right time'', \ie $40-60$ \efolds before $\phi_{\epsilon_{1}}$.] Initial conditions on $v_{k}, {v_{\uT}}_{k}$ and their derivatives are imposed such that each $k$ mode is in its adiabatic vacuum deep inside the Hubble radius.\\
Finally, from the knowledge of $v_{k}, {v_{\uT}}_{k}$ and their corresponding $N_{k}$, one can directly calculate the spectral amplitudes and indices on large scales. We know from the slow roll solutions to Eqs.~(\ref{eq:modeequation}) and (\ref{eq:vtensor}) and their limits in the super-Hubble regime that the perturbations cease to oscillate once outside the Hubble radius\footnote{More precisely, the spectra approach their late-time constant value some time after their Hubble crossing. Using $v_{k}, {v_{\uT}}_{k}$ exactly at Hubble exit introduces a factor $\order{2}$ \cite{Kinney:2009vz}.}. 

At this point, let us recall that, as we mentioned when considering the horizon problem, at the time of decoupling the Hubble radius $H^{-1}(t_\udec)$ corresponded to directions within $\sim 1^{\circ}$ of angular separation in the sky. 
As a consequence, only these $k$ have been affected by the cosmological evolution between the end of inflation and decoupling, and the smaller the scale, the longer the time period it spent under the evolution of gravitational instability inside the Hubble radius. At angles $>1^{\circ}$, we always probe larger scales than the Hubble length at decoupling, which have retained the primordial form of their spectra \cite{liddle:inflation}.

\begin{quotation}
{\bf Example: slow roll power spectrum and index for large field inflation}\\
For large field potentials of the form (\ref{eq:largefield}), we obtained the trajectory (\ref{eq:largefieldtrajectory}), which gives the time dependence of the slow roll Hubble parameter when inserted into Eq.~(\ref{eq:H-approx}). If we use Eq.~(\ref{eq:Ntot-largefield}) to replace $\phiin$ in Eq.~(\ref{eq:eps1ofN}), and since $\epsilon_{2}=(4/n)\,\epsilon_{1}$ during slow roll large field inflation, we finally find for the scalar spectral amplitude $\mathscr{A}_\uS$ and $n_\uS$ from Eqs.~(\ref{eq:scalarspectrum}) and (\ref{eq:indices}):
\bea
\mathscr{A}_\uS&=&\frac{8}{3\pi^{n/2}}\left(\frac{\Lambda_{\mathrm{lf}}}{\mpl}\right)^{4}\left(\frac{n}{4}\right)^{(n-2)/2}\left(\Delta N_{k}+1\right)^{(n+2)/2}\label{eq:lfamplitudeofN}\\
n_\uS-1&=&-\frac{n+2}{2}\,\frac{1}{\Delta N_{k}+1}\label{eq:lfindexofN}
\eea
For the ratio $r=\calP_\uT/\calP_\uS$ it follows that $r=4n/(\Delta N_{k}+1)$. In these expressions, one then varies $\Delta N_{k}=N_{\mathrm{tot}}-N_{k}=40\dots60$ for observable scales. Note that from Eq.~(\ref{eq:lfindexofN}) it follows that the scalar power spectrum always has a red tilt, $n_\uS-1<0$, in large field models. For the simple case of a pure mass term $V(\phi)=(m^{2}/2)\,\phi^{2}$, one obtains from Eq.~(\ref{eq:lfamplitudeofN}) and (\ref{eq:lfindexofN}) that
\bea
n_\uS-1&\approx&[-0.049,-0.033]\,,\qquad r\approx[0.132,0.196]\,,\label{eq:ns-chaotic}\\
\mathscr{A}_\uS&=&\frac{4}{3\pi}\,\frac{m^{2}}{\mpl^{2}}\,\left(\Delta N_{k}+1\right)^{2}=\frac{4}{3\pi}\,\frac{m^{2}}{\mpl^{2}}\,\frac{4}{(n_\uS-1)^{2}}\,,\label{eq:A-chaotic}
\eea
where the numerical values in Eq.~(\ref{eq:ns-chaotic}) are for $\Delta N_{k}=40\dots60$, respectively. Using Eq.~(\ref{eq:A-chaotic}) one can therefore observationally determine the mass scale $m$, if the amplitude of the scalar power spectrum is measured.
\end{quotation}

Hence, the primordial spectra (\ref{eq:tensorspectrum}) and (\ref{eq:scalarspectrum}) of scalar and tensor perturbations during inflation can be calculated either numerically, or in certain cases analytically in their slow roll approximation. They then have to be propagated through reheating and the radiation dominated epoch until the release of the CMB. From the surface of last scattering to today, the perturbations are then subject to the \emph{integrated Sachs Wolfe effect} explained below, before they can eventually be compared with temperature fluctuation measurements taken today. We now briefly comment on the transition from quantum to classical perturbations before we turn to the issue of relating primordial spectra to CBM observations.

\subsubsection{Transition to classical perturbations, non-Gaussianity}
The continuous production of perturbations around an inflating FLRW background is a decidedly quantum phenomenon: any classical perturbations are diluted away during inflation, while perturbations of quantum origin can develop a non-zero super-Hubble amplitude. On the other hand, it is evident that they must turn classical eventually 
because their observational consequences such as the CMB spectra and galaxy distributions do not have quantum mechanical features. From the experimental point of view, the perturbations are turned classical by replacing the standard quantum average in the temperature two-point correlation function by a spatial average over the celestial sphere \cite{Grishchuk:1997pk}. The difference between these averaging procedures gives rise to ``cosmic variance'', as we discuss in the next Section. The details process of the classicalization process of the perturbations are much more involved \cite{Martin:2007bw,Polarski:1995jg}. 
In summary, the statistical properties of quantum fluctuations during inflation are interpreted as the statistical behavior of a classical random field after inflation. In particular, the perturbation spectra at linear order preserve their Gaussian character (which is due to their quantum origin, and the conservation of Gaussianity by linear evolution), and hence a two-point correlation function suffices to describe them. Small non-Gaussian corrections can be accounted for by the three-point function, and may be of primordial origin (\eg due to features in the potential leading to sudden changes of $V$ and its derivatives), second order perturbation effects or non-linear evolution (``secondary'' non-Gaussianity).

\section{From Inflation to CMB Observations}\label{sec:CMB}
We showed that once a choice of potential $V(\phi)$ for the inflaton is made, 
the spectra of both tensor and scalar perturbations are straightforward to calculate. We also know that the scales observable today left the Hubble radius about $40-60$ \efolds before the end of inflation, and re-entered the comoving Hubble radius at some point afterwards during the SBBM evolution, which is when they became observable to us. The missing piece in this puzzle is a way to relate (\ie transfer) the primordial $\calP_\uT(k),\calP_\uS(k)$ to today's density perturbation spectra $\calP_{\rho}(k)$ obtained from observations of the CMB and large scale structure. Schematically, one may write this as
\beq
\calP_{\rho}(k,t_{0})=\left[\calP_{\mathrm{po}}(k)\right]_{t=t_\uend}\,\mathscr{T}(k, t_\uend, t_{0})\,
\eeq
where $t_\uend$ is the end of inflation, the index ``po'' stands for ``primordial'' and $t_{0}$ is today. The function $\mathscr{T}(k, t_\uend, t_{0})$ is called the transfer function, 
and in principle, it depends on all the physics that occurred in the Universe ever since the end of inflation. Luckily, our understanding of the SBBM allows to narrow its description down to a few parameters, and the calculation of $\mathscr{T}(k, t_\uend, t_{0})$ can be broken up into several pieces: from the end of inflation until the release of the CMB photons, perturbations have evolved through reheating, the radiation dominated and the beginning of the matter dominated epoch. From the time of decoupling until they deposit information about the primordial spectra in our detectors, these photons were essentially free-streaming, their frequencies being redshifted by the (decelerated) universal expansion. The tiny inhomogeneities in their temperature are due to the \emph{Sachs Wolfe effect} \cite{Sachs:1967er} we discuss below. Though complicated and involved, the physical processes during the SBBM evolution are well understood, and the description of the reheating phase terminating inflation can also be significantly simplified. From a practical point of view, the calculation of CMB observables is automated to a large extent today, and computer codes (\eg the \texttt{CAMB} code\footnote{\texttt{http://camb.info/}}) are available from which the CMB spectrum corresponding to the primordial input of a given inflationary model is readily obtained.

\subsection{The Sachs Wolfe Effect}
The link between primordial cosmological fluctuations and temperature fluctuations on the last scattering surface is provided by the Sachs Wolfe effect \cite{Sachs:1967er}, which allows the calculation of a photon's energy change while propagating from its release at the time of decoupling $t_\udec$ to a present-day detector. This change can be expressed in terms of the ratio of its temperatures (energies) at detection (index ``d'', today) and emission (index ``e'', at the time of last scattering) $T_\ud/T_\ue$. At zeroth order, one simply has $T_\ud=[a(t_\udec)/a_{0}]\,T_\ue$. At first order in perturbations, photons can pick up a delay $\delta t_\udec$ depending on their position because they have to climb out of gravitational ``potential wells'' of different depth on the surface of last scattering. Hence they acquire a slightly different redshift, leading to fluctuations in their temperature $\delta T_\ud$. The depths of these wells in turn is a consequence of the inflationary perturbations propagated through reheating and the radiation dominated epoch.\\
In calculating the Sachs Wolfe effect, one establishes $\delta t_\udec(t_\udec,\vec{x}_\ue)$, which requires the definition of a surface of emission\footnote{One possibility is the surface of constant photon density, \ie $\rho_{\gamma}(t_\udec)=\const$, in which case $\delta t_{\udec}$ is proportional to the density contrast of photons $\delta\rho_{\gamma}/\rho_{\gamma}$.}, on which $\vec{x}_\ue$ is the position of the photon. Then it can be shown (see \eg \cite{Martin:2004um}) that the relative temperature fluctuation today (we now drop the detection index ``d'') consists of three contributions,
\beq\label{eq:dTcontributions}
\frac{\delta T}{T}=\left(\frac{\delta T}{T}\right)_{\mathrm{dipole}}+\left(\frac{\delta T}{T}\right)_\uS+\left(\frac{\delta T}{T}\right)_\uT\,.
\eeq
The first term is the dipole caused by the relative motion of our galaxy with respect to the reference frame of the CMB\footnote{The amplitude of the dipole component is $(\delta T/T)_{\mathrm{dipole}}\approx 1.2\cdot 10^{-3} \cite{Durrer:CMB}.$}. The terms ``S'' and ``T'' are perturbations of scalar and tensor type, respectively, and they are directly related to the corresponding fluctuations of inflationary origin. Exact calculations show that $\left(\delta T/T\right)_\uS$ contains three different terms \cite{Durrer:CMB}: \emph{i)} a Doppler-like term (photon velocity at the time of decoupling), \emph{ii)} an integral over the change in the gravitational potential along the photon's trajectory from $t_\udec$ to $t_{0}$ (the so-called \emph{integrated Sachs Wolfe effect}) and \emph{iii)} the photon density contrast $\delta\rho_{\gamma}/\rho_{\gamma}$ and the gravitational (Bardeen) potential $\Phi$ evaluated at $t_\udec$. On large scales, only the third contribution persists and its two parts can be combined to give
\beq\label{eq:Tfluct}
\left(\frac{\delta T}{T}\right)_\uS\simeq\frac{1}{3}\,\Phi\left(t_\udec,\vec{x}_\ue\right)\,.
\eeq
During inflation, we previously derived the evolution equation Eq.~(\ref{eq:eofmPhi}) for $\Phi$ from the Einstein equations. For the purposes of quantization, $\Phi$ was replaced by the comoving curvature perturbation $\mathscr{R}$, whose inflationary power spectrum is given by Eq.~(\ref{eq:scalarspectrum}). Hence, if we find a way to propagate this spectrum from the time inflation ends at $t_\uend$ to decoupling, Eq.~(\ref{eq:Tfluct}) relates it to the observed CMB temperature fluctuations.

\subsection{Conserved Quantity}
A general consideration of the Einstein equations in a universe filled with a fluid obeying $p=w\,\rho$ (where $w$ does \emph{not} have to be constant) allows us to identify a convenient tracing variable for perturbations \cite{Martin:2004um,Burgess:2007pz,Durrer:CMB}, commonly denoted by $\zeta$, where
\beq
\zeta=\Phi+\frac{2}{3}\,\frac{\Phi'+\calH\,\Phi}{\calH (1+w)}\,.
\eeq
This can be understood as the first integral of Eq.~(\ref{eq:eofmPhi}), and one can show that $\zeta$ fulfills
\beq\label{eq:zetaprime}
\zeta'=\frac{2}{3\calH}\,\frac{1}{1+w}\,\nabla^{2}\Phi\,.
\eeq
The physical interpretation of $\zeta$ is the curvature perturbation on uniform density hypersurfaces. From Eq.~(\ref{eq:zetaprime}), two very important conclusions can be drawn: firstly, after Fourier decomposition and in cosmic time $t$, it follows from Eq.~(\ref{eq:zetaprime}) that $\dot{\zeta}_{k}\propto k^{2}/(a^{2}H^{2})$ and therefore on scales large compared to the comoving Hubble radius, $\zeta_{k}$ is conserved. Secondly, from the general gauge-invariant definition of the comoving curvature perturbation $\mathscr{R}$ \cite{Mukhanov:1990me} [which we used in its longitudinal, single field form in Eq.~(\ref{eq:intrinsic})], it can be shown that the difference $\Delta=-\zeta_{k}-\mathscr{R}_{k}$ obeys $\Delta\propto k^{2}/(a^{2}H^{2})$ as well, and hence, on large scales, they coincide, 
with $\mathscr{R}\approx-\zeta$.\\
Recall that the scalar power spectrum (\ref{eq:scalarspectrum}) is essentially given by $\left|\mathscr{R}_{k}\right|^{2}$, and approaches a constant for scales outside the Hubble radius. Therefore, the relation $\mathscr{R}\approx-\zeta$ provides the ``missing link'' to connect inflationary perturbations to a universe where the equation of state parameter is $w$. In particular, this renders the calculation of spectra independent of our microphysical understanding (or lack thereof) of reheating: we can simply model the transfer of energy from the inflaton field to relativistic particles by an effective (time-dependent) equation of state $w_\ueff$ which smoothly interpolates between the end of inflation (with an energy density $\rho_\uend$) and the radiation dominated epoch where $w_\urad=1/3$. Finally, we can use Eq.~(\ref{eq:Tfluct}) to establish the connection to today's temperature fluctuations in the CMB.

\subsection{Multipole Moments}
From an observational point of view, we measure the relative excess in temperature\footnote{We consider only the scalar contribution to $\delta T/T$ in the following [see Eq.~(\ref{eq:dTcontributions})], but drop the index ``S''.} $\delta T\left(\vec{e}\right)$ of a photon arriving from a direction $\vec{e}$ in the sky with respect to the average temperature of all CMB photons $T$.  On the celestial sphere, these may be decomposed into spherical harmonics as
\beq
\frac{\delta T(\vec{e})}{T}=\sum_{\ell=2}^{\infty}\sum_{m=-\ell}^{\ell}a_{\ell m}\,Y_{\ell m}(\theta,\phi)\,.
\eeq
While the temperature $T$ by itself is not an operator, the quantity $\left(\delta T/T\right)$ is directly related to the (quantized) Bardeen potentials, therefore the $a_{lm}$ in this expression should be considered as operators. Calculating their two-point correlation function for two directions $\vec{e}_{1}$ and $\vec{e}_{2}$ then corresponds to calculating a vacuum expectation value,
\beq\label{eq:vac-dT}
\left\langle\Omega\left|\frac{\delta T(\vec{e}_{1})}{T}\,\frac{\delta T(\vec{e}_{2})}{T}\right|\Omega\right\rangle=\frac{1}{4\pi}\sum_{\ell=2}^{\infty}(2\ell+1)\,C_{\ell}\,P_{\ell}(\cos\theta)\,,
\eeq
where the $P_{\ell}$ denote Legendre polynomials and $\theta$ is the angle between $\vec{e}_{1}$ and $\vec{e}_{2}$. Moreover, in Eq.~(\ref{eq:vac-dT}) the multipole moments $C_{\ell}$, which are independent of the index $m$, are used\footnote{Note that the dipole component of $\delta T/T$ due to the proper motion of our galaxy has been separated off.}, where
\beq
C_{\ell}\equiv\frac{1}{2\ell+1}\sum_{m=-\ell}^{\ell}\left|a_{\ell m}\right|^{2}\,.
\eeq
In statistical terminology, Eq.~(\ref{eq:vac-dT}) denotes an \emph{ensemble average}. 
Since we can only perform measurements in \emph{one} realization of the Universe, the resulting $C_{\ell}$ are plagued by the so-called ``cosmic variance'',
\beq\label{eq:cosmicvariance}
\frac{\delta C_{\ell}}{C_{\ell}}=\frac{1}{\sqrt{2\ell+1}}\,,
\eeq
which is small for large $\ell$ (\ie on small scales, for which we can take many measurement on our one celestial sphere), but important for the lowest multipole moments. Therefore, after choosing the inflaton potential $V(\phi)$ and calculating the corresponding primordial tensor and scalar power spectra $\calP_\uT(k)$ and $\calP_\uS(k)$ numerically, these spectra are propagated through the reheating phase with an effective equation of state parameter $w_\ueff$. At the end of reheating, they are carried through the radiation and part of the matter dominated epoch and translated into temperature anisotropies $\delta T/T$ of the CMB photons via the Sachs Wolfe effect. A present-day measurement of $\delta T/T$ is then expanded over the celestial sphere to obtain the multipole moments $C_{\ell}$. The largest observed scale corresponds to $C_{2}$ ($180^{\circ}$ separation of the photon arrival directions on the sky), while the smallest scale (hence largest moment $\ell$) is determined by the resolution of the experiment.

\subsection{Polarisation of the CMB}
At this point, let us make a few comments about polarization of the CMB, \ie an observable ``vector field'' on the celestial sphere. Prior to decoupling, the frequently scattering photons in the Universe are unpolarized. The process of recombination, however, is not instantaneous, therefore a quadrupole anisotropy is gradually produced by both scalar and tensor perturbations \cite{Mukhanov:2005sc,Durrer:CMB}. In the presence of a quadrupole, Thomson scattering of photons off electrons then leads to linearly polarized radiation on the scales of multipoles with $\ell\geq100$, or about $\sim1^{\circ}$ on the celestial sphere, since this is the Hubble size at the time of recombination. (Later in the history of the Universe, during the epoch of reionization, some more Thomson scattering between free electrons and photons can occur.)\\
On top of their temperature $T$, polarized CMB photons require two additional parameters to describe them, typically called the $E$ and $B$ modes. (If circular polarization were also present, a third parameter would be required, but circular polarization is not excited by Thomson scattering.) Given that the cross-correlations $\mean{TB}$ and $\mean{EB}$ are forbidden by parity invariance \cite{Durrer:CMB}, there are in principle four measurable spectra, $\mean{TT}, \mean{EE}, \mean{BB}$ and $\mean{TE}$. Above, we have only considered the $\mean{TT}$ spectrum, of which the WMAP satellite has provided a measurement of unprecedented accuracy (see below). Though not designed to observe polarization, WMAP also put upper bounds on the $\mean{TE}$ component. From the Planck satellite, a more detailed polarization map of the sky is expected, and notably one may hope for a detection of the $B$ mode: since the $\mean{BB}$ spectrum is only sourced from tensor (but not from scalar) perturbations, this would correspond to observational evidence for primordial gravitational waves.

\subsection{Experimental Results}\label{subsec:WMAP5}
\begin{figure}[t]
\begin{center}
\includegraphics[width=\textwidth]{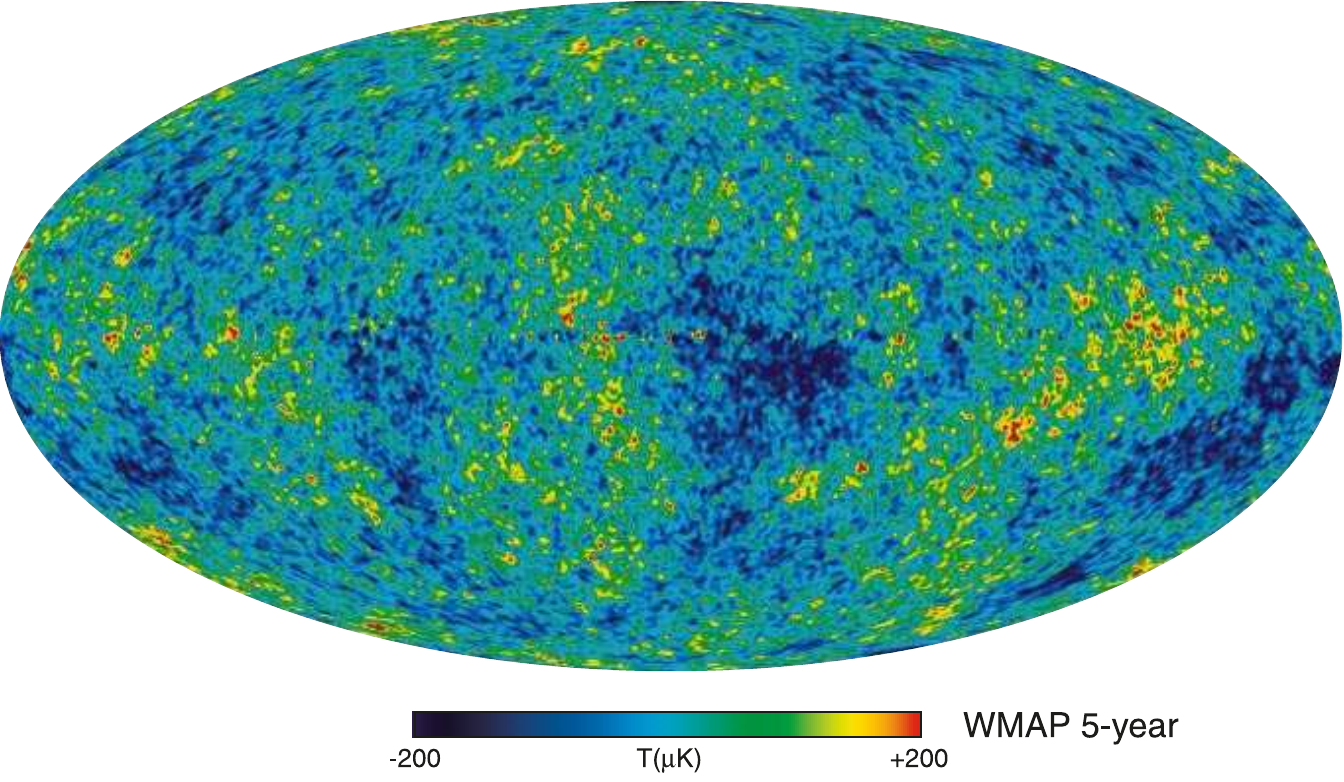}
\caption[WMAP5 map of CMB temperature fluctuations. (Source: \cite{Hinshaw:2008kr})]{\small Foreground-reduced map (with the galactic plane removed) of CMB temperature fluctuations in the sky based on the five-year results of the WMAP satellite. (Figure from \cite{Hinshaw:2008kr}).}
\label{fig:wmap5-map}
\end{center}
\end{figure}

\begin{figure}[t]
\begin{center}
\includegraphics[width=\textwidth]{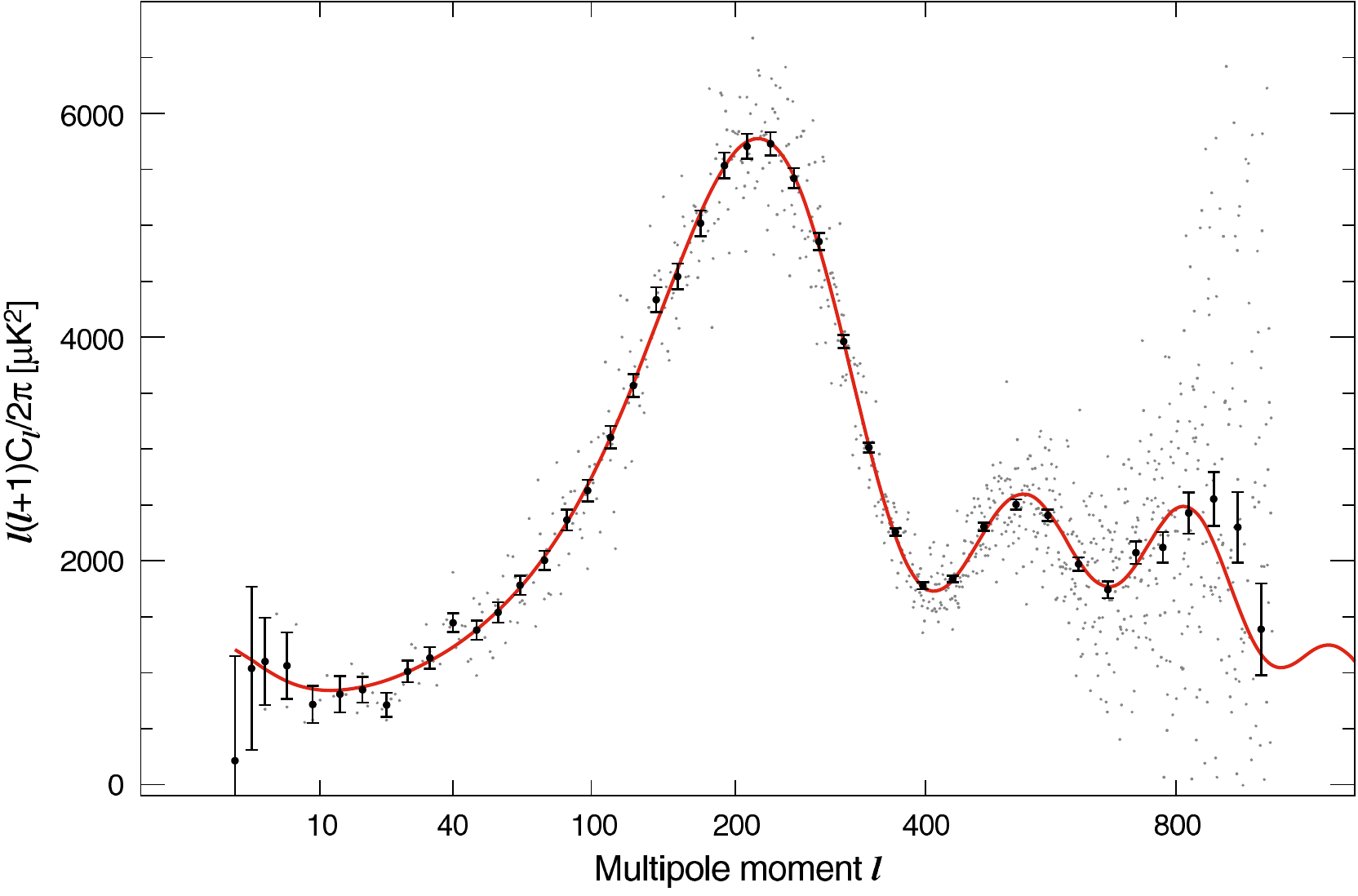}
\caption[WMAP5 plot of multipole moments $C_{\ell}$. (Source: \cite{Dunkley:2008ie})]{\small Temperature angular power spectrum obtained from WMAP5. Grey dots are unbinned data, while the black points are the binned data with $1\sigma$ error bars calculated from both noise and cosmic variance. (At low multipoles, cosmic variance is largely dominant.) The solid red line is the WMAP-only best fit \Lcdm\, model with the parameters given in the text. (Figure from \cite{Dunkley:2008ie})}
\label{fig:wmap5-Cls-dunkley}
\end{center}
\end{figure}

The most recent measurement of the CMB radiation was performed by the ``Wilkinson Microwave Anisotropy Probe'' satellite, which released the results of five years of data taking in March 2008 \cite{Dunkley:2008ie,Gold:2008kp,Hill:2008hx,Hinshaw:2008kr,Komatsu:2008hk,Nolta:2008ih,Wright:2008ib}. Their best map of differential temperature fluctuations 
in the sky after removal of foregrounds\footnote{The largest contaminant in the foreground are point sources. Other effects that can generate or amend anisotropies are the \emph{Sunyaev Zel\textquoteright dovic effect} and gravitational lensing.} and instrumentation noise, as well as having the galactic plane subtracted, is shown in Fig.~\ref{fig:wmap5-map}. The latest data release confirmed and refined the standard ``\Lcdm\, plus inflation'' model of the Universe that was established using the one- and three-year results \cite{Spergel:2003cb,Spergel:2006hy}. For WMAP, the main scientific tool for extracting cosmological information from maps such as the one shown in Fig.~\ref{fig:wmap5-map} is the temperature two-point correlation power spectrum $\mean{TT}$ plotted in terms of its (conveniently normalized) multipole moments $\ell(\ell+1)C_{\ell}/2\pi$ over $\ell$ (see Fig.~\ref{fig:wmap5-Cls-dunkley}). This plot shows two distinct features.

\paragraph{Sachs Wolfe plateau} For small values of $\ell$ in the range of roughly $2\leq l\leq20$, the amplitude of the $C_{\ell}$'s becomes almost independent of $\ell$. This is called the \emph{Sachs Wolfe plateau}. (Note also that at small $\ell$ the effects of cosmic variance as calculated from Eq.~(\ref{eq:cosmicvariance}) are largely dominant over any measurement errors.) Information about the amplitude and spectral index of the primordial inflationary spectra is encoded in the height and slope of this plateau because its $\ell$ correspond to scales $>1^{\circ}$ that were still outside the Hubble radius (and therefore ``frozen in'') at the time of decoupling. This part of the spectrum plotted in Fig.~\ref{fig:wmap5-Cls-dunkley} is therefore not very sensitive to other cosmological parameters (except the cosmological constant $\Lambda$).

\paragraph{Acoustic peaks} After the Sachs Wolfe plateau, the $C_{\ell}$ amplitude rises towards the first of a series of ``acoustic peaks'' due to baryon acoustic oscillations in the plasma at the time of last scattering. These $\ell$ correspond to scales inside the Hubble radius (and therefore dependent on cosmological parameters) when the CMB was released. After its Hubble re-entry, a given perturbation scale in the tightly coupled plasma of baryons and photons develops a standing wave oscillation of the baryon density (\ie a standing acoustic wave). At decoupling, this acoustic oscillation ends because no more Thomson scattering takes place, cutting off the pressure support for the density oscillation. The peaks and troughs in the CMB spectrum (see Fig.~\ref{fig:wmap5-Cls-dunkley}) capture the oscillations just before they came to an end: the first acoustic peak, for example, corresponds to the comoving wavelength just entering the Hubble radius, which had enough time to perform only one oscillation and therefore was at its maximum amplitude (\ie not damped yet) at decoupling. The finite thickness of the last scattering surface eventually causes a cutoff in the peak spectrum due to damping out of the anisotropies around $l\geq 800$: this effect is called \emph{Silk damping} and describes the fact that coupling of photons to the baryon and electron plasma is still present at these scales but is no longer perfect.

This structure in itself --a plateau followed by a sequence of peaks-- confirms that the spectrum is predominantly scale invariant and adiabatic. As stated earlier, the height and slope of the Sachs Wolfe plateau can be interpreted in terms of the amplitude and spectral index of the primordial perturbation spectra. On the other hand, the position and relative height of the peaks provide a wealth of information about the early Universe, and the best-fit theory spectrum corresponding to the red line in Fig.~\ref{fig:wmap5-Cls-dunkley} found from WMAP observations only is characterized by the parameter set $\{\OmegaB h^{2},\OmegaM h^{2},\Delta_{\mathscr{R}}^{2},\nS,\tau,H_{0}\}=\{0.0227,0.131,2.41,0.961,0.089,72.4\}$ \cite{Dunkley:2008ie,Nolta:2008ih}. As before, $h$ is the reduced Hubble parameter, related to $H_{0}$ by Eq.~(\ref{eq:Hubble-today}), $\OmegaB$ and $\OmegaM$ are the baryon and matter densities, respectively, and $\nS$ the spectral index of the scalar perturbations. The parameter $\tau$ is the so-called reionization depth (where the optical depth to an event taking place at redshift $z$ is the scattering probability of a photon integrated from $z$ until today) \cite{Durrer:CMB}. The amplitude of the scalar perturbation spectrum at leading order here is denoted by $\Delta_{\mathscr{R}}^{2}=k^{3}\calP_\uS/(2\pi^{2})$. This minimal set of six \Lcdm\, parameters characterizes a flat Universe dominated by a cosmological constant with adiabatic and nearly scale invariant Gaussian fluctuations. The sensitivity of the peaks and troughs to various cosmological parameter is discussed in detail \eg in \cite{Mukhanov:2005sc,Durrer:CMB}, and many details on how the WMAP5 values are obtained are given in \cite{Dunkley:2008ie,Komatsu:2008hk}. These references also show how more stringent bounds can be placed on the parameters, if more datasets than just the WMAP results are included.

\begin{figure}[t]
\begin{center}
\includegraphics[width=0.3\textwidth]{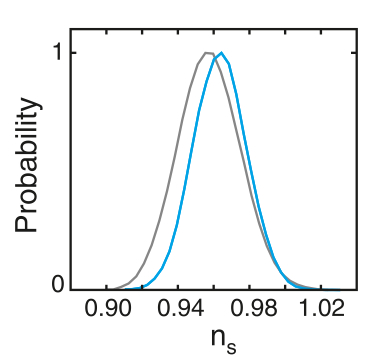}
\hfill
\includegraphics[width=0.6\textwidth]{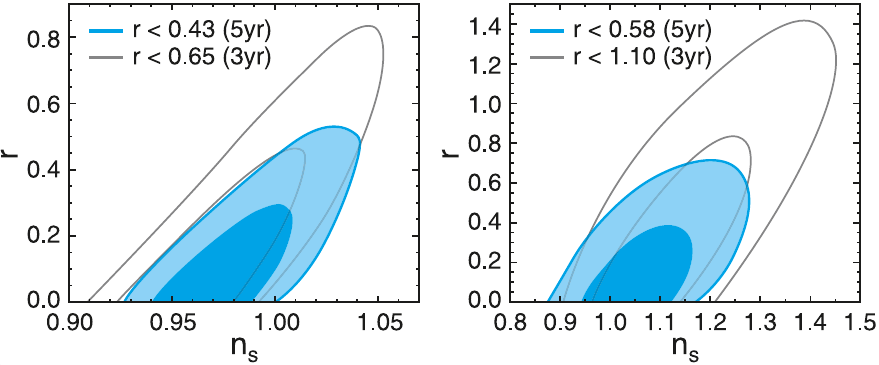}
\caption[One-dimensional probability distribution for the scalar spectral index $\nS$ and two-dimensional constraints on $r$ and $n_\uS$ obtained from WMAP5. (Source: \cite{Dunkley:2008ie})]{\small \emph{Left:} WMAP5 constraint (blue) on the scalar spectral index $\nS$ from the marginalized one-dimensional probability distribution.  The grey line is the previous result from WMAP3. \emph{Center and right:} Two-dimensional marginalized constraints on $r$ and $\nS$ obtained from WMAP5. The darker shaded region is the 95\% confidence level region, the lighter shading shows 68\% confidence level. (The grey lines show the previous regions from WMAP3.) The pivot scale here is chosen at $k_{*}=0.002/\Mpc$. In the central plot, no running $\alpha_\uS$ of the spectral index is assumed, while the right plot allows for a non-zero $\alpha_\uS$. (Figures from \cite{Dunkley:2008ie})}
\label{fig:wmap5-proba-ns}
\end{center}
\end{figure}

\begin{figure}[t]
\begin{center}
\includegraphics[width=0.45\textwidth]{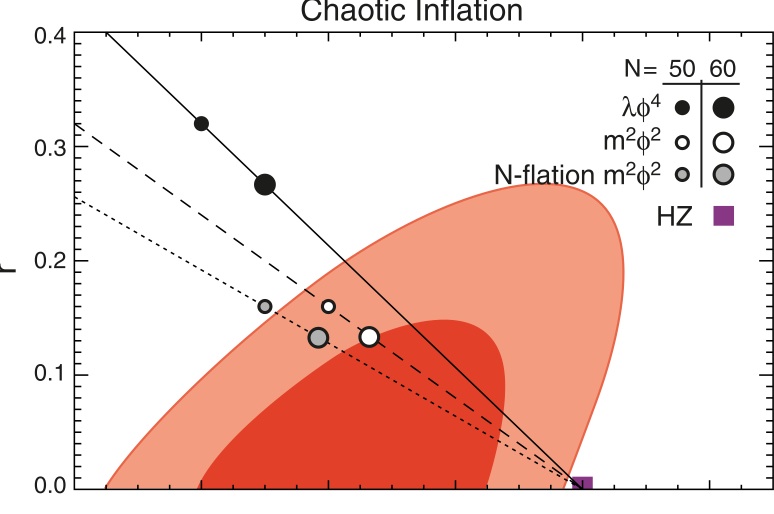}
\hfill
\includegraphics[width=0.45\textwidth]{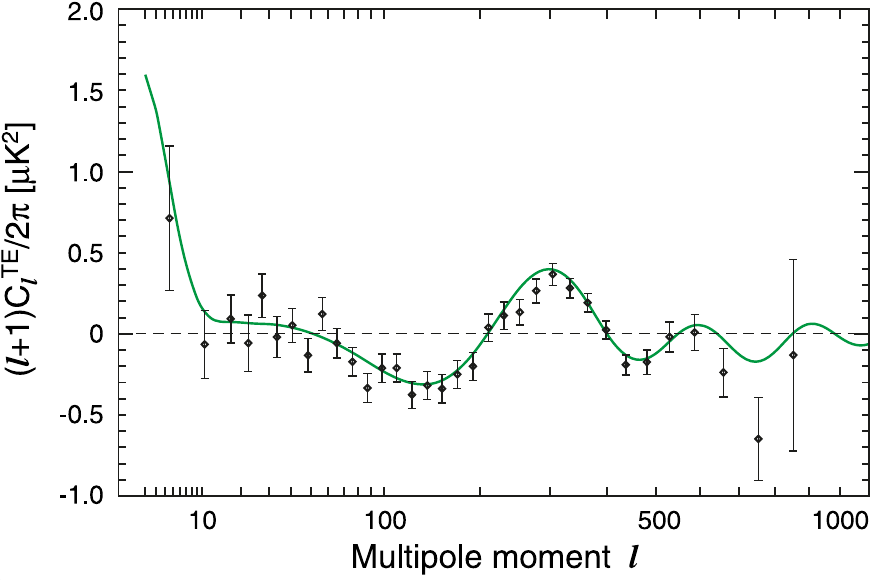}
\caption[Chaotic inflation models compared to WMAP5 measurements of $r$ and $\nS$. (Source: \cite{Komatsu:2008hk}) TE spectrum from WMAP5. (Source: \cite{Nolta:2008ih})]{\small \emph{Left:} Localization of various inflaton potentials of the chaotic type in the $(\nS,r)$ parameter plane (see legend for the identification of the symbols). The horizontal axis is the spectral index $\nS$, with a scaling as in Fig.~\ref{fig:wmap5-proba-ns}. The darker shaded region is the WMAP5 95\% confidence level contour, while the lighter shading indicates the 68\% confidence level region. Note that the simple $V(\phi)=(m^{2}/2)\phi^{2}$ model (white circles) still fits the data very well, while the quartic potential $\propto\phi^{4}$ is well outside the 68\% contour. (Figure from \cite{Komatsu:2008hk}) \emph{Right:} Though not originally designed as a polarimeter, WMAP has been able to measure the cross polarization spectrum $\mean{TE}$, plotted here in terms of its multipole moments $C_{\ell}^{\mathrm{TE}}$. Again, a peak-like structure is obtained; note that the vertical axis here is $(l+1)C_{\ell}^{\mathrm{TE}}/(2\pi)$. (Figure from \cite{Nolta:2008ih})}
\label{fig:wmap5-chaoticinf}
\end{center}
\end{figure}

Tensor perturbations in principle show the same structure in the multipole moments $C_{\ell}$ \cite{Durrer:CMB}, but their amplitude falls off rapidly for $\ell\geq 60$ and they are therefore largely subdominant in the region of the acoustic peaks. They do, however, contribute to the amplitude of the Sachs Wolfe plateau, \ie they add power to the spectrum on smaller scales $\ell<60$. Therefore, their presence is degenerate with a red-tilted spectrum $n_\uS<1$, equally leading to an excess of power for small $\ell$. As an example, the likelihood distribution obtained for the scalar spectral index is plotted in Fig.~\ref{fig:wmap5-proba-ns}, and its degeneracy with the tensor to scalar ratio $r$ (which, in single field inflation, is $r=-8\epsilon_{1}$) is shown. The quantities $r$ and $\nS$ can be calculated numerically for a given inflationary model, and one may characterize an inflaton potential $V(\phi)$ by its ``coordinates'' $(\nS,r)$ in parameter space. For models of the large field type, this is done in Fig.~\ref{fig:wmap5-chaoticinf} (see the caption of that figure for a discussion). Similar plots for other generic classes of models using the WMAP5 data can be found \eg in \cite{Komatsu:2008hk,Alabidi:2008ej}.

\begin{quotation}
{\bf Example: scalar spectrum normalization in chaotic inflation}\\
For simple models such as the chaotic potential $V(\phi)=(m^{2}/2)\phi^{2}$ with only one free parameter, \ie the mass $m$, one can use the slow roll approximation to establish the link with observations ``by hand'' via the WMAP quadrupole $Q_{\mathrm{rms-PS}}/T$, since $k^{3}\calP_\uS\propto \left(Q_{\mathrm{rms-PS}}/T\right)^{2}$. Taking into account numerical prefactors [like \eg the factor $1/3$ in Eq.~(\ref{eq:Tfluct})], using the slow roll spectral amplitude (\ref{eq:scalarspectrum}) at leading order and the Hubble parameter from Eq.~(\ref{eq:H-approx}), one can establish that \cite{Martin:2006rs}
\beq
\qquad\frac{V_{k}}{\mpl^{4}}\simeq\frac{45\epsilon_{1}^{(k)}}{2}\,\QQoverTT\,,
\eeq
where $V_{k}$ and $\epsilon_{1}^{(k)}$ are the potential and the first parameter evaluated when the observable scale $k$ left the Hubbe radius. Using the slow roll trajectory of Eq.~(\ref{eq:largefieldtrajectory}) for $n=2$, one finds
\beq
\frac{m^{2}}{\mpl^{2}}\simeq\frac{45\pi}{\left(\Delta N_{k}+1\right)^{2}}\,\QQoverTT\,.
\eeq
With the measured value $Q_{\mathrm{rms-PS}}/T\simeq6\times10^{-6}$ \cite{Hinshaw:2008kr} and for $\Delta N_{k}=50$, this determines the mass of the chaotic inflaton as $m/\mpl\simeq1.4\times10^{-6}$.
\end{quotation}

\section{Inflationary Predictions}\label{eq:infl-predictions}
Above, we explored the consequences of an early inflationary era in the Universe at both the background and the perturbation level. We argued that while the microphysics at play during inflation and at the transition to the radiation dominated epoch (reheating) are complicated and largely unknown, they can be expressed in terms of a few generic parameters, which allow one to broadly distinguish different classes of scenarios. At the top level, we may isolate some cast-iron predictions of inflation independently of its concrete realizations.

\textsc{Spatially flat homogeneous and isotropic Universe}\\
Inflation dilutes away all previous classical inhomogeneities in the observable patch, \ie ``our'' Hubble volume. Therefore, the observable Universe must be described by a FLRW metric. Though inflation only last for a tiny fraction of a second, the violent exponential expansion renders the Universe so close to spatial flatness that all of SBBM evolution cannot push it away again.

\textsc{Almost scale invariant scalar and tensor power spectrum}\\
The spectrum of both scalar and tensor perturbations may be calculated from inflation, and is found to be very close to the scale invariant Harrison Zel\textquoteright{}dovich case. 
The scalar perturbations couple to matter and are hence observable as temperature fluctuations in the CMB. Tensor perturbations (decoupled from matter at linear order) lead to a background of gravitational waves, whose amplitude is suppressed and therefore difficult to detect. If it were observed, however, it would allow a direct determination of the energy scale of inflation.

\textsc{Almost Gaussian perturbations}\\
The statistics of the perturbations are, at linear order, unaffected by the expansion that occurs after their generation. Since their spectra are calculated from expectation values taken in the vacuum $\ket{\Omega}$, inflationary perturbations obey Gaussian statistics, while small deviations from Gaussianity may be induced \eg by features in the inflaton potential, non-canonical dynamics and backreaction.

\textsc{Adiabatic, coherent perturbations}\\
Inflation (with a single scalar field) predicts that the perturbations are adiabatic\footnote{We discuss multifield inflation and perturbations of isocurvature type in Chapter \ref{chapter:ext-alt}.} and generated in a coherent manner. Though a lot of complicated particle physics enters into the calculation of the observable CMB spectra from the primordial inflationary perturbations, coherence is important to obtain the ``acoustic peak'' features of the CMB multipole moments $C_{\ell}$. This is a crucial difference between perturbations generated by a scalar field like the inflaton, and those from residual defects such as cosmic strings because in the latter case, coherence is absent and hence no peak structure in the CMB produced.

Observations so far lend ample justification to the hypothesis of an early inflationary phase. Since the $\epsilon_{i}$ constrained from CMB measurements are a calculational output of a concrete scenario of inflation [see Eq.~(\ref{eq:sr-eps})], they put limits on the shape of the inflaton field's interactions. To some extent, this is already possible with today's WMAP5 data (see Section \ref{subsec:WMAP5}), and the Planck mission will provide much more stringent bounds. Therefore, if we can derive  these interactions from a theoretical framework, as we discuss in the following Chapter, observations may be able to discriminate between them.

\section{Beyond the Standard Treatment}
In this Section, we discuss two aspects of inflation related to physics at the smallest and the largest scales, respectively. From a conceptional point of view, these issues have to be addressed within the inflationary framework, though their consequences may not be accessible to us experimentally.

\subsection{Physics of the Very Small: Trans-Planckian Effects}
We previously mentioned that the minimum amount of inflation is bounded by SBBM problem solutions. Most models, however, predict a number of \efolds much bigger than this requirement. In terms of the mode labels $k$ we introduced when studying perturbations, the number of \efolds measures the ``stretching'' of a given $k$ (or of the wavelength $\lambda$, respectively), since the \emph{physical} length associated with it is proportional to the scale factor, $\lambda_\uphys=a\,\lambda$. Scales observed today were smaller in the past and one finds that, if the amount of inflationary expansion is large, certain modes $k$ should have originated below the Planck scale $\ell_{\uPl}$. We know, however, that at the Planck scale at the very latest we must touch onto the realm of new physics.\\
The issue of trans-Planckian physics in inflation\footnote{Previously, similar questions had been studied in the context of black hole radiation \cite{Unruh:1980cg,broutetal,Corley:1997ef}: following a black hole photon back into its past closer and closer to the event horizon, its wavelength undergoes a blueshift. However, for the type of dispersion relations studied, it was found that any trace of high energy physics is erased thermodynamically from the observed photon spectrum of a black hole.} was first raised in \cite{Brandenberger:1999sw,Martin:2000xs} and one may study it from two different vantage points: on the one hand, one must hope that the trans-Planckian past of observable scales does not affect the predictions of inflation too severely because these predictions were derived using General Relativity and quantum field theory, both of which must be cast into doubt close to the Planck scale. In this sense, one is interested in the robustness of inflation's predictions to trans-Planckian effects \cite{Brandenberger:2000wr,Starobinsky:2001kn}. On the other hand, the prospect that, via the exponential expansion of the Universe, these high energy scales may lie within the grasp of observations is tantalizing. Inflation can then be understood as a ``Planck scale microscope'', and the more we know about physics close to the Planck scale, the more concrete predictions we can test using precision cosmological measurements.

Inspired by the strategy pursued for a similar black hole problem, it was argued that the dispersion relation in the perturbations' equation of motion Eq.~(\ref{eq:modeequation}) may be modified in order to mimic trans-Planckian effects \cite{niemeyer01,niemeyerparentani01,brandenbergermartin02,lemoineetal}. This is reminiscent of condensed matter physics where, at wavelengths comparable to the lattice spacing in crystals, the dispersion relation departs from the linear relation \(\omega_{k}\propto{}k\). Hence, in the inflationary context, if a mode's wavelength is comparable to the ``spacing of spacetime'' at the Planck scale, it may experience effects of this discretization. Therefore, at short distances, the time-dependent frequency of the harmonic oscillator of Eq.~(\ref{eq:modeequation}) changes, and one replaces $k$ by an effective wavenumber $k_\ueff$.\\
This modification, however, is engineered by hand and hence lacks fundamental justification. Starting from first principles assumptions about a theory of quantum gravity, one may therefore take a different approach \cite{kempf00,kempfniemeyer01,greeneetal0104,greeneetal0110,Kempf:2006wp}: a finite minimum length (corresponding to the discretization length of spacetime) is introduced via quantum gravity correction terms to the commutation relations of quantized perturbation theory. Such corrections arise from a high energy modified Heisenberg uncertainty relation, $\Delta x\Delta p\geq \frac{1}{2}\left[1+\beta(\Delta p)^{2}+\dots\right]$. Here \(\beta\) is a positive constant related to the ``spacetime spacing''  $\Delta x_{\mathrm{min}}$ by $\Delta x_{\mathrm{min}}=\sqrt{\beta}$.\\
Independent of the framework chosen to describe them, once a scale $k$ has grown sufficiently for its mode function to reach the standard form of Eq.~(\ref{eq:modeequation}), the trans-Planckian effects manifest themselves as a vaccum choice different from the usual Eq.~(\ref{eq:vacuumchoice}). Traces of the other (decaying) branch of the two-dimensional solution space of Eq.~(\ref{eq:modeequation}) should persist, meaning that the mode is not in the perfectly adiabatic vacuum in the far past. In the CMB spectrum, this non-adiabaticity of the vacuum takes the form of super-imposed oscillations, which are already tightly constrained \cite{Martin:2003sg,Martin:2004yi}.

\subsection{Physics of the Very Large: Eternal Inflation}\label{subsec:eternalinf}
\begin{figure}[t]
\begin{center}
\includegraphics[width=0.4\textwidth]{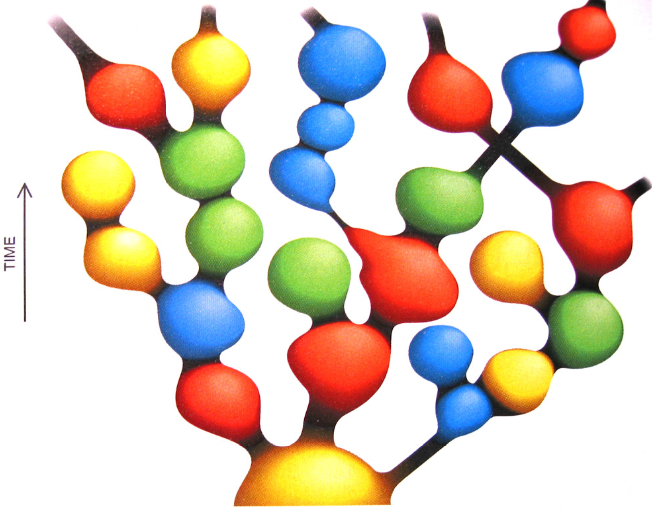}
\hfill
\includegraphics[width=0.5\textwidth]{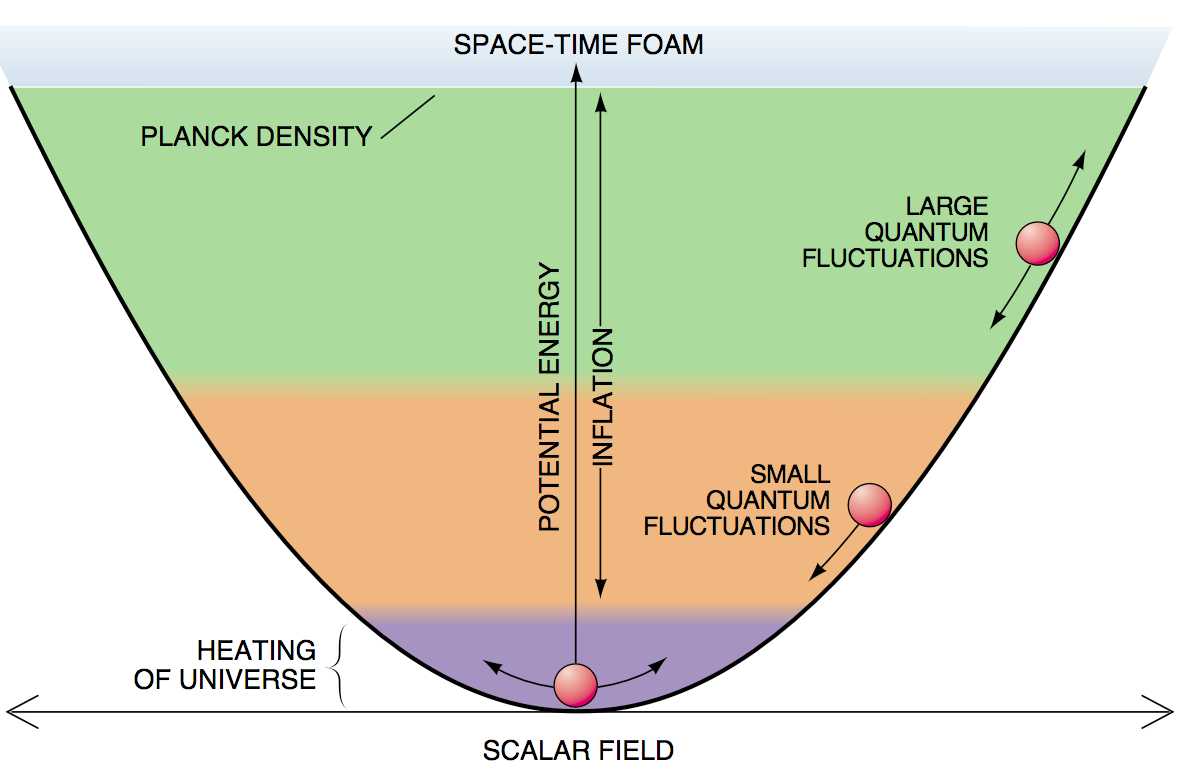}
\caption[Eternally inflating self-reproducing Universe and chaotic inflation potential with the regions or large and small quantum fluctuations. (Source: \cite{Linde:1994eq})]{\small \emph{Left:} The eternally inflating self-reproducing Universe with its inhomogeneous structure of separate ``Hubble bubbles'' on very large scales. The different colors of the bubbles indicate different realizations of the laws of physics, arising from the initial field value in the bubble. Whenever the field fluctuates to a value where the Universe can start inflating in the slow roll regime, a new bubble begins to form. \emph{Right:} Sketch of a typical large field (``chaotic'') inflation potential with the regions of small and large quantum fluctuations. While the field value is big enough, the field can move upwards on the potential through quantum jumps, in the opposite direction than its slow roll classical trajectory dictates. (Figures from \cite{Linde:1994eq})}
\label{fig:eternalinf}
\end{center}
\end{figure}

The observation of homogeneity and isotropy on scales $\order{\sim100\Mpc}$ justifies the use of the FLRW metric to describe our observable patch of the Universe, which is $\order{\sim3000\Mpc}$. However, we have no means of knowing if the Universe remains homogeneous and isotropic on even larger scales: it is possible that globally, spacetime is highly inhomogeneous and made of distinct ``Hubble bubbles'', each corresponding to a separate observational patch with possibly different laws of physics (see Fig.~\ref{fig:eternalinf}) \cite{Vilenkin:1983xq,Linde:1986fd}. This immediately raises new questions: how did this global structure arise, and how can we develop a measure of probability to account for it, given that we are trapped within our own bubble?

Quantum fluctuations of the inflaton field $\delta\phi$ were previously treated as small corrections to the homogeneous background $\phi_{0}$. However, the typical size of quantum fluctuations for a massless free scalar field in a de Sitter universe is given by 
$\delta\phi\simeq H/2\pi$, and hence $\delta\phi$ depends on the value of the Hubble parameter, which in turn is a function of the background scalar field value $\phi_{0}$. During slow roll inflation, $H$ is given by the potential only, see Eq.~(\ref{eq:H-approx}). Therefore, depending on where $\phi_{0}$ is located on its potential, quantum fluctuations can have different amplitude. For illustration, let us assume a potential of large field type, \ie the inflaton starts at a large field value $\phi_\uin$ (in Planck units) and rolls towards smaller ones. Therefore, the classical field motion described by Eq.~(\ref{eq:dphi-approx}) is ``to the left'' (see Fig.~\ref{fig:inflation-bg}), and during a time interval $\Delta t$, the field typically moves by [see Eq.~(\ref{eq:dphi-approx})]
\beq
\Delta\phi=-\frac{V'}{3H}\,\Delta t=-\frac{V'}{3H^{2}}\,,
\eeq
where we used one Hubble time $H^{-1}$ for the time interval $\Delta t$ along with the slow roll expression Eq.~(\ref{eq:H-approx}). This classical movement ``to the left'' has to be compared with the quantum jumps $\delta\phi\simeq H/2\pi$ that occur randomly to the right and to the left. Classical and quantum fluctuations are of equal amplitude if
\beq
\abs{\Delta \phi}\simeq\abs{\delta\phi},\qquad\textnormal{hence}\quad\abs{-\frac{V'}{3H^{2}}}\simeq\abs{\frac{H}{2\pi}}\,,
\eeq
Using Eq.~(\ref{eq:largefield}) and Eq.~(\ref{eq:H-approx}), we find that for the large field case this holds true for
\beq\label{eq:phistar}
\frac{\phi_{*}}{\mpl}\simeq\left[\left(\frac{3}{8\pi}\right)^{1/2}\frac{n}{4}\left(\frac{\mpl}{\Lambda_{\mathrm{lf}}}\right)^{2}\right]^{2/(n+2)}\,.
\eeq
Therefore, in the region where $\phi\geq\phi_{*}$, quantum effects can exceed the amount of classical motion occurring during one Hubble time, \ie $\phi$ can largely compensate its slow rolling downwards  Eq.~(\ref{eq:dphi-approx}) by a quantum jump upwards on the potential. Extrapolating this argument, we see that at the next step (given that it is still in the regime $\phi\geq\phi_{*}$), the inflaton again has a chance of moving upwards instead of rolling down \etc It could therefore forever stay in the region where quantum effects are dominant. Only once a quantum jump in the same direction as classical motion allows the field to reach values $\phi<\phi_{*}$ can the slow roll evolution take over \cite{Vilenkin:1983xq,Linde:1986fd}.\\
Recall that we previously also determined the scale $\phi_{\mathrm{qg}}$ [see Eq.~(\ref{eq:Ntot-largefield})] where quantum gravity effects become important because $V(\phi_{\mathrm{qg}})\simeq\mpl^{4}$. For large field potentials of the form Eq.~(\ref{eq:largefield}), it is related to $\phi_{*}$ by
\beq\label{eq:phistar-phiqg}
\frac{\phi_{\mathrm{qg}}}{\mpl}=\left(\frac{8\pi}{3}\right)^{1/n}\left(\frac{4}{n}\right)^{2/n}\left(\frac{\phi_{*}}{\mpl}\right)^{(n+2)/2}\,,
\eeq
\ie there exists a regime where quantum jumps $\delta\phi$ are important, but still $\phi<\phi_{\mathrm{qg}}$.\\
Classically, the field value in large field inflation continuously decreases from $\phi_\uin$, which is bounded from below by the requirement of $N_{\mathrm{tot}}=40-60$, but can be \emph{much} larger (as long as the regime of quantum gravity is avoided). One may consider the region inside each Hubble radius $H^{-1}$ as a separate universe [one of the bubbles in Fig.~(\ref{fig:eternalinf})] since it is causally disconnected from the rest of spacetime (which can comprise many other $H^{-1}$ size regions). We can therefore think of the initial field value $\phi_\uin$ as a number ``pulled out of a hat'' (with a certain probablity) for every Hubble volume separately: those regions where $\phi_\uin$ is at least large enough to create $N_{\mathrm{tot}}$ \efolds of inflation are \apriori candidates for our part of the Universe. However, if $\phi_\uin$ is $\order{\phi_{*}}$ as calculated in Eq.~(\ref{eq:phistar}), the field may undergo an erratic sequence of jumps $\delta\phi$ first before its value has dropped enough to start inflating according to Eq.~(\ref{eq:dphi-approx}). In this context, one may speak of an ``eternally self-reproducing inflationary universe'' \cite{Linde:1986fd}, since each region where inflation does set in immediately produces a volume large enough to fit our own Hubble bubble, and continuous quantum jumps ensure that there are always regions left in which $\phi$ still exceeds $\phi_{*}$. The consequences of this are the subject of the ``stochastic inflation'' approach \cite{Starobinsky:1986fx,Starobinsky:1994bd,
Martin:2005ir,Kuhnel:2008yk,Finelli:2008zg,Kuhnel:2008wr}.

\begin{quotation}
{\bf Example: stochastic effects in chaotic inflation}\\
For $n=2$, one can write the potential as a simple mass term  $V(\phi)=(m^{2}/2)\,\phi^{2}$. Therefore, we have from Eqs.~(\ref{eq:phistar}) and (\ref{eq:phistar-phiqg}) that
\beq
\frac{\phi_{*}}{\mpl}=\left(\frac{3}{16\pi}\right)^{1/4}\left(\frac{\mpl}{m}\right)^{1/2}\approx10^{3},\qquad\frac{\phi_{\mathrm{qg}}}{\mpl}=\left(\frac{32\pi}{3}\right)^{1/2}\left(\frac{\phi_{*}}{\mpl}\right)^{2}\approx 10^{6}\,,
\eeq
where the numerical values are for $m/\mpl\approx 10^{-6}$, a typical value found from normalizing perturbation spectra to the CMB, see Section \ref{subsec:WMAP5}. Recall that the required minimum initial field value $\phi_\uin$ is $\order{\mpl}$ in this model, see Eq.~(\ref{eq:Ntot-largefield}). Clearly, if one were to choose $\phi_\uin\approx\order{\phi_{*}}$, the number of \efolds produced is $N_{\mathrm{tot}}\approx\order{10^{6}}$ [see Eq.~(\ref{eq:Ntot-largefield})], \ie \emph{much} larger than the required amount. 
We also see that quantum fluctuations dominate the motion of the inflaton field in chaotic inflation long before the scale of quantum gravity.
\end{quotation}

\subsubsection{Stochastic inflation: quantum fluctuations interpreted as ``noise''}
These considerations motivate the idea of ``coarse graining'' the inflaton field over the entire Universe at a scale of roughly the Hubble volume, \ie each separate $H^{-1}$ region is assigned a background field value $\phicl$ averaged over its volume (see Fig.~\ref{fig:eternalinf}), and all fluctuations on smaller scales are summarily considered as ``quantum noise''. Technically, this is achieved by dividing the field into its long and short wavelength modes and ``smearing out'' the latter ones. 
The classical field $\phicl$ (comprising only the long wavelength modes) by definition exactly obeys the slow roll equation of motion Eq.~(\ref{eq:dphi-approx}). The field $\phi$, re-interpreted as a stochastic quantity, instead obeys a \emph{Langevin equation},
\beq\label{eq:standardlangevin}
\frac{\drm \phi}{\drm t}=-\frac{2}{\kappa}\,H'+\frac{H^{3/2}}{2\pi}\,\xi(t)\,,
\eeq
where the stochastic function $\xi(t)$ describes the quantum noise produced by the short wavelength modes. By $H'$ we denote the derivative of the Hubble parameter with respect to $\phi$ (not with respect to conformal time $\eta$, as it was the case for $\calH'$ earlier). It is important to note, however, that the derivation of Eq.~(\ref{eq:standardlangevin}) is straightforward only in a de Sitter background (where the Hubble parameter is constant, $H'=0$, and $\phi$ corresponds to a test field in a fixed background) \cite{Starobinsky:1994bd}, and especially the normalization factor in front of $\xi(t)$ is obtained by considering jumps of a test field in de Sitter spacetime. It is then generalized to time-dependent backgrounds, but one must be careful because $\phi$ no longer behaves like a test field, but back-reacts on the spacetime.\\
The (Gaussian) noise is completely characterized by its statistical properties\footnote{For simplicity, we only consider the case of ``white Gaussian noise'' here. ``Colored'' noise functions are related to more complicated window functions used for separating long and short wavelength modes.}
\beq\label{eq:noiseproperties}
\mean{\xi(t)}=0,\qquad\mean{\xi(t)\,\xi(t')}=\delta(t-t')\,.
\eeq
It is well known \cite{Risken} that a Langevin equation of the general form
\beq\label{eq:Langevinform}
\frac{\drm \phi}{\drm t}=\alpha(\phi)+\beta(\phi)\,\xi(t)\,,
\eeq
where for Eq.~(\ref{eq:standardlangevin}) we have
\beq
\alpha(\phi)=-\frac{2}{\kappa}\,H'\,,\quad\beta(\phi)=\frac{H^{3/2}}{2\pi}\,,
\eeq
is related to a \emph{Fokker Planck equation} which describes the evolution of the probability distribution $P(\phi,t)$ for finding a given field value $\phi$ inside a Hubble domain at time $t$. This Fokker Planck equation is written as  \cite{Risken}
\beq\label{eq:FPform}
\frac{\partial P(\phi, t)}{\partial t}=\frac{1}{2}\,\frac{\partial^{2}\left[b(\phi) P(\phi, t)\right]}{\partial \phi^2}-\frac{\partial \left[a(\phi)P(\phi,t)\right]}{\partial \phi}\,,
\eeq
where the coefficients $a(\phi)$ and $b(\phi)$ are related to the $\alpha(\phi),\beta(\phi)$ of Eq.~(\ref{eq:Langevinform}) by
\bea
a(\phi)&=&\alpha(\phi)+\frac{1}{4}\,\frac{\drm}{\drm \phi}\,\left[\beta(\phi)\right]^{2}=-\frac{2}{\kappa}\,H'+\frac{3}{4}\,\frac{H^{2}}{(2\pi)^{2}}\,H'\,,\\
b(\phi)&=&\beta^{2}(\phi)=\frac{H^{3}}{(2\pi)^{2}}\,.
\eea
It should be noted that there is an ambiguity in defining the Fokker Planck equation for a given Langevin equation which is related to the question of whether, in the limit of continuous time, the coefficients in Eq.~(\ref{eq:Langevinform}) are evaluated before or after the quantum jump has occurred \cite{Risken}. Here, we exclusively use the \emph{Stratonovich rule} for deriving the Fokker Planck equation, which corresponds to the usual laws of differentiation. (Alternatively, one may use the \emph{It\^{o} rule}.) The inflationary Langevin equation for the stochastic field $\phi$ Eq.~(\ref{eq:standardlangevin}) hence corresponds to a Fokker Planck equation obeyed by the probability distribution $P(\phi,t)$ which reads\footnote{One may also write Eq.~(\ref{eq:standardFP}) in the form \cite{Linde:1993xx}
\beq\label{eq:LindeFP}
\frac{\partial}{\partial t}\,P(\phi,t)=\frac{1}{2}\,\frac{\partial}{\partial \phi}\left\{\frac{H^{3/2}}{2\pi}\,\frac{\partial}{\partial \phi}\left[\frac{H^{3/2}}{2\pi}\,P(\phi,t)\right]\right\}+\frac{\partial}{\partial \phi}\left[\frac{V'}{3H}\,P(\phi,t)\right]\,.
\eeq}  \cite{Linde:1993xx,Winitzki:2006rn}
\beq\label{eq:standardFP}
\frac{\partial}{\partial t}\,P(\phi,t)=\frac{1}{2}\,\frac{\partial^{2}}{\partial\phi^{2}}\left[\frac{H^{3}}{(2\pi)^{2}}\,P(\phi, t)\right]-\frac{\partial}{\partial\phi}\left\{\left[-\frac{2}{\kappa}\,H'+\frac{3}{4}\,\frac{H^{2}}{(2\pi)^{2}}\,H'\right]\,P(\phi, t)\right\}\,.
\eeq
A stationary (\ie equilibrium) solution of this equation for which $\partial P_\ueq(\phi)/\partial t=0$ would be of particular interest since the field then should be distributed according to $P_\ueq(\phi)$ at late times $t\rightarrow\infty$. From Eq.~(\ref{eq:standardFP}) or Eq.~(\ref{eq:LindeFP}), respectively, one can show that \cite{Linde:1993nz}
\beq
P_\ueq(\phi)\propto\exp\left[\frac{8\pi^{2}V(\phi)}{3H^{4}}\right]=\exp\left[\frac{3\mpl^{4}}{8V(\phi)}\right]\,,
\eeq
where Eq.~(\ref{eq:H-approx}) has been used to replace $H^{2}$.\\
Let us now consider the probability that the inflaton field takes a certain value $\phi$ at time $t$ under the condition that its value at some earlier time $t'<t$ was \eg equal to $\chi$ (in particular for $t'=t_\uin$ and $\chi=\phi_\uin$). This distribution $P(\phi,t|\chi)$ then obeys a \emph{backward Kolmogorov equation} \cite{Risken} and it can be shown that a stationary solution to both this backward equation and Eq.~(\ref{eq:standardFP}) is
\beq
P_\ueq(\phi,\chi)\propto\exp\left[\frac{3\mpl^{4}}{8V(\phi)}\right]\,\exp\left[-\frac{3\mpl^{4}}{8V(\chi)}\right]\,.
\eeq
This distribution cannot be normalized if the potential $V(\phi)$ vanishes at its minimum (which, following its classical motion, $\phi$ approaches for late times $t$). Intuitively, one can understand this as the absence of quantum jumps as soon as the field reaches $\phi<\phi_{*}$: it is then only subject to the classical force that pushes it downwards on the potential, there is no counteraction trying to increase $\phi$ again and therefore no stationary distribution can be established.\\
The way out of this puzzle leads to the realization that $P(\phi,t)$ does not account for the relative size of the Hubble domain that contains the value $\phi$ at time $t$: the larger $\phi$, the larger the Hubble parameter $H$ for a large field potential. Therefore, the \emph{physical} probability distribution $P_{\uphys}(\phi,t)$ should obey \cite{Linde:1993nz}
\beq
\frac{\drm}{\drm t}\,P_{\uphys}(\phi,t)=\frac{1}{2}\,\frac{\partial}{\partial \phi}\left[\frac{H^{3/2}}{2\pi}\,\frac{\partial}{\partial \phi}\left(\frac{H^{3/2}}{2\pi}\,P_{\uphys}(\phi,t)\right)\right]+\frac{\partial}{\partial \phi}\left(\frac{V'}{3H}\,P_{\uphys}(\phi,t)\right)+3HP_{\uphys}(\phi,t)\,,
\eeq
where the extra term accounts for the growth of the corresponding volume during a given time interval. Again, a corresponding backward Kolmogorov equation for $P_{\uphys}(\phi,t|\chi)$ holds and it can be shown that in this case there indeed exists a normalizable stationary solution.

\subsubsection{Perturbative solution in the noise}
A different approach for obtaining the probability distribution of the inflaton field $\phi$ once it has been turned into a stochastic quantity $\phi[\xi]$ consists of using the expansion \cite{Martin:2005ir,Martin:2005hb}
\beq\label{eq:noiseexpansion}
\phi(t)\simeq\phicl(t)+\delta\phi_{1}(t)+\delta\phi_{2}(t)+\dots\,,
\eeq
in the Langevin equation (\ref{eq:standardlangevin}), where $\delta\phi_{1}(t)$ is of first order in the noise $\xi(t)$, $\delta\phi_{2}(t)$ is of second order \etc. 
A consistent expansion of the coefficients in Eq.~(\ref{eq:standardlangevin}) leads to differential equations for $\delta\phi_{1}(t)$ and $\delta\phi_{2}(t)$, which can be solved in terms of integrals of $H$ and its derivatives, multiplied by one or two noise functions $\xi(t)$, respectively. The idea then is to exploit the properties of the noise (\ref{eq:noiseproperties}) to obtain the mean values $\mean{\delta\phi^{2}_{1}}$ and $\mean{\delta\phi_{2}}$ in terms of the classical field $\phicl(t)$ [which exactly solves Eq.~(\ref{eq:dphi-approx})] only.

To tie the perturbative approach of Eq.~(\ref{eq:noiseexpansion}) to the probability distribution $P(\phi,t)$ satisfying a Fokker Planck equation, 
note that the probability distribution $P(\phi,t)$ may also be defined as
\beq\label{eq:P-as-delta}
P(\phi,t)=\mean{\delta(\phi-\phi[\xi])}\,,
\eeq
where $\phi[\xi]$ here is the stochastic field that is expanded around $\phicl$ in Eq.~(\ref{eq:noiseexpansion}).
The mean $\mean{\dots}$ in Eq.~(\ref{eq:P-as-delta}) is calculated with respect to the functional probability distribution of the noise,
\beq
\mathcal{P}[\xi]=\mathcal{N}_{0}\,\exp\left[-\frac{1}{2}\,\xi^{T}{\bf C}^{-1}\xi\right]\,,
\eeq
where $\mathcal{N}_{0}$ is a normalization coefficient given by $\mathcal{N}_{0}=\left[\mathcal{D}\xi\,\exp\left(-\frac{1}{2}\,\xi^{T}{\bf C}^{-1}\xi\right)\right]^{-1}$ and we use the shorthand notation $f^{T}g\equiv\int\drm\tau\,f(\tau)g(\tau)$. One can then show that the approximation Eq.~(\ref{eq:noiseexpansion}) in the Langevin equation (\ref{eq:standardlangevin}) gives an approximate probability distribution solution $P(\phi,t)$ to the Fokker Planck equation (\ref{eq:standardFP}):
\beq\label{eq:distri_approx}
P(\phi,t)\simeq\frac{1}{\sqrt{2\pi\mean{\delta\phi_{1}^{2}}}}\,\exp\left[-\frac{\left(\phi-\phicl-\mean{\delta\phi_{2}}\right)^{2}}{2\mean{\delta\phi_{1}^{2}}}\right]\,.
\eeq
For the case of standard inflation, this distribution has been studied for different types of potentials in \cite{Martin:2005ir}. In this thesis, we present (Chapter \ref{chapter:stochastic-project}) the generalization of the perturbative approach of Eq.~(\ref{eq:noiseexpansion}) to the case of Langevin equations originating from an inflaton equation of motion with a non-canonical kinetic term.

\bigskip
In this Chapter, we pushed the description of a FLRW universe filled with scalar field matter to first order in perturbations, and calculated the resulting spectra for scalar and tensor perturbations. The former leave an imprint on the CMB radiation reaching us today from the surface of last scattering, and provide the seeds for structure formation during the SBBM epoch. Having firmly established benefits and predictions of inflation, we now address the question of the inflaton's identity.

\chapter{The Origin of the Inflaton}\label{chapter:infl-guts}
\begin{quotation}
\emph{Searching for an inflaton candidate in high energy extensions of the Standard Model\footnote{of particle physics (as opposed to the SBBM)} (SM) seems promising because the energy scale of inflation is possibly close to the realm of Grand Unified Theories (GUTs). In this Chapter, we present a concise overview of attempts to embed inflation within a GUT framework, along with the difficulties generally encountered.}
\end{quotation}

\section{From the Standard Model to Grand Unified Theories}
\subsection{The Standard Model of Particle Physics}

\begin{figure}[t]
\begin{center}
\includegraphics[width=0.35\textwidth]{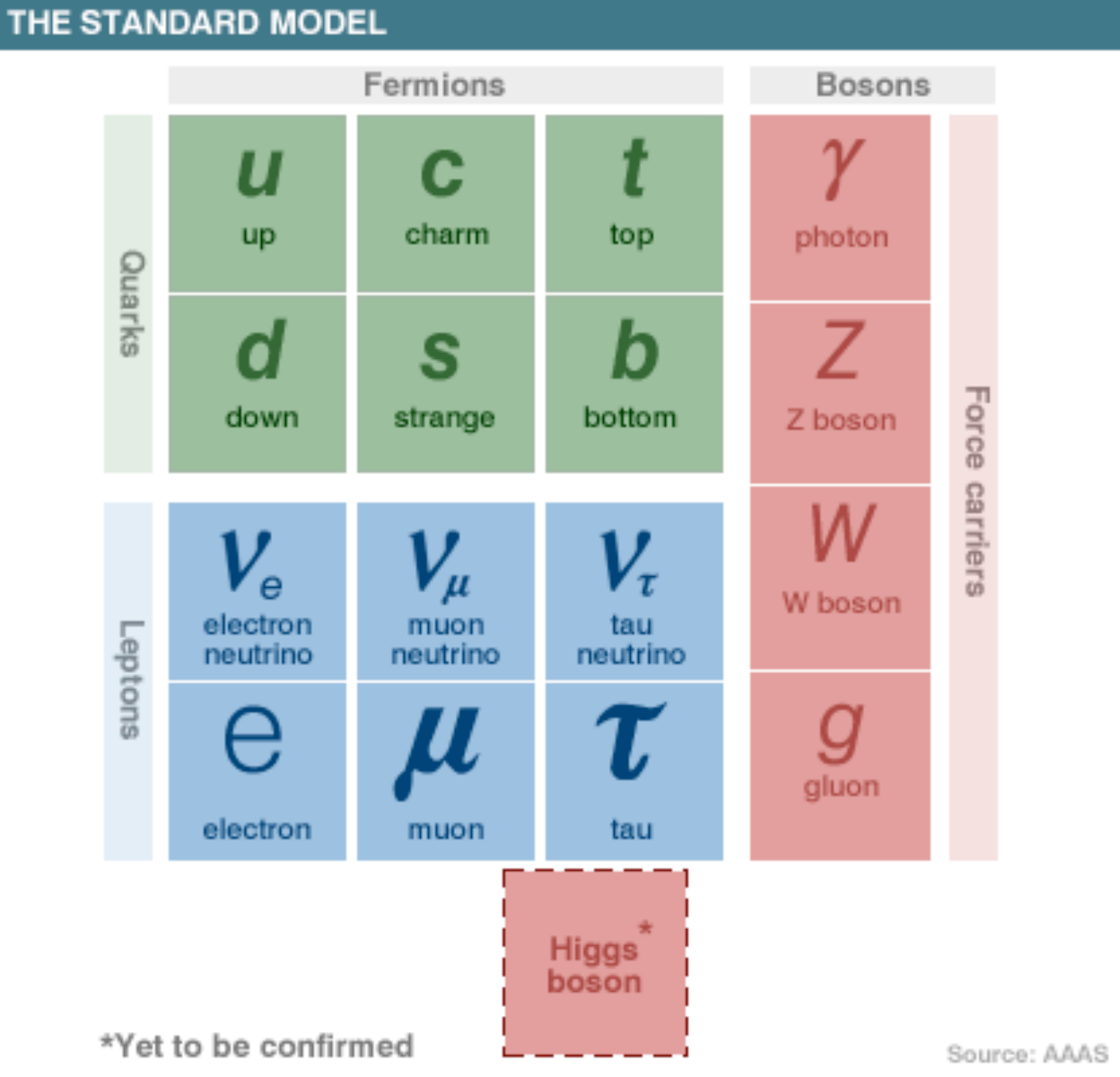}
\hfill
\includegraphics[width=0.6\textwidth]{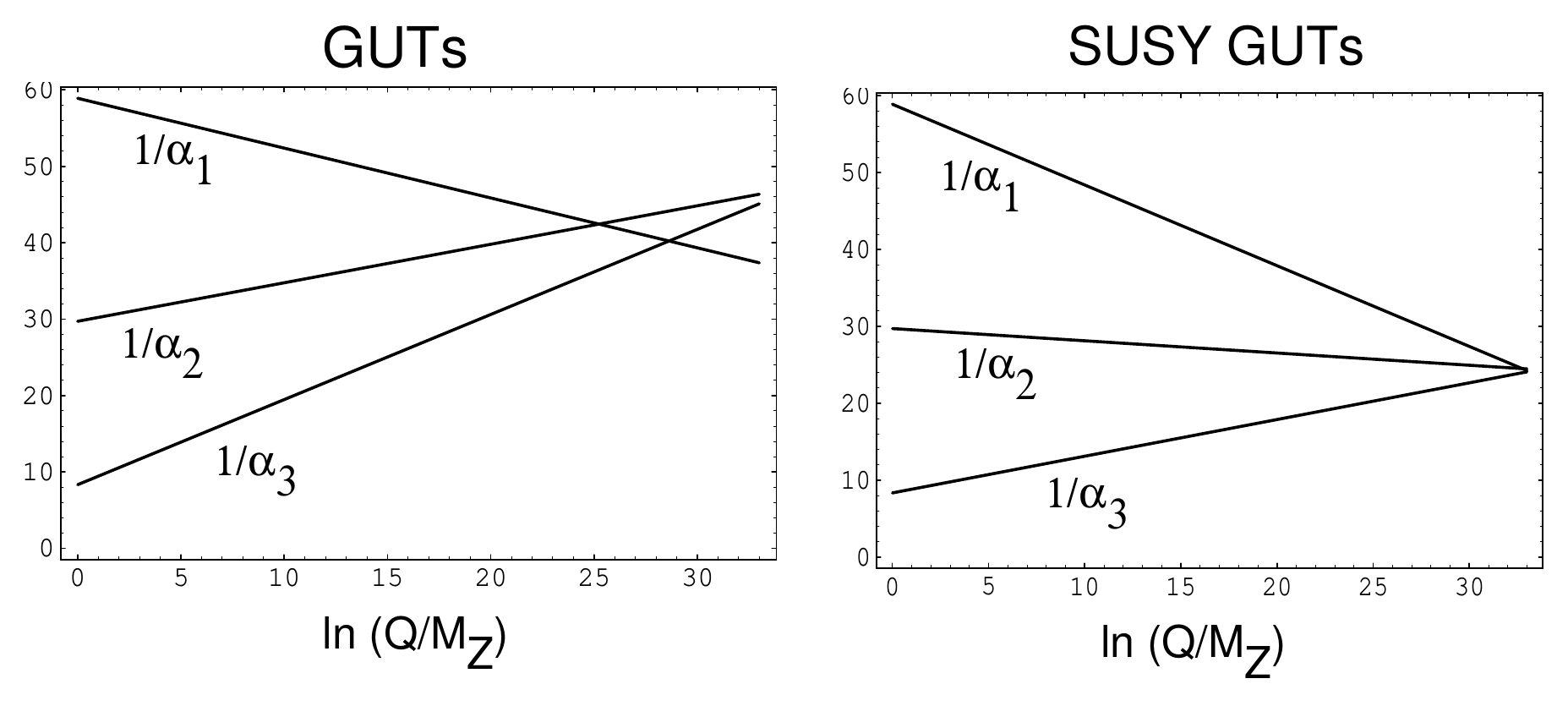}
\caption[Particle families in the Standard Model of particle physics (Source: American Association for the Advancement of Science). Running of Standard Model coupling constants with and without supersymmetry (Source: \cite{Lopez:1996gy}).]{\small \emph{Left:} The particle families of the Standard Model of particle physics. Each particle is accompanied has a corresponding anti-particle with opposite charges. (Figure by American Association for the Advancement of Science) \emph{Right:} Running of Standard Model coupling constants with and without supersymmetry.  $\alpha_{1}$ is the coupling constant of weak hypercharge $Y$, $\alpha_{2}$ that of the chiral $SU(2)$ interaction, and $\alpha_{3}$ the coupling of quantum chromodynamics. They seemingly unify around an energy scale of $\order{10^{15}\GeV}$. (The horizontal energy scale is expressed in terms of the $Z^{0}$ gauge boson mass.) The unification becomes exact when supersymmetry is assumed. (Figure from \cite{Lopez:1996gy})}
\label{fig:sm_particles}
\end{center}
\end{figure}

All experiments in the laboratory to-date confirm the predictions of the Standard Model of particle physics. Established over the second half of the twentieth century in close interplay between theoretical progress and accelerator discoveries, the SM explains the strong, weak and electromagnetic interactions from gauge theories based on the groups $SU(3)_{\mathrm{C}}\times SU(2)_{\mathrm{L}}\times U(1)_{\mathrm{Y}}$. Generally speaking, if a particle is charged under a certain symmetry group, it partakes in the corresponding interaction: $SU(3)_{\mathrm{C}}$ gives rise to quantum chromodynamics, describing processes between coloured particles (hence the subscript ``C''), while $SU(2)_{\mathrm{L}}\times U(1)_{\mathrm{Y}}$ stands for the unification of quantum electrodynamics and the weak interaction within the Glashow Weinberg Salam model \cite{Glashow:1961tr,Salam:1964ry,Weinberg:1967tq} at energies of $\order{200\GeV}$. The subscript ``L'' indicates that the electroweak interaction is left-handed, and ``Y'' stands for the weak hypercharge $Y$. The symmetries $SU(2)_{\mathrm{L}}\times U(1)_{\mathrm{Y}}$ are then broken down into the weak and electromagnetic forces observed at lower energies by virtue of the so-called Higgs mechanism \cite{Higgs:1964ia,Englert:1964et,Guralnik:1964eu}.\\
The Standard Model matter particles, called quarks and leptons, are fermions (\ie particles of half-integer spin) and can carry color (for quarks only), weak and electric charges. Forces are mediated by messenger particles (which are bosons, \ie their spin is integer): eight gluons $g$ for the strong force, three weak gauge bosons $W^{\pm},Z^{0}$ for the weak interaction and the photon $\gamma$ for electromagnetism. Photons are massless, which means that the electromagnetic force has infinite range\footnote{As does gravity; we shall see in later Chapters that in string theory, gravitational interaction is mediated by a massless spin-$2$ particle called the ``graviton''. This motivates the interpretation of string theory as a quantum gravity candidate.}, and reflects the fact that a residual electromagnetic $U(1)$ symmetry is unbroken after $SU(2)_{\mathrm{L}}\times U(1)_{\mathrm{Y}}\rightarrow U(1)_{\mathrm{em}}$ via the Higgs mechanism. Gluons are also massless [$SU(3)_{\mathrm{C}}$ is unbroken], but the strong interaction they mediate is confined to the \emph{very} short distance scales $\order{10^{-15}\mathrm{m}}$ inside atomic nuclei. Fermions and weak gauge bosons, on the other hand, are massive: their masses are generated in the process of electroweak symmetry breaking when the Higgs particle, the only elementary scalar (spin-$0$) field in the SM, acquires a non-zero vacuum expectation value.\\
Fermion and boson families of the Standard Model of particle physics are summarized in Fig.~\ref{fig:sm_particles}. With the exception of the elusive Higgs particle, they have been detected and their properties measured in high energy accelerator experiments \cite{Amsler:2008zzb}. The large discrepancy in mass between leptons and quarks is explained by their different coupling strength to the Higgs. In contrast to leptons, quarks are only observed in ``colorless'' combinations of three (baryons) or two (mesons), held together by gluons.

The SM is a renormalizable theory, meaning that it depends on a finite number (in this case 
19) of input parameters to be determined from experiment. Moreover, these parameters depend on the energy scale at which the interactions are probed: in the process of renormalization, fields and couplings are redefined to absorb counter terms (necessary to cancel divergent Feynman loop diagrams) and hence pick up a dependence on the high energy cutoff $\Lambda_\uUV$ of the theory. In this sense, the Standard Model is an effective theory that may be used to calculate processes up to $\Lambda_\uUV$, where it should be replaced with a more general framework. Indeed, measurements of the SM gauge coupling constants at current accelerator energies suggest that the electroweak and the strong interaction unify further around $\order{10^{15}\GeV}$ (see Fig.~\ref{fig:sm_particles}). This raises the question whether a fundamental theory can be found for all three, \ie strong, weak and electromagnetic forces, much like electroweak unification is achieved by the Glashow Weinberg Salam model.

\subsection{Supersymmetry and Supergravity}
Great effort has been dedicated to the construction of such ``Grand Unified Theories'', which may rely on an extended gauge group like $SO(10)$, broken at energies below $\order{10^{15}\GeV}$ to recover the Standard Model. An important building block of many GUTs is ``supersymmetry'', a symmetry that extends Lorentz invariance by introducing spin-$1/2$ group generators. At the most basic level, supersymmetry predicts a ``superpartner'' for each SM particle, \ie for each fermion in Fig.~\ref{fig:sm_particles} there should be a boson with the same mass, and \emph{vice versa}. In later Chapters, we discuss this formalism in more detail; textbook descriptions are given \eg in \cite{bailin:susy,binetruy:susy}. Any supersymmetric version of the SM therefore comprises at least twice its particle content. Immediate consequences of supersymmetry are that \emph{i)} the SM gauge coupling constants now unify precisely at one point (see Fig.~\ref{fig:sm_particles}), and \emph{ii)} there are now (at least) two Higgs doublets\footnote{Scalar fields always come in pairs in supersymmetry (which can be written \eg as the real and imaginary part of a complex scalar) since two degrees of freedom are needed for the superpartner for a spin-$1/2$ field.}. Conceptually, it is possible to introduce not only one, but a number $\calN\geq1$ of supersymmetries, which increases the particle content even further. However, to eventually recover the chiral character of the electroweak interaction (under which left-handed particles transform as doublets, while the right-handed ones are singlets), is is advisable to focus on the $\calN=1$ case. The simplest extension of the SM using supersymmetry is called the ``Minimally Supersymmetric Standard Model'' (MSSM).


Note that gravity is no integral part of the Standard Model picture. Indeed, GR is fundamentally different from the $SU(3)_{\mathrm{C}}\times SU(2)_{\mathrm{L}}\times U(1)_{\mathrm{Y}}$ framework: apart from the ``running'' of coupling constants, renormalizability also imposes restrictions on the form of allowed interaction terms in the SM Lagrangian. Only terms with dimensionless coupling constants are permitted, limiting to four the number of fields participating in any given vertex. In the case of gravity, however, we know from the action (\ref{eq:Sgrav}) that the coupling constant $\kappa$ is dimensionful. Therefore, gravity has to be added ``by hand'' to the SM to obtain a complete description of Nature.\\
\emph{A priori}, supersymmetry is an extension of the Standard Model only and hence does not describe gravity. However, when supersymmetry is turned from a global into a local symmetry (by making its infinitesimal Grassmann transformation parameter 
spacetime dependent), it includes gravity (\ie invariance under general reparametrizations of spacetime). The resulting theory is then called ``supergravity'', and in this process one passes from a potentially renormalizable theory (global supersymmetry) to a non-renormalizable one with an infinite number of parameters. [As a consequence, non-renormalizable terms in the ``superpotential'' $W({\boldmath \Phi})$, see below, may be allowed in supergravity.] A common strategy is to consider globally supersymmetric models as the limit of supergravity scenarios when $\mpl\rightarrow\infty$, though this approach has its limitations.\\
Since we do not observe an equal mass bosonic partner for \eg the electron, even $\calN=1$ supersymmetry must be broken (and the SM recovered from the MSSM) below a certain energy scale. In the theory's local form as supergravity, the breaking occurs spontaneously (\ie via a non-zero vacuum expectation value, like electroweak symmetry breaking by the Higgs), usually in a ``hidden sector'' of unobserved particles, from where it is communicated to the ``visible sector'' of SM particles. The resulting theory in the visible sector then looks like global supersymmetry with explicit, so-called\footnote{Soft supersymmetry breaking is characterized by the fact that it gives the supersymmetric partners of quarks and leptons masses within the ``correct'' range $\order{100\GeV-1\TeV}$, \ie such that they are heavy enough to have escaped detection until now, but light enough not to interfere with the SM Higgs mechanism.} ``soft'' supersymmetry breaking.

The study of inflationary scalar perturbations in the previous Chapter showed that the energy density during inflation can be close to the GUT scale. 
More precisely, the Universe could have been at this energy scale during the observable part (\ie the last $\sim60$ \efolds) of inflation, 
with the limit $\rho\leq10^{-8}\mpl^{4}$ derived from the non-detection of primordial gravitational waves \cite{Lyth:2007qh}. At these scales, supersymmetry (or supergravity) sets the stage. Moreover, in contrast to the SM with its lone Higgs scalar, supersymmetry makes an abundance of scalar fields available for inflationary model building.

\subsection{Supersymmetry Multiplets}\label{subsec:multiplets}
Let us briefly examine the minimal field content of a supersymmetric model. In $\calN=1$ supergravity, we have at our disposal \emph{i)} the chiral supermultiplet (from which supersymmetric matter is built) containing a Weyl spinor and a complex scalar field, \emph{ii)} the vector supermultiplet (for supersymmetric gauge particles) made of a Weyl spinor and a massless vector (spin-$1$) and \emph{iii)} the gravity supermultiplet with the spin-$2$ graviton and its superpartner, the gravitino (spin-$3/2$). With inflation (\ie scalar fields) in mind, we focus on the chiral supermultiplet, which we write as ${\boldmath \Phi}(x^{\mu},\theta,\bar{\theta})$. It may be expanded in ``superspace''\footnote{Supersymmetric fields are functions not only of the spacetime coordinates $x^{\mu}$, but also of anti-commuting Grassmann variables $\theta,\bar{\theta}$.} in component fields \cite{bailin:susy}
\bea\label{eq:chiralmultiplet}
{\boldmath \Phi}(x^{\mu},\theta,\bar{\theta})&=&\phi(x^{\mu})+\sqrt{2}\,\theta\,\psi(x^{\mu})+\theta^{2}\,F(x^{\mu})\\
&&+i\partial_{\mu}\phi(x^{\mu})\,\theta\sigma^{\mu}\bar{\theta}-\frac{i}{\sqrt{2}}\,\theta^{2}\,\partial_{\mu}\psi(x^{\mu})\,\sigma^{\mu}\bar{\theta}-\frac{1}{4}\,\partial_{\mu}\partial^{\mu}\phi(x^{\mu})\,\theta^{2}\,\bar{\theta}^{2}\,,\nonumber
\eea
where $\phi$ is a complex scalar, $\psi$ a spinor and $F$ an auxiliary scalar field. (Auxiliary fields like $F$ are introduced to make the supersymmetry algebra close off-shell, \ie without use of the equations of motion.) In the above expansion, we have not displayed the spinor indices, and the $\sigma^{\mu}$ denote the usual vector of $(2\times 2)$ matrices $\sigma^{\mu}=(\mathbb{I},\vec{\sigma})$, where the $\sigma^{i}$ are the Pauli matrices. The conjugate of the (left chiral) superfield ${\boldmath \Phi}$ is denoted by ${\boldmath \Phi}^{\dagger}$. We discuss the form of the general supersymmetric Lagrangian for the superfield ${\boldmath \Phi}(x^{\mu},\theta,\bar{\theta})$ in both its renormalizable and in its supergravity version below (see Section \ref{subsec:susylagrangians}).\\
The two-component field $\phi$ can be composed \eg into its real and imaginary part (we denote the complex conjugate of $\phi$ by $\bar{\phi}$), or, alternatively, into a radial and an angular mode (much like the decomposition of a symmetry-breaking field like the Higgs). Both its degrees of freedom are natural supersymmetric candidates for the inflaton, and there can be $n$ copies of the chiral multiplet of Eq.~(\ref{eq:chiralmultiplet}) in a supersymmetric theory, giving $\phi_{i},i=1,\dots,n$ complex fields. It should then in principle be possible to derive the effective inflaton potential $V(\phi_{i})$ from the chosen supersymmetric particle physics model, \eg the MSSM. Our strategy is the following: we first discuss conditions on the most general type of inflaton potential, before turning to the question how these can be satisfied by a supersymmetric candidate field.

\section{Inflation and Supersymmetry}\label{sec:infl-susy}
\subsection{Conditions on a General Inflaton Potential}
In a general effective field theory, the inflaton potential with an infinite number of terms reads
\beq\label{eq:generalpot-sf}
V(\phi)=V_{0}\pm\frac{m^{2}}{2}\,\phi^{2}\pm\frac{\lambda}{4}\,\phi^{4}+\sum_{n=5}^{\infty}\lambda_{n}\,\kappa^{(4-n)/2}\,\phi^{n}
\eeq
where the first (constant) term plays the r\^{o}le of vacuum energy, and the second term gives the scalar field its mass. A linear term is absent by a suitable choice of origin for the field $\phi$. Generically, the only renormalizable interaction term present is the quartic self-coupling. (A cubic term is allowed by renormalizability, but makes it necessary to check explicitly if the potential is positive definite.) One must have $\lambda<1$ to be in the perturbative regime. The couplings $\lambda_{n}$ of the non-renormalizable terms (in which the Planck mass appears explicitly via $\kappa$) are generically $\order{1}$ if the cutoff scale is Planckian $\Lambda_\uUV\approx\mpl$ [or $|\lambda_{n}|\simeq(\mpl/\Lambda_{\uUV})^{n}$ in general]. Their values should be calculable within the theory that replaces the effective framework at the cutoff. The $\lambda_{n}$ vanish in the limit where the cutoff scale $\Lambda_{\uUV}\rightarrow\infty$ (\ie, with respect to the above discussion, when supergravity is replaced by global symmetry for $\mpl\rightarrow\infty$). One must be careful when taking this limit during inflation because the coupling to gravity is essential to drive spacetime expansion, and hence the terms $\propto\lambda_{n}$ can be problematic when $\phi\geq\mpl$.\\
It is possible that the underlying theory beyond the cutoff prescribes the exact form of the potential  \cite{Linde:2007fr} and that, for example, only the quadratic or the cubic term are present in Eq.~(\ref{eq:generalpot-sf}), corresponding to the popular large field potentials $V(\phi)=(m^{2}/2)\phi^{2}$ and $V(\phi)=(\lambda/4)\phi^{4}$. If, however, we consider Eq.~(\ref{eq:generalpot-sf}) in its full generality, we see that, discarding accidental cancellations between the non-renormalizable higher order terms (note that the $\lambda_{n}$ can have either sign), a canonically normalized field $\phi$ inflates if the constant term $V_{0}$ dominates \cite{Lyth:2007qh}. This is usually expressed in terms of a potential slow roll parameter $\eta_{V}$ with $\eta_{V}\propto V''/V$, \ie a measure of curvature for the potential\footnote{In terms of the Hubble flow parameters evaluated in the slow roll limit, $\eta_{V}=2\epsilon_{1}-\epsilon_{2}/2$, see \cite{Schwarz:2001vv,Schwarz:2004tz}.}. The condition $|\eta_{V}|\ll1$ for prolonged slow roll inflation then puts upper bounds on $\lambda$ and the $\lambda_{n}$ \cite{Lyth:1998xn,Lyth:2007qh}, which are increasingly difficult to satisfy the larger the field value $\phi$ is compared to $\mpl$. One finds that, since the range of $\phi$ for which inflation takes place cannot be made arbitrarily short, the $\lambda_{n}$ are required to be unnaturally small, \ie fine-tuned. 

\subsection{Supersymmetric Lagrangians}
\subsubsection{Chiral supersymmetric Lagrangian}
From the superspace expansion (\ref{eq:chiralmultiplet}) of a superfield ${\boldmath \Phi}_{i}$ (and its analogue for the conjugate ${\boldmath \Phi}_{i}^{\dagger}$), one may determine the components of products like ${\boldmath \Phi}_{i}{\boldmath \Phi}_{j}$ and ${\boldmath \Phi}_{i}{\boldmath \Phi}_{j}^{\dagger}$. From the supersymmetry transformation properties of the former, it can be shown that any product of only ${\boldmath \Phi}_{i}$'s is again a (left) chiral superfield, while a combination ${\boldmath \Phi}_{i}{\boldmath \Phi}_{j}^{\dagger}$ behaves like a vector superfield (whose superspace expansion we have not written out here, but is readily available in the literature \cite{bailin:susy}). Moreover, one can exploit the integration over the Grassmann coordinates $(\theta,\bar{\theta})$ of superspace, for which $\int\dd\theta=0,\,\int\dd\theta\,\theta=1$ \etc, to ``extract'' components from these superfield expansions. The most general supersymmetric Lagrangian for chiral superfields can then be written as
\beq\label{eq:susyL}
\calL_{\mathrm{susy}}=\int\dd^{4}\theta\,\sum_{i}{\boldmath \Phi}_{i}^{\dagger}{\boldmath \Phi}_{i}+\left[\int\dd^{2}\theta\,W({\boldmath \Phi})+\mathrm{h.c.}\right]\,,
\eeq
where for the integrations we use the notation $\dd^{4}\theta=\dd^{2}\theta\,\dd^{2}\bar{\theta}$ and $\dd^{2}\theta=-\dd\theta\,\dd\theta/4$, the spinor indices again being suppressed. The function $W({\boldmath \Phi})$ is the superpotential we mentioned earlier and contains products of the superfields ${\boldmath \Phi}_{i}$ (and \emph{not} their conjugates ${\boldmath \Phi}_{i}^{\dagger}$) only. [The reverse applies to its hermitian conjugate $W^{\dagger}$, indicated by ``h.c.'' in Eq.~(\ref{eq:susyL}).] By renormalizability, only up to cubic powers (\ie at most a term ${\boldmath \Phi}_{i}{\boldmath \Phi}_{j}{\boldmath \Phi}_{k}$) are permitted. The integration $\int\dd^{2}\theta$ then projects out the so-called ``$F$ term'' of $W({\boldmath \Phi})$, \ie the coefficient of $\theta\theta$ in the superspace expansion of ${\boldmath \Phi}_{i}{\boldmath \Phi}_{j}$ and/or ${\boldmath \Phi}_{i}{\boldmath \Phi}_{j}{\boldmath \Phi}_{k}$, which contains the auxiliary fields $F_{i}$.\\
The first term in $\calL_{\mathrm{susy}}$ provides the kinetic terms for the component fields of the ${\boldmath \Phi}_{i}$, and its notation in Eq.~(\ref{eq:susyL}) is explained as follows: in the superspace expansion for a vector superfield [in analogy to Eq.~(\ref{eq:chiralmultiplet}) for the chiral superfield], an auxiliary pseudo-scalar field appears, which is usually denoted as $D$. Since the product ${\boldmath \Phi}_{i}{\boldmath \Phi}_{i}^{\dagger}$ transforms as a vector superfield, its expansion has a ``$D$ term'', which multiplies the product $\theta\theta\bar{\theta}\bar{\theta}$ of superspace coordinates. This term is projected out by the $\dd^{4}\theta$ integration in Eq.~(\ref{eq:susyL}). An alternative notation for the chiral supersymmetric Lagrangian $\calL_{\mathrm{susy}}$ of Eq.~(\ref{eq:susyL}) is therefore
\beq
\calL_{\mathrm{susy}}=\sum_{i}\left[{\boldmath \Phi}_{i}{\boldmath \Phi}_{i}^{\dagger}\right]_{D}+\left\{\left[W({\boldmath \Phi}_{i})\right]_{F}+\mathrm{h.c.}\right\}\,.
\eeq
Note that in global supersymmetry, the kinetic terms of the components $\phi_{i}$ and $\psi_{i}$ are canonical, \ie using the expansion (\ref{eq:chiralmultiplet}), one finds
\bea
\left[{\boldmath \Phi}_{i}{\boldmath \Phi}_{i}^{\dagger}\right]_{D}&=&F_{i}^{\dagger}F_{i}+\frac{1}{2}\,\partial_{\mu}\phi^{\dagger}_{i}\,\partial^{\mu}\phi_{i}-\frac{1}{4}\left(\phi^{\dagger}_{i}\,\partial_{\mu}\partial^{\mu}\phi_{i}+\mathrm{h.c.}\right)+\frac{1}{2}\left(i\bar{\psi}_{i}\,\bar{\sigma}^{\mu}\,\partial_{\mu}\psi_{i}+\mathrm{h.c.}\right)\,.
\eea

\subsubsection{Chiral supergravity Lagrangian}
When supersymmetry is made local, the supersymmetric Lagrangian of Eq.~(\ref{eq:susyL}) generalizes to
\beq\label{eq:sugraL}
\calL_{\mathrm{sugra}}=\int\dd^{4}\theta\,\mathscr{K}\left({\boldmath \Phi}^{\dagger},{\boldmath \Phi}\right)+\left[\int\dd^{2}\theta\,W({\boldmath \Phi})+\mathrm{h.c.}\right]\,,
\eeq
where $\mathscr{K}({\boldmath \Phi}^{\dagger},{\boldmath \Phi})$ is a general function of the superfields and their conjugates, and therefore non-minimal kinetic terms for the component fields are possible. Since the theory now includes gravity and is not renormalizable anyway, there is also no longer a restriction on the form of the superpotential $W({\boldmath \Phi})$.\\
It turns out that the scalar component fields $\phi_{i},\bar{\phi}_{i}$ of chiral multiplets only enter into the Lagrangian via a specific combination,
\beq\label{eq:defKaehler}
G(\phi,\bar{\phi})=K(\phi,\bar{\phi})+\log|W|^{2},\qquad K(\phi,\bar{\phi})=-3\log\left(-\mathscr{K}/3\right)\,,
\eeq
where $K(\phi,\bar{\phi}$) is often called the \emph{K\"ahler potential}, and $W$ and $\mathscr{K}$ are the superpotential $W({\boldmath \Phi})$ and the general function $\mathscr{K}({\boldmath \Phi}^{\dagger},{\boldmath \Phi})$ of Eq.~(\ref{eq:sugraL}), with the superfields replaced by their scalar components. For the purpose of inflation, our main interest is the bosonic part $\calL_{\mathrm{b}}$ of supergravity Lagrangians like Eq.~(\ref{eq:sugraL}). Up to a metric determinant, and in units where we set $\kappa=1$ for the moment, $\calL_{\mathrm{b}}$ can be obtained using the function $G(\phi,\bar{\phi})$ only,
\beq\label{eq:sugraLbosonic}
-\calL_{\mathrm{b}}\propto\frac{1}{2}\,R+{G^{i}_{}}_{\bar{\jmath}}\,D_{\mu}\phi_{i}\,D^{\mu}\bar{\phi}^{\bar{\jmath}}+e^{G}\left[G_{\bar{\imath}}\,{\left(G^{-1}\right)^{i}_{}}_{\bar{\jmath}}\,G^{j}-3\right]\,,
\eeq
where $R$ is the Ricci scalar, \ie the gravity part of the theory. The indices for $G$ indicate derivatives with $G_{\bar{\imath}}=\partial G/\partial\bar{\phi}^{\bar{\imath}},\,G^{i}=\partial G/\partial\phi_{i}$, and also ${G^{i}_{}}_{\bar{\jmath}}=\partial^{2}G/(\partial\phi_{i}\partial\bar{\phi}^{\bar{\jmath}})$. By $G^{-1}$ we then denote the inverse matrix of second derivatives. The second term in Eq.~(\ref{eq:sugraLbosonic}) then provides the kinetic terms for the scalar fields $\phi, \bar{\phi}$. Note that $W$ is a function of the $\phi$ only (not the $\bar{\phi}$), therefore it does not contribute to the second derivative ${G^{i}_{}}_{\bar{\jmath}}$, and one may write ${G^{i}_{}}_{\bar{\jmath}}={K^{\bar{\imath}}_{}}_{j}$. Consequently, a minimal kinetic term ${K^{i}_{}}_{\bar{\jmath}}=\delta^{\bar{\imath}}_{j}$ is recovered for a general function $\mathscr{K}=-3\exp(-{\boldmath \Phi}_{i}{\boldmath \Phi}^{i\dagger}/3)$ in Eq.~(\ref{eq:sugraL}), which, using Eq.~(\ref{eq:defKaehler}) gives $G=\phi_{i}\bar{\phi}^{\bar{\imath}}+\log|W|^{2}$. 
The last term in Eq.~(\ref{eq:sugraLbosonic}) provides the so-called tree level effective potential $V(\phi)$ for the scalar fields, 
\beq\label{eq:treeleveleffective}
V(\phi,\bar{\phi})=e^{G}\left[G_{\bar{\imath}}\,{\left(G^{-1}\right)^{i}_{}}_{\bar{\jmath}}\,G^{j}-3\right]=e^{K}\left(K^{\bar{\imath}j}\,\overline{D_{\bar{\imath}}W}\,D_{j}W-3\left|W\right|^{2}\right)\,.
\eeq
Here, $D_{i}W=W_{,i}+K_{,i}W$ is a the ``covariant derivative'' of the superpotential $W$. If the kinetic terms are minimal and only renormalizable terms are present in the superpotential $W({\boldmath \Phi})$, we recover the globally supersymmetric Lagrangian (\ref{eq:susyL}). We return to the potential (\ref{eq:treeleveleffective}) in detail in later Chapters within the context of string theory.

\subsubsection{No scale supergravity}
In the supergravity limit of the superstring theories that are studied in Part \ref{part:string} of this thesis, one usually deals with ``no scale'' supergravity, therefore here we briefly state the definition of these models.
In no scale supergravity, the Planck scale is the only mass scale in the picture, \ie there is no intermediate scale introduced (such as \eg the gravitino mass) by effects supersymmetry breaking \etc In these models, the K\"ahler potential has the form (if there is but one scalar field $\phi$)
\beq\label{eq:noscaleKaehler}
G=-3\,\log(\phi+\bar{\phi})\,,\qquad\frac{\partial G}{\partial\phi}=\frac{\partial G}{\partial\bar{\phi}}=-3\,(\phi+\bar{\phi})^{-1}\,,\qquad\frac{\partial^{2}G}{\partial\phi\,\partial\bar{\phi}}=3\,(\phi+\bar{\phi})^{-2}\,.
\eeq
(To make contact with the notation in later Chapters, note that $\phi$ is then replaced $\phi\rightarrow-i\phi$.) As a consequence, one has from Eq.~(\ref{eq:treeleveleffective}) that $V(\phi)=0$ for all values of $\phi$. Hence, at tree level, the field $\phi$ is massless, \ie without potential, and its value can be changed at no energy cost. Fields with the property that they do not enter into the tree level effective potential $V(\phi)$ obtained from supergravity are referred to as ``flat directions''.

\subsection{Inflaton Potentials from Supersymmetry}\label{subsec:susylagrangians}
\subsubsection{Lagrangians for scalar component fields}
Let us now put our understanding of the Lagrangian (\ref{eq:sugraLbosonic}) of complex scalar component fields $\phi_{i}$ of chiral supermultiplets to ``inflationary'' use. We saw that two functions $K(\phi_{m},\bar{\phi}_{m})$ and $W(\phi_{i})$ enter into the scalar Lagrangian $\Lphi=\calL_{\mathrm{kin}}-V(\phi_{i})$. Of these, $K(\phi_{m},\bar{\phi}_{n})$ is the K\"a{}hler potential from which the scalars' kinetic terms are derived,
\beq\label{eq:Kaehler}
\calL_{\mathrm{kin}}=\sum_{m,n}K_{m\bar{n}}\,\partial_{\mu}\phi_{m}\,\partial^{\mu}\bar{\phi}_{n}\,,\qquad K_{m\bar{n}}=\frac{\partial^{2}K}{\partial\phi_{m}\,\partial\bar{\phi}_{n}}\,.
\eeq
The derivatives $K_{m\bar{n}}$ therefore play the r\^ole of a ``metric in field space''. 
The function $W(\phi_{i})$ is obtained from the superpotential $W({\boldmath \Phi})$ by inserting the expansion (\ref{eq:chiralmultiplet}) in terms of component fields. Let us specialize to the case of canonical kinetic terms for the scalar fields for now (as it is the case in global supersymmetry, before it is turned into supergravity). The scalar interaction potential is then obtained from derivatives of the superpotential only [compare Eq.~(\ref{eq:treeleveleffective})],
\beq\label{eq:Ftermpot}
V(\phi_{i})=\sum_{i}\left|\frac{\partial W}{\partial\phi_{i}}\right|^{2}\,,
\eeq
and is called the ``$F$ term potential'' because after solving the supersymmetric equations of motion for the auxiliary field $F_{i}$ in each chiral multiplet, Eq.~(\ref{eq:Ftermpot}) is equivalent to $V(\phi_{i})=\sum_{i}\left|F_{i}\right|^{2}$. [Recall our discussion of the superpotential $F$ term projected out by integrating over $\int\dd^{2}\theta$ in Eq.~(\ref{eq:susyL}).] From this it follows that supersymmetry can be spontaneously broken, \ie the potential $V(\phi_{i})$ can have have a non-zero minimum, if one of the $F_{i}$ acquires a non-vanishing vacuum expectation value $\mean{F_{i}}\neq0$.

For definiteness, let us assume that the inflaton $\phi$ is a supersymmetric field whose effective potential generically has the form (\ref{eq:generalpot-sf}), where the $(V_{0},m^{2},\lambda)$ terms are present at tree-level. If the inflaton is one of the ``flat directions'' while supersymmetry is unbroken, the constant term $V_{0}$ dominates and at first sight inflation may proceed with a very small mass arising from supersymmetry breaking effects. However, at loop order, an additional mass term is generated via the inflaton's coupling to the fields running around the loop. 
At fixed particle content of the chosen theory, however, these loop corrections are exactly calculable (and consequently, this should also be the case in string theory). An extensive review of possible models is provided in \cite{Lyth:1998xn}, and short discussions can be found in \cite{Lyth:2007qh,peter:cosmo,binetruy:susy}. 
One may employ two strategies to render a quantum-corrected potential flat enough for slow roll inflation: the superpotential can be dominated either by an $F$ or by a $D$ term [note that the latter is \emph{not} the $D$ term that gave rise to the kinetic terms of the scalar fields in Eq.~(\ref{eq:susyL}), but a separate contribution to the superpotential which we discuss below]. These can be engineered to be flat, at least in a certain regime. We now consider one example each for two or three chiral multiplets, denoted ${\boldmath \Phi_{1},\Phi_{2}}$ and ${\boldmath \Phi_{0},\Phi_{+},\Phi_{-}}$, respectively, with their scalar components written as $\phi_{1},\phi_{2}$ and $\phi_{0},\phi_{+},\phi_{-}$.

\subsubsection{$F$ term inflation}
Consider the (renormalizable) superpotential for two chiral multiplets \cite{Copeland:1994vg},
\beq\label{eq:Ftermexample}
W_{F}(\phi_{1},\phi_{2})=\phi_{1}\left(\lambda\phi_{2}^{2}-M^{2}\right)\,,
\eeq
which, with the redefinition $|\phi_{1}|=\tilde{\phi}_{1}/\sqrt{2}$ and Eq.~(\ref{eq:Ftermpot}), leads to the scalar potential
\beq
V_{F}(\tilde{\phi}_{1},\phi_{2})=2\lambda^{2}\,\tilde{\phi}_{1}^{2}\,|\phi_{2}|^{2}+\left|\lambda\phi_{2}^{2}-M^{2}\right|^{2}\,,
\eeq
whose global minimum is at $\tilde{\phi}_{1}=0,\,\phi_{2}^{2}=M^{2}/\lambda$. However, if $\tilde{\phi}_{1}$ is held fixed at a value exceeding the critical $\tilde{\phi}_{1}^{\mathrm{(c)}}=M^{2}/\lambda$, there is another minimum (for vanishing real and imaginary part of $\phi_{2}$) with $V=M^{4}$ \cite{binetruy:susy}. Therefore, for $\tilde{\phi}_{1}>\tilde{\phi}_{1}^{\mathrm{(c)}}$, the field $\tilde{\phi}_{1}$ behaves like a flat direction with non-zero potential energy. Loop corrections generate a slope towards $\tilde{\phi}_{1}^{\mathrm{(c)}}$, and when $\tilde{\phi}_{1}$ reaches its critical value, the field $\phi_{2}$ rolls off towards the global minimum. Thus, the resulting inflationary model has the characteristics of hybrid inflation, where $\phi_{2}$ is the field fixed during inflation, and $\tilde{\phi}_{1}^{\mathrm{(c)}}$ the waterfall point.\\
However, in true supergravity the ``prescription'' for obtaining the scalar potential $V(\phi_{i})$ from the superpotential $W({\boldmath \Phi})$ is more complicated than Eq.~(\ref{eq:Ftermpot}), namely it should be replaced by Eq.~(\ref{eq:treeleveleffective}). In particular, the supergravity scalar potential has a prefactor $V\propto\exp\left(\kappa\,K\right)$, where $K$ is the K\"a{}hler potential of Eq.~(\ref{eq:Kaehler}). As a consequence, the potential slow roll parameter picks up a dependence $\eta_{V}\propto K_{i\bar{\imath}}$, where $K_{i\bar{\imath}}$ is the component of the field space metric in front of the $\phi_{i}$ kinetic term. In the true vaccum of supergravity, this term should be canonical, \ie $K_{i\bar{\imath}}\approx\order{1}$, and during inflation it is not much smaller. This is referred to as the ``$\eta$ problem'' of supergravity because the inflaton hence develops a mass (given by $V''$) of $\order{H^{2}}$, which prohibits inflation. A possible way out is to use a K\"a{}hler potential of non-minimal form that makes $\eta_{V}$ small enough \cite{Stewart:1994ts}.

\subsubsection{$D$ term inflation}
As mentioned earlier, one can write down a superspace expansion for a vector supermultiplet analogous to Eq.~(\ref{eq:chiralmultiplet}), making its spinor and vector components explicit. There is also another real pseudo-scalar auxiliary field $D$ in this multiplet. 
If a theory contains both chiral and vector supermultiplets, the scalar potential $V(\phi)$ has a so-called $D$ term $\propto D^{2}/2$  on top of the $F$ term discussed earlier. This term is obtained from $D\propto\bar{\phi}\phi$ plus a possible \emph{Fayet Iliopoulos term}, 
if a $U(1)$ symmetry --like \eg hypercharge $Y$-- is present. Unlike the $F$ term potential, the $D$ term does not receive an exponential prefactor involving the K\"ahler potential when supergravity corrections are introduced. It therefore does not suffer from the $\eta$ problem and can be flat enough for slow roll inflation.\\
A superpotential from global supersymmetry including an additonal $U(1)$ takes the form \cite{Halyo:1996pp,Binetruy:1996xj}
\beq
W_{D}=\lambda\Phi_{0}\Phi_{+}\Phi_{-}\,,
\eeq
where the three chiral superfields have charges $0,\pm1$ under the $U(1)$ symmetry and $\lambda$ is a coupling constant. The scalar potential then has the form
\beq\label{eq:Dtermpot}
V_{D}(\phi_{0},\phi_{+},\phi_{-})=\lambda^{2}\left(\left|\phi_{+}\phi_{-}\right|^{2}+\left|\phi_{0}\phi_{+}\right|^{2}+\left|\phi_{0}\phi_{-}\right|^{2}\right)+\frac{g^{2}}{2}\left(\left|\phi_{+}\right|^{2}-\left|\phi_{-}-\xi\right|^{2}\right)^{2}\,,
\eeq
where $g$ is the gauge coupling and $\xi$ is the parameter in the Fayet Iliopoulos term (here $\xi>0$). The unique supersymmetric vacuum (where gauge symmetry is broken) is at $\phi_{0}=\phi_{-}=0,\,\phi_{+}=\sqrt{\xi}$. However, while $\phi_{0}$ is in the regime $|\phi_{0}|>g\sqrt{\xi}/\lambda$, there is a minimum at $\phi_{+}=\phi_{-}=0$, and the potential is flat at tree-level in the direction of $\phi_{0}$ (with $V=g^{2}\xi^{2}/2$), while there is a ``valley'' in the $(\phi_{+},\phi_{-})$ directions. At tree-level the $F$ terms vanish in the inflationary field space direction. Loop corrections generate a small slope along $\phi_{0}$, leading to slow roll towards $\phi_{0}^{\mathrm{(c)}}=g\sqrt{\xi}/2$. The potential (\ref{eq:Dtermpot}) therefore is a supersymmetric realization of hybrid inflation. A generalization of the superpotential (\ref{eq:Dtermpot}) is obtained when the $U(1)$ symmetry is pseudo-anomalous as \eg in string theory \cite{binetruy:susy,Green:1984sg}.

\bigskip
We close this Chapter with two remarks. First, the fine-tuning constraints on the inflaton couplings $(\lambda,\lambda_{n})$ can be lifted by relieving the inflaton from the duty of generating the primordial curvature perturbations. [Fine-tuning is necessary to ensure the predicted perturbation amplitude does not exceed the observed one of $\order{10^{-5}}$, which is guaranteed by restricting inflation to the slow roll regime.] Instead, the perturbation spectrum may be generated by a different field, called the ``curvaton'', at the end of inflation \cite{Enqvist:2001zp,Moroi:2001ct,Lyth:2002my}.\\
In supersymmetry, many weakly (\ie only gravitationally) coupled scalars are available for this mechanism. If the curvaton decays late enough, its inhomogeneous perturbations can lead to fluctuations in the radiation density seeding the CMB temperature fluctuations. However, the decay has to be fast enough (within the first second) not to jeopardize SBBM evolution, which tightly constrains the possible mass range. Moreover, the resulting density perturbation is highly dependent on the evolution of a given mode $k$ outside the Hubble scale. While inflationary model building is less constrained in the presence of a curvaton field, this mechanism also restricts the observational window on the microphysics of inflation.\\
Second, there are (at least) two more cosmological puzzles one would like to see resolved in a GUT theory \cite{Burgess:2007pz,Mukhanov:2005sc}: it should contain feasible candidates for Dark Matter (of which there are many in supersymmetry), and it must be able to explain baryogenesis. A successful realization of reheating at the end of inflation together with baryogenesis can be used as a powerful criterion to discriminate between GUT candidates.

\chapter{Extensions and Alternatives}\label{chapter:ext-alt}
\begin{quotation}
\emph{In this Chapter, we introduce two generalizations of the inflationary perturbation theory presented in Section \ref{sec:perturbations}: the possibility of multiple dynamic scalar fields during inflation makes it necessary to consider the evolution of both adiabatic (curvature) and entropy (isocurvature) perturbations. We also discuss the scenario of $k$-inflation (Section \ref{subsec:k-inflation}) at the perturbative level, which allows the calculation of fluctuation spectra for an inflaton field with non-canonical kinetic term. Finally, we briefly comment on alternatives to inflation.}
\end{quotation}

\section{Multifield Inflation and Entropy Perturbations}\label{sec:multifieldpert}
In Chapters \ref{chapter:cosmo-infl} and \ref{chapter:toolkit} we focused on scenarios with only one dynamic scalar $\phi$ during inflation, 
but attempts to identify the inflaton among the degrees of freedom of a (supersymmetric) Grand Unified Theory usually confront us with a multitude of fields $\phi_{i}$. Several of them can contribute to the accelerated expansion if they are light compared to the Hubble scale. Each field undergoes quantum fluctuations $\delta\phi_{i}$, and therefore a qualitatively new form of perturbations arises: while a single field constitutes the energy density of the Universe, its perturbations are essentially equivalent to fluctuations in the energy density $\delta\rho$, and hence of adiabatic (curvature\footnote{Recall that $\delta\phi$ and the metric variable $\Phi$ determine the comoving curvature perturbation $\mathscr{R}$ of Eq.~(\ref{eq:intrinsic}).}) type. However, in the multifield case where $\rho$ depends on all $\phi_{i}$, different field perturbations $\delta\phi_{i}$ can combine in such a way that the overall density perturbation vanishes, $\delta\rho=0$, but the relative contribution of each $\phi_{i}$ to the total energy density $\rho$ fluctuates. These perturbations are hence of entropy (or isocurvature) type. We briefly show how this new perturbation component can be calculated from the background evolution in multifield models. Details are readily available in the literature, see \eg \cite{Gordon:2000hv,Bartolo:2001rt,Bassett:2005xm,Wands:2007bd}.

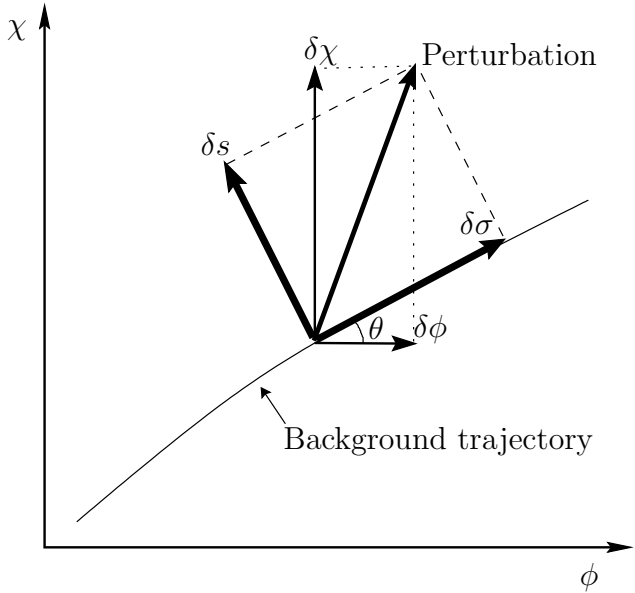
\begin{figure}[t]
\begin{minipage}{0.45\textwidth}
\begin{picture}(0,0)%
\includegraphics{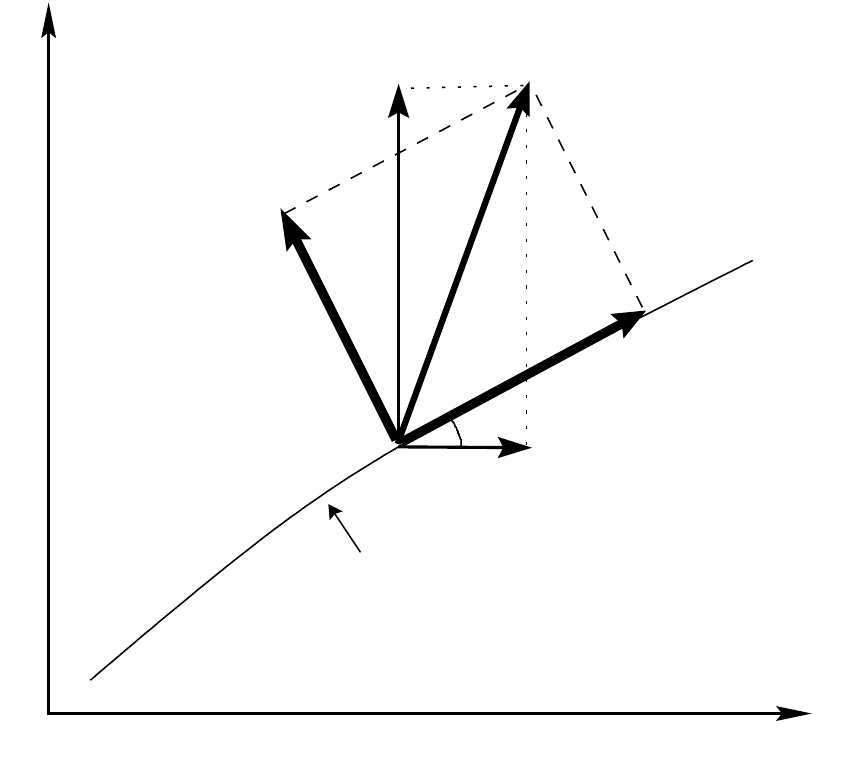}%
\end{picture}%
\setlength{\unitlength}{3947sp}%
\begingroup\makeatletter\ifx\SetFigFont\undefined%
\gdef\SetFigFont#1#2#3#4#5{%
  \reset@font\fontsize{#1}{#2pt}%
  \fontfamily{#3}\fontseries{#4}\fontshape{#5}%
  \selectfont}%
\fi\endgroup%
\begin{picture}(3897,3705)(1271,-3424)
\put(4085,-1171){\makebox(0,0)[lb]{\smash{\SetFigFont{12}{14.4}
{\rmdefault}{\mddefault}{\updefault}$\delta \sigma$}}}
\put(2990,-2517){\makebox(0,0)[lb]{\smash{\SetFigFont{12}{14.4}
{\rmdefault}{\mddefault}{\updefault}Background trajectory}}}
\put(3847,-113){\makebox(0,0)[lb]{\smash{\SetFigFont{12}{14.4}
{\rmdefault}{\mddefault}{\updefault}Perturbation}}}
\put(3116,-76){\makebox(0,0)[lb]{\smash{\SetFigFont{12}{14.4}
{\rmdefault}{\mddefault}{\updefault}$\delta \chi$}}}
\put(2476,-676){\makebox(0,0)[lb]{\smash{\SetFigFont{12}{14.4}
{\rmdefault}{\mddefault}{\updefault}$\delta s$}}}
\put(3798,-1813){\makebox(0,0)[lb]{\smash{\SetFigFont{12}{14.4}
{\rmdefault}{\mddefault}{\updefault}$\delta \phi$}}}
\put(3517,-1825){\makebox(0,0)[lb]{\smash{\SetFigFont{12}{14.4}
{\rmdefault}{\mddefault}{\updefault}$\theta$}}} \put(1271,
84){\makebox(0,0)[lb]{\smash{\SetFigFont{12}{14.4}{\rmdefault}
{\mddefault}{\updefault}$\chi$}}}
\put(4831,-3366){\makebox(0,0)[lb]{\smash{\SetFigFont{12}{14.4}
{\rmdefault}{\mddefault}{\updefault}$\phi$}}}
\end{picture}
\end{minipage}
\hfill
\begin{minipage}{0.45\textwidth}
\caption[Decomposition of the two-dimensional field space trajectory into an adiabatic and an entropy direction. (Source: \cite{Gordon:2000hv})]{\small In a two-field inflationary model with scalars $(\phi,\chi)$ (both of which are light compared to the Hubble scale), the curved trajectory in field space can be decomposed into an adiabatic component $\sigma$ (along the trajectory) and an entropy component $s$ (orthogonal to the trajectory). Both $\sigma$ and $s$ have fluctuations, which give rise to the adiabatic and isocurvature perturbation modes, respectively. The angle $\theta$ parametrizes the curvature of the background trajectory. (Figure from \cite{Gordon:2000hv})}
\end{minipage}
\label{fig:twofieldtrajectory}
\end{figure}

For a collection of scalar fields $\phi_{i}, i=1,2,\dots n$, we have [instead of Eq.~(\ref{eq:KleinGordon})] a set of $i$ Klein Gordon equations,
\beq\label{eq:KG-multi}
\ddot{\phi}_{i}+3H\dot{\phi}_{i}+\frac{\partial}{\partial\phi_{i}}\,V(\phi_{1},\phi_{2},\dots)=0\,,
\eeq
where the potential $V$ is a function of all scalar fields at once, which can contain mass terms for certain fields, their self-interactions and notably cross-couplings between different $\phi_{i}$. 
Even in the absence of explicit interactions, however, the fields are coupled gravitationally via the multifield Hubble parameter, $H^{2}=(\kappa/3)\,\left[\sum_{i}(\dot{\phi}_{i}^{2}/2)+V\right]$, where it has been assumed that all $\phi_{i}$ have canonical kinetic terms. Via the sum over all fields in the Hubble parameter, slow roll inflation may be possible even if the individual field potentials are steep. This mechanism is used, for example, in the ``assisted inflation'' scenario \cite{Liddle:1998jc} as well as in ``N-flation'' \cite{Dimopoulos:2005ac}.\\
With the field space now being $n$-dimensional, it is useful to assign a ``weight factor'' $f_{i}$ to each field, which is defined as \cite{Gordon:2000hv}
\beq
f_{i}=\frac{\dot{\phi_{i}}}{\sqrt{\sum_{j}\dot{\phi}_{j}^{2}}}\,,
\eeq
measuring its contribution to the inflationary direction in field space. The direction $\sigma$ is obtained from integrating $\sigma=\int\dd t\,\sum_{i}f_{i}\dot{\phi}_{i}$, and it can be shown from Eqs.~(\ref{eq:KG-multi}) that it evolves as
\beq\label{eq:KG-sigma}
\ddot{\sigma}+3H\dot{\sigma}+V_{,\sigma}=0,\qquad V_{,\sigma}=\sum_{i}f_{i}\,\frac{\partial V}{\partial\phi_{i}}\,.
\eeq
There are, however, $(n-1)$ more directions in field space which we may denote by $s_{j},j=1,\dots, n-1$, and which can be chosen orthogonal to $\sigma$ (and relative to each other). In the following, we concentrate on the case $n=2$ with the scalar fields $\phi$ and $\chi$ (and write $s_{1}\equiv s$), for which the situation is illustrated in Fig.~ \ref{fig:twofieldtrajectory}.

We have hence exchanged the field space basis $(\phi,\chi)$ for the more convenient pair $(\sigma,s)$ along and orthogonal to the inflating direction. These new fields evidently can be decomposed again into a background and perturbations, $\sigma(t,\vec{x})=\sigma_{0}(t)+\delta\sigma(t,\vec{x})$ and $s(t,\vec{x})=\delta s(t,\vec{x})$, where $\sigma_{0}$ obeys Eq.~(\ref{eq:KG-sigma}) and the entropy background field vanishes \cite{Gordon:2000hv}. The perturbations $\delta s$ precisely give the isocurvature mode that the two-field scenario exhibits on top of the adiabatic mode\footnote{For the general case of $n$ fields, the adiabatic perturbation is constructed from $\delta\sigma=\sum_{i}f_{i}\delta\phi_{i}$, while for the entropy perturbation $\delta s_{i}=\sum_{j}e_{ij}\delta\phi_{j}$, where $e_{ij}$ are coefficients such that $\sum_{i}e_{ji}f_{i}=0$..} $\delta\sigma$.\\
In straightforward analogy with the single field case, the inherent gauge dependence of the adiabatic perturbation $\delta\sigma$ requires a redefinition $\delta\sigma^{\uPgiP}=\delta\sigma+(\dot{\sigma}_{0}/H)\,\psi$, while $\delta s$ is automatically gauge independent \cite{Stewart:1974uz}. It can then be shown that the evolution of $\delta\sigma^{\uPgiP}$ and $\delta s$, respectively, may schematically be written as
\bea
\delta\ddot{\sigma}^{\uPgiP}+3H\,\delta\dot{\sigma}^{\uPgiP}+\left(\frac{k^{2}}{a^{2}}-m^{2}_{\sigma}\right)\delta\sigma^{\uPgiP}&=&\mathscr{C}\,\dot{\theta}\delta s+2\frac{\dd}{\dd t}\,(\dot{\theta}\delta s)\,,\label{eq:adiabaticmode}\\
\delta \ddot{s}+3H\,\delta\dot{s}+\left(\frac{k^{2}}{a^{2}}-m^{2}_{s}\right)\delta s&=&\frac{4\dot{\theta}}{\kappa\dot{\sigma}_{0}}\,\frac{k^{2}}{a^{2}}\,\Psi\,,\label{eq:entropymode}
\eea
where $m_{\sigma},m_{s}$ are the (time-dependent) adiabatic and entropic ``mass'' of the respective perturbations (calculated from weighted derivatives of the potential as well as other contributions), and $\mathscr{C}$ is a time-dependent factor (also consisting of background quantities). The angle $\theta$ describes the curvature of the inflationary trajectory in field space, see Fig.~\ref{fig:twofieldtrajectory}. The detailed derivation of these equations is given elsewhere \cite{Gordon:2000hv,Wands:2007bd}; here, we emphasize the following two observations: note that the only source term on the right hand side of Eq.~(\ref{eq:entropymode}) is $\propto\Psi$ [where $\Psi$ is the Bardeen potential of Eq.~(\ref{eq:bardeenpots})], and that it dies out on large scales when $k/a\rightarrow 0$. On the other hand, Eq.~(\ref{eq:adiabaticmode}) is sourced by terms $\propto\dot{\theta}\,\delta s$, hence if the inflationary trajectory is nontrivial in field space ($\dot{\theta}\neq0$), a non-vanishing entropy mode can feed the adiabatic curvature perturbation. As a consequence, the adiabatic mode no longer becomes constant on large scales (outside the Hubble radius), which is easily seen from the time derivative of the comoving curvature perturbation [compare the conservation law (\ref{eq:zetaprime}) for $\zeta$ derived in the single field case]:
\beq\label{eq:Rdot}
\dot{\mathscr{R}}=\frac{H}{\dot{H}}\,\frac{k^{2}}{a^{2}}\,\Psi+\frac{2H}{\dot{\sigma}_{0}}\,\dot{\theta}\,\delta s
\eeq
Another way to see this is that for a single inflaton field $\phi$, the dominant Hubble friction term in its Klein Gordon equation during slow roll usually ensures that $\phi$ is quickly driven towards a unique attractor in phase space $(\phi,\dot{\phi})$, which is independent of the field's initial conditions \cite{Goldwirth:1991rj}. However, for multifield inflation the Hubble parameter depends on all fields $\phi_{i}$ simultaneously, and hence there are several possible trajectories in phase space depending on the set of initial field values. This ambiguity is responsible for the survival of non-adiabatic perturbations.

Generalizing the slow roll regime to the two field case allows to obtain first order differential equations for $\delta\sigma^{\uPgiP}$ and $\delta s$ on large scales \cite{Wands:2002bn,Byrnes:2006fr}, instead of the second order equations (\ref{eq:adiabaticmode}), (\ref{eq:entropymode}) . The two modes of these first order equations can be described in terms of the curvature and isocurvature perturbations \cite{Gordon:2000hv,Amendola:2001ni}
\beq\label{eq:adiabatic-entropy-modes}
\mathscr{R}=\frac{H}{\dot{\sigma}_{0}}\,\delta\sigma^{\uPgiP}\,,\qquad \mathscr{S}=\frac{H}{\dot{\sigma}_{0}}\,\delta s\,,
\eeq
respectively. 
Since over many Hubble times, the slow roll approximation becomes less and less reliable, the evolution of $\mathscr{R}$ and $\mathscr{S}$ after Hubble crossing has to be taken into account by a transfer matrix,
\beq\label{eq:transfer}
\left(\begin{matrix}\mathscr{R}\\\mathscr{S}\end{matrix}\right)=\left(\begin{matrix}1&T_{\mathscr{RS}}\\0&T_{\mathscr{SS}}\end{matrix}\right)\left(\begin{matrix}\mathscr{R}\\\mathscr{S}\end{matrix}\right)_{\mathrm{hc}}\,.
\eeq
The first column in this matrix reflects the fact that \emph{i)} adiabatic perturbations remain adiabatic if $\mathscr{S}=0$ and \emph{ii)} entropy perturbations are not sourced by adiabatic ones [see Eq.~(\ref{eq:entropymode})].\\
In principle, instead of the sole comoving curvature spectrum obtained in single field inflation, three spectra have to be calculated in the multifield case: curvature and isocurvature perturbations as well as their correlation. They all contribute to the scalar spectrum $\calP_\usssS$ and hence the observed temperature fluctuation $\delta T/T$ in the CMB, see \eg \cite{Langlois:1999dw}. 
Notably, pure isocurvature perturbations would lead to CMB anisotropies six times bigger than their adiabatic counterpart. Therefore, observations exclude a pure isocurvature origin of the temperature fluctuations, but with their correlation taken into account, a mixed adiabatic and isocurvature origin of the observed $\delta T/T$ can be feasible.

Note that we have considered scalar perturbations only so far. Indeed, since they decouple from the matter sector at linear order, tensor perturbations are not affected by the presence of several scalar fields, and their spectrum remains unchanged. However, the ratio $\calP_{\usssT}/\calP_{\usssS}$ has changed and we now have
\beq
r=\frac{\calP_{\usssT}}{\calP_{\usssS}}\simeq-8n_{\usssT}\,\sin^{2}\Delta\,,
\eeq
where the additional factor measures for the cross-correlation due to $T_{\mathscr{RS}}$ in Eq.~(\ref{eq:transfer}), with $\cos\Delta=T_{\mathscr{RS}}/\left(1+T_{\mathscr{RS}}^{2}\right)^{1/2}$. Other observational consequences of isocurvature modes during inflation are, for example, non-Gaussianities, though they are of detectable size only if generated at the end of inflation \cite{Bernardeau:2002jy,Bernardeau:2002jf} or from inhomogeneous reheating \cite{Kolb:2005ux,Byrnes:2006fr}.

\section{Perturbations in $k$-Inflation}\label{sec:kinflationpert}
In Section \ref{subsec:k-inflation} we discussed the possibility of inflation from a generalized scalar field Lagrangian written as $\calL_{\phi}=p(X, \phi)$, where $X=(g_{\mu\nu}\,\partial^{\mu}\phi\,\partial^{\nu}\phi)/2$. The function $p(X, \phi)$ plays the r\^ole of pressure, and the corresponding energy density was calculated in Eq.~(\ref{eq:rho-kinf}). Note that the GR sector of the theory is kept unchanged: we still use the Einstein equations (\ref{eq:Einstein}), but their ``right hand side'', \ie the energy momentum tensor, now reads \cite{ArmendarizPicon:1999rj,Garriga:1999vw}
\beq\label{eq:Tmunu-kinf}
T_{\mu\nu}=\frac{p+\rho}{\sqrt{2X}}\,\partial_{\mu}\phi\,\partial_{\nu}\phi-p\,g_{\mu\nu}
\eeq
with $p$ and $\rho$ given by their ``$k$-inflationary'' expressions. It is then cumbersome but straightforward to again push the Einstein equations to first order in perturbations with a generalized $\delta T_{\mu\nu}$ obtained from Eq.~(\ref{eq:Tmunu-kinf}). Introducing the quantity\footnote{If $\cS^{2}<0$, the theory is violently unstable already at the background level and therefore these models do not have physical significance \cite{ArmendarizPicon:1999rj,Bruneton:2007si}.}
\beq
\cS^{2}=\frac{\left(\partial p/\partial X\right)}{\left(\partial \rho/\partial X\right)}
=\frac{\left(\partial p/\partial X\right)}{2X\left(\partial^{2}p/\partial X^{2}\right)-\left(\partial p/\partial X \right)}\,,
\eeq
one can show that the second order perturbed action fit for quantizing scalar perturbations in a universe filled with a $k$-inflaton reads \cite{Garriga:1999vw}
\begin{equation}\label{eq:perturbedaction-kinf}
\delta_{2}S=\frac{1}{2}\int\dd^{4}x\,\left(v'^{2}-\cS^{2}\,v_{,i}v_{,i}+\frac{z''}{z}\,v^{2}\right)\,.
\end{equation}
Here, we have assumed a flat universe $\mathcal{K}=0$ and the perturbations' ``mass'' term $z''/z$ is now calculated with $z=\cS^{-3/2}\,(a\phi'/\calH)$. Eq.~(\ref{eq:perturbedaction-kinf}) suggests interpretation of $\cS$ as the ``speed of sound'' for the perturbations (where in standard perturbation theory $\cS=1$), which is further confirmed by the equation of motion for the modes $v_{k}$ after Fourier decomposition:
\beq\label{eq:modeeq-kinflation}
v_{k}''+\cS^{2}\,k^{2}v_{k}-\frac{z''}{z}\,v_{k}=0
\eeq
It immediately follows that, provided the second term can still be neglected on large scales when $-k\eta\rightarrow0$, the solution of Eq.~(\ref{eq:modeeq-kinflation}) obeys $v_{k}\propto z$ in this limit. It is equally possible to calculate the perturbation spectra in an analogous fashion to the standard case of Section \ref{sec:perturbations}, however, the ``slow roll approximation'' now includes the assumption of an adiabatically varying speed of sound $\cS$. Moreover, the freezing of modes now occurs when a given scale $k$ satisfies $\cS k=aH$ (which can be understood as ``sound horizon crossing''). The derivation of $k$-inflationary power spectra and their indices using the so-called uniform approximation is part of the results presented in Chapter \ref{chapter:kinf-WMAP5-paper}, and will be discussed there, along with its consequences for recent CMB observations. At zeroth order, the amplitude of the scalar perturbation spectrum picks up an additional factor of $1/\cS$,
\beq
k^{3}\calP_{\usssS}=\left(\frac{H^{2}}{\pi\mpl^{2}\epsilon_{1}\cS}\right)_{k=\kstar}\,,
\eeq
where the quantities on the right hand side are evaluated at the sound horizon crossing of the pivot scale $\kstar$. Since tensor perturbations are decoupled from matter at linear order, the tensor modes still obey Eq.~(\ref{eq:vtensor}). Therefore, their spectrum does not change with respect to the standard case, which has an important consequence for the single field consistency condition: it now reads $r=-8\,\cS\,n_{\usssT}$. Recall that in the last Section, we also concluded that multifield scenarios can be distinguished observationally from the standard single field case by their prediction for $r$. A measurement of tensor perturbations therefore promises to discriminate between the minimal inflationary perturbation theory of Chapter \ref{chapter:toolkit} and its extensions presented in this and the previous Section.

We saw in Chapter \ref{chapter:infl-guts} that supersymmetry inspired models of inflation with several scalar fields can have a non-trivial metric in field space $K_{m\bar{n}}$, depending on the form of their K\"a{}hler potential $K(\phi,\bar{\phi})$. In the following, we shall find that multiple fields with non-canonical kinetic terms are also common in string theory. The derivation of inflationary spectra in these scenarios is quite involved and requires the combination of the extensions presented in Sections \ref{sec:multifieldpert} and \ref{sec:kinflationpert}. For recent work in this direction, see \cite{Langlois:2008mn,Langlois:2008wt,Langlois:2008qf} and references therein.

\section{Alternatives to Inflation}\label{sec:alternatives}
Generic predictions of the inflationary scenario (see Section \ref{eq:infl-predictions}) pass all observational tests to date and have made it the major paradigm for the very early Universe. There are however, loose ends to the inflationary story: despite many attempts, it remains difficult to identify a good candidate for the inflaton field from viable particle physics theories. Moreover, in some scenarios the problem of initial conditions rears its ugly head: by design, inflation solves the initial conditions issues behind the SBBM's flatness and horizon problems, but \eg within the class of small field models, fine tuning of the initial field value and velocity of the inflaton can be necessary to achieve enough expansion.\\ 
Most importantly, however, inflation is not a solution to the ``initial singularity problem'': in an expanding universe described by GR and filled with matter sources that obey the weak energy condition, a singularity in the far past is inevitable \cite{Hawking:1973uf,Borde:1993xh}. When approaching this singularity, our description of spacetime must be radically different. An inflationary period may emerge from this (like the SBBM emerges after the end of inflation), but other options can be considered. We conclude this Chapter by briefly citing some alternatives to inflation.

\textsc{String gas cosmology}\\
The string gas scenario for the early Universe \cite{Brandenberger:1988aj,Tseytlin:1991xk,Alexander:2000xv} uses ideas from superstring theory (to be discussed in the following Chapters), which predicts a ten-dimensional spacetime. Six spatial dimensions therefore have to be compactified to recover our four-dimensional Universe, and string gas cosmology resolves the question ``Why three extended spatial dimensions?'' by studying the annihilation of string winding modes around dimensions that are initially curled up with radius $R$.\\
Due to the property of ``T-duality'' (which relates the compactification on radius $R$ with the one of inverse radius $1/R$ and is discussed in Section \ref{subsec:dualities}), strings in thermal equilibrium have a maximum temperature $T_{\mathrm{H}}$, the Hagedorn temperature. Therefore, even when a gas of strings is further and further compressed while we follow the early Universe backwards in time, a temperature singularity is avoided. 
For large $R$, only the momentum string modes contribute to the energy density\footnote{There are three types of modes, momentum, oscillating and winding ones. Momentum modes correspond to degrees of freedom also available to point particles (the center of mass string motion), while oscillatory (corresponding to fluctuations of the string) and winding (wrapped around extra dimensions) modes are intrinsically stringy.}, and they behave like radiation, hence the initial phase of SBBM evolution is recovered. It must be checked whether the flatness and horizon problems are still solved for the resulting universe (which directly passes from the Hagedorn to the radiation phase without an intermediate stage of inflation). Other key issues to address in this framework are the stabilization of moduli fields (see below), and the generation of perturbation spectra \cite{Nayeri:2005ck,Brandenberger:2006vv,Brandenberger:2008nx}.

\textsc{Ekpyrotic / cyclic universes}\\
In an ekpyrotic universe \cite{Khoury:2001wf,Khoury:2001zk}, the horizon problem is addressed by a period of slow contraction in the far past (before the Big Bang). The perturbation spectra are generated during this contraction leading to a ``Big Crunch'', from which the Universe then emerges into the SBBM epoch. (In its extended version of this model, several subsequent contraction and expansion phases are considered, leading to the terminology of a cyclic universe \cite{Steinhardt:2001st}.) These scenarios can be motivated from the collision of branes in string theory (see following Chapters). Again, it is crucial to check whether the resulting perturbation spectra can have the correct (\ie nearly scale invariant) shape to match observations.

\textsc{Pre Big Bang and bouncing universe}\\
In the pre Big Bang scenario, one assumes that the Universe never experienced a singularity, but initially was in a state of finite maximum curvature \cite{Gasperini:2002bn}. Like the ekpyrotic model, this can be motivated from string theory, and it is found that the Universe  underwent a contracting phase in the past, passing through a bounce before it emerged into the present period of expansion \cite{Martin:2003bp,Abramo:2007mp,Peter:2008qz}. The minimum radius of the Universe at the bounce (\ie the maximum curvature) then introduces another length scale into the problem. Perturbation spectra generated before the bounce and propagated through it generically experience a mixing of the pre-bounce $k$ modes in the following expanding phase \cite{Martin:2003sf}.

\newpage
\thispagestyle{empty}
\mbox{}
\newpage
\part{String-Inspired Cosmology}\label{part:string}
\chapter{Elements of String Theory}\label{chapter:stringelements}
\begin{quotation}
\emph{We discussed in Chapter \ref{chapter:infl-guts} that inflation may be connected with unifying theories of high energy physics. A promising contender for a description of all interactions including gravity is string theory, according to which the fundamental constituents of Nature are oscillating strings instead of pointlike particles. In this Chapter, we introduce the basic concepts of string theory and establish its connection with a supersymmetric effective field theory in ten dimensions.}
\end{quotation}
 
String theory originally emerged in the 1960s as a framework for the strong interaction prior to the development of quantum chromodynamics. It was then realized that the massless spin-$2$ state in the string theoretic particle spectrum can be interpreted as the graviton, and therefore string theory could be seen as a quantum theory of gravity. Today five supersymmetric formulations of string theory in ten-dimensional spacetime are known, closely related among themselves by so-called duality transformations. These relations suggest that they are not five distinct theories, but rather different branches on the tree of a single underlying framework.\\
String theory explores the consequences of the bold assumption that, instead of elementary pointlike particles, the Universe is filled with tiny strings of characteristic length $\ells$. Just like vibrations of a violin string produce different sounds, the various particles we observe in Nature then are associated with distinct oscillations of the fundamental string. However, there is only one string type, and therefore all particles are described by the same theory \cite{zwiebach:strings}. The string length $\ells$ is the only (dimensionful) input parameter in string theory. All other (dimensionless) parameters like coupling constants \etc are determined by the vacuum expectation values of scalar fields. This is in sharp contrast to the situation in the Standard Model with its $\order{20}$ adjustable dimensionless parameters, and is commonly taken as a sign that string theory can provide a fundamental (not just an effective) description of physics. Many textbooks discuss string theory from the introductory to the expert level \cite{zwiebach:strings,bailin:susy,binetruy:susy,kiritsis:nutshell,Becker:2007zj,Polchinski:1998rq,Polchinski:1998rr,ortin:strings}.

\section{The Bare Necessities}
\subsection{The Bosonic String}\label{subsec:bosonicstring}
\begin{figure}[t]
\begin{center}
\includegraphics[width=0.6\textwidth]{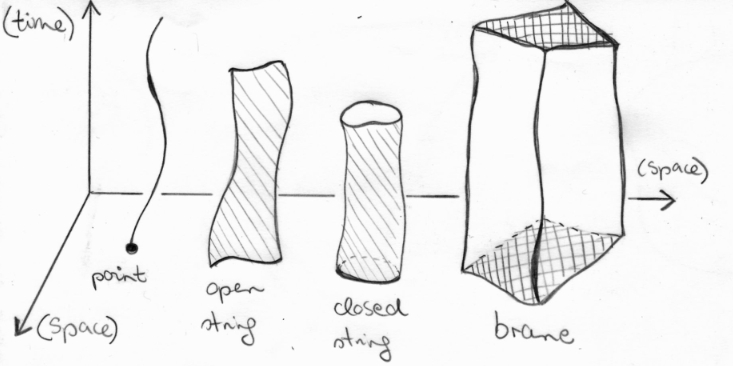}
\hfill
\includegraphics[width=0.3\textwidth]{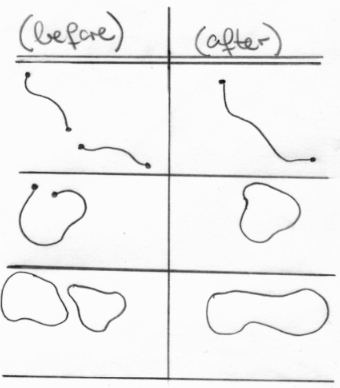}
\caption[World line of particles, world sheets of open and closed strings, and world volume of a $p$-brane. Overview of closed and open string combinatory options.]{\small \emph{Left:} A pointlike particle traces out a $d=1$ trajectory (its ``world line'') in spacetime, while a one-dimensional object like a fundamental string has a $d=2$ ``world sheet''. (Consequently, a closed string has a ``world tube''.) This can be generalized to higher dimensional objects of string theory like $p$-branes (with $p$ spatial dimensions) whose action is proportional to their $(p+1)$ dimensional ``world volume''. \emph{Right:} The ends of open strings can join to either form a longer open string, or (if two ends of the same string meet) a closed string. Therefore, a theory of only open strings cannot be consistent and closed strings must always be included. Closed strings can join to form larger string loops.}
\label{fig:worldlines}
\end{center}
\end{figure}

A relativistic particle following a (one-dimensional) trajectory $T$ in $d$-dimensional spacetime has the degrees of freedom $x^{m}(\xi)$ (where $m=0,1,\dots,d-1$, and $\xi$ is a convenient parameter along $T$) and is described by the action $\mathcal{S}_{\mathrm{particle}}=-m\int_{T}\dd s$. (Recall that we set $c=1$.) The parameter $m$ is the mass of the particle, and $\dd s$ is the line element obtained from the $d$-dimensional generalization of Eq.~(\ref{eq:linelement}). Note that one may write $\dd s$ as $\dd s=\dd \xi\,\sqrt{g_{mn}\left(\dd x^{m}/\dd \xi\right)\left(\dd x^{n}/\dd \xi\right)}$, where we have used $x^{0}=\xi$, the remaining $x^{i}$ (where now $i=1,\dots, d-1$) being spatial coordinates. Hence, for $d=4$ with $x^{\mu}=\left(t,\vec{x}\right)$ in Minkowski space (\ie with the metric $g_{\mu\nu}=\eta_{\mu\nu}$), we clearly obtain the familiar point particle action of Special Relativity, $\mathcal{S}_{\mathrm{particle}}=-m\int_{T}\dd t\, \sqrt{1-\vec{v}^{2}}$.\\
A one-dimensional object like a fundamental string traces out a (two-dimensional) world sheet $\Sigma$ in spacetime (see Fig.~\ref{fig:worldlines}), hence we need two parameters $\left(\xi^{1},\xi^{2}\right)$ to describe it. Suppose the string action is proportional to the area\footnote{As is conventional in string theory, we now use capital letters $X^{m}$ for the spacetime coordinates in Eq.~(\ref{eq:linelement}) since they really are ``mapping functions'' $X^{m}(\xi^{1},\xi^{2})$ \cite{zwiebach:strings}, \ie they describe the spacetime embedding of the string.} of $\Sigma$,
\beq\label{eq:NambuGoto}
\mathcal{S}_{\mathrm{string}}=-T\int_{\Sigma}\dd\xi^{1}\dd\xi^{2}\,\sqrt{-\det\left(g_{mn}\frac{\partial X^{m}}{\partial\xi^{a}}\frac{\partial X^{n}}{\partial\xi^{b}}\right)}\,,
\eeq
where $T$ is a (dimensionful) parameter called the string tension, and the determinant under the square root runs over the coordinates on the world sheet, \ie it may be understood as the ``induced world sheet metric'' $\gamma_{ab}$,
\beq\label{eq:inducedmetric}
\gamma_{ab}=g_{mn}\frac{\partial X^{m}}{\partial\xi^{a}}\frac{\partial X^{n}}{\partial\xi^{b}}\,.
\eeq
This $(2\times2)$ metric is also referred to as the ``pullback'' of the spacetime metric $g_{mn}$ onto the world sheet. Until further notice, we implicitly set $g_{mn}=\eta_{mn}$, with $\eta_{mn}$ the $d$-dimensional Minkowski metric.

The coordinates $X^{m}\left(\xi^{1},\xi^{2}\right)$ are the (bosonic) degrees of freedom of string theory. For definiteness, we fix the world sheet coordinate system as $\xi^{1}=\tau,\,\xi^{2}=\sigma$, where $\sigma$ is a spatial coordinate running along the string and $\tau$ is the string's proper time. Without loss of generality, one may restrict $\sigma$ to the interval $\sigma=[0,2\pi]$, while $\tau$ runs between some initial and final instants of world sheet time, $\tau=[\tau_{i},\tau_{f}]$. Therefore, we may rewrite Eq.~(\ref{eq:NambuGoto}) as
\beq
\mathcal{S}_{\mathrm{string}}=-T\int_{\tau_{i}}^{\tau_{f}}\dd\tau\int_{0}^{2\pi}\dd\sigma\,\sqrt{-\gamma}\,,
\eeq
where $\gamma=\det \gamma_{ab}$. In the above form the string action is called the \emph{Nambu Goto action}. The string tension $T$ (with dimensions of mass per unit length, \ie mass squared in our units) is but a rewriting of the stringy length scale $\ells$,
\beq
T=\frac{1}{2\pi\ells^{2}}=\frac{1}{2\pi\alphas}\,,
\eeq
where $\alphas=\ells^{2}$ is called the \emph{Regge slope}\footnote{Historically, the parameter $\alphas$ described the relation between the angular momentum of a rigidly rotating open string (in units of $\hbar$) and its mass squared, \ie $J/\hbar\propto\alphas\,m^{2}$. Hence, in plots of $J/\hbar$ \emph{vs.} $m^{2}$ used to study hadronic excitations (recall that string theory started out as a theory of the strong interaction), the parameter $\alphas$ measured the slope.}. Quantization of the string action in the form (\ref{eq:NambuGoto}) is difficult. At least at the classical level, there exists the alternative \emph{Polyakov formulation}, where the $(2\times2)$ world sheet metric is used as an auxiliary field $h_{ab}$,
\beq\label{eq:stringaction}
\mathcal{S}_{\mathrm{string}}=-\frac{T}{2}\int_{\tau_{i}}^{\tau_{f}}\dd\tau\int_{0}^{2\pi}\dd\sigma\sqrt{-h}\,h^{ab}g^{mn}\,\partial_{a}X_{m}\,\partial_{b}X_{n}\,.
\eeq
Varying Eq.~(\ref{eq:stringaction}) with respect to $X^{m}$ and $h_{ab}$, we obtain the equations of motion and the world sheet energy momentum tensor $T_{ab}$, respectively:
\begin{eqnarray}
\partial_{a}\left(\sqrt{-h}\,h^{ab}\partial_{b}X^{m}\right)&=&0\label{eq:eofm_X}\\
T_{ab}=-\partial_{a}X^{m}\,\partial_{b}X_{m}+\frac{1}{2}\,h_{ab}h^{cd}\,\partial_{c}X^{m}\,\partial_{d}X_{m}&=&0\label{eq:eofm_wsmetric}
\end{eqnarray}
Note that as an auxiliary field, $h_{ab}$ has no kinetic term in Eq.~(\ref{eq:stringaction}), hence its equation of motion corresponds to the vanishing of the energy momentum tensor, $T_{ab}\propto\delta\mathcal{S}_{\mathrm{string}}/\delta h^{ab}$. By construction, the action (\ref{eq:stringaction}) has Poincar\'{e} invariance in $d$ spacetime dimensions and is invariant under local reparametrizations of $\Sigma$, \ie the world sheet experiences two-dimensional gravity. There is, however, an additional conformal invariance 
on $\Sigma$, corresponding to a coordinate-dependent rescaling 
of $h_{ab}$ (also called Weyl rescaling). Together, world sheet reparametrization and conformal invariance allow us to fix a gauge such that $h_{ab}=\eta_{ab}=\diag(1,-1)$, \ie the world sheet is flat. This is called the ``conformal gauge'', in which the equations of motion Eq.~(\ref{eq:eofm_X}) take a particularly simple form:
\beq\label{eq:stringwave}
\partial_{a}\partial^{a}X^{m}=\left(\frac{\partial^{2}}{\partial\tau^{2}}-\frac{\partial^{2}}{\partial\sigma^{2}}\right)X^{m}=0
\eeq
Note that, since we have already fixed the world sheet metric, the vanishing of the energy momentum tensor (\ref{eq:eofm_wsmetric}) now amounts to constraint equations imposed on the solutions of Eq.~(\ref{eq:stringwave}). Moreover, when we derived Eq.~(\ref{eq:eofm_X}) from $\delta \mathcal{S}_{\mathrm{string}}/\delta X^{m}=0$, we quietly set surface terms to zero, which amounts to boundary conditions for the $X^{m}$. These can, separately for each $m=0,1,\dots d-1$, be of two distinct types:
\begin{itemize}
\item For closed strings, $X^{m}(\tau,\sigma+2\pi)=X^{m}(\tau,\sigma)$, so-called \emph{Dirichlet boundary conditions}.
\item For open strings, $\partial X^{m}/\partial\sigma=0$ at $\sigma=0,\sigma=2\pi$, so-called \emph{Neumann boundary conditions}.
\end{itemize}
Only Neumann boundary conditions allow to conserve Poincar\'{e} invariance for open strings and ensure that no momentum is flowing through the ends of the string. Later, however, it was realized that open strings can also have Dirichlet boundary conditions if they are attached to dynamical objects called D-branes. We come back to this in Section \ref{subsec:Dbranes}.

\subsubsection{Closed strings}\label{subsubsec:closed}
The classical solution of Eq.~(\ref{eq:stringwave}) in the closed string case $X^{m}_\uclosed(\tau,\sigma)$ can always be written as a sum of ``left-movers'' [functions of $(\tau+\sigma)$ only] and ``right-movers'' [functions of $(\tau-\sigma)$ only],
\beq\label{eq:leftandright}
X^{m}_\uclosed(\tau,\sigma)=X^{m}_\uleft(\tau+\sigma)+X^{m}_\uright(\tau-\sigma)\,.
\eeq
As is readily demonstrated in the literature, $X^{m}_\uleft$ and $X^{m}_\uright$ contain a constant and a linear term, plus an infinite tower of oscillatory modes with hermitian coefficients $\alpha_{n}^{m}$ and $\tilde{\alpha}_{n}^{m}$, respectively\footnote{The index $n$ here and in the following runs over the modes, while $m$ still denotes the dimensions of spacetime, $m=0,1,\dots d-1$.}:
\bea
X^{m}_\uleft(\tau+\sigma)&=&\frac{x_{0}^{m}}{2}+\frac{\alphas p^{m}}{2}\,(\tau+\sigma)+i\sqrt{\frac{\alphas}{2}}\sum_{n\neq0}\frac{\alpha_{n}^{m}}{n}\,e^{-in(\tau+\sigma)}\label{eq:closedleft}\\
X^{m}_\uright(\tau-\sigma)&=&\frac{x_{0}^{m}}{2}+\frac{\alphas p^{m}}{2}\,(\tau-\sigma)+i\sqrt{\frac{\alphas}{2}}\sum_{n\neq0}\frac{\tilde{\alpha}_{n}^{m}}{n}\,e^{-in(\tau-\sigma)}\label{eq:closedright}
\eea
With these solutions in hand, we proceed to imposing the constraints of Eq.~(\ref{eq:eofm_wsmetric}) in the case of flat world sheet metric. Given the separation of Eq.~(\ref{eq:leftandright}), it is useful to define components $T_{++}=\frac{1}{2}\left(T_{00}+T_{01}\right)$ and $T_{--}=\frac{1}{2}\left(T_{00}-T_{01}\right)$ of the energy momentum tensor, which depend only on $X^{m}_\uleft$ and $X^{m}_\uright$, respectively. The closed string solutions (\ref{eq:leftandright})-(\ref{eq:closedright}) then must guarantee $T_{++}=T_{--}=0$. (The mixed componentes $T_{+-}$ and $T_{-+}$ vanish identically because the energy momentum tensor is traceless.)

Classically, there is no obstacle to these constraints. However, to obtain a quantum theory, the coefficients $\left(\alpha_{n}^{m},\,\tilde{\alpha}_{n}^{m}\right)$ are promoted to the status of (hermitian) operators with the standard harmonic oscillator commutation relations (see Section \ref{subsec:linearpert}). 
To impose Eq.~(\ref{eq:eofm_wsmetric}), one then uses the Fourier components of $T_{++}$ and $T_{--}$ related to the oscillator creation and annihilation operators as (for now, the index $k\neq0$)
\bea
L_{k}&=&2T\int_{0}^{\pi}\dd\sigma\,e^{ik(\tau-\sigma)}T_{--}=\frac{1}{2}\sum_{n=-\infty}^{\infty}\alpha^{m}_{k-n}\alpha_{m\,n}\,,\label{eq:Lk}\\
\tilde{L}_{k}&=&2T\int_{0}^{\pi}\dd\sigma\,e^{ik(\tau+\sigma)}T_{++}=\frac{1}{2}\sum_{n=-\infty}^{\infty}\tilde{\alpha}^{m}_{k-n}\tilde{\alpha}_{m\,n}\,\label{eq:Lk_tilde}.
\eea
Here, the oscillator operators are hermitian with $\left(\alpha_{n}^{m}\right)^{\dagger}=\alpha_{-n}^{m},\,\left(\tilde{\alpha}_{n}^{m}\right)^{\dagger}=\tilde{\alpha}_{-n}^{m}$. The $L_{k},\tilde{L}_{k}$ constructed from them are called the \emph{Virasoro operators}, and the rationale behind their definition is that $T_{++}=T_{--}=0$ holds true if the physical states $\ket{\varphi}$ of the theory obey\footnote{For negative $k$, using $L_{k}=L_{-k}^{\dagger}$, one still has $\bra{\varphi}L_{n}\ket{\varphi}=0$, meaning that states of zero norm occur but decouple from the positive norm physical states.}
\beq
L_{k}\ket{\varphi}=\tilde{L}_{k}\ket{\varphi}=0,\qquad k>0\,.
\eeq
For $k, k'\neq 0$, one may obtain the \emph{Virasoro algebra} from Eqs.~(\ref{eq:Lk}) and (\ref{eq:Lk_tilde}) and the $\left(\alpha_{n}^{m},\,\tilde{\alpha}_{n}^{m}\right)$ commutation relations to read
\beq\label{eq:Virasoro}
\left[L_{k},L_{k'}\right]=(k-k')L_{k+k'},\qquad\left[\tilde{L}_{k},\tilde{L}_{k'}\right]=(k-k')\tilde{L}_{k+k'},\qquad \left[L_{k},\tilde{L}_{k'}\right]=0\,.
\eeq
For $k=0$, there is an ordering ambiguity in the definitions (\ref{eq:Lk}) and (\ref{eq:Lk_tilde}), and if one defines normal ordered operators as $L_{0}^{\mathrm{(no)}}\equiv\frac{1}{2}\sum_{n=-\infty}^{\infty}:\alpha^{m}_{-n}\alpha_{m\,n}:$ (and likewise for $\tilde{L}_{0}$), this ordering ambiguity is illustrated by a (formally infinite) constant $a$ appearing in the Hamiltonian,
\beq
H=\left(L_{0}^{\mathrm{(no)}}+\tilde{L}_{0}^{\mathrm{(no)}}-2a\right)\,.
\eeq
Likewise, the generalization of the algebra (\ref{eq:Virasoro}) to arbitrary $k,k'$ is
\beq
\left[L_{k},L_{k'}\right]=(k-k')L_{k+k'}+\frac{d}{12}\,k(k^{2}-1)\delta_{k+k',0}\,,
\eeq
where we have simplified the notation setting $L_{0}^{\mathrm{(no)}}:=L_{0}$, and an analogous replacement for the second commutator in Eq.~(\ref{eq:Virasoro}).

Two steps allow to understand the r\^{o}le of $a$ and $d$ and determine their values. Firstly, one can show that the conformal gauge $h_{ab}=\eta_{ab}$ still left some residual gauge freedom, \ie not all of degrees of freedom $X^{m}, m=0,1,\dots d-1$, really are physical. In order to limit ourselves to propagating degrees of freedom only, we define two so-called light cone coordinates from the combinations
\beq\label{eq:lightcone}
X^{\pm}=\frac{1}{\sqrt{2}}\left(X^{0}\pm X^{d-1}\right)\,,
\eeq
while the remaining $X^{i}, i=1,\dots d-2$, are kept as before and shall now be called the transverse degrees of freedom. [When we defined $T_{++}$ and $T_{--}$ before, these were the components of the energy momentum tensor in the light cone gauge of Eq.~(\ref{eq:lightcone}).] The residual gauge freedom permits to fix $X^{+}$ in an essentially trivial way (\ie without oscillator components), $X^{+}=x^{+}_{0}+\alphas p^{+}\tau$. As a consequence, $X^{-}$ may be expressed as a function of the transverse $X^{i}$ only, and we hence only need to find solutions of the type (\ref{eq:leftandright})-(\ref{eq:closedright}) to the remaining $(d-2)$ equations of motion for the $X^{i}$.\\
The bosonic closed string state space can be built from the vacuum $\ket{\Omega,k}$ at given momentum $k$, which is annihilated by both the (transverse) $\alpha_{n}^{i}$ and the $\tilde{\alpha}_{n}^{i}$ (where $n>0$) and for which
\beq
p^{+}\ket{\Omega,k}=k^{+}\ket{\Omega,k},\qquad p^{i}\ket{\Omega,k}=k^{i}\ket{\Omega,k},
\eeq
with the ``light cone momentum'' $p^{+}$ and transverse momentum $\vec{p}$ (whose components are the $p^{i}$). We act on $\ket{\Omega,k}$ with left- and right-mover creation operators $\alpha_{-n}^{i}$ and $\tilde{\alpha}_{-n}^{i}$, which leads to physical states (in the light cone gauge) with masses
\beq\label{eq:closedmass}
\alphas m^{2}_\uclosed=4\left(\sum_{n=1}^{\infty}:\alpha_{-n}^{i}\alpha_{i\,n}:-a\right)\,,\qquad i=1,2,\dots d-2\,.
\eeq
In this expression $a$ plays the r\^ole of the ``vacuum energy'' that arises when normal ordering is carried out, and one has 
\beq\label{eq:aofd}
a=-\frac{d-2}{2}\,\sum_{n=1}^{\infty}n=-\frac{d-2}{2}\,\zeta(-1)=\frac{d-2}{24}\,,
\eeq
where the analytic continuation of the $\zeta$ function, $\zeta(s)=\sum_{n=1}^{\infty}n^{-s}$, to $s=-1$ has been used, with $\zeta(-1)=-1/12$. This procedure is called ``$\zeta$ function regularization'', and it has allowed us to express $a$ in terms of the spacetime dimension $d$ in Eq.~(\ref{eq:aofd}).\\
Secondly, it turns out that $d$ in bosonic string theory is not a free parameter, but must be fixed to avoid quantum anomalies in the commutators. These anomalies would destroy spacetime Lorentz invariance after quantization of the theory, and it can be shown that they are avoided if $d=26$, which consequently gives $a=1$.

We now turn to the mass spectrum predicted by Eq.~(\ref{eq:closedmass}). For the lowest-lying level of the closed bosonic string it follows that $m_\uclosed^{2}<0$. We come back to the r\^ole of this closed string tachyon (and its open string analogue) below. For now let us note that as a consequence of shift invariance along the $\sigma$ direction on the ``world tube'' (see Fig.~\ref{fig:worldlines}) left- and right-movers give equal contributions to the mass, $m_\uleft^{2}=m_\uright^{2}$, and hence states with $m_\uclosed=0$ are only obtained if both $m_\uleft$ and $m_\uright$ are zero. (These massless states are of special interest since they appear in the effective supergravity description of string theory, see below.) A general massless state may be written as
\beq\label{eq:masslessstate}
\sum_{ij}R_{ij}\,\alpha_{-1}^{i\,\dagger}\tilde{\alpha}_{-1}^{j\,\dagger}\ket{\Omega,k}\,.
\eeq
The matrix $[R]_{ij}$ may be decomposed into a symmetric and traceless part $[S]_{ij}$, an anti-symmetric part $[A]_{ij}$ and a trace $t$ multiplied by the identity matrix $\mathbb{I}$. The states described by Eq.~(\ref{eq:masslessstate}) accordingly belong to three different categories: 
$[S]_{ij}$ gives rise to the spin-2 field $h_{ij}$, which is interpreted as the $d$-dimensional graviton $g_{mn}$. Note that this state has emerged without putting in any spacetime gravity at the classical level. 
$[A]_{ij}$ describes the \emph{Kalb Ramond anti-symmetric tensor field} $B_{mn}$ in $d$ dimensions. This two-index field $B_{mn}$ may be thought of as the stringy generalization of the four-vector potential $A_{\mu}$ in electrodynamics, which couples to the world line element $\dd x^{\mu}$. (Indeed, we shall see that $B_{mn}$ couples to the two-dimensional world sheet element $\dd\tau\,\dd\sigma$ in an analogous way.) 
The trace $t$ is a scalar degree of freedom called the \emph{dilaton} $\Phi$. 
We come back to these two companion states $(B_{mn},\Phi)$ of the graviton below.

In distinguishing left- and right-movers for the closed string, we have assumed that its (tube-like) world sheet is oriented. However, closed strings can be made unoriented using the world sheet parity operation $\Omega$, which takes $\Omega:\sigma\rightarrow 2\pi-\sigma,\,\tau\rightarrow\tau$ (\ie it exchanges the left- and right-moving modes). In a theory of unoriented closed strings, only states invariant under $\Omega$ are kept. In the case of the massless degrees of freedom, the graviton as well as the dilaton survive in the unoriented theory, while the Kalb Ramond tensor field is projected out.

\subsubsection{Open strings}\label{subsubsec:open}
For open strings, due to the Neumann boundary conditions, left- and right-moving parts in $X^{m}_\uopen(\tau,\sigma)$ are not independent, hence there is no separation like in Eq.~(\ref{eq:leftandright}), and only one set of oscillator coefficients $\alpha_{n}^{m}$ is needed. The mode expansion replacing Eqs.~(\ref{eq:closedleft}) and (\ref{eq:closedright}) in this case reads
\beq
X^{m}_\uopen(\tau,\sigma)=x_{0}^{m}+2\alphas p^{m}\tau+i\sqrt{2\alphas}\sum_{n\neq0}\frac{\alpha^{m}_{n}}{n}\,e^{-in\tau}\,\cos(n\sigma)\,.
\eeq
A similar quantization procedure involving the definition of (one set of) Virasoro operators goes through, and notably anomalies are again avoided in $d=26$ dimensions. Using the light cone gauge, one may again obtain a mass formula
\beq\label{eq:openmass}
\alphas m^{2}_\uopen=\sum_{n=1}^{\infty}:\alpha_{-n}^{i}\alpha_{i\,n}:-1\,.
\eeq
Once more we postpone the discussion of the tachyonic ground state for the moment. In this case, the first excited state $\alpha^{i}_{-1}\ket{\Omega,k}$ gives a massless vector field with 24 degrees of freedom. (We do not pay attention to the higher-lying massive states since they do not enter in the effective field theory description we shall use eventually.) 
It is intuitively clear that open strings also have an orientation since their endpoints are distinct. A world sheet parity transformation $\Omega$ therefore identifies the endpoints, making the open string unoriented. 

To summarize, the bosonic string action (\ref{eq:stringaction}) describes the coupling of $d=26$ scalar fields $X^{m}(\tau, \sigma)$ to two-dimensional gravity on the world sheet. The resulting particle spectrum after quantization contains only spacetime bosons, and among the low-lying (closed string) states, we encountered the graviton along with the Kalb Ramond field and the dilaton. However, both the spectra of the closed and of the open string contain a state of negative mass. While the open string tachyon has been understood in terms of the decay of D-branes (see Section \ref{subsec:Dbranes}), the closed string tachyon remains pathologic. Moreover, we cannot be satisfied with a theory of only bosons, given that matter in the real world is made of fermions. This leads to the introduction of superstrings, which we now discuss.

\subsection{The Superstring}\label{subsec:superstring}
In Chapter \ref{chapter:infl-guts}, we gave arguments in favour of an underlying ``supersymmetry'' between bosons and fermions in the realm of GUT energies, though this symmetry must be broken in the Universe at present. Since string theory is aimed at describing physics beyond the Standard Model, supersymmetry may be part of it. This suggests the introduction of word sheet spinors $\Psi^{m}(\tau, \sigma)$ as superpartners for the $X^{m}(\tau, \sigma)$ coordinates in a supermultiplet\footnote{In terms of an expansion in component fields analogous to Eq.~(\ref{eq:chiralmultiplet}), this superfield reads
\beq\label{eq:superstringsuperfield}
Y^{m}(\sigma^{\alpha},\theta)=X^{m}(\sigma^{\alpha})+\bar{\theta}\Psi^{m}(\sigma^{\alpha})+\frac{1}{2}\,\bar{\theta}\theta\,B^{m}(\sigma^{\alpha})\,,
\eeq
where $\sigma^{\alpha}=(\tau,\sigma)$ and the $(\bar{\theta},\theta)$ are the superspace coordinates on the world sheet. The field $B^{\mu}$ is again an auxiliary field that allows to close the world sheet supersymmetry algebra off-shell.}. Both fermions and bosons should then be coupled to two-dimensional (super-)gravity on the world sheet. Note that the theory obtained in such a way has world sheet supersymmetry by construction. Spacetime supersymmetry is only obtained after a suitable projection (see below), and is never explicit in the formalism (called \emph{Ramond Neveu Schwarz}) presented here. Spacetime supersymmetry can, however, be made manifest using the alternative \emph{Green Schwarz formalism}.

A reasonable starting point is the action
\beq\label{eq:superaction}
\mathcal{S}_{\mathrm{superstring}}^{(0)}=-\frac{1}{2\pi}\int\dd\xi^{1}\dd\xi^{2}\,\sqrt{-h}\left(h^{ab}\partial_{a}X^{m}\,\partial_{b}X_{m}+i\Psi^{m}\rho^{a}\partial_{a}\Psi_{m}\right)\,,
\eeq
where the matrices $(\rho^{0},\rho^{1})$ are the $\gamma$-matrices in two dimensions with $\{\rho^{a},\rho^{b}\}=2\eta^{ab}\mathbb{I}$. Indeed, this action has the familiar world sheet reparametrization invariance from before\footnote{In terms of the superstring superfield of Eq.~(\ref{eq:superstringsuperfield}), this action may be written as
\beq
\mathcal{S}_{\mathrm{superstring}}^{(0)}=\frac{i}{4\pi}\int\dd^{2}\sigma\,\dd^{2}\theta\,\bar{D}^{A}Y^{m}\,D_{A}Y_{m},\qquad D_{A}=\frac{\partial}{\partial\bar{\theta}_{A}}+\left(\rho^{\alpha}\theta\right)_{A}\partial_{\alpha}\,.
\eeq
The auxiliary field $B^{\mu}$ is found to obey the equation of motion $B^{\mu}=0$ and can be eliminated.}, plus an additional (on-shell) global world sheet supersymmetry relating $\Psi^{m}$ and $X^{m}$. One can show that the commutator of two successive such transformations acts as a world sheet translation on $\Psi^{m}$ and $X^{m}$, respectively. However, when supersymmetry is made local by taking the infinitesimal two-component Majorana spinor parameter $\zeta$ as world sheet dependent, $\zeta\rightarrow\zeta(\tau, \sigma)$, a supercurrent appears when varying the action $\mathcal{S}_{\mathrm{superstring}}^{(0)}$ of Eq.~(\ref{eq:superaction}). It may be cancelled by introducing a two-dimensional ``gravitino'' $\chi_{a}$ (the superpartner of the graviton, with appropriate transformation properties),
\beq\label{eq:superS1}
\mathcal{S}_{\mathrm{superstring}}^{(1)}=-\frac{1}{\pi}\int\dd\xi^{1}\dd\xi^{2}\,\sqrt{-h}\,\bar{\chi}_{a}\rho^{b}\rho^{a}\Psi^{m}\,\partial_{b}X_{m}\,.
\eeq
However, to make the total superstring action invariant under local supersymmetry transformations at first order, still one more term is required:
\beq\label{eq:superS2}
\mathcal{S}_{\mathrm{superstring}}^{(2)}=-\frac{1}{\pi}\int\dd\xi^{1}\dd\xi^{2}\,\sqrt{-h}\,\bar{\Psi}_{m}\Psi^{m}\bar{\chi}_{a}\rho^{b}\rho^{a}\chi_{b}\,
\eeq
Then, the sum of Eqs.~(\ref{eq:superaction}), (\ref{eq:superS1}) and (\ref{eq:superS2}) is invariant with respect to local world sheet supersymmetry transformations at first order, \ie we have introduced two-dimensional supergravity on the world sheet.

The conformal invariance that, together with world sheet reparametrizations, allowed us to choose a flat world sheet metric survives the generalization from bosonic to superstrings and we may still take $h_{ab}=\eta_{ab}$. What is more, the two-dimensional world sheet supersymmetry together with an additonal ``superconformal'' invariance of the superstring action may be used to set the gravitino to zero, $\chi_{a}=0$. This is the analogue of the conformal gauge for the superstring, in which the equations of motion for $\Psi^{m}$ and $X^{m}$ simply read
\beq\label{eq:eofm_super}
\partial_{a}\partial^{a}X^{m}=0\,,\qquad i\rho^{a}\partial_{a}\Psi^{m}=0\,.
\eeq
Again, the remaining equations of motion (for $h_{ab}$ and $\chi_{a}$) turn into the constraint equations $T_{ab}=0,\,J^{a}=0$ once the supercovariant gauge choice is made, where the energy momentum tensor 
 $T_{ab}$ and the supercurrent $J^{a}$ are given by
\bea
T_{ab}&=&-\partial_{a}X^{m}\,\partial_{b}X_{m}-\frac{i}{4}\,\bar{\Psi}^{m}\left(\rho_{a}\partial_{b}+\rho_{b}\partial_{a}\right)\Psi_{m}\nonumber\\
&&+\frac{\eta_{ab}}{2}\left(\partial^{g}X^{m}\,\partial_{g}X_{m}+\frac{i}{2}\,\bar{\Psi}^{m}\rho^{g}\partial_{g}\Psi_{m}\right)\,,\label{eq:superTab}\\
J^{a}&=&\frac{1}{2}\,\rho^{b}\rho^{a}\Psi^{m}\,\partial_{b}X_{m}\,.\label{eq:supercurrent}
\eea
Boundary conditions for the solutions of Eqs.~(\ref{eq:eofm_super}) are again obtained from the requirement of vanishing surface terms when varying the superstring action. For the bosonic degrees of freedom $X^{m}$, these are unchanged from before. For the newly introduced spinors, note that one may write
\beq\label{eq:Psiansatz}
\Psi^{m}=\left(\begin{matrix}\Psi^{m}_\uright\\\Psi^{m}_\uleft\end{matrix}\right),\qquad\Psi^{m}_\uright=\Psi^{m}_\uright(\tau-\sigma),\,\Psi^{m}_\uleft=\Psi^{m}_\uleft(\tau+\sigma)\,,
\eeq
where for now the separation into left- and right movers is a tool: using the ansatz on the left in Eq.~(\ref{eq:Psiansatz}) in the equation of motion (\ref{eq:eofm_super}), we see that Eq.~(\ref{eq:eofm_super}) reduces to
\beq
\left(\frac{\partial}{\partial\tau}+\frac{\partial}{\partial\sigma}\right)\Psi^{\mu}_\uright=0\,,\qquad \left(\frac{\partial}{\partial\tau}-\frac{\partial}{\partial\sigma}\right)\Psi^{\mu}_\uleft=0\,.
\eeq
This justifies the expressions on the right in Eq.~(\ref{eq:Psiansatz}). 
There are then also the same two boundary choices, \ie open or closed fermionic strings.

For closed strings, one may impose periodic (referred to as \emph{Ramond}) and anti-periodic (referred to as \emph{Neveu Schwarz}) boundary conditions on $\Psi^{m}_\uleft$ and $\Psi^{m}_\uright$ separately, \ie choose one sign each in
\bea
\Psi^{m}_\uleft(\tau, \sigma)&=&\pm\Psi^{m}_\uleft(\tau, \sigma+2\pi)\,,\\
\Psi^{m}_\uright(\tau, \sigma)&=&\pm\Psi^{m}_\uright(\tau, \sigma+2\pi)\,.
\eea
Hence, there are four possible combinations of boundary conditions for closed fermionic strings, and the state space can be divided accordingly into the Ramond-Ramond (R-R), Ramond-Neveu Schwarz (R-NS), Neveu Schwarz-Ramond (NS-R) and the Neveu Schwarz-Neveu Schwarz (NS-NS) sectors. It can be shown that in spacetime, the R-R and NS-NS sectors give rise to bosons eventually, while the mixed sectors contain spacetime fermions.\\
For open strings, the boundary conditions are
\bea
\Psi^{m}_\uleft(\tau, \sigma=0)&=&\Psi^{m}_\uright(\tau, \sigma=0)\,\\
\Psi^{m}_\uright(\tau, \sigma=2\pi)&=&\pm\Psi^{m}_\uright(\tau, \sigma=2\pi)\,.\label{eq:otherend}
\eea
The overall relative sign in the first line, \ie for $\sigma=0$, is  a matter of convention, but the relative sign at the other end (the second line, $\sigma=2\pi$) is meaningful. If the ``$+$'' sign is chosen (periodic, \ie Ramond boundary conditions), it can be shown that the resulting states are spacetime fermions. If one opts for relative ``$-$'' sign at the other end (that is, anti-periodic or Neveu Schwarz boundary conditions), one recovers spacetime bosons.

The equations (\ref{eq:eofm_super}) may then again be solved using an oscillator expansion and imposing the above boundary conditions. As an example, let us cite the mode expansion for the right-moving part of the closed superstring with Ramond and Neveu Schwarz boundary conditions, respectively:
\beq
\left[\Psi^{m}_\uright\right]_{\mathrm{R}}=\sum_{k\,\in\,\mathbb{Z}}\gamma^{m}_{k}\,e^{-ik(\tau-\sigma)}\,,\qquad\left[\Psi^{m}_\uright\right]_{\mathrm{NS}}=\sum_{r\,\in\,\mathbb{Z}+1/2}\beta^{m}_{r}\,e^{-ir(\tau-\sigma)}
\eeq
Note that $k$ in the left expression runs over all integers, while the index $r$ on the right runs over all half-integers. For the left-movers, corresponding expansion coefficients $(\tilde{\gamma}^{m}_{k},\,\tilde{\beta}^{m}_{r})$ are introduced. Once promoted to operators, these coefficients now satisfy \emph{anti-}commutation conditions
\bea
\left\{\gamma^{m}_{k},\,\gamma^{n}_{l}\right\}=\left\{\tilde{\gamma}^{m}_{k},\,\tilde{\gamma}^{n}_{l}\right\}=\delta_{k+l,0}\,\eta^{mn}\,,\\
\left\{\beta^{m}_{r},\,\beta^{n}_{s}\right\}=\left\{\tilde{\beta}^{m}_{r},\,\tilde{\beta}^{n}_{s}\right\}=\delta_{r+s,0}\,\eta^{mn}\,,
\eea
for the Ramond and the Neveu Schwarz cases, respectively. The (half-)integer character of the indices $(r,s)$ and $(k,l)$, respectively, suggests that, as mentioned above, states from the R-R and NS-NS sectors are spacetime bosons, while fermions arise from R-NS and NS-R. For the open string, left- and right movers are not independent (as it was the case for the bosonic string), and therefore one set each of $\gamma^{m}_{k}$ (Ramond) and $\beta^{m}_{r}$ (Neveu Schwarz) suffices.\\
In our analysis of both the bosonic and the fermionic string, we encountered the property that closed strings --due to their subdivision into left-movers and right-movers-- required two copies of the oscillators $\alpha_{n}^{m}$ and $(\beta_{r}^{m},\gamma_{k}^{m})$, respectively, as opposed to one copy for open strings. In fact, one may think of the state space of closed strings as a direct product of left- and right-movers, each of which has the same structure as the open string spectrum. We therefore restrict our considerations to open superstrings with operators $\alpha_{n}^{m}$ for the bosonic $X^{m}$ and $(\beta_{r}^{m},\gamma_{k}^{m})$ for the fermionic $\Psi^{m}$ (with its two sectors R and NS) in the rest of this Section.

To impose the resulting constraints of Eqs.~(\ref{eq:superTab})-(\ref{eq:supercurrent}), it is again useful to introduce Virasoro operators as the Fourier components of $T_{++},\,T_{--}$ and $J_{+},\,J_{-}$, respectively\footnote{The latter are defined as $J_{+}=\Psi_\uleft^{m}\,\partial_{+}X_{m\,\uleft}$ and $J_{-}=\Psi_\uright^{m}\,\partial_{-}X_{m\,\uright}$, where $\partial_{\pm}=\partial_{\tau}\pm\partial_{\sigma}$.}. The superconformal invariance on the world sheet then demands that these components as well as the supercurrents vanish, $T_{++}=T_{--}=J_{+}=J_{-}=0$. The details are readily available in the literature \cite{bailin:susy,Becker:2007zj}. It is again possible to prohibit ghosts (\ie states of negative norm) by postulating conditions for physical states using the super-Virasoro operators. The fermionic analogues ($a_{\mathrm{R}}$ and $a_{\mathrm{NS}}$, one for each sector) of the bosonic ordering constant $a$ may be expressed as functions of the spacetime dimension $d$. From the consideration of quantum anomalies, it follows in this case that Lorentz invariance is preserved if $d=10$ (and $a_{\mathrm{R}}=0$, $a_{\mathrm{NS}}=1/2$). Superstring theories  therefore live in ten dimensions. Depending on the boundary conditions (\ie in its different sectors introduced above), they can describe spacetime fermions or bosons. The construction of the spectrum from the NS sector vacuum state $\ket{\Omega,k}_{\mathrm{NS}}$ (which is a scalar and annihilated by both $\alpha_{n}^{i}$ and $\beta_{r}^{i}$ for $n,r>0$) and $\ket{\Omega,k}_{\mathrm{R}}$ in the R sector (annihilated by $\alpha_{n}^{i}$ and $\gamma_{r}^{i}$ for $n,r>0$ and of spinor type) is discussed in detail \eg in \cite{Becker:2007zj}. 
Here, we state the resulting mass formulas in the light cone gauge, \ie using the transverse oscillators $i$ only (normal ordering is implied):
\bea
\alphap m^{2}_{\mathrm{NS}}&=&\sum_{n=1}^{\infty}\alpha_{-n}^{i}\alpha_{i\,n}+\sum_{r=1/2}^{\infty}r\beta_{-r}^{i}\beta_{i\,r}-\frac{1}{2}\label{eq:massNS}\\
\alphap m^{2}_{\mathrm{R}}&=&\sum_{n=1}^{\infty}\alpha_{-n}^{i}\alpha_{i\,n}+\sum_{n=1}^{\infty}n\gamma_{-n}^{i}\gamma_{i\,n}\label{eq:massR}
\eea
In the NS sector, the lowest-lying state again is a tachyon, as we see from Eq.~(\ref{eq:massNS}). Note that the index $r$ is half-integer valued and therefore the first excited (massless) state is obtained by acting on $\ket{\Omega,k}_{\mathrm{NS}}$ with $\beta_{-1/2}^{i}$. It is a spacetime vector because it arises from a vector operator acting on a scalar ground state. (The bosonic oscillators $\alpha_{-n}^{i}$ only contribute to the higher excitations we do not consider.) The lowest state in the R sector is a massless spinor.

\subsubsection{The GSO projection}
We saw that a Lorentz invariant, ghost free superstring theory of open (and closed, see below) strings can be formulated in ten dimensions. The resulting particle spectrum contains both spacetime bosons and fermions. There are, however, two points we glossed over so far: firstly, supersymmetry on the world sheet does not imply spacetime supersymmetry. Secondly, as is readily seen from Eqs.~(\ref{eq:closedmass}), (\ref{eq:openmass}) and (\ref{eq:massNS}), massless states are not the lowest ones in the spectrum: both bosonic and superstring theory contain tachyonic states of negative mass, which means that their vacuum state is unstable. 
The persistent closed string tachyon lead us to abandon bosonic string theory, but luckily the situation is different for the superstring: it turns out that the superstring NS tachyon may be removed, and moreover spacetime supersymmetry in ten dimensions achieved, by applying a projection due to \emph{Gliozzi, Scherk and Olive} (GSO).\\
The GSO projection operators for the NS and the R sectors are defined from
\beq
P_{\mathrm{NS}}=\frac{1}{2}\left[1+\left(-1\right)^{F_{\mathrm{NS}}+1}\right]\,,\qquad P_{\mathrm{R}}=\frac{1}{2}\left[1+\eta\left(-1\right)^{F_{\mathrm{R}}+1}\right]\,,\label{eq:P_R}
\eeq
where $F_{\mathrm{NS}}=\sum_{r=1/2}^{\infty}\beta_{-r}^{i}\beta_{i\,r}$ counts the $\beta$ oscillator excitations (\ie it is the world sheet fermion number). Likewise, $F_{\mathrm{R}}=\sum_{n=1}^{\infty}\gamma_{-n}^{i}\gamma_{i\,n}$ in the Ramond sector. The additional parameter $\eta$ in the Ramond operator $P_{\mathrm{R}}$ can be chosen $\pm1$, depending on the chirality of the spinor ground state $\ket{\Omega,k}_{\mathrm{R}}$. In the NS sector, the GSO projection therefore removes all states with an even number of world sheet fermions, and notably the NS tachyonic ground state is projected out. In the R sector it is a matter of convention whether an even or odd number of $\gamma$ oscillations are kept, and the two choices give rise to different superstring theories, as we discuss below. It can be shown that after the GSO projection, the number of fermionic and bosonic degrees of freedom is equal at each excitation level. This is necessary for, but not a proof of spacetime supersymmetry. One may, however, show explicitly that the theory is supersymmetric in ten spacetime dimensions after the GSO projection if the alternative Green Schwarz formalism is employed (see \eg \cite{Becker:2007zj}).

\subsection{Five Theories}\label{subsec:five}
The last part of the above discussion was carried out for open superstrings only. However, the endpoints of open strings can meet and form a closed string (see Fig.~\ref{fig:worldlines}), which is why a consistent theory cannot be built from open strings (and their R and NS sectors) alone. Combining two copies of the open string state space (one for the left- and one for the right-movers), we can construct the closed string spectrum with its R-R, R-NS, NS-R and NS-NS sectors. Note that in Eq.~(\ref{eq:P_R}) we have the choice of either projecting onto states with odd or even $F_{\mathrm{R}}$ (depending on the chirality of the ground state, which we write $\ket{+}_{\mathrm{R}}$ or $\ket{-}_{\mathrm{R}}$, respectively). This choice can be made independently for left- and right-movers, and their R sectors then have the same or opposite chirality. The former choice gives rise to the so-called type IIB superstring theory\footnote{The chirality in this case is chosen positive for definiteness.}, while the latter defines type IIA superstring theory.\\
This distinction along with two other criteria --whether a theory contains open strings on top of closed ones, and whether its fundamental strings are oriented\footnote{Recall that in Section \ref{subsec:bosonicstring} we the notion of a world sheet parity transformation $\Omega$ that turns oriented into unoriented strings.} or not-- allows the identification of five distinct superstring theories referred to as type I, the above-mentioned types IIA and IIB, and the heterotic $SO(32)$ and $E_{8}\times E_{8}$ theories.

\textsc{Theories of oriented closed superstrings: type IIA and type IIB}\\
Their spectra are obtained by taking the direct product of one open string spectrum each for the left and the right movers of the closed string, giving rise to the four sectors R-R, R-NS, NS-R and NS-NS. As before, left-mover operators carry a tilde, and since the effective field theory we are ultimately interested in only contains the massless states, and we restrict ourselves to these here. The state space then is summarized as \cite{Becker:2007zj}:

\begin{center}
\begin{tabular}{p{2cm}p{5cm}p{5cm}}
&\textsc{type IIA}&\textsc{type IIB}\\
&&\\
R-R&$\ket{-}_{\mathrm{R}}\otimes\ket{+}_{\mathrm{R}}$&$\ket{+}_{\mathrm{R}}\otimes\ket{+}_{\mathrm{R}}$\\
&&\\
R-NS&$\ket{-}_{\mathrm{R}}\otimes\beta^{i}_{-1/2}\ket{\Omega}_{\mathrm{NS}}$&$\ket{+}_{\mathrm{R}}\otimes\beta^{i}_{-1/2}\ket{\Omega}_{\mathrm{NS}}$\\
&&\\
NS-R&$\tilde{\beta}^{i}_{-1/2}\ket{\Omega}_{\mathrm{NS}}\otimes\ket{+}_{\mathrm{R}}$&$\tilde{\beta}^{i}_{-1/2}\ket{\Omega}_{\mathrm{NS}}\otimes\ket{+}_{\mathrm{R}}$\\
&&\\
NS-NS&$\tilde{\beta}^{i}_{-1/2}\ket{\Omega}_{\mathrm{NS}}\otimes\beta^{j}_{-1/2}\ket{\Omega}_{\mathrm{NS}}$&$\tilde{\beta}^{i}_{-1/2}\ket{\Omega}_{\mathrm{NS}}\otimes\beta^{j}_{-1/2}\ket{\Omega}_{\mathrm{NS}}$
\end{tabular}
\end{center}
Note that there are two chiralities in the type IIA case, but only one in type IIB. We now list the fields in the different sectors.
\begin{itemize}
\item {\bf NS-NS sector}\\
It is easily seen from the above table that this sector is shared by both theories. Its massless states are bosonic and give the symmetric graviton $g_{mn}$, the anti-symmetric Kalb Ramond tensor gauge field $B_{mn}$ and the dilaton $\Phi$.
\item {\bf NS-R and R-NS sectors}\\
Note that the first of these is again the same for both theories, but in the R-NS sectors the chiralities are different. The (fermionic) fields in these sectors are a spin-3/2 ``gravitino'' and a spin-1/2 ``dilatino'' each. The presence of two gravitinos means that type IIA and type IIB superstring theory have $\calN=2$ supersymmetry in ten dimensions. In type IIA these gravitinos have opposite chirality.
\item {\bf R-R sector}\\
This sector again contains bosons. In the type IIA case, one obtains a $U(1)$ gauge field $A_{m}$ and a three-form $A_{lmn}$. For type IIB, there is a zero-form (\ie a scalar) $C_{0}$, a two-form $C_{mn}$ and a four-form field $C_{klmn}$. Note that each of these fields (with $n$ indices) is associated with a field strength $F_{(n+1)}^{\mathrm{IIA}}=\dd A_{(n)}\,(n=1,3)$, or $F_{(n+1)}^{\mathrm{IIB}}=\dd C_{(n)}\,(n=0,2,4)$, respectively, and they are summarily referred to as the ``R-R gauge potentials''. We shall see later that the number of indices of the fields present determines the dimensionality $(p+1)$ of each theory's stable D$p$-branes.
\end{itemize}
Given their state spectrum, both type IIA and IIB have $\calN=2$ supergravity in ten dimensions, but type IIB is non-chiral on the world sheet. (Another way to see this is that the world sheet parity transformation $\Omega$ is a symmetry of type IIB because $\Omega$ exchanges left- and right-movers, and in type IIB they have the same chirality.) Our focus in later Chapters is on cosmological models living in type II theories, therefore our account of the remaining three superstring theories below is shorter.

\textsc{Theory of open and closed unoriented superstrings: type I}\\
In this theory, unoriented open strings are combined with an ``unoriented'' version of type IIB theory of closed strings, \ie the world sheet parity symmetry of type IIB is gauged away. The NS-R and R-NS sectors transform into each other under the $\Omega$ operation, and only one copy remains in the unoriented theory. It can be shown that type I has the field content of a chiral $\calN=1$ supergravity multiplet: the dilaton $\Phi$, the graviton $g_{mn}$ with its gravitino, and an anti-symmetric tensor field $C_{mn}$ [as well as the gauge boson and the gaugino of a $\calN=1$ super Yang Mills theory of $SO(32)$]. The Kalb Ramond field $B_{mn}$ is absent from the spectrum because it does not survive the ``unorientation'' procedure.

\textsc{Theories of heterotic closed oriented strings: heterotic $SO(32)$ and $E_{8}\times E_{8}$}\\
In these theories, the decoupling of left- and right-movers is used to make only the latter ones supersymmetric, while the left-moving degrees of freedom are described by bosonic string theory. However, the bosonic string is consistent with Lorentz invariance in 26 dimensions, while $d=10$ for the superstring. Hence, 16 of the left-movers must be compactified, and this can only be done on a special type of internal lattice. The two heterotic theories differ by the choice of lattice, which can either be $E_{8}\times E_{8}$ or $SO(32)$. When the supersymmetric right-movers and the bosonic, compactified left-movers are combined, one obtains $\calN=1$ supergravity in ten spacetime dimensions with the dilaton $\Phi$, the graviton $g_{mn}$ plus gravitino, the anti-symmetric $B_{mn}$ [again together with $\calN=1$ super Yang Mills theory of either $E_{8}\times E_{8}$ or $SO(32)$].

\begin{figure}[t]
\begin{center}
\includegraphics[width=0.65\textwidth]{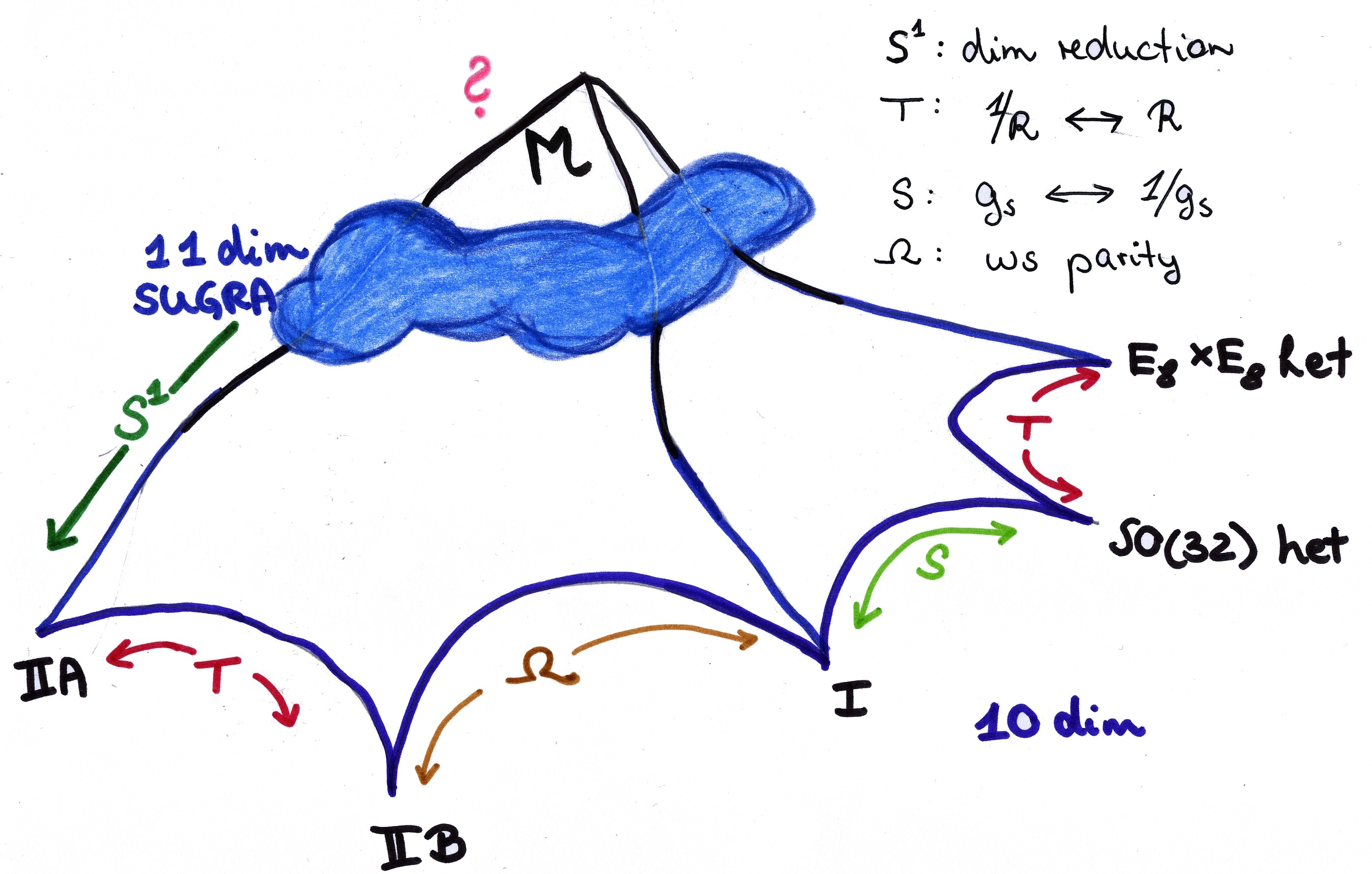}
\hfill
\begin{minipage}[b]{0.3\textwidth}
\includegraphics[width=\textwidth]{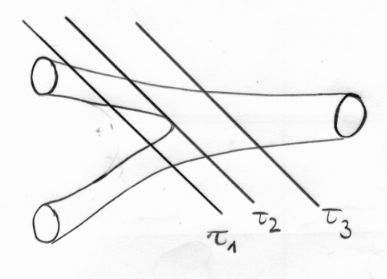}\\
\includegraphics[width=\textwidth]{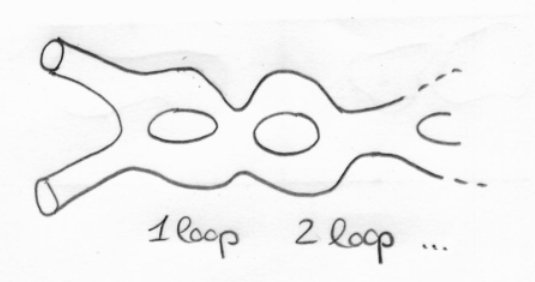}
\end{minipage}
\caption[The web of dualities for superstring theories, and the ``mountain'' of M-theory with its eleven-dimensional supergravity limit at low energies. Feynman diagrams for closed strings. (Following examples from \cite{PhD_Anke}, \cite{bailin:susy} and \cite{Becker:2007zj}, respectively.)]{\small \emph{Left:} The five different superstring theories (type IIA and IIB, type I, heterotic $SO(32)$ and heterotic $E_{8}\times E_{8}$) are related by a web of dualities: these are T- and $S$-duality (exchanging compactification radii $R\rightarrow 1/R$ and the dilaton $\Phi\rightarrow-\Phi$, see Section \ref{subsec:dualities}), and the ``unorientation'' procedure $\Omega$. By compactification on a circle $S^{1}$, type IIA superstring theory in ten dimensions can be obtained from eleven-dimensional supergravity, which is strongly believed to be the low-energy limit of M-theory. \emph{Top right:} Closed string Feynman diagram at tree order: because of the extended nature of the strings, interactions no longer take place at a fixed point in spacetime, and the theory is therefore UV finite. The lines are ``cuts'' of the tree level diagram on different time hypersurfaces $\tau_{1}<\tau_{2}<\tau_{3}$. \emph{Bottom right:} Closed string Feynman diagram at loop order: for closed oriented surfaces, the number of handles counts the number of loops.}
\label{fig:stringdualities}
\end{center}
\end{figure}

These five different formulations of superstring theory are related among each other by a close web of so-called dualities, which is sketched in Fig.~\ref{fig:stringdualities} and explored further in Section \ref{subsec:dualities}. As a consequence, it is strongly believed that they are but distinct realizations of one underlying theory, currently dubbed ``M-theory''. In contrast to superstring theories, M-theory lives in eleven dimensions, and its fundamental objects are not strings, but membranes. M-theory is still very much under development, but it is known that in its low energy limit, it most likely reduces to $\calN=1$ supergravity for $d=11$. We do not venture further into the realm of M-theory here, but we shall use its eleven-dimensional supergravity description shortly when establishing the effective $d=10$ supersymmetric action of type IIA. First, however, we pause for a brief discussion of the dilaton field $\Phi$.

\section{The Special R\^ole of the Dilaton}\label{sec:dilaton}
Taking closer look at the dilaton $\Phi$, one can appreciate a remarkable property of string theory as opposed to the quantum field theory of point particles: in the latter, \emph{many} Feynman diagrams have to be calculated at each loop order for a given interaction process, and a cutoff $\Lambda_\uUV$ has to be imposed on the momenta running around the loop to make the theory finite in the UV. In string theory, on the other hand, there is a single diagram at any loop order $n_{\mathrm{loop}}$, and there are no UV divergencies. An illustrative explanation of this is given in Fig.~\ref{fig:stringdualities}.

Let us Euclideanize the (so far) Lorentzian world sheet metric $h_{ab}$ by taking $\tau\rightarrow -i\tau$, and define complex coordinates $(z,\bar{z})$ in local patches by
\beq\label{eq:definez}
z=e^{\tau-i\sigma},\qquad \bar{z}=e^{\tau+i\sigma}\,.
\eeq
The world sheet can now be regarded as a \emph{Riemann surface}, \ie a one-dimensional complex manifold. The Polyakov action (\ref{eq:stringaction}) we wrote down in Section \ref{subsec:bosonicstring} is at lowest order in and expansion in terms of  $\alphap$ (visible in the prefactor, \ie the string tensions $T\propto1/\alphap$). At the next order $\order{\alphap^0}$, Eq.~(\ref{eq:stringaction}) is supplemented by a term $\tilde{\action}\propto\int\dd^{2}\xi\,\sqrt{-h}\,\Ricci^{(2)}$, where $\Ricci^{(2)}$ is scalar curvature on the two-dimensional world sheet with metric $h_{ab}$. This term serves as the ``string coupling constant'' in the following way: using the $z$ coordinates defined in Eq.~(\ref{eq:definez}), the contribution of the dilaton $\Phi$ in the spectrum of the closed string to the world sheet action up to order $\alphap^{0}$ is \cite{Becker:2007zj}
\bea\label{eq:dilatonaction}
\action_{\Phi}&=&\frac{1}{4\pi}\int_{\Sigma}\dd^{2}z\,\sqrt{-h}\,\Phi\,\Ricci^{(2)}\,.
\eea
If the dilaton is constant, $\Phi=\Phi_{0}$, the integrand in Eq.~(\ref{eq:dilatonaction}) is a total derivative, and therefore $\action_{\Phi}$ is determined by the global topology of the world sheet. 
More precisely, it corresponds to the \emph{Euler characteristic} of the surface $\Sigma$,
\beq\label{eq:Euler}
\chi(\Sigma)=\frac{1}{4\pi}\int_{\Sigma}\dd^{2}z\,\sqrt{h}\,\Ricci^{(2)}=2-2n_{\mathrm{h}}-n_{\mathrm{b}}-n_{\mathrm{c}}\,.
\eeq
Here, $n_{\mathrm{h}}$ (the genus of $\Sigma$) counts the number of ``handles'', $n_{\mathrm{b}}$ the number of boundaries, and $n_{\mathrm{c}}$ the number of cross-caps of the Riemann surface in question. In a theory of only closed strings (as it is the case for type IIA and type IIB), the world sheet has no boundaries (which would be created by the ends of open strings). The type IIA/B fundamental strings are also oriented, therefore $\Sigma$ is necessarily orientable and has no cross-caps. With $n_{\mathrm{b}}=n_{\mathrm{c}}=0$, and the only possible world sheet topologies are closed and oriented Riemann surfaces, which are uniquely characterized by $n_{\mathrm{h}}$, and this is also the number of loops $n_{\mathrm{loop}}$ in the corresponding Feynman diagram (see Fig.~\ref{fig:stringdualities}). The Euler characteristic then is $\chi=2-2n_{\mathrm{h}}$, and from Eq.~(\ref{eq:dilatonaction}) we find $\action_{\Phi}=\Phi_{0}(2-2n_{\mathrm{h}})$.

When string theoretic Feynman diagrams are calculated using the path integral formalism,
\beq
Z\propto\int\left[\mathcal{D} h_{ab}\right]\,\int\left[\mathcal{D}X^{m}\right]\cdots\,e^{-(\action_{\mathrm{string}}+\action_{\Phi})}\,,
\eeq
the first integral runs over all possible world sheet topologies, with the world sheet diffeomorphisms and Weyl scalings that leave the string action $\action_{\mathrm{string}}$ invariant ``taken out''. Therefore, $Z$ is found from a perturbative expansion in the number of loops (or handles) of the world sheet, $Z=\sum_{n_{\mathrm{h}}=0}^{\infty}Z_{n_{\mathrm{h}}}$. For the tree level diagram with $n_{\mathrm{h}}=0$, the dilaton contribution to the action $\action_{\Phi}$ then reads $e^{-2\Phi_{0}}\equiv\gs^{-2}$, and each handle gives an additional factor of $\gs^{2}$. In this sense, the ``string coupling'' $\gs$ is the expansion parameter in the number of string loops, and one must have $\gs<1$ for a perturbative regime. Note that we considered the dilaton constant, but in principle it is a dynamical scalar field, which must be stabilized at a value $\Phi_{0}$ to allow for the above reasoning. This is a first example of a dimensionless parameter in string theory (here, the coupling $\gs$) determined by the vacuum expectation value of a scalar field, as we mentioned at the very beginning of this Chapter.

\section{Effective Supergravity Field Theories}\label{sec:effectivesugra}
Apart from the number of string loops, parametrized by $\gs$, there is a second expansion parameter: the (dimensionful) Regge slope $\alphap=\ells^{2}$ is a measure of ``stringiness'' since it describes the difference between a one-dimensional string and a point particle, \ie the limit $\alphap\rightarrow0$. Note that this is also the low energy limit because (in a time-independent background) $\alphap E^{2}$ is the only dimensionless combination with $\alphap$. In Section \ref{subsec:five}, we restricted ourselves to massless fields in the spectrum because when turning off the ``stringiness'' by taking $\alphap\rightarrow0$, all but the massless particles become infinitely heavy, see Eqs.~(\ref{eq:massNS}) and (\ref{eq:massR}). The field content of the different superstring theories listed in Section \ref{subsec:five} suggested that they have $\calN=2$ (or $\calN=1$) local supersymmetry in ten dimensions, and hence we may expect a $d=10$ supergravity form for their low energy effective actions. 
We only consider the bosonic part of supergravity actions in the following since we are looking for classical solutions without fermionic degrees of freedom.

\subsection{Type IIA Supergravity}\label{subsec:typeIIAsugra}
The supergravity description of type IIA theory is the most straightforward to obtain: we stated earlier that the low energy limit of M-theory (see Fig.~\ref{fig:stringdualities}) is eleven-dimensional supergravity, and it is known that type IIA theory follows from it by simple ``dimensional reduction'' (see below) on a circle of radius $R$. At the effective supergravity level, the action in eleven dimensions is constrained from the requirements of level matching between fermionic and bosonic degrees of freedom (to preserve supersymmetry), general coordinate invariance and local Lorentz invariance, and the bosonic part of this unique action reads
\beq\label{eq:11daction}
\action_{11d}=\frac{1}{2\kappa_{11}}\int\dd^{11}x\,\sqrt{-G}\left(R^{(11)}-\frac{1}{2}\left|F_{(4)}\right|^{2}\right)-\frac{1}{6}\int A_{(3)}\wedge F_{(4)}\wedge F_{(4)}\,.
\eeq
The eleven-dimensional gravitational coupling $\kappa_{11}$ is related to the $d=11$ Planck length $\ell_{_\uPl}^{(11)}$ by $2\kappa_{11}=\left[2\pi\ell_{_\uPl}^{(11)}\right]^{9}/2\pi$, and $R^{(11)}$ is the Ricci scalar in eleven dimensions (obtained from the metric\footnote{Indices with primes run over eleven (not ten) dimensions in this Section, \ie $m',n'=0,\dots,10$, while $m,n=0,\dots,9$.} $G_{m'n'}$ with determinant $G$). The field strength $F_{(4)}=\dd A_{(3)}$ is obtained from the three-form gauge potential introduced to match the number of fermionic degrees of freedom (whose action we do not display). 
Dimensional reduction is a procedure where one dimension is ``curled up'' on a circle, and only the zero modes of the corresponding Fourier expansion in that dimension are kept for each of the fields in the original higher-dimensional theory. (For compactification, the entire tower of modes survives in the lower dimensionality, see Section \ref{subsec:compactify}.) 
In the case at hand, reducing Eq.~(\ref{eq:11daction}) down to ten dimensions produces the $d=10$ graviton $g_{mn}$, a gauge field $A_{m}$ and the dilaton $\Phi$ from the eleven-dimensional metric $G_{m'n'}$. Studying the eleven-dimensional line element $\dd s^{2}_{11}$ in this decomposition $G_{m'n'}\rightarrow(g_{mn},A_{m},\Phi)$, one obtains the relation $\ell_{\uPl}^{(11)}=\gs^{1/3}\ells$ between the eleven-dimensional Planck and the fundamental string scale \cite{Becker:2007zj}. The three-form $A_{(3)}$ in $d=11$ leads to a three-form and a two form, $A_{mnr}$ and $B_{mn}$, at the ten-dimensional level. One also finds that the four-form field strength has to be replaced by the gauge-invariant combination
\beq
\tilde{F}_{(4)}=\dd A_{(3)}+A_{(1)}\wedge H_{(3)}\,,
\eeq
where $H_{(3)}=\dd B_{(2)}$. (This is necessary to preserve invariance under supersymmetry variations.) The integration over the compact eleventh coordinate is performed with the radius of the circle\footnote{Following our remarks about $\gs$ being determined by the vacuum expectation value of the dilaton (Section \ref{sec:dilaton}), the compactification radius $R$ is therefore again in principle a dynamical quantity. In this sense, one sometimes speaks fo the eleventh dimension ``opening up'' in the limit of strong (non-perturbative) coupling $\gs\gg1$.} set to $R=\gs\ells$. For the gravitational coupling constant in ten dimensions, this gives $2\kappa_{10}=\left(2\pi\ells\right)^{8}\gs^{2}/(2\pi)=2\tilde{\kappa}_{10}\,\gs^{2}$ (note that a factor of $\gs^{2}$ has been taken out in the last definition). 
The type IIA supergravity action can then schematically be written as $\action^{\mathrm{IIA}}=\action_\uNS+\action_\uR^{\mathrm{IIA}}+\action_\uCS^{\mathrm{IIA}}$, where the three contributions (in the ``string frame'', see below) read:
\bea
\action_\uNS&=&\frac{1}{2\tilde{\kappa}_{10}}\int\dd^{10}x\,\sqrt{-g}\,e^{-2\Phi}\left(R+4\,\partial_{\mu}\Phi\,\partial^{\mu}\Phi-\frac{1}{2}\,\left|H_{(3)}\right|^{2}\right)\label{eq:actionNS}\\
\action_\uR^{\mathrm{IIA}}&=&-\frac{1}{4\tilde{\kappa}_{10}}\int\dd^{10}x\,\sqrt{-g}\left(\left|F_{(2)}\right|^{2}+\left|\tilde{F}_{(4)}\right|^{2}\right)\\
\action_\uCS^{\mathrm{IIA}}&=&-\frac{1}{4\tilde{\kappa}_{10}}\int B_{(2)}\wedge F_{(4)}\wedge F_{(4)}
\eea
Note that the NS part in the first line does not carry a superscript ``type IIA'' because, as we saw at the level of fundamental massless states in Section \ref{subsec:five}, type IIA and type IIB superstring theory have the same NS-NS sector, therefore we will equally use Eq.~(\ref{eq:actionNS}) in type IIB supergravity below. The prefactor $e^{-2\Phi}$ of the Ricci scalar 
reminds us that these actions are written in the string frame, and ``plain'' ten-dimensional General Relativity is obtained by transforming to the Einstein frame, with the two metrics related by $g_{mn}^{\mathrm{(string)}}=e^{\Phi/2}g_{mn}^{\mathrm{(E)}}$. Re-writing the NS part of the action in the latter frame gives
\beq
\action_\uNS^{\mathrm{(E)}}=\frac{1}{2\tilde{\kappa}_{10}}\int\dd^{10}x\,\sqrt{-g^{\mathrm{(E)}}}\left(R^{\mathrm{(E)}}-\frac{1}{2}\,\partial_{\mu}\Phi\,\partial^{\mu}\Phi-\frac{1}{2}\,e^{-\Phi}\left|H_{(3)}\right|^{2}\right)\label{eq:actionNS-Einstein}\\
\eeq
In the following, we drop the superscript ``(E)'' and always work in the Einstein frame where gravity takes its canonical form.

\subsection{Type IIB Supergravity}\label{subsec:typeIIBsugra}
The most prominent difference between type IIA and type IIB arises from the different R-R gauge potentials $C_{(n)}$ they contain, with $n^{\mathrm{IIA}}=1,3$ and $n^{\mathrm{IIB}}=0,2,4$. We already mentioned that the field strength $F_{(5)}=\dd C_{(4)}$ then must be self-dual in ten dimensions. This is one of the obstructions to formulating a manifestly supersymmetric action for type IIB supergravity. One possible strategy is to work directly with the equations of motion, which can be written in a covariant way, and moreover may be obtained one from the other by supersymmetric variation. Alternatively, one may resort to writing an action which is not by itself invariant under local supersymmetry, but has to be supplemented with the self-duality constraint $\tilde{F}_{(5)}=*\tilde{F}_{(5)}$ on the gauge-invariant field strength combination, see below. (This allows the elimination of superfluous bosonic degrees of freedom on top of those needed to match the fermionic ones.) The NS part of the action is the same as in type IIA, see Eq.~(\ref{eq:actionNS-Einstein}). Its other parts in the Einstein frame read
\bea
\action_\uR^{\mathrm{IIB}}&=&-\frac{1}{4\tilde{\kappa}_{10}}\int\dd^{10}x\,\sqrt{-g}\left(e^{2\Phi}\left|F_{(1)}\right|^{2}+e^{\Phi}\left|\tilde{F}_{(3)}\right|^{2}+\frac{1}{2}\left|\tilde{F}_{(5)}\right|^{2}\right)\label{eq:actionR-typeIIB}\,,\\
\action_\uCS^{\mathrm{IIB}}&=&-\frac{1}{4\tilde{\kappa}_{10}}\int C_{(4)}\wedge H_{(3)}\wedge F_{(3)}\,.\label{eq:actionCS-typeIIB}
\eea
Here the notation used is $F_{(n+1)}=\dd C_{(n)},\,H_{(3)}=\dd B_{(2)}$, and tildes indicate combinations
\bea
\tilde{F}_{(3)}&=&F_{(3)}-C_{(0)}H_{(3)}\,,\label{eq:F3tilde}\\
\tilde{F}_{(5)}&=&F_{(5)}-\frac{1}{2}\,C_{(2)}\wedge H_{(3)}+\frac{1}{2}\,B_{(2)}\wedge F_{(3)}\,.\label{eq:F5gi}
\eea
On the solutions to the equations of motion derived from the complete type IIB action $\action^{\mathrm{IIB}}=\action_\uNS+\action_\uR^{\mathrm{IIB}}+\action_\uCS^{\mathrm{IIB}}$, one has to impose the self-duality condition for $\tilde{F}_{(5)}$. 

Note that type IIB supergravity has two scalar fields, $C_{(0)}$ and the dilaton $\Phi$, as well as two two-forms, the Kalb Ramond field $B_{(2)}$ and the R-R potential $C_{(2)}$. It turns out that the action $\action^{\mathrm{IIB}}$ built from Eqs.~(\ref{eq:actionNS-Einstein}), (\ref{eq:actionR-typeIIB}) and (\ref{eq:actionCS-typeIIB}) has a symmetry under transformations by the special linear group $SL(2,\mathbb{R})$, 
which is not manifest in the above notation. To make this symmetry explicit, one may regroup the dilaton and the $C_{(0)}$ field (often called the ``axion'') into the complex axion-dilaton field $\tau$, and assemble the two-form potentials into a two-component vector as in
\beq\label{eq:axiondilaton}
\tau=C_{(0)}+i\,e^{-\Phi}\,,\qquad \mathcal{B}_{(2)}=\left(\begin{array}{c}B_{(2)}\\C_{(2)}\end{array}\right)\,.
\eeq
Note that the field strength $\mathcal{H}_{(3)}$ that follows from $\mathcal{B}_{(2)}$ then also has two components. The type IIB action in the Einstein frame can then be rewritten as
\bea
\action^{\mathrm{IIB}}&=&\frac{1}{2\tilde{\kappa}_{10}}\int\dd^{10}x\,\sqrt{-g}\left[R-\frac{1}{12}\,\mathcal{H}_{(3)}^{T}\,\mathcal{M}\,\mathcal{H}^{(3)}+\frac{1}{4}\,\mathrm{tr}\left(\partial^{\mu}\mathcal{M}\,\partial_{\mu}\calM\right)\right]\nonumber\\
&&-\frac{1}{8\tilde{\kappa}_{10}}\left(\int\dd^{10}x\,\sqrt{-g}\,\left|\tilde{F}_{5}\right|^{2}+\int\varepsilon_{ij}\,C_{(4)}\wedge \calH_{(3)}^{i}\wedge\calH_{(3)}^{j}\right)\,,\label{eq:SL2R-typeIIB}
\eea
where the matrix $\calM$ and the field strength $\tilde{F}_{5}$ are given by
\beq
\calM=e^{\Phi}\left(\begin{array}{cc}|\tau|^{2}&-C_{(0)}\\-C_{(0)}&1\end{array}\right)\,,\qquad \tilde{F}_{5}=F_{5}+\frac{1}{2}\,\varepsilon_{ij}\,\mathcal{B}_{(2)}^{i}\wedge\calH_{(3)}^{j}\,,
\eeq
using the definitions of Eqs.~(\ref{eq:axiondilaton}). The manifestly $SL(2,\mathbb{R})$ invariant notation of Eq.~(\ref{eq:SL2R-typeIIB}) is useful to find solutions for all the bosonic background fields of type IIB supergravity. To this end and for later use, we also define the complex three-form
\beq\label{eq:G3form}
G_{(3)}=F_{(3)}-\tau\,H_{(3)}\,,
\eeq
which is (in the applications we will consider) imaginary self-dual, \ie it obeys $*G_{(3)}=iG_{(3)}$. We consider the background equations of motion derived from this type IIB supergravity action in Section \ref{subsec:typeIIB-eofm}.

\bigskip
In this Chapter, we established effective ten-dimensional supergravity actions as the low-energy limits of both type IIA and type IIB superstring theory. The fields $g_{mn},B_{mn},\Phi$ as well as the R-R gauge potentials $C_{(n)}$ make up the bulk field content of these theories. There are, however, additional ingredients to the picture: for example, it is possible to embed hypersurfaces, called ``$p$-branes'', into the supergravity backgrounds. In the next Chapter, we study a particular class of branes, so-called \emph{Dirichlet} or \emph{D-branes}, for short. We also address the question of compactifying this Chapter's ten-dimensional actions down to the four-dimensional world of our everyday experience.

\chapter{Branes, Compactifications and Dualities}\label{chapter:branes}
\begin{quotation}
\emph{The bulk supergravity actions obtained in Chapter \ref{chapter:stringelements} do not cover the full range of string theoretic building blocks available for the construction of a unified theory. In this Chapter, we discuss non-perturbative string objects called D-branes. We also show how one can establish contact between the $d=10$ picture of superstring theory and our low-energy world where we observe but four dimensions. This also sheds light on the relations (``dualities'') between the five different formulations of superstring theory.}
\end{quotation}

\section{D-Branes}\label{subsec:Dbranes}
\begin{figure}[t]
\begin{center}
\includegraphics[width=0.45\textwidth]{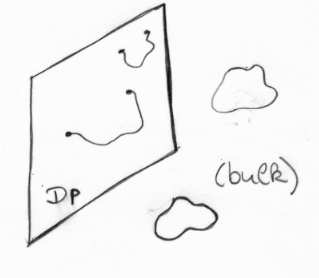}
\hfill
\includegraphics[width=0.45\textwidth]{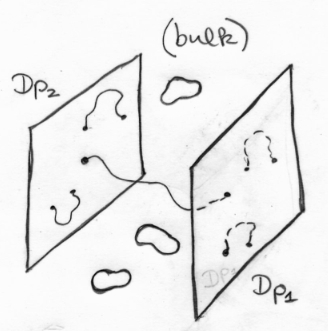}
\caption[Open strings attached to D-branes and closed strings propagating in the bulk. D-brane interactions due to open strings stretched between them, and exchange of closed strings between branes.]{\small \emph{Left:} D-branes are dynamical objects to which the ends of open strings can attach if they have Dirichlet boundary conditions. On the world volume of the branes, (supersymmetric) gauge theories can live, whose fields correspond to the massless modes of the open strings attached to the brane. The excitation spectrum for strings with both ends on the brane is tachyon-free. Closed strings can propagate the entire ten-dimensional bulk. \emph{Right:} Branes interact through the exchange of closed string modes that penetrate the bulk, and open strings can also stretch between different branes. In a brane--anti-brane system, these strings have tachyonic modes and announce the onset of mutual brane annihilation into closed string annihilation.}
\label{fig:Dbranes}
\end{center}
\end{figure}

When we solved the equations of motion for open strings in Section \ref{subsubsec:open} for the bosonic and in Section \ref{subsec:superstring} for the superstring case, we only considered Neumann boundary conditions: they are reasonable because they conserve momentum at the string endpoints. It was later realized, however, that open strings can have Dirichlet boundary conditions provided that their ends in these dimensions are attached to dynamical objects called D$p$-branes, where the ``D'' stands for Dirichlet, and $p$ denotes the number of spatial dimensions of the brane, see Fig.~\ref{fig:Dbranes}. [In spacetime, D$p$-branes then trace out a $(p+1)$ dimensional world volume, see Fig.~\ref{fig:worldlines}.] The momentum flowing out of the string endpoints is conserved by the brane.  Apart from the spacetime filling D-brane with $p=d-1$, the presence of these objects breaks $d$-dimensional Poincar\'e invariance.

\subsection{Chern Simons (Charge) Term}
Recall that we stated earlier the analogy between the electromagnetic world line coupling for charged point particles, $q\int_{T} \dd x^{\mu}\,A_{\mu}$, and a coupling of the anti-symmetric Kalb Ramond tensor field $B_{mn}$ to the string world sheet with the ten-dimensional field $B_{mn}$ ``pulled back'' onto $\Sigma$. [Compare the pullback of the metric $g_{mn}$ in Eq.~(\ref{eq:inducedmetric}).] In this sense, fundamental strings are charged under the field $B_{mn}$. It turns out that, analogously, D-branes carry charges corresponding to the R-R gauge potentials we encountered in the state spectra of type IIA/B superstring theories. There is a natural coupling between a R-R form $C_{(n)}$ with $n=p+1$ indices and the world volume $\Sigma_{p+1}$ of a D$p$-brane,
\beq\label{eq:electriccoupling}
\action_\uCS=\mu_{p}\int_{\Sigma_{p+1}}C_{(p+1)}=\frac{\mu_{p}}{(p+1)!}\int_{\Sigma_{p+1}}\dd^{p+1}\xi\,A_{\nu_{1}\dots\nu_{p+1}}\,\frac{\partial x^{\nu_{1}}}{\partial\xi^{0}}\cdots\frac{\partial x^{\nu_{p+1}}}{\partial\xi^{p}}\,,
\eeq
where the $\xi^{i}\,(i=0,\dots ,p)$ are the coordinates on the world volume of the brane. The index ``CS'' stands for ``Chern Simons'', and the proportionality constant $\mu_{p}$ will be explained shortly. To be precise, the gauge potentials entering into the Chern Simons term of D-brane actions may differ by field redefinitions from the ones in the bulk supergravity actions of Chapter \ref{chapter:stringelements}. This is because the supergravity theories established previously with the use of symmetry arguments are unique up to those redefinitions, which means that the identification between the supergravity fields and the string modes is ambiguous \cite{ortin:strings}.

Coupling as in Eq.~(\ref{eq:electriccoupling}) is called ``electric'' because of the analogy with $\dd x^{\mu}A_{\mu}$. Since the $n$-forms have $n=1,3$ (\ie odd) in type IIA and $n=0,2,4$ (\ie even) in type IIB superstring theory, we immediately see that $p$ must be even in the former, and $p$ odd in the latter case. The D$p$-branes then serve as (electric) sources for the $C_{(p+1)}$ gauge potentials in each theory. Apart from Eq.~(\ref{eq:electriccoupling}), there is also ``magnetic'' coupling: in this case, the electromagnetic analogy is with the field strength $F_{\mu\nu}\propto\partial_{\mu}A_{\nu}-\partial_{\nu}A_{\mu}$, which creates a flux through a two-dimensional surface $S^{2}$, as is described by Gauss' law. In $d=10$ dimensions, the R-R field strength $F_{(n)}=\dd C_{(n-1)}$ creates a flux through a $(d-2-n)=(8-n)$ dimensional hypersurface, written as $S^{8-n}$: therefore, a $C_{(n-1)}$ potential experiences electric coupling to D$p$-branes with $p=n-2$, and magnetic coupling to a brane with $p=8-n$. [From the point of view of the branes, a D$p$-brane is an electric souce for the same potential for which a D$(6-p)$ brane is a magnetic source.] Mathematically, this is described by the \emph{Hodge duality} or \emph{``star'' operator} $*$ (see Appendix \ref{app:geotopo} and \cite{Nakahara:1990th,Becker:2007zj}). The higher-dimensional generalization of Gauss' law then allows one to calculate the electric brane charge $\mu_{p}$ from
\beq
\mu_{p}=\int_{S^{8-p}}*F_{(p+2)}\,.
\eeq
This charge, and the one of the magnetically sourcing D$(6-p)$-brane, are subject to the \emph{Dirac quantization condition} with $\mu_{p}\,\mu_{6-p}\in2\pi\mathbb{Z}$. Note that in type IIB, the D3-brane is special because it carries a self-dual charge: the (five-index) field strength of the four-form potential $F_{(5)}=\dd C_{(4)}$ must be self-dual in ten dimensions, $F_{(5)}=*F_{(5)}$. [Note that the type IIB bulk supergravity action of Section \ref{subsec:typeIIBsugra} required a gauge-invariant redefiniton of this field strength, see Eq.~(\ref{eq:F5gi}).]\\
Branes with $p$ even in type IIA and $p$ odd in type IIB are stable, and they preserve half of the supersymmetry of the background. [Branes of the other dimensionalities are unstable in the respective theory because there are no gauge potentials for them to couple to, see Eq.~(\ref{eq:electriccoupling}). If present, these branes break all of the supersymmetry.] If more than one brane is embedded into the same background, they can interact among themselves by exchanging (closed string) graviton and R-R modes. Gravity is described by the so-called Dirac Born Infeld (DBI) term of the brane world volume action, which we discuss in the next Section. In principle, closed string dilaton modes are also exchanged, but in the applications we consider, the dilaton is usually a constant, therefore we do not take them into account here.\\
Like the repelling force between two particles of the same charge $q$, the R-R interaction tries to drive two D$p$-branes (with charges $\mu_{p}$) apart, but for parallel branes, the attractive gravitational force exactly balances this. (For the same reason, a single D-brane put in a supergravity background does not experience any force: its CS and the DBI terms have opposite signs, see below, and cancel exactly.) A system of D-branes only is therefore stable and also preserves half the supersymmetry.\\
If anti-D-branes (with opposite R-R charge $-\mu_{p}$, written as $\overline{\mathrm{D}}p$) are introduced, they attract D-branes and supersymmetry is broken. (Put a simple way, anti-D-branes can only be introduced into the theory at the quantum level, therefore one has necessarily left the realm of classical supergravity solutions obtained from the bosonic part of the action in Chapter \ref{chapter:stringelements}.) Since for an anti-brane, the sign of the CS term is reversed, the cancellation with the DBI action vanishes for a an anti-brane: even by itself, an anti-brane in a supergravity background will feel a force and try to minimize its energy. Combinations of D- and $\overline{\mathrm{D}}$-branes are of particular interest in cosmology, as we shall see later.

\subsection{Dirac Born Infeld (Dynamical) Term}\label{subsec:DBIterm}
\begin{figure}[h]
\begin{center}
\includegraphics[width=0.6\textwidth]{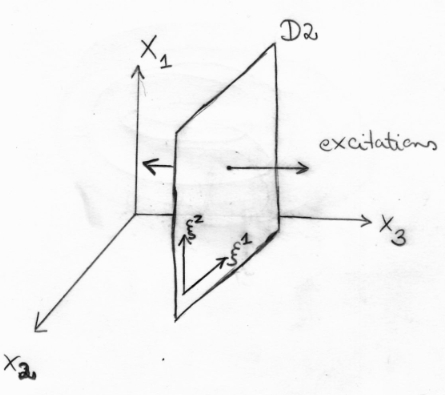}
\hfill
\includegraphics[width=0.3\textwidth]{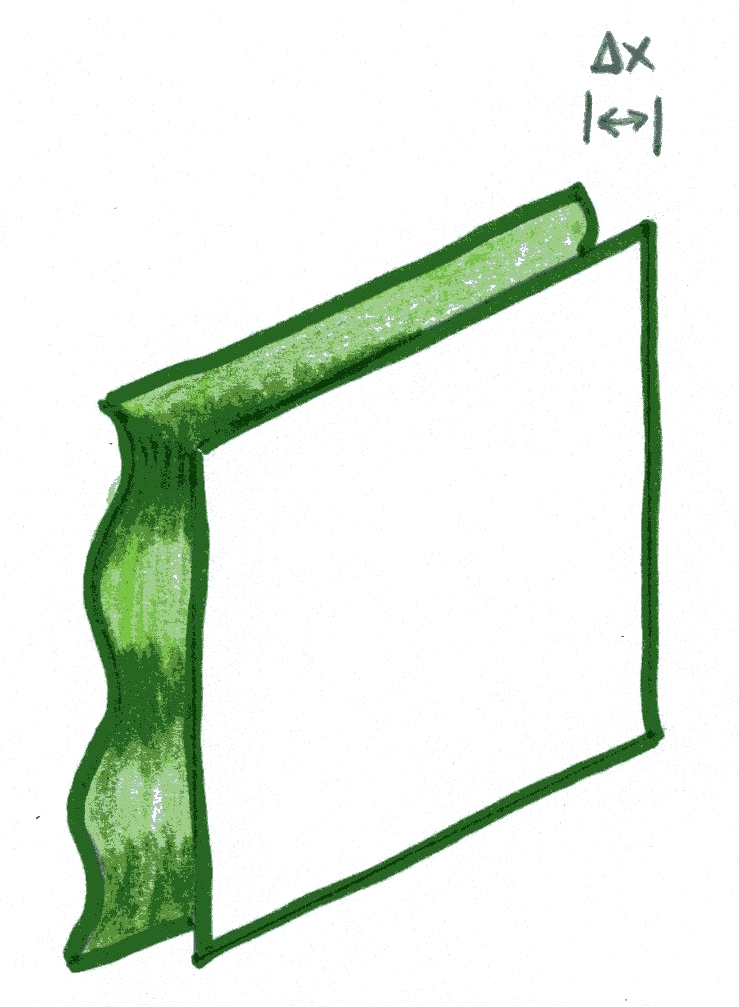}
\caption[D-branes aligned in the static gauge, and transverse oscillations of the brane.]{\small \emph{Left:} When D-branes are embedded into the higher-dimensional spacetime, their world volume directions can be aligned with certain coordinates of spacetime in the so-called static gauge. \emph{Right:} In the spacetime directions orthogonal to its volume, the D-brane can oscillate since it is a dynamical object.}
\label{fig:braneembeddings}
\end{center}
\end{figure}

A theory with D-branes must also contain open strings (that end on the branes) apart from closed ones, because a string loop hitting a D-brane can break into an open string with both ends attached to the brane, see Fig.~\ref{fig:Dbranes}. When the branes carry conserved charges [that is, if they are of a dimensionality $p$ that is stable in the corresponding theory], one can show that there are no tachyons in the spectrum of open strings starting and ending on branes.\\
On the world volume of D-branes, gauge theories can reside (which can possibly be non-abelian for multiple coincident branes, see below), and these theories have as much supersymmetry as the background after introduction of the branes. The fields in the spectrum of the gauge theory correspond to the massless modes of the open strings attached to them. Put another way, the dynamics of a D$p$-brane are described by the $(p+1)$-dimensional effective theory of massless fields living on the brane,
\beq\label{eq:DBIbraneaction}
S_\uDBI=-T_{p}\int_{\Sigma_{p+1}}\dd^{p+1}\xi\,\left\{-\det\left[\gamma_{ab}+B_{ab}+2\pi\alphap\, F_{ab}\right]\right\}^{1/2}\,,
\eeq
where the dimensionful parameter $T_{p}$ is the D-brane tension, where in our units
\beq\label{eq:Dbranetension}
T_{p}=\left[\gs\,(2\pi)^{p}\,\alphap^{(p+1)/2}\right]^{-1}\,.
\eeq
The subscript ``DBI'' in Eq.~(\ref{eq:DBIbraneaction}) stands for ``Dirac Born Infeld'', and the two-form $B_{(2)}$ is the pullback onto the brane of the two-form $B_{mn}$ in ten-dimensional spacetime, in straight analogy to the pullback of the metric $\gamma_{ab}$, compare Eq.~(\ref{eq:inducedmetric}). The second two-form $F_{(2)}$ arises from a $U(1)$ gauge potential associated with the strings with both ends on the brane. 
A derivation of this dynamical brane action can be found \eg in \cite{Polchinski:1998rq,Becker:2007zj}. In the so-called static gauge, diffeomorphism symmetry is used to align the $(p+1)$ world volume coordinates $\xi^{i}$ with the first $(p+1)$ components of the bosonic spacetime coordinates $X^{m}$, while the other $(9-p)$ coordinates become scalar fields on the world volume of the brane describing transverse excitations, see Fig.~\ref{fig:braneembeddings}. 
Very often, it is reasonable to consider the case of vanishing pullback $B_{(2)}$ and gauge field $F_{(2)}$ in Eq.~(\ref{eq:DBIbraneaction}). For completeness, however, let us note that if these two-forms are non-zero on a D$p$-brane, this brane can also carry induced charges of D$(p-2n)$-type, where $n=0,1,\dots$. 
In these cases, the Chern Simons term of the brane action instead of Eq.~(\ref{eq:electriccoupling}) is written as $\action_\uCS=\mu_{p}\int\left[C\,\exp(B+2\pi\alphap\,F)\right]_{(p+1)}$, with the $(p+1)$-dimensional piece extracted, and $C=\sum_{n=0}^{8}C_{(n)}$. (In this sum it is understood that only the values of $n$ permitted for the corresponding theory are counted.) A case of interest we will encounter later is a D5-brane, which, given its orientation in the background geometry (see below) also carries charges of D3- and D1-brane type. Note that a D1-brane, while geometrically a string (``D-string''), must not be confused with the fundamental strings of the theory (``F-strings''): the former is charged under the R-R two-form potential $C_{(2)}$ of type IIB theory, while the fundamental F1-strings are charged under the NS-NS Kalb Ramond anti-symmetric tensor field $B_{(2)}$.

In the right units and if the dilaton is constant, the charge and tension of a brane are equal in the Einstein frame (in the anti-brane case, up to a sign), $T_{p}=\pm\mu_{p}$. This makes it possible for the CS and DBI terms to cancel for D-branes (hence they can be moved around in a supergravity background without energy cost), and add for anti-branes (which will seek to minimize their energy at a preferred position in the background). From Eq.~(\ref{eq:Dbranetension}), one sees that $T_{p}\propto\gs^{-1}$, meaning that D$p$-branes become infinitely heavy (\ie non-dynamical) in the weak coupling limit $\gs\rightarrow0$. In this sense, they are non-perturbative excitations of string theory.

\subsection{Other Extended Objects}
A complete ten-dimensional supergravity action describing the low-energy limit of type IIA/B superstring theory consists of the bulk terms derived in Sections \ref{subsec:typeIIAsugra} and \ref{subsec:typeIIBsugra}, respectively, plus contributions of the form of Eqs.~(\ref{eq:electriccoupling}) and (\ref{eq:DBIbraneaction}) for any type of (stable) D$p$-brane allowed by the background. There exist, however, also a variety of other (non-perturbative) string theoretic objects that may be included, if certain certain restrictions are respected. Here, we only comment on so-called \emph{orientifold planes}, denoted by O$p$, where $p$ is again the number of spatial dimensions, and the rules for their dimensionality in type IIA/B are the same as for D$p$ branes. These O$p$-planes carry charges opposite to the D$p$-branes, but they are not dynamic, since they are basically defined by the points left invariant under the orientifold projections $\Omega$ mentioned in the last Chapter. (Recall that D$p$ branes are surfaces on which fundamental strings can end, and therefore dynamic.) Other objects are \eg ``instantons'' [or D$(-1)$-branes in type IIB theory], and they will occasionally be mentioned later on. A thorough discussion of extended string objects can be found, for example, in \cite{ortin:strings}.

\section{From Ten to Four Dimensions}\label{subsec:compactify}
We so far accepted without bewilderment that a string theoretic spacetime can have more than four dimensions. The embedding of $p$-dimensional hypersurfaces like D-branes opens up the interesting possibility of identifying our observable Universe with \eg the (3+1)-dimensional world volume of a D3-brane in type IIB superstring theory (or with a stack of coincident branes, which is phenomenologically attractive from the gauge theory point of view). Models like this are often called ``brane world scenarios'', see \eg \cite{Lidsey:2003ws} for a review. The gauge theories of the Standard Model could then be described by the open strings attached to the brane (to which the SM interactions therefore are confined), while gravitons (closed string modes) propagate in the entire $d=10$ spacetime. This suggests an intuitive explanation why gravity is much weaker than the other forces of Nature \cite{Randall:1999vf,Randall:1999ee}.\\
However, setting aside the ``brane world'' approach for the moment, we can also obtain a four-dimensional theory from the $d=10$ supergravity actions of Section \ref{sec:effectivesugra} by compactifying six out of the nine spatial dimensions. Indeed, there is a precedent that the attempt to unify interactions may proceed using additional space dimensions, the so-called \emph{Kaluza Klein compactification procedure} \cite{Kaluza:1921tu,Klein:1926tv}: starting from five-dimensional General Relativity with spacetime coordinates $(t,\vec{x},y)$, one obtains in the effective four-dimensional theory the usual $d=4$ gravity, but also a four-vector field $A_{\mu}$ (which maybe associated with the gauge field of electromagnetism), plus a scalar field (the ancestor of the string dilaton $\Phi$). The additional dimension $y$ is made compact, \ie one identifies it with a circle by setting $y\sim y+2\pi R$. All fields in the theory are then periodic functions of $y$, and may hence be developed in a Fourier series in this dimension, giving rise to the so-called \emph{Kaluza Klein modes}. (Note that when, in search for the type IIA supergravity action in Section \ref{subsec:typeIIAsugra}, we applied dimensional reduction to eleven-dimensional supergravity, we only kept the zero-modes in this Fourier expansion.) At the quantum level, an infinite tower of additional energy states is found, with masses $m_{\mathrm{KK},n}$ quantized in terms of the inverse radius $1/R$ of the extra dimension. If $R$ is small (\ie the compactification scale $m_{\sss{c}}=1/R$ is large), the new energy levels lie far above those of the effective four-dimensional theory and go unnoticed in low energy experiments \cite{zwiebach:strings}. Let us also state the effect of a Kaluza Klein compactification on the Ricci scalar term of an Einstein frame supergravity action in $d$ dimensions, $m^{d-2}_{\uPl,d}\int\dd^{d}x\sqrt{g}\,R^{(d)}$, which after compactification reads
\beq\label{eq:Planckmassrelations}
V_{d-4}(\scrM)\,m^{d-2}_{\uPl,d}\int\dd^{4}x\sqrt{g}\,R^{(4)}+\dots\,.
\eeq
By $m_{\uPl,d}$ we have denoted the Planck mass in $d$ dimensions, and $\scrM$ is the $(d-4)$-dimensional space into which all but four dimensions have been compactified. Its volume therefore enters into the (observed) four-dimensional Planck mass $\mpl$ as $\mpl^{2}=m_{\uPl,d}^{d-2}\, V_{d-4}(\scrM)$. The compactification (or Kaluza Klein) mass scale $m_{\sss{c}}$ is just the inverse of the compact volume, \ie we have in our units $m_{\sss{c}}^{d-4}=1/ V_{d-4}(\scrM)$. Using this relation, one can establish an upper bound on the size of the compact space since the tower of Kaluza Klein excitations in the particle spectrum has not been observed yet. It also follows that, for the four-dimensional Planck mass to be constant, the compactification volume must be fixed, which, as we show below, is usually non-trival in string theory.

\subsection{Internal and External Spaces}\label{subsec:int-and-ext}
In the $d=10$ superstring theories, one has to compactify not one, but six spatial dimensions to (potentially) get to our observed four-dimensional world. In these remaining four dimensions, we wish to preserve Poincar\'e invariance, therefore the ten-dimensional metric is split into a direct product of a $(d_{\mathrm{ext}}=4)$ external and a $(d_{\mathrm{int}}=6)$ internal manifold, $\scrM_{10}=\scrM_{4}\times \scrM_{6}$. In the external space, the coordinates read $x^{\mu}\,(\mu=0,\dots,3)$, and on the compact internal space $y^{M}\,(M=4,\dots,9)$. The internal metric $g_{MN}$ does not depend on the external coordinates, and we specify to Minkowski space with $\eta_{\mu\nu}$ in four dimensions, so that the ten-dimensional line element reads
\beq\label{eq:10d-notwarped}
\dd s^{2}_{10}=\eta_{\mu\nu}\,\dd x^{\mu}\,\dd x^{\nu}+g_{MN}(y)\,\dd y^{M}\,\dd y^{N}\,.
\eeq
Note that, as a consequence, the ten-dimensional Ricci tensor $R_{mn}$ splits into a separate external piece $R_{\mu\nu}$, and an internal piece $R_{MN}$, while the mixed components vanish. Let us also remark that the requirement of maximal symmetry in the extended dimensions restricts the possible ans\"atze one can make for the other background fields such as $B_{mn}$ and the R-R gauge potentials: their field strength components with one or more indices along the extended dimensions vanish.

In a straightforward generalization of Kaluza Klein, each of the six extra dimensions can be curled up in a circle, corresponding to toroidal compactification\footnote{Another simple idea is to compactify on ``orbifolds'' (\ie manifolds with certain points taken out) because the equations of motion for the superstring Eq.~(\ref{eq:eofm_super}) then stay very simple.}. 
We consider the important example of the two-dimensional torus $\mathscr{T}$ below, but in general the resulting $d=4$ theory after toroidal compactification has too much supersymmetry. For example, in the case of type IIB theory, one would obtain $\calN=8$ supersymmetry in four dimensions. Requesting that only a ``reasonable'' amount of supersymmetry be preserved in four spacetime dimensions (more precisely $\calN=2$ for type IIA/B, and $\calN=1$ for type I and heterotic theories) suggests a compactification on more complicated spaces known as \emph{Calabi Yau (CY) manifolds}. (This can be shown by imposing that the internal manifold admits a covariantly constant spinor.) In general, CY compactifications preserve 3/4 of the original supersymmetry: the original $\calN=8$ for type IIB is reduced to $\calN=2$, which may be broken down further to $\calN=1$ \eg by adding D-branes to the background. 
A vast selection of CY manifolds with three complex (hence six real) dimensions exists \cite{Hubsch:1992nu}, and the choice made to compactify a given theory is of crucial importance for the resulting four-dimensional physics at low energies: as mentioned earlier, dimensionless physical parameters in string theory are determined by the expectation values of scalar fields, and these expectation values in turn are a consequence of the particular higher-dimensional geometry. One may hence schematically write $\mathcal{S}_{10d}[\mathcal{C}]\rightarrow\mathcal{S}_{4d}$, where $\mathcal{C}$ stands for the chosen compactification \cite{Baumann:2009ni}. Moreover, the supergravity approximation is justified when the size of the compact internal manifold is large compared to the string scale, \ie when the mass $m_{\sss{c}}$ is much smaller than $\ms$.

\begin{figure}[t]
\begin{center}
\includegraphics[width=0.45\textwidth]{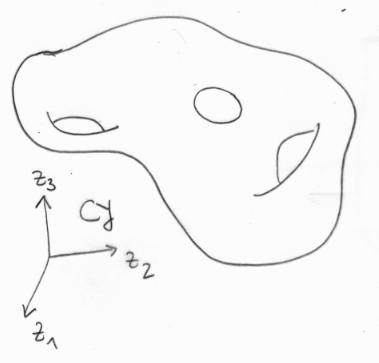}
\hfill
\includegraphics[width=0.45\textwidth]{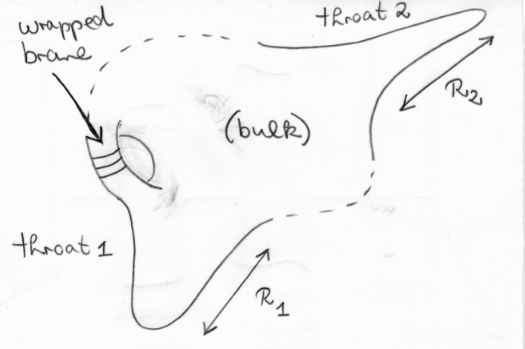}
\caption[Calabi Yau manifolds, and the ``patchwork'' geometry of warped compactifications.]{\small \emph{Left:} The six internal dimensions of a ten-dimensional supergravity description of string theory are compactified on a Calabi Yau manifold, which breaks 3/4 of the original supersymmetry in the background. CY manifolds are complex manifolds with coordinate $z_{1},z_{2},z_{3}$ and a K\"ahler metric. \emph{Right:} More general compactification schemes are possible, \ie one may include a warp factor into the ten-dimensional metric. As we show later, these geometries typically have the form of a cone. The metric on these (non-compacts) cones is usually known, but they have to be attached to a (compact) CY bulk manifold at their UV end, to make the total compactification volume (and hence the Planck constant in four dimensions) a constant.}
\label{fig:compactifications}
\end{center}
\end{figure}

\subsection{Calabi Yau Manifolds}
A Calabi Yau $n$-fold is a \emph{K\"ahler manifold} in $n$ complex dimensions with $SU(n)$ holonomy (see Appendix \ref{app:geotopo}). In one complex dimension, these are the complex plane $\mathbb{C}$ (which is non-compact), and the torus $\scrT$ (which is compact). For $n=3$, many thousand CY manifolds are known, and their number may even be infinite. A compact Calabi Yau $n$-fold is \emph{Ricci flat}, \ie $R_{MN}=0$ (here, $M,N$ are indices running over the $2n$ real coordinates). 
If the CY has a metric, the \emph{Betti numbers} $b_{k},\,k=0,\dots,n$ count the number of linearly independent harmonic $p$-forms on the manifold. For K\"ahler manifolds, the Betti numbers can be decomposed in terms of the \emph{Hodge numbers} $h^{p,q}$ which give the number of harmonic $(p,q)$-forms (see Appendix \ref{appsec:complexmanif}), with $b_{k}=\sum_{p=0}^{k}h^{p,k-p}$. Note that there is no requirement to know the exact form of the metric on the Calabi Yau; indeed it is not known for any non-trivial compact example.\\
One can characterize Calabi Yau manifolds topologically by their Hodge numbers $h^{p,q}$, where $p,q=0,\dots,n$, and $h^{p,q}=h^{q,p}$ from complex conjugation. One also has $h^{p,0}=h^{n-p,0}$ from Poincar\'e duality. Since for any complex manifold $h^{0,0}=1$ (which means that constant functions can exist on the manifold), and $h^{1,0}=0$ for simply-connected manifolds\footnote{The torus, which we consider in Section \ref{subsec:torus} as an example of two-dimensional compactification, is not simply connected and has $h^{1,0}=h^{0,1}=1$.}, the characterization of the $n=3$ Calabi Yau manifold only requires $h^{1,1}$ and $h^{2,1}$. (This is illustrated by grouping the Hodge numbers together in the \emph{Hodge diamond} for CY $n$-folds, see \eg \cite{Nakahara:1990th,Becker:2007zj}. Note that $h^{3,0}=1$ by Poincar\'e duality.) The Euler characteristic [compare Eq.~(\ref{eq:Euler})] of the CY then is
$\chi=\sum_{p=0}^{6}(-1)^{p}\,b_{p}=2(h^{1,1}-h^{2,1})$.

However, CY manifolds with specified Hodge numbers are not unique: inequivalent Calabi Yaus can have the same $h^{1,1}$ and $h^{2,1}$, and be smoothly related among themselves by deformations of the parameters characterizing their size and shape, the so-called (scalar) ``moduli fields''. The moduli fields with their respective possible values span the ``moduli space'' of parameters that may be changed without affecting the topology. 
In Eq.~(\ref{eq:10d-notwarped}), it would therefore be more accurate to write $g_{mn}(y,\omega)$, indicating that the metric on the internal space is not only a function of the coordinates $y$, but also of all other parameters $\omega$ required to fix the geometry completely \cite{Burgess:2007pz}. Once compactified to four dimensions (\ie at energies below the compactifiction scale $m_{\sss{c}}$), we must expect a massless $4d$ scalar field for each modulus of the extra-dimensional metric (and more from the compactification of the $C_{(p)}$ forms in the respective theory). The fact that they are massless means that these fields do not enter into the effective scalar potential (at lowest order at least), and have \apriori no reason to be stabilized at a given value.

\subsection{Warped Compactifications}\label{subsec:warpedcompactifications}
Compactification on compact Calabi Yau manifolds is consistent with the background fields and R-R gauge potentials ($B_{mn}$ and $C_{(p)}$) in the bulk supergravity actions of Section \ref{sec:effectivesugra} set to zero. Note, however, that a supergravity background can include non-perturbative objects like D-branes. This has two important consequences: firstly, these objects source the R-R gauge potentials (a D$p$ is an electric source of $C_{(p+1)}$), which (along with their associated ``fluxes'', \ie the field strengths $F_{(p+2)}$) should then have non-zero background values. The fluxes are often said to ``thread'' cycles of the $\scrM_{6}$ manifold. (Recall that the gauge potentials' field strengths are forbidden to have ``legs'' along the extended dimensions by the requirement of maximal symmetry.) If some of the $p$ spatial dimensions of a D$p$-brane lie along the internal manifold $\scrM_{6}$, the brane is ``wrapped'' along the corresponding dimension (in type IIB, this will be the case for D-branes with $p=5,7,9$ dimensions\footnote{The D9-brane is not dynamical because it is spacetime-filling in ten dimensions.}). This is precisely a case where \eg the induced $B_{(2)}$ field that enters into the DBI brane action (\ref{eq:DBIbraneaction}) no longer vanishes. As we mentioned in Section \ref{subsec:DBIterm}, such a brane then carries also induced charges of lower $p$-type: for example, a D5-brane with two wrapped dimensions, it then also contributes a so-called ``fractional'' D3-charge.\\
Secondly, D$p$-branes embedded into the ten spacetime dimensions give rise to additional scalar moduli fields associated with their position within the internal manifold (see Figure \ref{fig:braneembeddings}), on top of the geometric moduli $\omega$ that arise from the choice of compactification. (Note that $\overline{D}$-branes do not have world volume moduli because they minimize their energy at special positions in the background.) We shall see in the next Chapter that it is of crucial importance in string-cosmological model building that (almost all) moduli be fixed. Since our ultimate goal is to study models of inflation derived from string theory, let us rephrase this condition in ``inflationary'' terminology: the moduli fields (or at least most of them) should be heavy compared to the Hubble scale, $m^{2}_\umod\gg H^{2}$, leaving only a manageable number of dynamic scalar fields during inflation.

The metric ansatz of Eq.~(\ref{eq:10d-notwarped}) therefore must be generalized to a larger class of compactification manifolds allowing for the presence of branes and non-zero background values for NS-NS and R-R fields. The ten-dimensional space in this case is a so-called ``warped'' product of a Minkowski space and an internal manifold. Its metric can be written as
\beq\label{eq:10d-warped}
\dd s^{2}_{10}=h^{-1/2}(y)\,\eta_{\mu\nu}\,\dd x^{\mu}\,\dd x^{\nu}+h^{1/2}(y)\,g_{MN}(y)\,\dd y^{M}\,\dd y^{N}\,,
\eeq
where the function $h(y)$ (which depends on the internal coordinates $y^{M}$) is called the warp factor. The internal manifold with metric $g_{MN}(y)$ 
is no longer necessarily Calabi Yau and can be non-compact. (It does not even have to be a K\"ahler or a complex manifold.) A well-known example of a non-compact CY used in ``flux compactifications'' is the so-called conifold we study below. In this particular case, even the explicit metric on the manifold is known.\\
There are two main advantages in choosing a ten-dimensional metric of the form of Eq.~(\ref{eq:10d-warped}) over one like Eq.~(\ref{eq:10d-notwarped}). Firstly, the warp factor $h(y)$ [which can have a very strong, \ie exponential dependence on (proper) distances within the internal manifold] can be used to explain the hierarchy between the strength of physical interactions. This is analogous to the \emph{Randall Sundrum models} \cite{Randall:1999vf,Randall:1999ee} (the first of the ``brane world'' scenarios mentioned earlier), apart from the fact that these live in five instead of ten dimensions. Secondly, as it will be discussed at length in later Chapters, the presence of fluxes allows for the stabilization of certain moduli fields, a feature much sought after in models of string cosmology. In particular, the non-compact conifold mentioned above comes in different varieties (called simple, deformed or resolved) depending on which of its internal cycles (if any) is threaded by fluxes. Intuitively, the moduli stabilization mechanism provided by fluxes (or branes wrapping internal cycles) may be understood as follows: without fluxes, cycles in the internal manifold can be deformed at will without energy cost. If there is a flux along the cycle, changing its size modifies the energy distribution of the background, and the cycle is therefore stabilized at the position of least energy. In terms of the effective scalar potential of the lower-dimensional theory, it has acquired a mass.\\
The conifold example, however, also points to a major problem of warped compactifications: the manifolds employed can be non-compact. At the technical level, this means that at least the modulus field corresponding to the overall compactification volume $V_{6}$ remains unfixed. This is problematic because $V_{6}$ determines the four-dimensional Planck scale via Eq.~(\ref{eq:Planckmassrelations}). Therefore, the metric (\ref{eq:10d-warped}) is usually considered to be valid only in some ``corner'' of the overall six-dimensional space, and should be glued smoothly into a compact Calabi Yau bulk of the form (\ref{eq:10d-notwarped}) somewhere. On the one hand, this ``patchwork'' structure of the compactified dimensions is useful in string theoretic model building because one can attach different corners with \eg different warp factors $h_{1}$ and $h_{2}$ [as well as different local metrics $g_{MN}^{(1)},g_{MN}^{(2)}$] in Eq.~(\ref{eq:10d-warped}) to the same bulk. One of these then might produce inflation in the early Universe, while another one can contain the Standard Model of particle physics. On the other hand, the metric of the compact bulk connecting the two corners is unknown.

\section{Duality Relations}\label{subsec:dualities}
In Section \ref{subsec:five} we discussed the five superstring theories and mentioned that they are considered as different realizations of one underlying framework. This hope is fueled by the fact that, as indicated in Fig.~\ref{fig:stringdualities}, the distinct formulations can be obtained one from the other by certain transformations, on which we now comment.

\textsc{T-duality}\\
The ``T'' refers to ``target space'', because under this duality, compactification of a dimension on a circle with radius $R$ is exchanged with compactification on $\tilde{R}=\alphap/R$. (The string scale is often set to unity, $\alpha'=1$, hence T-duality takes $R\rightarrow1/R$.) The resulting physics in each case is identical, which illustrates the fact that spacetime geometry is ``seen'' differently by extended objects like strings than it would be by point particles.\\
A T-duality transformation takes the two type II string theories one into the other, as well as the two heterotic theories. (If more than one direction is toroidially compactified, an even number of transformations gives back the same theory on the dual torus.) More precisely, $R$ and $\tilde{R}$ are the two limits of a continuous change of the compactification radius, which is a dynamical quantity. Note that in this sense, the string coupling $\gs$ measures the compactification radius of the eleventh dimension that was curled up (with $\gs=\sqrt{\alpha'}\,R$) to obtain type IIA superstring theory from $d=11$ supergravity. 
T-duality moreover justifies the existence of D-branes as fundamental objects of string theory because it can be shown that by taking $R\rightarrow\alphap/R$, Neumann boundary conditions in the direction of the compactified dimension map into Dirichlet boundary conditions.

\textsc{$S$-duality}\\
This duality relates the limits of small and strong string coupling by exchanging $\gs\rightarrow 1/\gs$. Again, recall that $\gs$ is fixed by the vacuum expectation value of the dilaton\footnote{The dilaton, together with its superpartner the axion, belongs to a chiral superfield usually denoted by $S$.}, $\gs=e^{\Phi_{0}}$, therefore $S$-duality is a field transformation taking the dilaton $\Phi\rightarrow-\Phi$ (and is in this sense again continuous). Remarkably, it is possible to study this duality at the level of the supergravity effective actions for the five theories (of which we only discussed the type IIA and type IIB cases in Section \ref{sec:effectivesugra}). This can be understood as supersymmetry ``protecting'' string quantities as the extrapolation from weak to strong coupling is carried out. $S$-duality relates type I to $SO(32)$ heterotic theory, while type IIB is self-dual under it. When type IIA theory is taken to the strong coupling limit, it ``grows'' an eleventh dimension (see above), leading to the low-energy supergravity limit of M-theory.

\textsc{Mirror symmetry and geometric transitions}\\
Mirror symmetry is a property relating certain Calabi Yau manifolds, and is most easily understood as an analogue of T-duality: two CY three-folds $\mathscr{Y}$ and $\mathscr{W}$ have mirror symmetry\footnote{In the Hodge diamond, this looks like a mirror transformation along its central axis.} if their cohomology groups (see Appendix \ref{app:geotopo}) satisfy $\mathscr{H}^{p,q}(\mathscr{Y})=\mathscr{H}^{3-p,q}(\mathscr{W})$. As a consequence, their Hodge numbers are interchanged, \ie $h^{1,1}(\mathscr{Y})=h^{2,1}(\mathscr{W})$, and \emph{vice versa}. It turns out that type IIA theory compactified on $\mathscr{Y}$ describes the same physics as type IIB compactified on $\mathscr{W}$. Their moduli spaces then are the same, and we come back to this in the next Chapter. (It is, however, possible that two CY manifolds have the same Hodge numbers, yet disjunct moduli spaces.)\\
Going beyond the supergravity approximation, one can also have ``geometric transitions'', which describe a smooth change in topology. These can even affect the Hodge numbers of manifolds, and one example is the ``conifold transition'' relating the singular, deformed and resolved conifolds we encountered earlier in the discussion of warped compactifications. Geometric transitions establish a link between backgrounds that contain D-branes (\ie sources of $C_{(n)}$ gauge potentials), and warped backgrounds with fluxes only (\ie with non-zero field strength $F_{(n+1)}=\dd C_{(n)}$, but no localized sources). This allows both to understand the presence and the quantization of flux in the latter backgrounds: the sources of the flux are explicit in the dual background, where they come in integer units because of the Dirac condition.

\bigskip

We therefore made it plausible that the five different formulations of superstring theory are related by duality transformations and describe but one underlying framework. However, it turns out that there is a vast number of possible compactifications from ten down to four dimensions, both on Calabi Yau manifolds and on more general ``warped'' backgrounds. Each of these gives rise to a different string theory vacuum (in which, at this point of our discussion, supersymmetry in four dimensions is still conserved). Therefore, we may have a unique theory, but infinitely many realizations of low-energy physics depending on the properties of each vacuum. [Note that this situation is not unlike General Relativity, where the Einstein equations also admit many solutions (including higher dimensional ones) that do not describe our Universe. In the context of cosmology, arguments of simplicity and symmetry help us to select a solution.] In the context of string theory, this is often referred to as the ``landscape'' \cite{Susskind:2003kw,Susskind:2005bd}. In the absence of a good criterion\footnote{other than our own existence, an argument known as the ``anthropic principle''} for choosing one among these vacua, one can consider the statistical probability for finding \eg a positive cosmological constant in one of them \cite{Douglas:2003um,Douglas:2004qg,Kumar:2006tn}.\\
With the elementary and advanced string theoretic tools of Chapter \ref{chapter:stringelements} and the present one in hand, we are ready to embark on the adventure of searching for string cosmological models. Before we proceed, let us add both a general \emph{caveat} and a precise motivation. It is partly by construction that inflation hides from our view most of its GUT scale origin because only a small window of scales (leaving the Hubble horizon towards the end of inflation) is accessible. The slow roll mechanism further reduces information to very few generic parameters constrained from observations. This makes detailed knowledge of the inflaton's interactions an accessory detail rather than a necessity. From a cosmologist's point of view, one may therefore ask what is to be gained from finding inflation's underlying theory -- apart from intellectual satisfaction? Among the reasons to continue and extend the search for the (string theoretic or other) origin of the inflaton, let us cite the following two \cite{Burgess:2007pz}: we will see that \eg the geometric interpretation of stringy inflaton candidates lends justification to seemingly arbitrary quantities like the range of field values. On the other hand, since string theory is also aimed at describing the Standard Model of particle physics, it offers the hope of a complete understanding of reheating, since both ends of the theory (the inflaton and the degrees of freedom after inflation) are known.

\chapter{Moduli and Their Stabilization}\label{chapter:stringcosmo}
\begin{quotation}
\emph{So far, we encountered several classes of scalars fields in inflation, among them the vast amount of moduli fields describing the (compactification of) the higher-dimensional geometry. In this Chapter we take a closer look at moduli fields, which can be of the complex structure or the K\"ahler structure type, and lay the groundwork for understanding their respective stabilization mechanisms.}
\end{quotation}

\section{Moduli Space}
We mentioned in Section \ref{subsec:compactify} that the ten-dimensional supergravity formulations of string theory, when  compactified down to $d=4$ on Calabi Yau three-folds, lead to an effective theory with a large number of massless scalars. These fields span the so-called moduli space and describe the size and shape of the compactification geometry. More precisely, moduli fields come in two different varieties, those related to the complex structure of the compactification, and the so-called \emph{K\"ahler structure moduli}. 
An important property of moduli space is that it is (locally) a direct product of these two components, $\scrM_\umod=\scrM_\uC\times\scrM_\uK$. For a CY manifold $\mathscr{Y}$, the dimension of the two parts is related to the Betti number $b_{2}$ and $b_{3}$ of the CY by $\dim (\scrM_\uC)=(b_{3}/2)-1$ and $\dim(\scrM_\uK)=b_{2}$ \cite{Douglas:2006es}. In terms of Hodge numbers, $\dim_\uc(\scrM_\uC)=h^{2,1}(\mathscr{Y})$ and $\dim_\uc(\scrM_\uK)=h^{1,1}(\mathscr{Y})$, where $\dim_\uc$ is the complex dimension \cite{Becker:2007zj}. It follows that under the operation of mirror symmetry, discussed in Section \ref{subsec:dualities}, the two parts are exchanged, but the two mirror manifolds have the same product moduli space $\scrM_\umod$.\\
Moving from a point $P_{1}$ in moduli space to $P_{2}$ corresponds to a continuous deformation of the parameters of the CY manifold $\mathscr{Y}_{1}$ into those of a second Calabi Yau $\mathscr{Y}_{2}$ with the same Hodge numbers (but interchanged). This means that the topology is not affected because the two Euler characteristics are related by $\chi(\mathscr{Y}_{1})=-\chi(\mathscr{Y}_{2})$. Fluctuations around a given CY $\mathscr{Y}_{1}$ therefore parametrize ways in which the geometry can be deformed without changing the topology, and one may (at least locally) use them to define a metric on moduli space\footnote{Note that this should not be confused with the metric \emph{on} a given Calabi Yau manifold, which, as we mentioned earlier, is not known for non-trivial compact examples.}. Because of the local product structure $\scrM_\umod=\scrM_\uC\times\scrM_\uK$ of moduli space, one may study separately both the complex structure and the K\"ahler structure deformations around their values for a given CY.

\subsection{Compactification on the Torus}\label{subsec:torus}
As an example, it is instructive to consider compactifying two dimensions in bosonic string theory on a torus $\mathscr{T}$, which is a compact CY one-fold with Hodge numbers $h^{0,0}=h^{1,1}=h^{1,0}=h^{0,1}=1$ and Euler characteristic $\chi(\mathscr{T})=0$. We first focus on deformations of the complex structure moduli space only: the scalar components arising from the metric $g_{mn}$ after compactification of two dimensions $(x,y)$ are
\beq
g=\left(\begin{array}{cc}g_{11}&g_{12}\\g_{12}&g_{22}\end{array}\right)\,, \qquad \sqrt{\det g}=1\,,\qquad\tau=\tau_{1}+i\tau_{2}=\frac{g_{12}}{g_{22}}+i\,\frac{1}{g_{22}}\,,
\eeq
\ie after fixing the overall volume (by setting $\sqrt{\det g}=1$), there are two real parameters left. They can be combined into the so-called complex structure $\tau$ of $\mathscr{T}$. The metric on the torus can then be written as
\beq\label{eq:toruslineelement}
\dd s_{\mathscr{T}}^{2}=\frac{1}{\tau_{2}}\left[\left(\tau_{1}^{2}+\tau_{2}^{2}\right)\dd x^{2}+2\tau_{1}\,\dd x\,\dd y+\dd y^{2}\right]\,,
\eeq
and one can define local complex coordinates $\dd z=\dd y+\tau\dd x$ (and the complex conjugate), in terms of which the torus line element (\ref{eq:toruslineelement}) reads $\dd s_{\mathscr{T}}^{2}=2g_{z\bar{z}}\,\dd z\,\dd \bar{z}$ with $g_{z\bar{z}}=1/2\tau_{2}$. 
[The $(\dd z,\dd\bar{z})$ define a basis of holomorphic (anti-holomorphic) one-forms on $\mathscr{T}$, see Appendix \ref{app:geotopo}.]\\
It can be shown (see \eg \cite{Becker:2007zj}) that the complex structure moduli space of the torus (parameterized by $\tau$) with one complex dimension is itself a K\"ahler manifold, which means that its metric $G_{\tau\bar{\tau}}$ is hermitian ($G_{\tau\tau}=G_{\bar{\tau}\bar{\tau}}=0$) and admits a K\"ahler potential (see Appendix \ref{app:geotopo})
\beq\label{eq:Kaehler-torus}
K(\tau,\bar{\tau})=-\log\left(i\int\dd z\wedge\dd \bar{z}\right)=-\log\left(2\tau_{2}\right),\qquad G_{\tau\bar{\tau}}=\partial_{\tau}\partial_{\bar{\tau}}K=\frac{1}{4\tau^{2}_{2}}\,.
\eeq
Therefore, the line element on the complex structure moduli space is $\dd s_{\mathscr{M_\uC}}^{2}=\dd\tau\,\dd\bar{\tau}/2\tau_{2}^{2}$. 
Note that we have held the total volume of the torus fixed by setting $\sqrt{\det g}=1$. If we drop this restriction, there is additional scalar degree of freedom (the ``radial modulus'') in the total moduli space of $\mathscr{T}$.

We know that the spectrum of the bosonic string at the massless level contains the Kalb Ramond anti-symmetric tensor field $B_{mn}$ on top of the metric $g_{mn}$. Let us see what moduli arise after toroidal compactification when both of these fields are considered. \emph{A priori}, the resulting scalar components after compactification now are
\beq
g=\left(\begin{array}{cc}g_{11}&g_{12}\\g_{12}&g_{22}\end{array}\right),\qquad B=\left(\begin{array}{cc}0&B_{12}\\-B_{12}&0\end{array}\right)\,,
\eeq
\ie there are four real parameters, which can be redefined as
\bea
\tau&=&\tau_{1}+i\tau_{2}=\frac{g_{12}}{g_{22}}+i\,\frac{\sqrt{\det g}}{g_{22}}\,,\\
\rho&=&\rho_{1}+i\rho_{2}=B_{12}+i\,\sqrt{\det g}\,.
\eea
As before $\tau$ parametrizes the complex structure moduli space in this example, and since $\rho_{2}=\sqrt{\det g}$, $\mathrm{Im}(\rho)$ is a measure for the total volume of the torus. 
Let us simplify to a rectangular torus (with radii $R_{1}$ and $R_{2}$ in the respective directions) for the moment. Then we have $g_{12}=0$, and the diagonal elements of $g$ are given by $g_{11}=R_{1}^{2},\,g_{22}=R_{2}^{2}$. Therefore the complex structure parameter is calculated as $\tau=i\,(R_{1}/R_{2})$, and $\rho=B_{12}+i\,R_{1}R_{2}$.\\
We know from Section \ref{subsec:dualities} that mirror symmetry exchanges the complex and the K\"ahler structure parts of moduli space, \ie here it takes\footnote{On the torus, mirror symmetry is just T-duality, which exchanges $R_{1,2}\rightarrow1/R_{1,2}$.}
 $\tau\leftrightarrow\rho$. Therefore it follows that $\rho$ must be purely imaginary if $\tau$ is, and $B_{12}=0$ in this case. However, for general tori at an angle, \ie with off-diagonal metric elements $g_{12}\neq0$, there is a non-zero component $B_{12}$, hence the fields $g$ and $B$ must be considered together.
 
\begin{figure}[t]
\begin{center}
\includegraphics[width=0.6\textwidth]{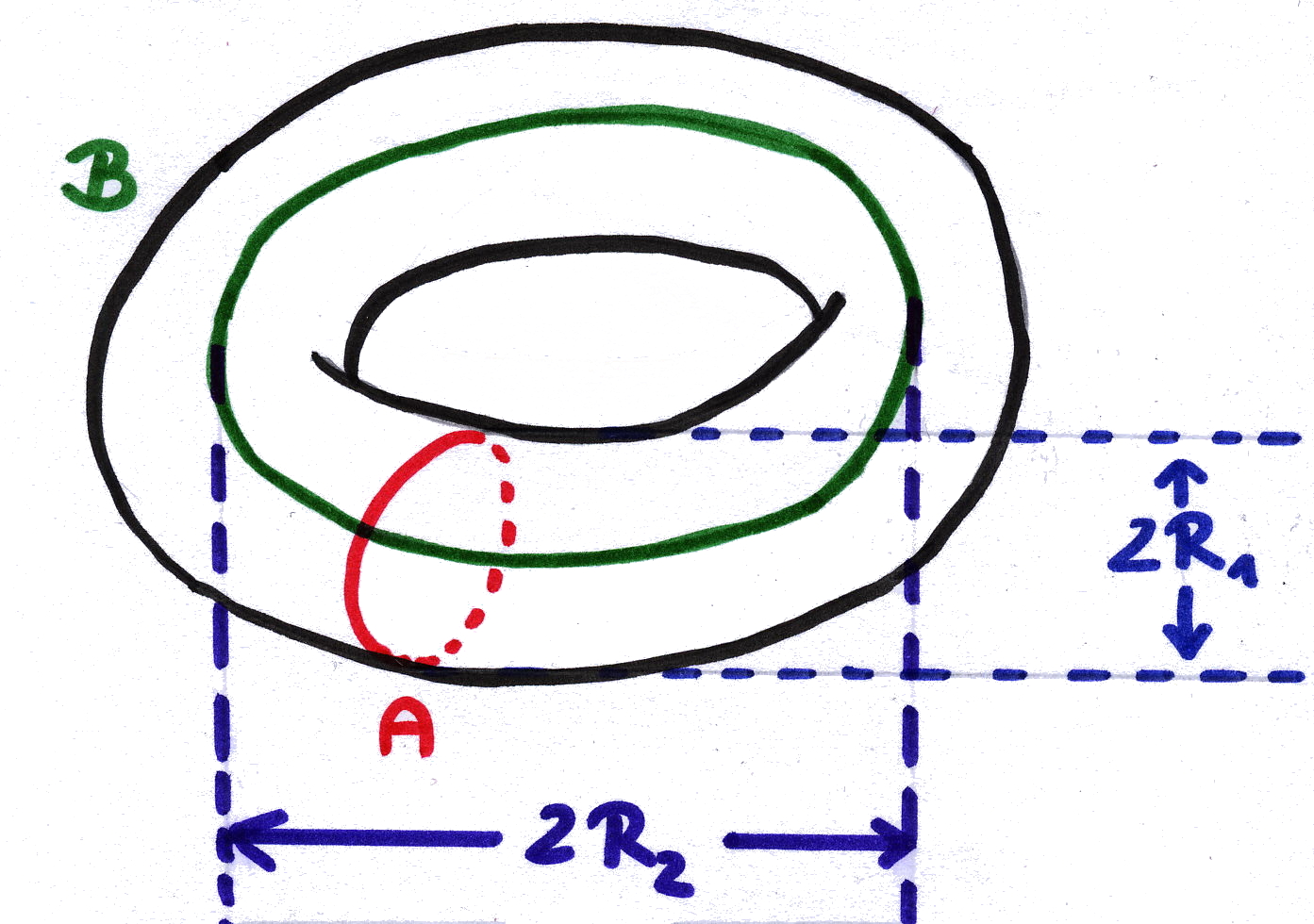}
\hfill
\begin{minipage}[b]{0.3\textwidth}
\includegraphics[width=\textwidth]{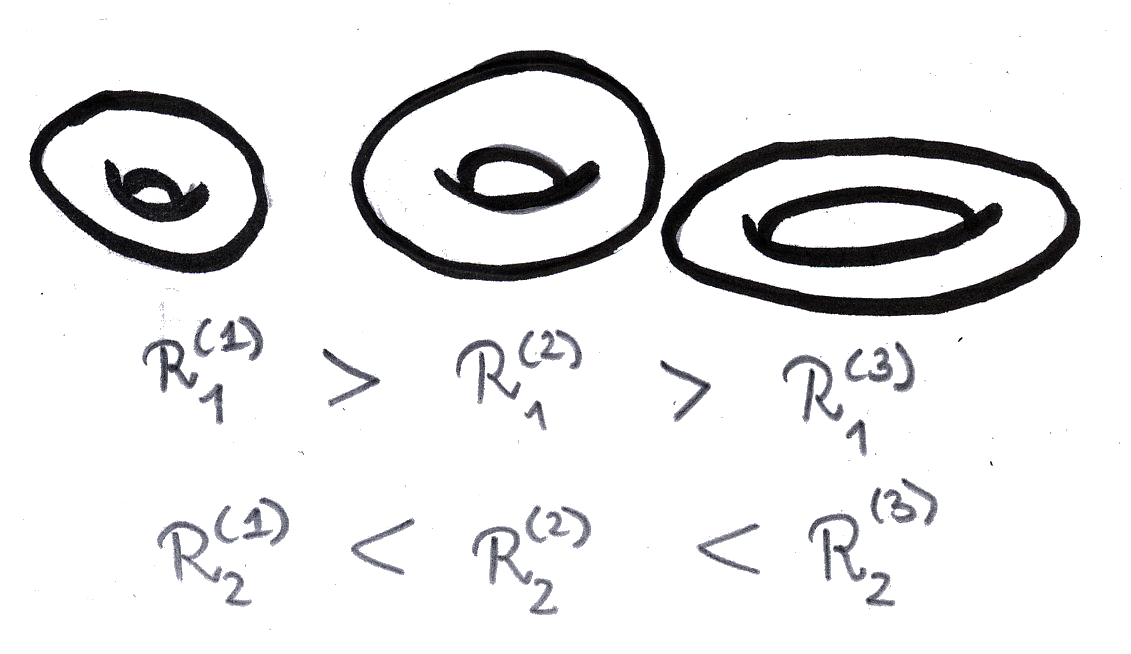}
\includegraphics[width=\textwidth]{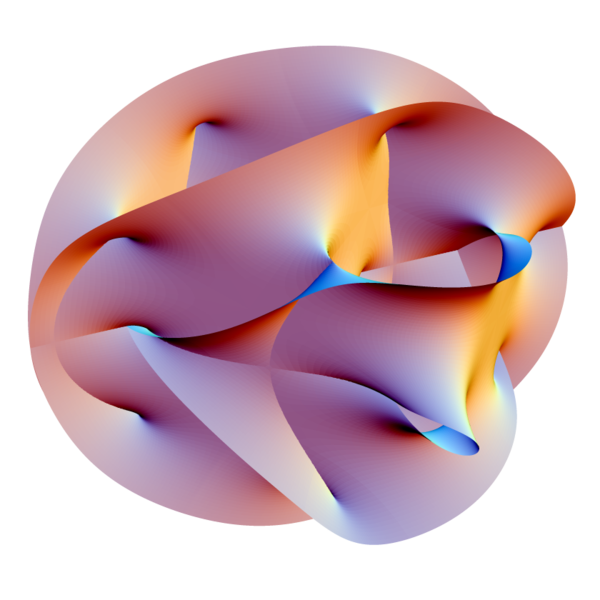}
\end{minipage}
\caption[Dual circles on the torus and their deformation, symbolic representation of a Calabi Yau manifold. (Source: Simple English Wikipedia website)]{\small \emph{Left:} A torus with its two dual cycles $\mathscr{A}$ and $\mathscr{B}$, with radii $R_{1}$ and $R_{2}$. \emph{Top right:} With the overall volume of the torus is kept constant, one can still change the radii of the cycles $\mathscr{A}$ and $\mathscr{B}$ without changing the topology: the radii are moduli fields for the simple toroidal compactification. \emph{Bottom right:} A symbolic representation of a Calabi Yau manifold typically used to compactify the internal spaces of superstring theory. Such a manifold can have many hundred moduli fields describing the size and shape. (Figure from Simple English Wikipedia)}
\label{fig:torus}
\end{center}
\end{figure}

\subsection{Moduli From Calabi Yau Three-Folds}\label{subsec:CYmoduli}
For compactification of the  $d_{\mathrm{int}}=6$ internal dimensions of superstring theory on a Calabi Yau manifold $\mathscr{Y}$, we again start with a simplified discussion of metric fluctuations $\delta g_{ab}$ (though we know that the components of the Kalb Ramond field must be considered on an equal footing). These fluctuations parametrize the complex structure part $\mathscr{M}_\uC$ of moduli space. For a Calabi Yau three-fold, we are interested in deformations under which the topological property of Ricci flatness in the extra dimensions is preserved, \ie one imposes both $R_{AB}(g)=0$ and $R_{AB}(g+\delta g)=0$ for indices $A,B$ running over the internal space. (One additionally demands that both the perturbed and unperturbed metric are K\"ahler manifolds.) This leads to the \emph{Lichnerowicz equation} for the perturbations $\delta g_{mn}$, in which the ten-dimensional operator $\Delta_{10d}$ appears.\\
The existence for (possibly many hundred) moduli fields in four dimensions is seen when the Lichnerowicz equation is decomposed into the four-dimensional external and the six-dimensional internal piece. Metric fluctuations $\delta g_{AB}(x^{\mu},y^{M})$ in the extra dimensions only then can be written as
\beq
\delta g_{AB}(x^{\mu},y^{M})=\sum_{k}\phi_{k}(x^{\mu})\,h_{AB}^{k}(y^{M})\,,
\eeq
where $h_{AB}(y^{M})$ are the tensor eigenfunctions of the \emph{Lichnerowitz operator} in six dimensions $\Delta_{6d}$, with $\Delta_{6d}h_{AB}(y^{M})=-m_{k}^{2}\,h_{AB}(y^{M})$. It then follows from the ten-dimensional equations that the mode functions $\phi_{k}$ (which only live in four dimensions, \ie they depend on the $x^{\mu}$) must satisfy $\left(\Box_{4d}-m_{k}^{2}\right)\phi_{k}=0$, which is the equation of motion of a scalar field in four dimensions with mass $m_{k}$. Moduli fields are those modes for which the mass vanishes, \ie a deformation of the Einstein equations in their direction comes without energy cost.

It can be shown, however, that under T-duality/mirror symmetry, the fields $g_{mn}$ and $B_{mn}$ mix (as it was the case for $\tau$ and $\rho$ in the simple example of the two-dimensional torus), hence at the same time as the perturbations $\delta g_{mn}$, one has to consider $\delta B_{mn}$. Using the strategy of perturbing around a given Calabi Yau $\mathscr{Y}$ to obtain the metric on moduli space $\scrM_\umod=\scrM_\uC\times\scrM_\uK$, one writes the most general variation of the moduli space line element as
\beq\label{eq:modulispacemetric}
\dd s^{2}_{\umod}=\frac{1}{2\,V_{6}(\mathscr{Y})}\int\dd^{6}x\sqrt{g}\,g^{a\bar{b}}g^{c\bar{d}}\left[\delta g_{ac}\,\delta g_{\bar{b}\bar{d}}+\left(\delta g_{a\bar{d}}\,\delta g_{c\bar{b}}-\delta B_{a\bar{d}}\,\delta B_{c\bar{b}}\right)\right]\,.
\eeq
By $V_{6}(\mathscr{Y})$, we denote the volume of the Calabi Yau three-fold. Note that the background metric elements $g_{ac}$ and $g_{\bar{b}\bar{d}}$ vanish because $\mathscr{Y}$ is a K\"ahler manifold, but the perturbations in these components have to be considered. The line element (\ref{eq:modulispacemetric}) is then rewritten in terms of the fundamental $(p,q)$-forms admitted by the CY (which are counted by the Hodge numbers $h^{p,q}$). Luckily, the exact form of the Calabi Yau metric $g_{mn}$ itself is not needed. One finally finds that both components of moduli space, the complex structure part $\scrM_\uC$ and the K\"ahler structure moduli space $\scrM_\uK$, are themselves K\"ahler manifods with K\"ahler potentials [compare Eq.~(\ref{eq:Kaehler-torus}) for $\mathscr{T}$] \cite{Becker:2007zj}
\beq\label{eq:moduliKaehlerpotentials}
K^{2,1}=-\log\left(i\int_{\scrM_{6}}\Omega\wedge\bar{\Omega}\right)\,,\qquad K^{1,1}=-\log\left(\frac{4}{3}\int_{\scrM_{6}} J\wedge J\wedge J\right)\,.
\eeq
Here, the superscripts ``2,1'' and ``1,1'' translate the fact that the dimension of the respective component is given by the Hodge numbers $h^{2,1}(\mathscr{Y})$ and $h^{1,1}(\mathscr{Y})$, respectively. By $\Omega$ we denote the (3,0)-form on $\mathscr{Y}$ (which is unique up to a prefactor, and the wedge product $\Omega\wedge\bar{\Omega}$ gives the volume), and $J$ is the so-called \emph{K\"ahler form} of $\mathscr{Y}$ (see Appendix \ref{app:geotopo}). 
It can be shown that also the $K^{1,1}$ potential can be expressed in terms of the volume, with $\exp(-K^{1,1})=\frac{4}{3}\int_{\scrM_{6}} J\wedge J\wedge J=8\,V_{6}$. Note that via $J$, the second potential $K^{1,1}$ contains the K\"ahler modulus field $\rho$, to which we return in Chapter \ref{chapter:kaehlerstabilized}.

\subsubsection{Integrating out heavy fields}
Let us return to a schematic notation and assume that we can expand each scalar field $\tilde{\varphi}(x^{m},y^{M})$ arising from a higher dimensional $p$-form or the ten-dimensional metric as 
\beq\label{eq:modulimodeexpansion}
\tilde{\varphi}(x^{\mu},y^{M})=\sum_{m}\varphi_{m}(x^{\mu})\,f_{m}(y^{M})\,,
\eeq
where the $f_{m}(y^{M})$ are complete set of eigenfunctions of the appropriate wave operator\footnote{Let us illustrate this on the example of the torus $\mathscr{T}$ again: in this case, the internal operator $\Delta$ is two-dimensional and in terms of the vector ``counterparts'' $\partial_{z}=(i/2\tau_{2})\left[\bar{\tau}\,(\partial/\partial x)-(\partial/\partial y)\right]$ and $\partial_{\bar{z}}$ of the one-forms $(\dd z,\dd\bar{z})$, it is written as $\Delta=-2\partial_{z}\partial_{\bar{z}}$. (Note that we drop the restriction $\sqrt{\det g}=1$.) The eigenfunctions and -values of this operator are
\beq
f_{m,n}(x,y)=\exp\left[2\pi i\left(n\,y+m\,y\right)\right]\,,\qquad\lambda_{m,n}=\frac{2\pi^{2}}{\tau_{2}^{2}}\,(m-\tau n)(m-\bar{\tau}n)\,.
\eeq} on the internal space $\scrM_{6}$. In the ten-dimensional actions of Section \ref{sec:effectivesugra}, the full Lagrangian $\calL_{10d}$ is integrated  over all ten spacetime dimensions. To obtain the Lagrangian in four dimensions, let us integrate only over the internal coordinates, $\calL_{4d}[\varphi_{m}(x^{\mu})]=\int\dd^{6}y\,\calL_{10d}[\tilde{\varphi}(x^{\mu},y^{M})]$. This still depends on all $\varphi_{m}(x^{\mu})$ regardless of their mass. To identify Lagrangian for the four-dimensional moduli fields (which are by definition massless), we split $\tilde{\varphi}(x^{\mu},y^{M})$ up into light (\ie massless) and heavy contributions, depending on whether they have a zero or non-vanishing eigenvalue under the internal wave operator. 
The heavy modes $\varphi_{m}^{\mathrm{(h)}}$ are integrated out, but they cannot simply be ``set to zero''. Instead, one must extremize the action with respect to them while holding the light fields fixed. After integrating over the internal manifold, the heavy fields in the expansion (\ref{eq:modulimodeexpansion}) contribute to the effective potential in the resulting $d_{\mathrm{ext}}=4$ theory, but they are not dynamic. The light modes, called moduli, do not enter into the potential, but have (potentially non-canonical) kinetic terms in the four-dimensional Lagrangian.\\
The choice to compactify on Calabi Yau manifolds was motivated by obtaining a supersymmetric four-dimensional theory and we saw earlier in Section \ref{subsec:susylagrangians} that a (chiral) superfield Lagrangian is entirely fixed by prescribing a superpotential $W$ (a holomorphic function, which, in our present notation means that it depends on the $\phi_{m}$ only, and not on their complex conjugates) and a general function $K$ (the K\"ahler potential, which can contain $\phi_{m}$ as well as $\bar{\phi}_{m}$). The remaining step  from superstring theory towards a four-dimensional Lagrangian of the familiar supergravity form therefore consists in determining $W(\phi_{m})$ and $K(\phi_{m},\bar{\phi}_{m})$ after the compactification. Note that we have just completed this step for the latter of the two functions: we obtained the K\"ahler potentials for both the complex and the K\"ahler structure moduli in Eqs.~(\ref{eq:moduliKaehlerpotentials}). We will see how the superpotential $W$ is calculated in concrete backgrounds below. Before we turn to the question of moduli stabilization, we now briefly list the moduli field content of type IIA and type IIB supergravity.

\subsection{Moduli of Type IIA/B Superstring Theory}
The field content of the two theories in ten dimensions was listed in Sections \ref{subsec:typeIIAsugra} and \ref{subsec:typeIIBsugra}, and we know that compactification on a Calabi Yau manifold in this case leads to  $d_{\mathrm{ext}}=4,\,\calN=2$ supersymmetry. The ten-dimensional metric $g_{mn}$ as well as the $B_{mn}$ field and the R-R gauge potentials give rise scalar zero modes upon compactification, and (since the lower-dimensional theory is still supersymmetric) these moduli fields belong to supergravity vector and chiral multiplets analogous to the one discussed in Section \ref{subsec:multiplets}.\\
For type IIA superstring theory, these four-dimensional fields are $h^{1,1}(\mathscr{Y})$ abelian vector multiplets and $h^{2,1}(\mathscr{Y})+1$ hypermultiplets in the four-dimensional theory. (The product form of the moduli space inhibits mixiing between these sets of moduli). Each vector multiplet gives rise to two real scalar fields, therefore the (real) dimension of the K\"ahler structure moduli space is $\dim(\scrM_\uK)=2h^{1,1}$. Each hypermultiplet contains four real scalar fields, so $\dim(\scrM_\uC)=4(h^{2,1}+1)$. For type IIB, these dimensionalities are reversed: there are $h^{2,1}$ abelian vector multiplets and $h^{1,1}+1$ hypermultiplets. The additional moduli fields from the ``+1'' in both theories are due to the dilaton (and another axionic partner which comes from the four-dimensional Poincar\'e dual of the two-form $B_{\mu\nu}$). Note that the overall volume modulus corresponding to $V_{6}(\mathscr{Y})$ always survives down to the four-dimensional theory because of the scale-invariance of the supergravity equations of motion. (This modulus is sometimes called the ``breathing mode''.) If there are non-perturbative objects such as D-branes present in the background geometry, they will give rise to additional moduli fields.

\section{Stabilization Techniques}
We saw that even the simplest compactifications of superstring theory leave us with a proliferation of complex scalar fields $\phi_{m}$ in four dimensions. By definition they do not contribute to the effective potential $V$ in four dimensions at lowest order because a variation of the Einstein equations in the direction of a modulus has no energy cost. 
Therefore, the moduli fields generically are unfixed and not stabilized to a particular vacuum expectation value (which would make them massive) at tree level. It is expected that loop corrections, supersymmetry breaking and non-perturbative effects generate a potential for moduli fields, which is typically very shallow.\\
This description makes moduli fields sound like cut-out candidates for inflation. However, the obstacle to this is that for slow-roll inflation, the potential must be nearly flat in its steepest direction. In multidimensional moduli space this means that one must know the correction-induced dependence of $V$ on all moduli fields at higher order before one can tell whether a given modulus can serve as an inflaton. Any mechanism that lifts the ``flat directions'' of the other moduli in the potential generically will also lift the inflaton flat direction. In the terminology of Section \ref{subsec:CYmoduli}, one might say that integrating over the ``heavy'' fields among the modes of Eq.~(\ref{eq:modulimodeexpansion}) produces contributions to the potential for the remaining ``light'' degrees of freedom, which can render their potential directions too steep for inflation. It was therefore not until significant progress was made in the direction of ``moduli stabilization'', dramatically reducing the number of dynamical fields in the four-dimensional theory, that inflationary model building in string theory could proceed successfully.\\
Above, we argued that flux compactifications may allow to fix most or all of the complex structure moduli: the fluxes threading internal cycles bestow an energy cost on the variation of previously unfixed geometric parameters. In the next section, we lay the ground work for understanding the mechanism of flux stabilization techniques. However, the K\"ahler structure moduli still remain massless after turning on fluxes. Their stabilization requires non-perturbative effects, and we come back to them in Chapter \ref{chapter:kaehlerstabilized}.\\
Two examples of moduli of special importance are the dilaton $\Phi$ and the K\"ahler structure modulus $\rho$ fixing the overall size of the compactification manifold. Note that they parametrize the validity of the string loop and  supergravity approximations, respectively. These expansions rely on $\gs<1$ and $m_{\sss{c}}<\ms$, \ie the compactification manifold should be large with respect to the fundamental string scale. As long as $\Phi$ and $\rho$ are massless, they can be tuned to arbitrary precision. In the presence of fluxes, however, the stabilization of the other moduli is usually bought at the prize of a certain loss of control over the $\gs$ and supergravity expansions. Remarkably, there exists a flux compactification solution in which the supergravity expansion is justified in the $N\gg1$ limit, where $N$ is the amount of flux turned on. Moreover, in this solution, the dilaton  $\Phi$ is constant because it is ``protected'' by the remaining supersymmetry of the theory.

Proceeding from here is difficult because ``everything happens at once'', and all steps in the stabilization process are closely intertwined. We therefore take a step back and return to the original ten-dimensional supergravity action: by deriving the properties of fluxes enforced by the equations of motion derived from this action, we develop an understanding for the construction of concrete flux compactifications in the next Chapter. Because it is the best understood, we shall henceforth focus on the case of type IIB supergravity. (Usually, the type IIA dual to the constructions below is also known.) We postpone the stabilization of the ``second half'' of moduli space, \ie the K\"ahler structure moduli such as the total volume, until Chapter \ref{chapter:kaehlerstabilized}.

\section{Ten-dimensional Solutions of Type IIB Supergravity}\label{sec:typeIIBsolutions}
\subsection{Equations of Motion}\label{subsec:typeIIB-eofm}
To understand the rationale behind the scenarios we study later, it is useful to go back to the type IIB supergravity action and explore the consequences of its equations of motion as well as those of the self-duality condition of $\tilde{F}_{5}$. For the two scalars $\Phi$ and $C_{0}$, the equations of motion derived from Eq.~(\ref{eq:SL2R-typeIIB}) are
\bea\label{eq:dilaton-eofm}
\nabla_{m}\nabla^{m}\Phi+\frac{1}{2}\,e^{-\Phi}|H_{(3)}|^{2}-e^{2\Phi}|F_{(1)}|^{2}-\frac{1}{2}\,e^{\Phi}|\tilde{F}_{(3)}|^{2}&=&0\,,\\
\nabla_{m}\nabla^{m}C_{(0)}+2\,\nabla_{m}\Phi\,\nabla^{m}C_{(0)}-e^{\Phi}\left(F_{(3)}H^{(3)}+|H_{(3)}|^{2}C_{(0)}\right)&=&0\,.\label{eq:C0-eofm}
\eea
In terms of the combined fields $\tau$ and the three-form $G_{(3)}$ defined in Eqs.~(\ref{eq:axiondilaton}) and (\ref{eq:G3form}), these can be combined into the equation
\beq\label{eq:tau-eofm}
\nabla_{m}\nabla^{m}\tau=-ie^{\Phi}\,|\partial_{m}\tau|^{2}-\frac{i}{2}\,e^{\phi}G_{(3)}G^{(3)}\,,
\eeq
where the last factor of $G_{(3)}$ is \emph{not} a complex conjugate. In the following, the axion field $C_{(0)}$ is often be set identically to zero. From Eq.~(\ref{eq:C0-eofm}) we then see that in this case
\beq\label{eq:threeformsorthogonal}
F_{(3)}H^{(3)}=0\,,
\eeq
\ie the two three-form field strengths are orthogonal to each other. Recall that we know from the eventual splitting of the metric into internal and external pieces $\scrM_{4}\times\scrM_{6}$, with maximal symmetry preserved in the four extended dimensions, that fluxes must not have ``legs'' in the external $\scrM_{4}$. Therefore, the indices in Eq.~(\ref{eq:threeformsorthogonal}) run over the internal manifold only. For vanishing $C_{(0)}$ and constant dilaton field, Eq.~(\ref{eq:dilaton-eofm}), we also find that [note that $\tilde{F}_{3}=F_{3}$ for $C_{(0)}=0$, see Eq.~(\ref{eq:F3tilde})]
\beq\label{eq:threeformsequal}
|H_{(3)}|^{2}=e^{2\Phi}|F_{(3)}|^{2}\,.
\eeq
Next, let us investigate what we can learn from the intrinsic self-duality of the five-form flux $F_{(5)}$. We can make an ansatz with ``built in'' self-duality as
\beq\label{eq:F5ansatz}
\tilde{F}_{(5)}=(1+*)\,\dd\alpha(y)\wedge\dd x^{0}\wedge\dd x^{1}\wedge\dd x^{2}\wedge\dd x^{3}\,,
\eeq
where $\alpha(y)$ is a function of the extra dimensions only, and the ``$*$'' operator is the Hedge dual in ten dimensions. Recall that $\tilde{F}_{(5)}$ was defined as the gauge-invariant combination of Eq.~(\ref{eq:F5gi}), and that $\dd F_{(5)}=\dd^{2}B_{(3)}=0$. Then, the Bianchi identity for $\tilde{F}_{(5)}$ takes the form
\beq\label{eq:F5Bianchi}
\dd \tilde{F}_{(5)}=H_{(3)}\wedge F_{(3)}\,.
\eeq
(Because of the self-duality, the second Bianchi identity  $\dd *\tilde{F}_{(5)}$ gives the same equation.) If $H_{3}$ and $F_{(3)}$ obey Eq.~(\ref{eq:threeformsorthogonal}), we see immediately from Eq.~(\ref{eq:F5Bianchi}) that $\dd F_{(5)}$ must be proportional to the volume of the internal manifold.
 
Below we will study the other R-R and NS-NS gauge potentials using their Bianchi identities. As a last step for now, we content ourselves with writing down the ten-dimensional Einstein equations in the bulk (\ie without localized sources such as D-branes) \cite{Schwarz:1983qr,Frey:2003tf}:
\bea\label{eq:10d-Einstein}
R_{mn}&=&\frac{1}{2}\,g_{mn}\,\partial_{k}\Phi\,\partial^{k}\Phi-2\nabla_{m}\nabla_{n}\Phi-\frac{1}{4}\,g_{mn}\,\nabla^{2}\Phi\nonumber\\
&&+\frac{1}{4}\,e^{2\Phi}\,\tilde{F}_{m(4)}{\tilde{F}_{n}}^{(4)}+\frac{1}{2}\,e^{2\Phi}\,{G_{(m}^{}}^{pq}\,G_{n)pq}^{*}-\frac{1}{12}\,e^{2\Phi}\,g_{mn}\,G_{(3)}\,G^{*(3)}
\eea
Note that only the last four indices are bound up in the square of the five-form flux, and that there is a symmetrisation with respect to the first index in the first term involving the complexified three form flux $G_{(3)}=F_{(3)}-\tau\,H_{(3)}$. On the right hand side above, Eqs.~(\ref{eq:dilaton-eofm}) and (\ref{eq:C0-eofm}) have been used to write the first line in terms of the dilaton only. Therefore we see that, for a constant dilaton, the first line in Eq.~(\ref{eq:10d-Einstein}) vanishes and only fluxes from the $C_{(4)},C_{(2)}$ and $B_{(2)}$ gauge potentials appear as sources.\\
Recall that for a metric ansatz of the form (\ref{eq:10d-notwarped}), the Ricci tensor will cleanly split into a four-dimensional and a six-dimensional piece, and that we argued that fluxes can only extend within the dimensions of the internal manifold. We can now appreciate the property of Ricci flatness ($R_{MN}=0$) of CY manifolds: they are a solution for the internal compact piece of Eq.~(\ref{eq:10d-Einstein}) in the absence of fluxes, \ie when the right hand side vanishes. We now turn to more general compactifications of type (\ref{eq:10d-warped}), which allow a warp factor between $\scrM_{4}$ and $\scrM_{6}$.

\subsection{The No Go Theorem}\label{subsec:nogotheorem}
In a first step, we show that at leading order in the supergravity approximation of Section \ref{sec:effectivesugra}, any warp factor introduced into the metric via an ansatz like Eq.~(\ref{eq:10d-warped}) must be trivial in the absence of localized sources for the fluxes \cite{Gibbons:1984kp,Maldacena:2000yy,deWit:1987ph}.  
The trace of the Einstein equations (\ref{eq:10d-Einstein}), calculated with the self-dual ansatz (\ref{eq:F5ansatz}) for the five-form flux and the Minkowski metric $\eta_{\mu\nu}$ in extended spacetime, gives
\beq\label{eq:h-withoutsources}
\Delta_{6d} h^{-1}=\frac{h^{-2}}{2}\,e^{\Phi}|G_{3}|^{2}+h\,\left(|\partial\alpha|^{2}+|\partial h^{-1}|^{2}\right)\,,
\eeq
where $\Delta_{6d}$ here is the Laplacian on the internal manifold. If we were to integrate this equation over the internal compact space $\scrM_{6}$, the left hand side would vanish because it is a total derivative. However, the right hand side is a sum of positive-definite terns, which only vanishes if each of the individual terms vanishes. Then, the three-form flux $G_{3}$ must be zero, and the functions $h(y),\alpha(y)$ constant, \ie trivial. This ``no go theorem'' for a non-trivial warp factor and non-zero fluxes was shown here for a Minkowski metric in four dimensions, but it can be generalized to anti-de Sitter and de Sitter spacetimes (to which we shall return in a moment).\\
It is, however, possible to avoid the above no go theorem both by going beyond leading order in the supergravity approximation and by introducing sources (and possibly singularities) into the bulk background geometry \cite{Giddings:2001yu,Douglas:2006es}. We already know that D-branes carry charges under the R-R gauge fields, and we now show that they invalidate the no go theorem in another generalization from electromagnetism: the total charge enclosed in a compact space must vanish, because lines of field strength either have to go to sources, or to infinity, and the latter is impossible if the compactification volume is finite. In a compact space, there must therefore be an equal number of sources of opposite charge.

\subsubsection{D-brane sources}
If there are localized sources present in the background, the contribute a term on the right hand side of Eq.~(\ref{eq:h-withoutsources}) which is $\propto h^{-1/2}\mathcal{T}_{\mathrm{loc}}$, where $\mathcal{T}_{\mathrm{loc}}$ denotes the trace of the Einstein tensor describing the localized sources. We discussed the Chern Simons and Dirac Born Infeld world volume actions for D$p$ branes in Section \ref{subsec:Dbranes}, and we also mentioned the possibility that branes can be ``wrapped'' along some of the internal directions of the ten-dimensional spacetime. (In type IIB, for example, this will be the case if the number of spatial directions $p>3$.) 
We therefore write the action of a D$p$-brane wrapping a $(p-3)$-cycle $\Sigma$ (at leading order and for vanishing fluxes on the brane) as
\beq
\action_{\mathrm{D}_{p}}=-\int_{\scrM_{4}\times\Sigma}\dd^{p+1}\xi\,T_{p}\sqrt{-g}+\mu_{p}\int_{\scrM_{4}\times\Sigma} C_{(p+1)}\,.
\eeq
Our main interest is in branes with $p=3$, since they can couple to the gauge potential $C_{(4)}$ behind the $F_{(5)}$ flux. Note, however, that to the gauge invariant flux $\tilde{F}_{(5)}$, also the $H_{(3)}$ and $F_{(3)}$ fluxes contribute, see Eq.~(\ref{eq:F5gi}). As we mentioned in Section \ref{subsec:DBIterm} and will become important below, branes of higher $p$ can also carry charges of the D3-brane type if they are wrapped (so-called fractional D3 charges). In type IIB theory we should generically also expect contributions from D5-branes wrapping two-cycles, and D7-branes wrapping four-cycles.\\
Apart from an additional term on the right hand side of the Einstein equations, sources also change the Bianchi identity (\ref{eq:F5Bianchi}) for $\tilde{F}_{5}$ as
\beq
\dd \tilde{F}_{5}=H_{3}\wedge F_{3}+2\tilde{\kappa}_{10}\, T_{3}\,\rho_{3}^{\mathrm{(loc)}}\,,
\eeq
where we have specialized to D3-brane (type) charges with tension $T_{3}$ and density $\rho_{3}^{\mathrm{(loc)}}$. The charge density $\rho_{3}^{\mathrm{(loc)}}$ then typically contains $\delta$-functions that specify the location of the D-branes within the compact dimensions. If this equation is integrated over the internal space $\scrM_{6}$, then it leads to the type IIB ``tadpole cancellation condition'':
\beq\label{eq:tadpole}
\frac{1}{2\tilde{\kappa}_{10} T_{3}}\int_{\scrM_{6}} H_{(3)}\wedge F_{(3)}+Q_{3}=0
\eeq
Here, $Q_{3}$ is the total charge obtained from integrating $\rho_{3}$ over $\scrM_{6}$. It can be shown that the requirement for $G_{(3)}$ to be imaginary self-dual means that fluxes $H_{(3)}$ and $F_{(3)}$ are only induced if $Q_{3}$ is negative. Recall that negative charges can be provided by anti-D3 branes or O3 orientifold planes.\\
The condition (\ref{eq:tadpole}) can also be rewritten in a purely geometric way if type IIB theory is ``lifted'' to its so-called ``F-theory'' description, where it is compactified on an elliptically fibered CY four-fold $X$. (The base of that fibration then is the original type IIB, and the fibration describes the running of the axion-dilaton.) Then, the tadpole condition becomes
\beq\label{eq:tadpoleF}
\frac{\chi(X)}{24}=N_{\mathrm{D}3}+\frac{1}{2\tilde{\kappa}_{10} T_{3}}\int_{\scrM_{6}} H_{(3)}\wedge F_{(3)}\,,
\eeq
where $\chi$ is the Euler characteristic of $X$, and $N_{\mathrm{D}3}$ is the D3 brane charge present in the compactification. (The left hand side of this equation can be interpreted as the negative of the D3-brane charge induced by the curvature of wrapped D7-branes, which we mentioned earlier.)

\subsubsection{Fluxes from D-branes}
We can now derive the form of the fluxes sourced by D3-branes: these branes couple to the $C_{(4)}$ potential, therefore we expect a non-zero $F_{(5)}$. Let us set the other fluxes $F_{(3)}$ and $H_{(3)}$ to zero for the moment. (This is possible while there are only ``proper'' D3-branes, but needs to be revisited if there are fractional D3-charges in the background such as wrapped D5-branes.) Under these assumptions, insert the self-dual ansatz for the five-form flux (\ref{eq:F5ansatz}) into the Bianchi identity (\ref{eq:F5Bianchi}) and subtract the result from the trace of the  Einstein equation (\ref{eq:h-withoutsources}), now taking into account the contribution of localized branes. This gives the constraint
\beq
\Delta (h^{-1}-\alpha)=\frac{1}{6}\,e^{\Phi}\,h^{-2}|iG_{3}-*G_{3}|^{2}+h|\partial(h^{-1}-\alpha)|^{2}+2\tilde{\kappa}_{10}\,h^{-1/2}\left(\mathcal{T}_{\mathrm{loc}}-T_{3}\rho_{3}^{\mathrm{(loc)}}\right)\,.
\eeq
The solutions to this system of equations are then characterized by
\beq\label{eq:fluxsources}
*G_{3}=iG_{3},\qquad h^{-1}(y)=\alpha(y),\qquad \mathcal{T}_{\mathrm{loc}}= T_{3}\rho_{3}^{\mathrm{(loc)}}\,,
\eeq
where the last equation means that the sources involved should exactly saturate the ``BPS-like'' bound $\mathcal{T}_{\mathrm{loc}}\geq T_{3}\rho_{3}^{\mathrm{(loc)}}$. (This bound describes a general property of tension and charge for localized sources in string theory.) The BPS-like bound is exactly satisfied by D3-branes, and it is satisfied by anti-D3 branes (but not saturated, given that their charge is negative). D7-branes wrapped on four-cycles and O3 planes (other possible sources for D3-type charge) can saturate it, while D5-branes on wrapped two-cycles satisfy, but do not saturate the bound.

\bigskip
The property of imaginary self-duality for the complexified flux $G_{(3)}$ is the key to determining the missing superpotential $W$ we need for a complete description of the four-dimensional supersymmetric theory. We now show how this along with the form of the warp factor $h(y)$ can be found in a concrete model: starting from the intuition of the so-called AdS/CFT correspondence, we consider a stack of $N$ D3-branes embedded into a ten-dimensional type IIB supergravity background and study the spacetime geometry in its vicinity.

\chapter{From AdS/CFT to Fluxes on the Conifold}\label{chapter:AdSCFT}
\begin{quotation}
\emph{We saw how D-branes as sources of R-R flux can help avoid the no-go theorem for warped compactifications, and we now explain how this is related to the AdS/CFT correspondence and its generalizations. Of particular interest are compactifications on so-called conifolds, in which the fluxes backreact on the geometry to deform it in a singularity-avoiding way. On these backgrounds, all complex structure moduli can be stabilized while still preserving supersymmetry in four dimensions.}
\end{quotation}

\section{The AdS/CFT Correspondence}\label{sec:AdSCFT}
In Section \ref{subsec:int-and-ext}, we split the ten-dimensional metric into a four-dimensional external and a six-dimensional internal piece [see Eq.~(\ref{eq:10d-notwarped})], using the Minkowski metric $\eta_{\mu\nu}$ for the former. However, the requirement of maximal symmetry (Lorentz invariance) in four dimensions could, apart from Minkowski space, also be met by a de Sitter (dS) or an anti-de Sitter space (AdS), \ie four-dimensional spacetimes with positive or negative cosmological constant, respectively. De Sitter geometry is what one may hope for in a phenomenologically successful construction (given that a small but non-zero positive cosmological constant is observed in our Universe), but it has proven notoriously difficult to obtain in string theory. For AdS space, however, a remarkable relation with conformal field theory (CFT) on $p$-branes has been found, which is referred to as the ``AdS/CFT correspondence''. The case important to us concerns type IIB superstring theory with D3-branes embedded in the ten-dimensional background.\\
Consider $N$ coincident D3-branes in a higher-dimensional spacetime with structure $\mathrm{AdS}_{5}\times S^{5}$,\ie five dimensions compactified on a sphere, plus five-dimensional anti-de Sitter spacetime. As mentioned before, open strings with both ends attached to a D$p$-brane give rise to gauge theories living on the brane world volume. In the case of a stack of $N$ D$p$-branes, the gauge theory can be shown to be maximally supersymmetric $SU(N)$ Yang Mills\footnote{To be precise, in the absence of background fields and at lowest order in $\alphap$, the low-energy effective action on the brane is obtained from dimensional reduction of supersymmetric $SU(N)$ gauge theory in ten dimensions down to the $(p+1)$ dimensional world volume.}. The AdS/CFT correspondence then states that the low-energy world volume theory of these branes is dual to the string theory in the near-horizon (close to the branes) geometry of the bulk, in the sense that the two theories describe the same physics in different limits of ``coupling'': when the gauge theory living on the branes is weakly Yang Mills coupled, the background geometry is strongly curved and \emph{vice versa}.\\
D$p$-branes have tension, \ie mass, and they carry charges under the $C_{(p+1)}$ gauge fields, which means that they can source fluxes and curvature in ten dimensions. Such backgrounds then have the ``warped'' form of Eq.~(\ref{eq:10d-warped}), and we discussed them in Section \ref{sec:typeIIBsolutions}. For our stack of D3-branes, sourcing the four-form potential $C_{(4)}$, $N$ units of five-form flux are threading the internal $S^{5}$ (recall that fluxes cannot have legs in the extended dimensions). Again, this can be interpreted as a generalization of Gauss' law: the $N$ D3-branes (each of which carries unit charge) are inside the compact $S^{5}$, where they can be moved around at random since they do not experience a force in this background (their R-R interaction and gravity cancel). In this sense, their charge is ``smeared out'' over the entire $S^{5}$, and they are not properly ``localized''. Let us now calculate the warp factor in this setup.

\subsection{Warp Factor in an $\mathrm{AdS}_{5}\times S^{5}$ Background}
The ten-dimensional ansatz for the metric in an $\mathrm{AdS}_{5}\times S^{5}$ background [compare Eq.~(\ref{eq:10d-warped})] is
\begin{equation}\label{ansatz}
{\rm d}s^{2}_{10d}=\frac{1}{\sqrt{h(r)}}\,\eta_{\mu\nu}\,{\rm d}x^{\mu}\,{\rm d}x^{\nu}+\sqrt{h(r)}\,{\rm d}r^{2}+\sqrt{h(r)}\,r^{2}{\rm d}\Omega_{5}^{2}\,,
\end{equation}
where \({\rm d}\Omega_{5}^{2}\) is the angular metric on the five-sphere, and we take the warp factor $h$ to be a function of the ``radial'' coordinate $r$ only, independent of the angular position on the sphere. Because of the large amount of symmetry both in the internal and the external space, there are only two independent non-zero components of the ten-dimensional Ricci tensor,
\begin{eqnarray}
R_{00}&=&\,\,\frac{1}{4h^{2}}\left[\frac{1}{h}\left(\frac{\partial}{\partial r}h\right)^{2}-\left(\frac{\partial^{2}}{\partial r^{2}}h\right)-\frac{5}{r}\left(\frac{\partial}{\partial r}h\right)\right]\,,\label{eq:R00sphere5}\\
R_{44}&=&-\frac{1}{4h}\left[\frac{1}{h}\left(\frac{\partial}{\partial r}h\right)^{2}+\left(\frac{\partial^{2}}{\partial r^{2}}h\right)+\frac{5}{r}\left(\frac{\partial}{\partial r}h\right)\right]\,,\label{eq:R44sphere5}
\end{eqnarray}
and the Ricci scalar is found to read
\begin{equation}\label{eq:AdSRicci}
R=-\frac{1}{2h^{3/2}}\left[\left(\frac{\partial^{2}}{\partial r^{2}}h\right)+\frac{5}{r}\left(\frac{\partial}{\partial r}h\right)\right]\,.
\end{equation}
If we add Eqs.~(\ref{eq:R00sphere5}) and (\ref{eq:R44sphere5}), and consequently their respective right hand sides in the Einstein equations (\ref{eq:10d-Einstein}), we can eliminate the $\propto (\partial h/\partial r)^{2}$ terms, while the right hand (source) sides exactly cancel. (This is true in general for the ``bulk'' terms, without assumptions about the dilaton or fluxes; since we do not consider the $N$ D3-branes as localized sources, we did not introduce an energy momentum tensor $\mathcal{T}^{\mathrm{(loc)}}_{mn}$ for them.) We obtain for $h(r)$ that
\begin{equation}\label{eq:definewarpAdS}
\left(\frac{\partial^{2}}{\partial r^{2}}h\right)+\frac{5}{r}\left(\frac{\partial}{\partial r}h\right)=0\,,
\end{equation}
which is exactly the combination appearing in the Ricci scalar (\ref{eq:AdSRicci}), hence our chosen internal manifold of a $S^{5}$ basis and a $r$ radial coordinate is still Ricci flat. We can now exploit Eq.~(\ref{eq:h-withoutsources}), which was obtained from the trace of the Einstein equations and the five-form self-dual ansatz (\ref{eq:F5ansatz}). With the form of the Laplacian on the five-sphere, and for vanishing three-form fluxes, we find by combination with Eq.~(\ref{eq:definewarpAdS}) that
\beq
\frac{2}{h^{3}}\left(\frac{\partial}{\partial r}\,h\right)^{2}=h\,\left(\left|\frac{\partial \alpha}{\partial r}\right|^{2}+\left|\frac{\partial h^{-1}}{\partial r}\right|^{2}\right)\,.
\eeq
This is solved by setting $\alpha(r)=h^{-1}(r)$, as it was anticipated in Eq.~(\ref{eq:fluxsources}). [Note, however, that we have worked with Eq.~(\ref{eq:definewarpAdS}), arguing that the D3-branes are not localized, but their charge smeared out, causing the $F_{(5)}$ flux.] Finally, the solution of Eq.~(\ref{eq:definewarpAdS}) takes the form
\begin{equation}\label{eq:hAdSsolution}
h(r)=C_{1}+\frac{C_{2}}{r^{4}}\,,
\end{equation}
where $C_{1,2}$ are integration constants to be fixed by boundary conditions and by integrating the five-form flux over the $S^{5}$ sphere, which should give the total charge $N$, $\int_{S^{5}}F_{(5)}\propto N$, 
where the ansatz (\ref{eq:F5ansatz}) with $\alpha(r)=h^{-1}(r)$ should be used. Therefore, we see that the flux, in a generalization of the Dirac condition, is quantized. In particular, it can be tuned only in discrete units.

We can think of the radial coordinate $r$ roughly as the distance to the stack of D3-branes. The constant term $C_{1}$ leads to a plateau at large $r$ (\ie far away from the stack of branes) where the second term dies out. In this limit where $r\gg C_{2}/C_{1}$, the warped ``corner'' of the overall CY should join into the bulk. Here, we are interested in the other, ``near-horizon limit'' of small $r\ll C_{2}/C_{1}$, where the metric takes the form\footnote{This is obtained by treating the D3's as ``black branes'', the string theoretic analogues of black holes \cite{Horowitz:1996fn}.}
\beq\label{eq:blackmetric}
\dd s^{2}_{10d}\simeq \frac{r^{2}}{R^{2}}\,\eta_{\mu\nu}\dd x^{\mu}\,\dd x^{\nu}+\frac{R^{2}}{r^{2}}\,\dd r^{2}+R^{2}\dd\Omega_{5}^{2}\,,
\eeq
where the $x^{\mu}$ are the coordinates on the parallel D3-branes (which are aligned with the extended spacetime dimensions). $R$ is the ``radius of horizon'', with $R^{4}=4\pi\gs N\alphap^{2}$, and therefore $C_{2}=R^{4}$ in the notation of Eq.~(\ref{eq:hAdSsolution}). 
The scale $R$ is the characteristic curvature scale both for the $\mathrm{AdS}_{5}$ and the five-sphere. On the branes, the $\mathrm{AdS}_{5}\times S^{5}$ corresponds to a $\calN=4, d=4$ super Yang Mills theory with gauge group $SU(N)$. We have $\calN=4$ (instead of $\calN=8$) because the presence of the branes breaks half of the supersymmetry. The $SU(N)$ on the branes is UV finite (corresponding to large values of the $r$ coordinate where the warp factor becomes constant), and conformally invariant. In the IR limit for $r\rightarrow0$, the warp factor seems to run into a singularity. We shall see shortly how this can be avoided if corrections to the geometry are properly taken into account. The Yang Mills coupling constant $g$ of the gauge theory is related to the string coupling $\gs$ by $g^{2}=4\pi\gs$, which illustrates the ``duality'' character of the AdS/CFT correspondence: since $R^{4}\propto g^{2}$, the supergravity theory is strongly curved (small characteristic curvature $R$) when the gauge theory is weakly coupled. \\
From Eq.~(\ref{eq:blackmetric}), we can also make explicit the equivalence of warped backgrounds to Randall Sundrum models that was mentioned earlier \cite{Burgess:2007pz}: if we re-define the $r$ variable as $r=r_{0}\,e^{\xi/R}$, the ten-dimensional metric using $\xi$ becomes 
\beq
\dd s^{2}_{10d}\simeq e^{2\xi/R}\,\eta_{\mu\nu}\,\dd x^{\mu}\,\dd x^{\nu}+\dd\xi^{2}+R^{2}\dd\Omega_{5}^{2}\,,
\eeq
where the constant $r_{0}$ has been absorbed into the four-dimensional coordinates. The new variable $\xi$ plays the r\^ole of proper distance along the throat. Setting aside the $\dd\Omega_{5}^{2}$ part, this metric is five-dimensional de Sitter space, and we see that the warp factor varies exponentially quickly, $h\propto \exp(-4\xi/R)$, in terms of proper distance. 

\subsection{Warping on Einstein Spaces}\label{subsec:warpEinstein}
The five-sphere is an exceedingly simple choice for a compact space in string theory. One can generalize Eq.~(\ref{eq:blackmetric}) as
\beq\label{eq:blackgeneralized}
\dd s^{2}=h_{3}^{-1/2}\,\eta_{\mu\nu}\,\dd x^{\mu}\,\dd x^{\nu}+h_{3}^{1/2}(\dd r^{2}+r^{2}\dd s_{5}^{2})\,,
\eeq
where $\dd s_{5}^{2}$ now stands for a so-called \emph{Einstein space} $X_{5}$, which is characterized by a Ricci tensor $R_{mn}\propto g_{mn}$ (with constant proportionality factor). Comparing with the ansatz of Eq.~(\ref{eq:10d-warped}) for a general warped spacetime metric, we see that the six compact dimensions form a cone, with radial direction $r$ and the Einstein space as its basis. (We have written the warp factor with a subscript ``3'' to make it explicit that we are working in the horizon limit of a stack of ``proper'' D3-branes only so far.) For any $\dd s_{5}\neq\dd\Omega_{5}$, there is a singularity at the ``tip'' of the cone where $r\rightarrow0$. The $N$ D3-branes then are are localized at this singularity, instead of being ``smeared out'' over the $S^{5}$ in the non-singular case.

The $X_{5}$ is referred a \emph{Sasaki Einstein space} if it is an Einstein space, and if together with the radial direction $r$ in Eq.~(\ref{eq:blackgeneralized}) it gives a non-compact Calabi Yau (note that there is a singularity at the tip, therefore this is not really an manifold). CY spaces break 3/4 of supersymmetry, therefore the dual gauge theory $SU(N)$ on the world volume now should have $\calN=1$ (instead of $\calN=4$ for the $S^{5}$). The formula for the curvature radius [provided one uses coordinates on the $X_{5}$ such that $R_{m'n'}=g_{m'n'}\,(m',n'=5,\dots,9)$] is
\beq\label{eq:RonEinstein}
R^{4}=4\pi\,\gs\,\alphap^{2}\,\frac{N}{v}\,,\qquad\,v=\mathrm{Vol}(X_{5})/\mathrm{Vol}(S^{5})\,.
\eeq
The parameter $v$ is a convenient dimensionless measure for the volume of the cone basis relative to a five-sphere. There is an infinite family of choices of spaces $X_{5}$, but the simplest non-trivial example is $T^{1,1}$, which is $SU(2)\times SU(2)/U(1)$ and has the topology of $S^{3}\times S^{2}$. (Hence, this is the topology of the cone at its basis.) Together with the coordinate $r$, the space $T^{1,1}$ is called the ``simple conifold''.\\
The $S^{3}\times S^{2}$ basis beautifully illustrates the importance of geometry in string theory: we mentioned earlier that D$p$ branes can wrap cycles if some of their $p$ spatial directions lie along the compact coordinates. Therefore, the general type IIB background with D3-branes [which create the warped geometry Eq.~(\ref{eq:blackgeneralized}) with $\dd s_{5}^{2}$ as the Sasaki Einstein space $T^{1,1}$] can, for example, also contain D5-branes wrapped on a two-cycle in the basis of the conifold. If there are $M$ D5-branes wrapped, they provide $M$ quantities of (fractional) D3-brane charge on top of the $N$ ``proper'' D3-branes already in the geometry. However, at the tip where $r=0$, this two-cycle shrinks to a point, making the conifold singularity manifest.\\
In geometrical terms, one can think of two ways to avoid the singular behavior for $r\rightarrow0$: either the $S^{3}$ or $S^{2}$ can stay finite at the tip. These are called the ``deformed'' and the ``resolved conifold'', respectively, and we now turn to compactifications on the three varieties of conifolds. Keep in mind though that ``compactification'' in this context is somewhat a misnomer, since the conifold geometries are non-compact: they are usually ``cut off'' at the characteristic radius $R$, where there are glued to CY bulk manifold, which, in turn, is compact. Deviations from the known conifold geometry can therefore occur in the UV end of the theory where the unknown bulk takes over.

\section{Conifold Compactifications}\label{sec:conifolds}
Inspired by the AdS/CFT duality, a series of papers \cite{Giddings:2001yu,Klebanov:1999rd,Klebanov:2000nc,Klebanov:2000hb,PandoZayas:2000sq} presented solutions to the equations of motion of type IIB superstring theory in a background with $N$ D3-branes and $M$ fractional D3-branes. We now follow the line of development in these papers: the strategy is to place the regular D3-branes at the singularity in the $\mathrm{AdS}_{5}\times X_{5}$ (where they act as localized sources of curvature and flux), and consider the $M$ fractional branes as a perturbation. In terms of the dual gauge theory description, the stack of $N$ branes would originally (by itself) have a conformally invariant $SU(N)$ with $\calN=1$ supersymmetry, but the fractional branes break the conformal invariance and lead to a renormalization group (RG) flow. With guidance from the behavior of this RG flow, one can draw conclusions for the effect of the fractional D3-branes on the dual string geometry.

\subsection{The Simple Conifold}\label{subsec:simple}
We saw in Section \ref{subsec:warpEinstein} how a stack of $N$ D3-branes gives rise to a conical background geometry with a characteristic radius $R$, the prime example being the simple conifold. A convenient way to describe the six-dimensional conifold geometry is in terms of four complex variables $w_{a}$ restricted by the equation \cite{Candelas:1989js}
\beq\label{eq:conifolddef}
\sum_{a=1}^{4}w_{a}^{2}=0\,.
\eeq
Note that this corresponds to two conditions, because the $w_{a}$ are complex, \ie $w_{a}=w_{a}^{(1)}+i\,w_{a}^{(2)}$. [This component notation will be used shortly to define cycles within the conifold described by Eq.~(\ref{eq:conifolddef}).] The basis of the conifold is the Einstein space $T^{1,1}$, whose metric reads
\beq\label{eq:conifoldmetric}
\dd s^{2}_{T^{1,1}}=\frac{1}{9}\left(\dd\psi+\sum_{i=1}^{2}\cos\theta_{i}\,\dd\phi_{i}\right)^{2}+\frac{1}{6}\,\sum_{i=1}^{2}\left(\dd\theta_{i}^{2}+\sin^{2}\theta_{i}\,\dd\phi_{i}^{2}\right)\,,
\eeq
using five angular coordinates $(\psi,\phi_1,\theta_{1},\phi_{2},\theta_{2})$. As we stated earlier, this is topologically a product of spheres $S^{2}\times S^{3}$. Note the symmetry between the to pairs of angular coordinates $(\phi_{1},\theta_{1})$ and $(\phi_{2},\theta_{2})$. For later use, let us define a basis of one-forms $g^{i},i=1,\dots,5$, in terms of which the $T^{1,1}$ Einstein space metric (\ref{eq:conifoldmetric}) reads
\beq\label{eq:conifold-g-metric}
\dd s_{T^{1,1}}^{2}=\frac{1}{9}\left(g^{5}\right)^{2}+\frac{1}{6}\sum^{4}_{i=1}(g^{i})^{2}\,.
\eeq
These one-forms are defined from
\bea
g^{1}=\frac{e^{1}-e^{3}}{\sqrt{2}}\,,&&\qquad g^{2}=\frac{e^{2}-e^{4}}{\sqrt{2}}\,,\qquad g^{5}=e^{5}\,,\label{eq:gbasis1}\\
g^{3}=\frac{e^{1}+e^{3}}{\sqrt{2}}\,,&&\qquad g^{4}=\frac{e^{2}+e^{4}}{\sqrt{2}}\,,\label{eq:gbasis2}
\eea
where the $e^{i},i=1,\dots,5$ stand for
\bea
e^{1}=-\sin\theta_{1}\,\dd\phi_{1}\,,\quad e^{2}=\dd\theta_{1}\,,&&\qquad e^{3}=\cos\psi\,\sin\theta_{2}\,\dd\phi_{2}-\sin\psi\,\dd\theta_{2}\,,\\
e^{4}=\sin\psi\,\sin\theta_{2}\,\dd\phi_{2}+\cos\psi\,\dd\theta_{2}\,,&&\qquad e^{5}=\dd\psi+\cos\theta_{1}\,\dd\phi_{1}+\cos\theta_{2}\,\dd\phi_{2}\,.
\eea
It can be shown that the components of the Ricci tensor $R_{mn}$ as well as the Ricci scalar calculated with the metric (\ref{eq:conifoldmetric}) are exactly the same as those for the five-sphere we obtained in Eqs.~(\ref{eq:R00sphere5}), (\ref{eq:R00sphere5}) and (\ref{eq:AdSRicci}). However, given that the conifold, in contrast to the five-sphere, has a singularity where the stack of D3-branes is localized, there is now an additional source term $\mathcal{T}^{\mathrm{(loc)}}_{mn}$ (containing a $\delta$-function) on the right hand side of the Einstein equations. We now show how this singularity can be avoided by an appropriate ``deformation'' of the simple conifold.

\subsection{The Deformed Conifold}\label{subsec:deformed}
As announced above, we now add $M$ fractional D3-branes (\ie D5-branes wrapped on a two-cycle) to the type IIB supergravity background. It can be shown that in terms of the Yang Mills gauge theory living on the stack of $N$ D3-branes, the $SU(N)$ changes to $SU(N+M)\times SU(N)$ (with Yang Mills gauge couplings $g_{1}$ and $g_{2}$), and while this theory is still supersymmetric, the conformal invariance is now broken. This means that it undergoes a renormalization group flow, the direction of which lies along the $r$-direction of the cone geometry.\\
Using the so-called \emph{Seiberg duality}, it can be shown that the gauge theory confines deep in the IR, \ie in the limit $r\rightarrow 0$. However, we saw that the supergravity theory compactified on the simple conifold had a singularity in its IR limit, where the warp factor $h\propto r^{-4}$ blows up. However, the gauge theory and the supergravity geometric background are related by the generalized AdS/CFT correspondence -- how can these different types of behavior be reconciled? In \cite{Klebanov:1999rd,Klebanov:2000nc,Klebanov:2000hb} both the supergravity background and the renormalization group flow of the $SU(N+M)\times SU(N)$ theory were studied closely together: in this way, physical intuition as well as calculational results obtained in one picture may be transferred and reinterpreted in the dual description. We now highlight the basic steps of this development.

\subsubsection{Fluxes from wrapped branes}
The $M$ fractional D5-branes give rise to an R-R three-form flux contribution through a three-cycle $\mathscr{C}^{3}$ of the $T^{1,1}$, which is quantized as $\int_{\mathscr{C}^{3}}F_{(3)}\propto M$. In this sense, the wrapped D5-branes are coupled to the $C_{(2)}$ R-R gauge potential. However, this R-R potential has a two-form NS-NS twin, namely $B_{(2)}$, with which it is grouped together in the vector $\mathscr{B}_{(2)}$ used in the $SL(2,\mathbb{R})$ invariant notation of the type IIB supergravity action in Eq.~(\ref{eq:SL2R-typeIIB}). If there is an $F_{(3)}$ flux, there must hence also be a non-zero NS-NS potential $B_{(2)}$ in the background, for which one can make the ansatz 
\beq\label{eq:B2ansatz}
B_{(2)}=e^{\Phi}\,f(r)\,\omega_{(2)}\,,\qquad \int_{\mathscr{C}^{2}}\omega_{(2)}=1\,,
\eeq
where $\omega_{(2)}$ is a two-form of the background we specify below. The second expression in Eq.~(\ref{eq:B2ansatz}) restricts this two-form, and $\mathscr{C}^{2}$ is the cycle dual to $\mathscr{C}^{3}$ on the (five-dimensional) $T^{1,1}$ basis. (These will be made explicit in Chapter \ref{chapter:braneinflation}.)

We focus on the case of constant axion-dilaton $\tau$, which from Eq.~(\ref{eq:tau-eofm}) tells us that $G_{(3)}G^{(3)}=0$. The Bianchi identities for the imaginary self-dual three-form $G_{(3)}$ then read
\beq\label{eq:G3-eofm}
\dd * G_{(3)}=i\tilde{F}_{(5)}\wedge G_{(3)}\,,\qquad \dd G_{(3)}=0\,.
\eeq
The flux $F_{(3)}$ created by the $M$ fractional branes should be proportional to the closed three-form on the $T^{1,1}$ (so that $\dd F_{(3)}=\dd C_{(2)}=0$ is respected)
\beq
F_{(3)}\propto M\,\tilde{e}^{\psi}\wedge\left(\tilde{e}^{\theta_{1}}\wedge \tilde{e}^{\phi_{1}}-\tilde{e}^{\theta_{2}}\wedge \tilde{e}^{\phi_{2}}\right)\,,
\eeq
where the new one-form basis $\tilde{e}^{i}$ here is
\beq\label{eq:oneformbasis}
\tilde{e}^{\psi}=\frac{1}{3}\left(\dd\psi+\sum_{i=1}^{2}\cos\theta_{i}\,\dd\phi_{i}\right)\,,\qquad \tilde{e}^{\theta_{i}}=\frac{1}{\sqrt{6}}\,\dd\theta_{i}\,,\qquad \tilde{e}^{\phi_{i}}=\frac{1}{\sqrt{6}}\,\sin\theta_{i}\,\dd\phi_{i}\,.
\eeq
The two-form $\omega_{(2)}$ defined by the second equation in Eq.~(\ref{eq:B2ansatz}) in terms of this basis reads
\beq\label{eq:omega2}
\omega_{(2)}\equiv\frac{1}{\sqrt{2}}\left(\tilde{e}^{\theta_{1}}\wedge \tilde{e}^{\phi_{1}}-\tilde{e}^{\theta_{2}}\wedge \tilde{e}^{\phi_{2}}\right)\,.
\eeq
The dilaton $\Phi$ is constant, therefore it follows from the ansatz (\ref{eq:B2ansatz}) that the NS-NS flux is given by $H_{(3)}\propto\dd f(r)\wedge\left(\tilde{e}^{\theta_{1}}\wedge \tilde{e}^{\phi_{1}}-\tilde{e}^{\theta_{2}}\wedge \tilde{e}^{\phi_{2}}\right)$. We now also set the $C_{0}$ scalar to zero; note that the constancy of $\tau$ then imposes the conditions (\ref{eq:threeformsorthogonal}) and (\ref{eq:threeformsequal}).

Because $\tilde{F}_{(5)}$ is self-dual, it has to be proportional to the sum of the volumes (\ie the five-forms) on the AdS space and the compact volume. [Earlier, we came to the same conclusion using the Bianchi identity (\ref{eq:F5Bianchi}).] With the ansatz (\ref{eq:F5ansatz}) used earlier, this was automatically respected by using the prefactor $(1+*)$. As a consequence, $\tilde{F}_{(5)}\wedge H_{(3)}=0$, which we can use in the first Bianchi identity in Eq.~(\ref{eq:G3-eofm}). It then follows from (the imaginary part of) this expression that the function $f(r)$ in Eq.~(\ref{eq:B2ansatz}) obeys
\beq\label{eq:function-fofr}
\frac{1}{r^{3}}\,\frac{\dd}{\dd r}\left[r^{5}\frac{\dd}{\dd r}\,f(r)\right]\propto M\,,\qquad f(r)\propto M\log r\,.
\eeq
Therefore, the components of $B_{(2)}$ grow logarithmically with $r$, hence its derivative $H_{(3)}\propto r^{-1}$, which diverges in the IR for $r\rightarrow0$. Again, there is an ``electromagnetic'' intuition which makes this behavior plausible: like $F_{(3)}$, the $H_{(3)}$ flux threads a three-cycle in $T^{1,1}$, and towards the ``tip'' of the conifold, this three-cycle shrinks to zero, resulting in the divergence.

Let us briefly comment on the order to which these solutions are valid: if $M$ is kept fixed, and $N$ goes to infinity, then the backreaction of the fluxes on the metric can be ignored to leading order in $N$. At first order in $M/N$, we therefore identified the $\mathrm{AdS}_{5}\times T^{1,1}$ with three-form fluxes turned on as the supergravity dual of the gauge theory $SU(N+M)\times SU(N)$ (which lives on a stack of $N$ D3-branes in the presence of $M$ wrapped D5-branes). [The dilaton only varies at order $\order{M^{2}/N^{2}}$.] At large $r$ (\ie in the UV limit), the renormalization group flow of the dual gauge theory is correctly reproduced, but not in the IR for $r\rightarrow0$. 
On the other hand, in the gauge theory description the behavior of the RG flow is known at all scales, and in particular, there is confinement in the far IR. This can be put to use to determine the non-singular supergravity background which replaces the simple conifold geometry in the $r\rightarrow0$ limit, removing \eg the divergence in the $H_{(3)}$ flux.

\subsubsection{Following the RG flow}
The direction $r$ lies along the RG flow and what is more, the function $f(r)$ calculated in Eq.~(\ref{eq:function-fofr}) denotes the supergravity dual of the scale dependence for the Yang Mills couplings $g_{1}$ and $g_{2}$. Along the RG flow, $\calN=1$ supersymmetry of the gauge theory is preserved, which should lead to a corresponding feature in the dual supergravity description in ten dimensions. To obtain solutions of the full type IIB equations of motion at all orders, 
it seems promising to start from an ansatz for the metric that keeps the ``AdS + compact space'' splitting as well as the structure of the $T^{1,1}$ at the basis of the cone, but allows for deformations,
\bea
\dd s_{10d}^{2}&=&R^{2}\left[e^{-(2/3)(B+4C)}\dd s_{\mathrm{AdS}}^{2}+\dd s_{5}^{2}\right]\,,\\
\dd s_{5}^{2}&=&\frac{1}{9}\,e^{2B}\left(\dd\psi+\sum_{i=1}^{2}\cos\theta_{i}\,\dd\phi_{i}\right)^{2}+\frac{1}{6}\,e^{2C}\sum_{i=1}^{2}\left(\dd\theta_{i}^{2}+\sin^{2}\theta_{i}\,\dd\phi_{i}^{2}\right)\,.\label{eq:d5ansatz}
\eea
The functions $B,C$ depend on the internal coordinates only, and the conformal prefactor in front of the AdS piece is chosen such that after compactification, the Einstein frame is preserved. $R$ is the characteristic scale of both the AdS and the compact space, as we saw earlier. Note that in Eq.~(\ref{eq:d5ansatz}), the symmetry between $(\phi_{1},\theta_{1})$ and $(\phi_{2},\theta_{2})$ (\ie between the two two-spheres inside the $T^{1,1}$, of which one is then fibered over the $S_{3}$) is still preserved. Let the metric in the AdS piece have the form
\beq\label{eq:AdSmetric}
\dd s_{\mathrm{AdS}}^{2}=\dd u^{2}+e^{2A(u)}\dd x_{\mu} \dd x^{\mu}\,.
\eeq
(Note that $u$ is not identical to the radial coordinate $r$ used before.) This ansatz for the metric is supplemented by corresponding expressions for the two-form $B_{(2)}$ and the fluxes. As an example, let us cite the 
self-dual five-form, which is written as $\tilde{F}_{(5)}=\mathcal{F}_{(5)}+*\mathcal{F}_{(5)}$ (still using built-in self-duality), and given that it must be proportional to the volume forms as before, we have
\bea
\mathcal{F}_{(5)}&=&K(u)\,e^{\psi}\wedge e^{\theta_{1}}\wedge e^{\phi_{1}}\wedge e^{\theta_{2}}\wedge e^{\phi_{2}}\,,\label{eq:F-Kansatz}\\
*\mathcal{F}_{(5)}&=&e^{4A-(8/3)(B+4C)}\,K(u)\,\dd u\wedge\dd x^{1}\wedge\dd x^{2}\wedge\dd x^{3}\wedge\dd x^{4}\,.
\eea
The Bianchi identities and the equations of motion obtained from the type IIB supergravity action then lead to a system of coupled differential equations of second order (in the coordinate $u$) for the functions $A,B,C,K$ in Eqs.~(\ref{eq:d5ansatz}), (\ref{eq:AdSmetric}) and (\ref{eq:F-Kansatz}), as well as the functions used to parametrize the other fields.\\
How can one hope to find solutions for all of these functions at once? In \cite{Klebanov:2000nc,Klebanov:2000hb} the gauge theory-inspired fact that the solutions must preserve supersymmetry in a certain sense was used to reduce the system of equations from second to first order in $u$-derivatives. One remarkable feature of the resulting solutions is that the dilaton $\Phi$ is constant at all orders in this background. Another other striking fact is that, under the influence of the IR confinement of the dual gauge theory, the conifold geometry undergoes a deformation, which we now describe.

\subsubsection{Deformation of the simple conifold}
\begin{figure}[h]
\begin{center}
\includegraphics[width=0.45\textwidth]{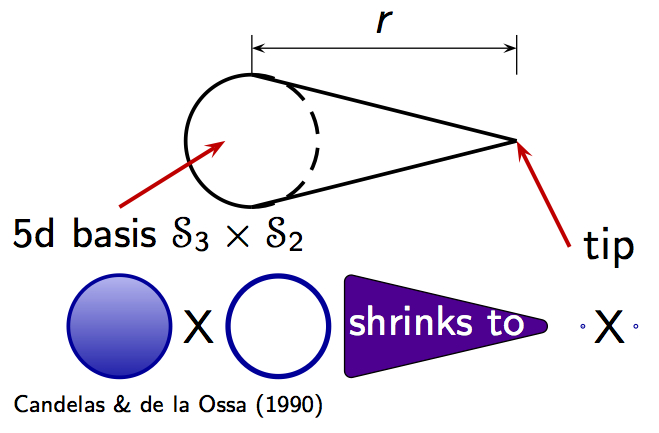}
\hfill
\includegraphics[width=0.45\textwidth]{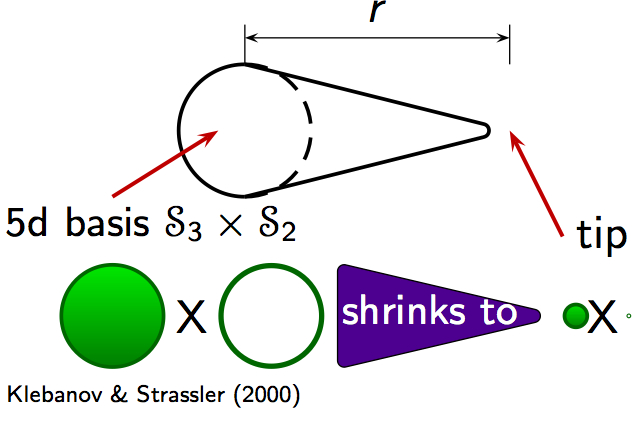}
\caption[Simple (singular) and deformed conifold geometries.]{\small \emph{Left:} The simple conifold of Section \ref{subsec:simple}: over a basis of spheres $S^{3}\times S^{2}$, the radial coordinate $r$ varies from its maximum value $r_\uUV$ to $r=0$ at the tip. The geometry is singular at the tip because both spheres shrink to zero size at $r=0$. \emph{Right:} The deformed conifold of Section \ref{subsec:deformed}: The basis still consists in the product $S^{3}\times S^{2}$, but the radial coordinate can now only take values in the range from $r_\uUV$ to a finite $r_{0}$ at the tip. The three-sphere stays finite at the tip, while the $S^{2}$ still shrinks to a point. The deformation is possible because the $S^{3}$ is stabilized by the three-fluxes of non-vanishing background fields threading it.}
\label{fig:conifolds}
\end{center}
\end{figure}

The simple conifold as defined by Eq.~(\ref{eq:conifolddef}) had a singularity at the point where all $w_{a}\equiv0$. In consequence of the additional three-form flux caused by the $M$ fractional D5-branes, the conifold is deformed such that now
\beq\label{eq:deformeddef}
\sum_{a=1}^{4}w_{a}^{2}=\epsilon^{2}\,,
\eeq
where $\epsilon^{2}$ is a dimensionful parameter. In terms of a conveniently defined ``radial'' coordinate $\tau$, the metric on the deformed conifold is written as
\bea
\dd s_{6}^{2}&=&\frac{1}{2}\,\epsilon^{4/3}\,\tilde{K}(\tau)\,\bigg\{\frac{1}{3\,\tilde{K}^{3}(\tau)}\,\left[\dd\tau^{2}+(g^{5})^{2}\right]\label{eq:deformedmetric}\\
&&\qquad\qquad\qquad+\cosh^{2}\left(\frac{\tau}{2}\right)\left[(g^{3})^{2}+(g^{4})^{2}\right]+\sinh^{2}\left(\frac{\tau}{2}\right)\left[(g^{1})^{2}+(g^{2})^{2}\right]\bigg\}\,,\nonumber
\eea
where we have used the $g^{i}$ basis defined in Eqs.~(\ref{eq:gbasis1}) and (\ref{eq:gbasis2}) and the function $\tilde{K}(\tau)$ reads
\beq
\tilde{K}(\tau)=\frac{[\sinh(2\tau)-2\tau]^{1/3}}{2^{1/3}\sinh\tau}\,.
\eeq
Note that at large $\tau$, we expect to recover the simple conifold because far from the tip of the cone the deformation should not play a r\^ole. Indeed, in this limit one can get back to the usual form of the $T^{1,1}$ in Eq.~(\ref{eq:conifoldmetric}) by setting $r^{3}\approx\epsilon^{2}e^{\tau}$. Let us now turn to the other limit, \ie the tip of the cone. At $\tau=0$, the metric (\ref{eq:deformedmetric}) degenerates into
\beq
\dd\Omega_{3}^{2}=\frac{1}{2}\,\epsilon^{4/3}\,(2/3)^{1/3}\left[\frac{1}{2}\,(g^{5})^{2}+(g^{3})^{2}+(g^{4})^{2}\right]\,.
\eeq
Using the expressions for the $g^{i}$, one can check that this is the metric of a three-sphere $S^{3}$. It is in this three-sphere that the $F_{(3)}$ flux due to the wrapped D5-branes lies at the apex of the cone, therefore we now know that the correct integral is $\int_{S^{3}}F_{(3)}\propto M$. As we did earlier for the $N$ D3-branes in the $\mathrm{AdS}_{5}\times S^{5}$ construction, one may think of the $M$ wrapped D5 branes ``smeared out'' over the $S^{3}$ of finite size at the tip of the cone.\\
The remaining two directions in Eq.~(\ref{eq:deformedmetric}), \ie the $S^{2}$ fibered over the $S^{3}$,  vanish quadratically as $(1/8)\,\epsilon^{4/3}(2/3)^{1/3}\tau^{2}\left[(g^{1})^{2}+(g^{2})^{2}\right]\propto\tau^{2}$. A convenient way to fix the parameter $\epsilon$ is to choose it such that the prefactor in this expression becomes 1, therefore $\epsilon=12^{1/4}$. Note that the symmetry of between the two two-spheres in the basis of the conifold [which went into the ansatz (\ref{eq:d5ansatz})] is still preserved. An illustration of the singular and the deformed conifold is provided by Fig.~\ref{fig:conifolds}.

\subsubsection{Corrections to the warp factor}
We know that in the UV limit, the deformed conifold smoothly turns into its simple cousin. Recall that the stack of $N$ coincident D3-branes is present in both cases (\ie non-zero $F_{(5)}$ flux sourced by the D3s), but the deformed conifold background on top has non-zero three-form fluxes because it contains $M$ D5-branes wrapped on a two-cycle. On the simple conifold, the warp factor $h(r)$ has the form (\ref{eq:hAdSsolution}). If we keep $r$ as a radial coordinate (instead of replacing it by $\tau$ as required in the apex of the deformed conifold), we can ask whether there still exists a function $\tilde{h}(r)$ which allows us to write the ten-dimensional metric as
\begin{equation}
{\rm d}s_{10}^{2}=\tilde{h}^{-1/2}(r)\,\eta_{\mu\nu}\,{\rm d}x^{\mu}\,{\rm d}x^{\nu}+\tilde{h}^{1/2}(r)\left({\rm d}r^{2}+r^{2}\,{\rm d}s^{2}_{T_{1,1}}\right)\,,
\end{equation}
where $\tilde{h}(r)$ defers from the simple warp factor $h(r)$ by a correction due to the $M$ D5-branes. Indeed it can be shown \cite{Klebanov:2000hb} that  Eq.~(\ref{eq:hAdSsolution}) in the presence of three-form fluxes is replaced by (see \cite{Herzog:2001xk} for a review)
\beq
\tilde{h}(r)=C_{1}+\frac{4\pi\,\alpha'^{2}}{r^{4}}\,\left[\gs\, N+a(\gs\,M)^{2}\,\ln\frac{r}{r_{0}}+a\,\frac{(\gs\,M)^{2}}{4}\right]\,,
\eeq
where $a$ is a constant of order 1, and $C_{1}$ again is the value of the plateau at large $r$. (One often sets $C_{1}=0$ if the focus is on the near-horizon limit close to the stack of D3-branes only.) The constant radius $r_{0}$ now corresponds to the IR end of the geometry, which is finite (instead of $r=0$ for the simple conifold). Remarkably, the deformation of the conifold by the three-form flux is modeled by a logarithmic correction to the D3-brane charge, \ie $N$ is replaced by  $N+a\gs\,M^{2}\,\ln(r/r_{0})+a\,(\gs\,M^{2}/4)$, and one may speak of an ``effective charge'' depending on the position in the radial direction. The deformed conifold is often referred to as the \emph{Klebanov Strassler (KS) throat}, and it is a crucial ingredient of the string cosmological scenarios we study in the following because all complex structure moduli are stabilized in this setup.

\subsection{The Resolved Conifold}
For completeness, let us briefly mention that the conifold family has a third member, the so-called resolved conifold. Geometrically, it defers from the deformed conifold in so far as now the $S^{2}$ (and not the $S^{3}$) stays finite at the tip of the cone. The approach to solving the supergravity equations of motion on this background is the same as before, but now instead of the ansatz (\ref{eq:d5ansatz}), the resolved conifold is ansatz is written such that it, while keeping $T^{1,1}$ structure, it allows for an asymmetry between the two $S^{2}$ contained in the basis \cite{PandoZayas:2000sq}. One may again exploit the preservation of supersymmetry to reduce the second order system of coupled equations to a first order one, which may be solved for all fluxes, the metric and the dilaton (which is still constant). The metric of the resolved conifold is exactly known and in suitably chosen coordinates (with a radial direction $\rho$) it reads
\begin{eqnarray}
{\rm d}s_{6}^{2}&=&\frac{1}{\tilde{k}(\rho)}\,{\rm d}\rho^{2}+\frac{1}{9}\,\tilde{k}(\rho)\,\rho^{2}\,\left({\rm d}\psi+\cos\theta_{1}\,{\rm d}\phi_{1}+\cos\theta_{2}\,{\rm d}\phi_{2}\right)^{2}\nonumber\\
&&+\frac{1}{6}\,\rho^{2}\,\left({\rm d}\theta_{1}^{2}+\sin^{2}\theta_{1}\,{\rm d}\phi_{1}^{2}\right)+\frac{1}{6}\left(\rho^{2}+6a^{2}\right)\left({\rm d}\theta_{2}^{2}+\sin^{2}\theta_{2}\,{\rm d}\phi_{2}^{2}\right)\,,\label{eq:resolvedmetric}
\end{eqnarray}
where the function \(\tilde{k}(\rho)\) is given by
\begin{equation}
\tilde{k}(\rho)=\frac{\rho^{2}+9a^{2}}{\rho^{2}+6a^{2}}\,.
\end{equation}
The parameter \(a\) is both a measure for the ``asymmetry'' between the two pairs of angles $(\theta_{1},\phi_{1})$ and $(\theta_{2},\phi_{2})$ and a singularity resolution parameter, as one can see from Eq.~(\ref{eq:resolvedmetric}) in the IR limit (\ie for $\rho\rightarrow 0$ close to the tip),
\beq
\left[{\rm d}s_{6}^{2}\right]_{\rho\rightarrow0}=\frac{2}{3}\,\dd\rho^{2}+\frac{1}{6}\,\rho^{2}\left(e^{2}_{\psi}+e^{2}_{\theta_{1}}+e^{2}_{\phi_{1}}\right)+\left(a^{2}+\frac{1}{6}\,\rho^{2}\right)\left(e^{2}_{\theta_{1}}+e^{2}_{\phi_{1}}\right)\,.
\eeq
The second term on the right hand side shows that this time the $S^{3}$ shrinks to zero size at the apex. As it was the case for the deformed conifold, in the limit of large $\rho$ the simple conifold metric (\ref{eq:conifoldmetric}) is recovered.

\begin{figure}[t]
\begin{center}
\includegraphics[width=\textwidth]{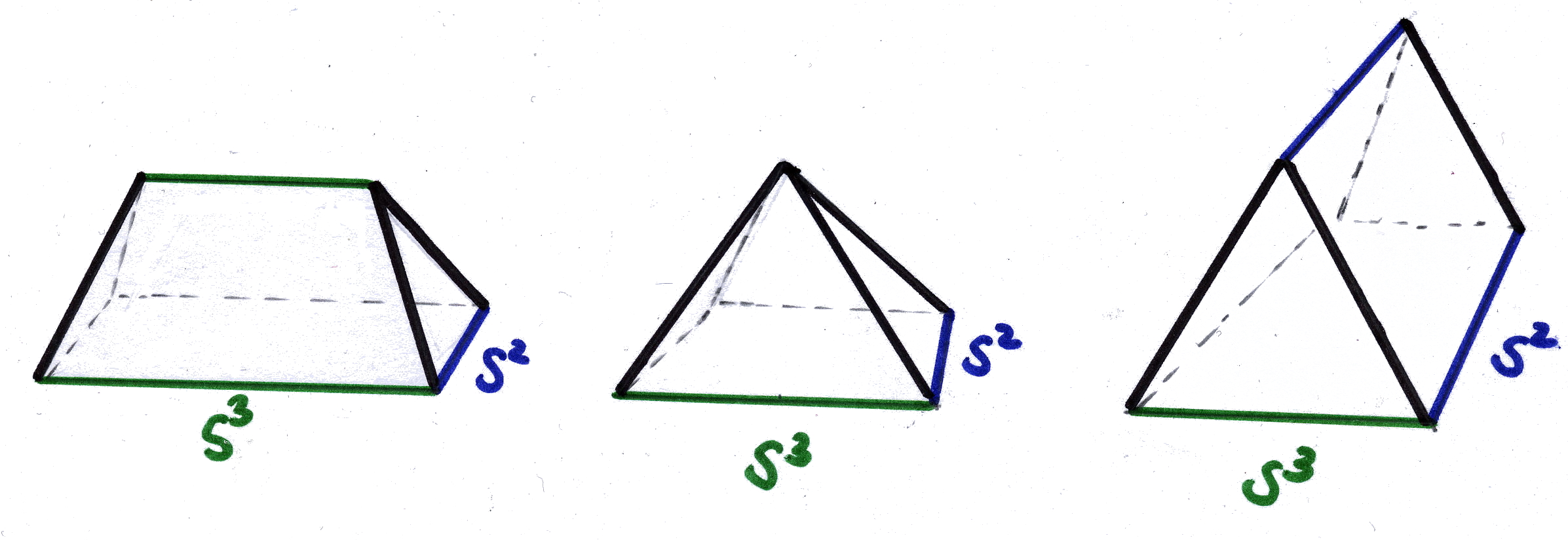}
\caption[The geometric transition relating deformed, simple and resolved conifold geometries. (Following the example of \cite{PhD_Anke}.)]{The ``confold transition'' relates the three geometries discussed in Section \ref{sec:conifolds}. All three conifolds have a product $S^{3}\times S^{2}$ at their basis, but they differ in the limit $r\rightarrow0$ towards the tip. In the deformed conifold (\emph{left}), the three-sphere of the basis stays of finite size, while the two-sphere shrinks to zero. The $S^{3}$ can be shrunk to zero to obtain the simple conifold (\emph{center}) with a singularity at its tip. From this, the $S^{2}$ then is blown up to reach the resolved conifold (\emph{right}).}
\label{fig:geomtrans}
\end{center}
\end{figure}

In Section \ref{subsec:dualities} we mentioned that the three distinct geometries describing the simple, deformed and resolved conifold are, in the exact string theory description beyond the leading supergravity order, related by so-called geometric transitions. Reviews can be found in \cite{PhD_Anke,Gwyn:2007qf}, and the general idea is illustrated in Fig.~\ref{fig:geomtrans}.

\bigskip
At the end of this rather technical Chapter, we pause to retain the key lessons learned from flux compactifications and the AdS/CFT correspondence. We saw that, respecting certain conditions resulting from the cancellation of tadpoles, one may embed D$p$-branes of suitable dimensionality into a the supergravity background. In the case of type IIB theory, these branes must have an odd number of spatial dimensions. They then act as ``pointlike'' sources for the R-R (and NS-NS) gauge potentials with an even number $(p+1)$ of indices, giving rise to $(p+2)$ fluxes. Branes wrapped along compact dimensions can provide ``fractional'' charges of lower $p$ type. If fluxes run through cycles of the compactification manifold, the shape of these cycles can no longer be changed at will (because it comes with an energy cost), and therefore the corresponding geometric modulus (of complex structure type) is stabilized. Moreover, these flux-enriched geometries typically have a warp factor $h(r)$, which means that energy scales such as \eg the tension of a brane depend on the position $r$. Expressed in the ``proper'' coordinate $\xi$, which measures distances inside the warped geometry, this dependence can be exponentially strong. The prime example of such a flux compactification is the Klebanov Strassler throat, also called the deformed conifold. In the UV limit, it resembles the simple conifold, but unlike the latter remains finite in the IR. In terms of the naive radial coordinate $r$, the warp factor in the KS throat receives logarithmic corrections, and moreover $r$ runs over a finite range of values, namely from the bottom of the throat $r_{0}$ to its edge $r_\uUV=R$. Both of these are functions of the chosen flux quantum numbers, the string scale, and the coupling $\gs$ [see Eq.~(\ref{eq:RonEinstein}) and Chapter \ref{chapter:braneinflation}]. In the next Chapter, we turn to the stabilization of the remaining part of moduli space, \ie the K\"ahler structure moduli.

\chapter{Non-Perturbative K\"ahler Moduli Stabilization}\label{chapter:kaehlerstabilized}
\begin{quotation}
\emph{We saw that after compactification on a suitable Calabi Yau space with NS-NS and R-R fluxes turned on, all compex structure moduli are fixed. In this Chapter, we discuss how the ``other half'', \ie the K\"ahler structure part of moduli space, can be stabilized, in particular the overall volume modulus: assuming it is the only field remaining massless in four dimensions, non-perturbative corrections to the superpotential are used to give it a (very heavy) mass.}
\end{quotation}

The decisive step towards fixing all moduli in the effective four-dimensional theory was presented in the work of {\bf K}achru, {\bf K}allosh, {\bf L}inde and {\bf T}rivedi \cite{Kachru:2003aw} and is called the KKLT stabilization procedure. In this setup, the last massless K\"ahler modulus --the overall compactification volume-- is stabilized by non-perturbative effects, leading to a supersymmetric anti-de Sitter minimum in four dimensions. Mildly breaking supersymmetry by adding anti-branes, this stable AdS minimum is then lifted to a metastable (but very long-lived) dS one. Let us now develop an intuition for this construction.

\section{Reminder: Supergravity Lagrangian}
For frequent use in this and the following Chapter, we briefly recall the relevant expressions for the K\"ahler and the superpotential that describe a locally supersymmetric theory in four dimensions. 
These two functions completely determine the scalar part of a supergravity Lagrangian\footnote{Here we ignore the $D$ term in the supergravity potential, and we do not consider the gauge kinetic function $f_{ab}$ for vector multiplets. It arises if there are low-energy gauge fields present and coupled to the scalars.}, and while the superpotential $W(\phi_{m})$ is holomorphic, the K\"ahler potential (responsible for kinetic terms) is a general function of both $\phi_{m}$ and $\bar{\phi}_{m}$, $K(\phi_{m},\bar{\phi}_{\bar{m}})$. The Lagrangian is then written as
\beq
\frac{\calL}{\sqrt{-g}}=-K_{\bar{\imath}j}(\phi,\bar{\phi})\,\partial^{\mu}\phi^{\bar{\imath}}\,\partial_{\mu}\bar{\phi}^{j}-V(\phi,\bar{\phi})\,,
\eeq
where $K_{\bar{\imath}j}$ is the ``K\"ahler metric'' on field space, obtained from second derivatives of the K\"ahler potential, see Eq.~(\ref{eq:Kaehler}). The $F$ term in the scalar potential (where we have suppressed the corresponding subscript) can be obtained from the super- and K\"ahler potentials as [compare Eq.~(\ref{eq:treeleveleffective})]
\beq\label{eq:VFagain}
V=e^{K}\left(K^{\bar{\imath}j}\,\overline{D_{\bar{\imath}}W}\,D_{j}W-3|W|^{2}\right)\,,\qquad D_{i}W=W_{,i}-K_{,i}W\,.
\eeq
We mentioned earlier that $D_{i}W$ is the order parameter for supersymmetry breaking, therefore, in a minimum of the potential $V_{F}$ where in addition $D_{i}W=0$, supersymmetry is preserved. Generally, the potential $V_{F}$ is (by definition) precisely flat in the direction of moduli $\phi$ at tree order (as it was the case in ``no scale'' supergravity), \ie these fields do not appear in Eq.~(\ref{eq:VFagain}). However, this situation is changed in the presence of the fluxes, which, as we argued before, result in an energy cost for the changing the size and shape of geometric details in the background, while the same topology is still preserved.

\section{K\"ahler and Superpotential in Flux Compactifications}\label{sec:kaehlersuper}
In Section \ref{subsec:CYmoduli}, we saw how to determine the two contributions $K^{2,1}$ and $K^{1,1}$ to the K\"ahler potential in the presence of background fluxes. 
The superpotential $W$ can also be calculated from the flux compactification of (in our case) type IIB string theory based on arguments from the equations of motion and (imaginary) self-duality. Recall that one of the conditions characterizing flux compactifications with D-brane sources (see Section \ref{subsec:nogotheorem}) was $*G_{(3)}=iG_{(3)}$, where $G_{(3)}=H_{(3)}-\tau F_{(3)}$ and $\tau$ the axion-dilaton. We also know that both $H_{(3)}$ and $F_{(3)}$ are restricted to lie within the compact manifold $\scrM_{6}$, therefore the ``$*$'' operation actually acts on the six-dimensional internal piece of spacetime. Its imaginary self-duality under ``$*_{6}$'' then restricts $G_{(3)}$ to have only pieces proportional to (2,1)- and (3,0)-forms on $\scrM_{6}$ \cite{Giddings:2001yu}. 
The constraint $*G_{(3)}=iG_{(3)}$ can also be derived from the superpotential 
\beq\label{eq:potW}
W=\int_{\mathscr{M}_{6}}G_{(3)}\wedge\Omega\,.
\eeq
with $\Omega$ the (3,0)-form of the compact internal space. Let us now examine the conditions for supersymmetry to be preserved with this superpotential \cite{Becker:2007zj}.

Suppose the CY manifold under scrutiny has (before fluxes are turned on) massless moduli fields corresponding to $h^{2,1}$ complex structure moduli $\phi_{a}$, the axion-dilaton $\tau$ and a superfield called $\rho$ which contains the (only) K\"ahler modulus. We then saw [compare Eq.~(\ref{eq:moduliKaehlerpotentials})] that the K\"ahler potential for the $\phi_{a}$ is
\beq\label{eq:K21again}
K^{2,1}(\phi_{a})=-\log\left(i\int_{\scrM_{6}}\Omega\wedge\bar{\Omega}\right)\,,
\eeq
to which we add those for $\tau$ and $\rho$, which are given by
\beq\label{eq:kaehlerrho}
K(\rho)=-3\log\left[-i\left(\rho-\bar{\rho}\right)\right]\,,\qquad K(\tau)=-\log\left[-i\left(\tau-\bar{\tau}\right)\right]\,.
\eeq
[Note the similarity with the no scale K\"ahler potentials of Eq.~(\ref{eq:noscaleKaehler}).] The total K\"ahler potential is given by the sum of the three terms in Eqs.~(\ref{eq:K21again}) and (\ref{eq:kaehlerrho}), $K=K^{2,1}(\phi_{a})+K(\rho)+K(\tau)$. The order parameter for unbroken supersymmetry is $D_{i}W=\partial_{i}W+K_{,i}\,W$, where $i$ runs over the $\phi_{a}$ as well as $\tau$ and $\rho$, and we need $D_{i}=0$ to preserve supersymmetry. 
The remarkable property of $\rho$ is that it does not appear in the superpotential, therefore $D_{\rho}W=0$, and more precisely
\beq
D_{\rho}W=\partial_{\rho}K\,W=-\left(\frac{3}{\rho-\bar{\rho}}\right)W=0\,,
\eeq
therefore also $W=0$ holds for supersymmetric configurations. Above, using the argument of self-duality, we already restricted the possible pieces of $G_{(3)}$ to (2,1)- and (3,0)-forms. But from Eq.~(\ref{eq:potW}) we see that to have $W=0$ identically, $G_{(3)}$ can only be of (2,1)-form type (see \eg \cite{Becker:2007zj} for a more detailed discussion). If, however, we had $W=W_{0}$ with $W_{0}$ a small constant calculated from the fluxes, this would tell us that \emph{i)} supersymmetry is mildly broken and \emph{ii)} that $W_{0}$ must be due to a (3,0)-form contribution in $W$. This will be the case of interest in the following Section. Let us one add one more remark on the ``no scale'' form for of the K\"ahler potential for $\rho$ observed in Eq.~(\ref{eq:kaehlerrho}): this is a manifestation of the property that the three conditions characterizing the flux compactification solutions, \ie the imaginary self-duality of $G_{(3)}$, the tadpole condition and the saturation of the BPS-like bound (relation between tension and charge) in Eq.~(\ref{eq:fluxsources}), are all invariant under rescaling by a constant. 

\section{The KKLT Procedure}
The starting point for KKLT were warped type IIB compactifications of the Klebanov Strassler type which have non-trivial fluxes for their NS-NS and R-R potentials, see Section \ref{subsec:deformed}. Let the fluxes be of such form that all complex structure moduli are stabilized, but supersymmetry still preserved. Then the resulting four-dimensional supergravity theory is of the no scale type, and hence the overall volume modulus $\rho$ remains unfixed.

As before, it is assumed that $\rho$ is the only unfixed K\"ahler modulus. [It is possible to construct explicit models with this property, and in terms of the Hodge numbers of the Calabi Yau it means that we have $h^{1,1}(\mathscr{Y})=1$.] One then includes corrections which violate the no scale structure, of which non-perturbative corrections to the superpotential are one example. 
Recall that the K\"ahler potential for $\rho$ is Eq.~(\ref{eq:kaehlerrho}), while its superpotential is constant, $W=W_{0}$. [As stated above, this arises from the (3,0)-part of the imaginary self-dual three-form flux $G_{(3)}$. Note that with the corrections we are about to include, supersymmetry can be preserved even if $W_{0}\neq0$.] Once can explicitly check from Eq.~(\ref{eq:VFagain}) that the no scale cancellation takes place and that hence $V=0$. 
This is true at leading order in both the $\alphap$ and the $\gs$ expansion, \ie in particular when the compactification manifold is large compared to the string scale.

\subsection{Fixing the Volume Modulus}
We are looking for corrections breaking the tree level no scale structure. Two known sources of non-perturbative corrections to this superpotential are \emph{i)} so-called instanton effects and \emph{ii)} ``gaugino condensation''. The former are caused by Euclidean D3-branes wrapped on four-cycles in the internal manifold, and the latter are due to non-Abelian gauge groups that can live an stacks of D7-branes also wrapping four-cycles. At special points in moduli space, these gauge theories undergo gaugino condensation, and one can assume to be at such a point. It can be shown that these two corrections enter into the superpotential in the same functional form (the gaugino condensate looks like a fractional instanton effect), and we treat them summarily. The remaining modulus $\rho$ is then fixed by a non-perturbative correction to the superpotential
\beq\label{eq:nonpert-correction}
W(\rho)=W_{0}+A\exp(-a\rho)\,,
\eeq
with $A,a$ constants. Note that we take the non-perturbative correction for the superpotential into account while keeping the K\"ahler potential only at tree level. This can be consistent if the size of the constant term $W_{0}$ is very small.

\subsection{The Resulting Vacuum}
We now try to understand the vacuum structure with tree level K\"ahler potential for $\rho$, and the non-perturbatively corrected superpotential. Let us also assume that the tadpole cancellation condition [which, in its F-theory formulation, was given in Eq.~(\ref{eq:tadpoleF})] has been solved by fluxes only and there are no localized D3-branes in the background so far. We know that in a supersymmetric vacuum, we have $D_{\rho}W=0$. Before, in Section \ref{sec:kaehlersuper}, we saw that using fluxes but \emph{without} the non-perturbative corrections we just introduced in Eq.~(\ref{eq:nonpert-correction}), this translates into $W_{0}$. Now we admit a small constant $W_{0}$, generated by a (3,0)-piece of the $G_{(3)}$ flux.\\
The volume modulus $\rho$ is a complex variable (as it was the case in our simple example of the torus in Section \ref{subsec:torus}), but let us set its real part, which is an axion decending from a four-form gauge potential, to zero. (We shall return to the r\^ole of this axion later.) This means that we have only $\rho=i\sigma$. Take the parameters $A,a,W_{0}$ to be real, and let $W_{0}$ be negative. Then we find from $D_{\rho}W=0$ with Eq.~(\ref{eq:nonpert-correction}) and Eq.~(\ref{eq:VFagain}) that
\beq\label{eq:AdSpotential}
W_{0}=-A\,e^{-a\sigma_\uc}\left(1+\frac{2}{3}\,a\sigma_\uc\right)\,,\qquad V=-\frac{a^{2}\,A^{2}\,e^{-2a\sigma_\uc}}{6\sigma_\uc}\,,
\eeq
which means that there is an anti-de Sitter vacuum at the critical value $\sigma_\uc$. As a consequence, $\rho$ (or $\sigma$, respectively) now has a potential and is stabilized, while supersymmetry is still preserved. Note that we need $\sigma_\uc$ to be large (compared to the string scale, where $\alphap$ has been set to one in our present units) for the supergravity expansion to be valid.

\subsection{Breaking Supersymmetry}
\begin{figure}[h]
\begin{center}
\includegraphics[width=0.45\textwidth]{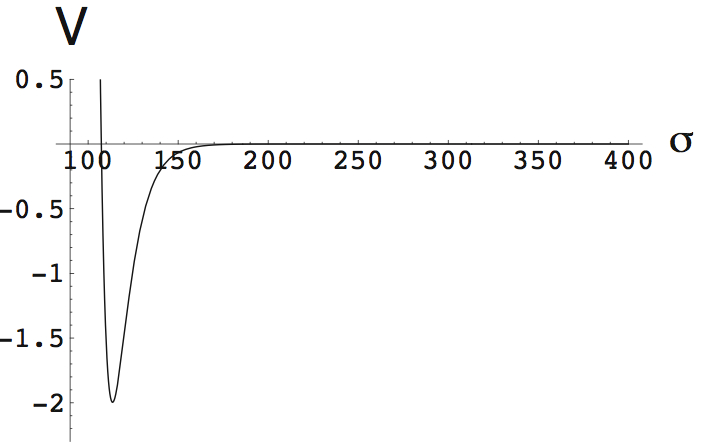}
\hfill
\includegraphics[width=0.45\textwidth]{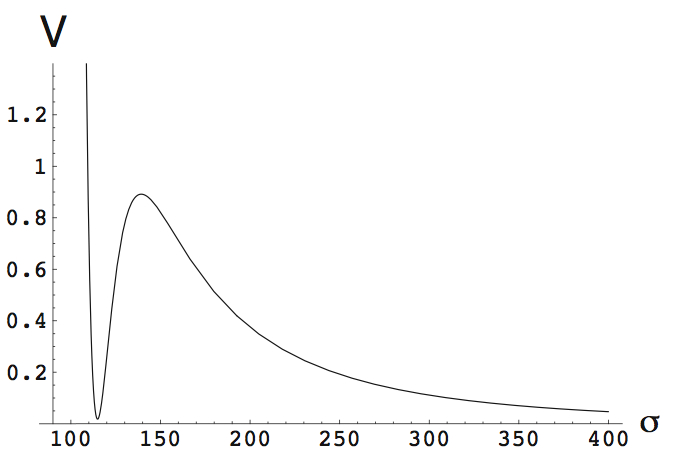}
\caption[Potentials with an AdS and a dS vacuum obtained after the KKLT procedure of K\"ahler moduli stabilization. (Source: \cite{Kachru:2003aw})]{\small \emph{Left:} The exemplary potential plotted in \cite{Kachru:2003aw} to illustrate Eq.~(\ref{eq:AdSpotential}). Here, $V$ has been multiplied by a factor of $10^{15}$ to show the behavior close to zero. The parameters are such that $W_{0}=-10^{-4},\, A = 1, \,a = 0.1$ in Eq.~(\ref{eq:AdSpotential}). The minimum at (a large value of) $\sigma_\uc$ is of AdS type. \emph{Right:} The same potential (again multiplied by $10^{15}$) after uplifting by a term $\delta V\propto D/\sigma^{2}$ with $D=3\times 10^{-9}$ in Eq.~(\ref{eq:dSpotential}). This turns the minimum at a new (still large) value of $\sigma_\uc$ into a dS one. Note that the potential is very steep around this potential and hence the $\sigma$ modulus very massive. The $\sigma_\uc$ minimum is metastable (but longlived), with the global minimum still being reached in the decompactification limit $\sigma\rightarrow\infty$. (Figures from \cite{Kachru:2003aw})}
\label{fig:KKLTpotentials}
\end{center}
\end{figure}

This sucessfully fixes the last of the moduli without breaking supersymmetry, and the geometry of the four extended dimensions is anti-de Sitter space, with negative vacuum energy density at $\sigma_\uc$, see Fig.~\ref{fig:KKLTpotentials}. Therefore, this system still needs additional ingredients to break supersymmetry and lift the vacuum energy to zero or positive value (\ie de Sitter space). However, this must be achieved in a way that does not ruin the stabilization we just engineered. The KKLT idea is to do so by adding an anti-D3 brane, or a small number thereof. (Note that anti-branes do not have world volume scalar moduli because there is a preferred position where $\overline{\mathrm{D}}$-branes minimize their energy.) Let us assume that \emph{too much} flux is turned on, and that therefore the tadpole cancellation condition (\ref{eq:tadpoleF}) is satisfied once we add the \Dbar.\\
An anti-brane breaks all of the supersymmetry, hence the resulting four-dimensional theory cannot be captured by supergravity anymore. There is then much less control over the corrections to the calculation. In particular, the anti-branes would normally create a runaway potential for the volume $\sigma$, the imaginary part of $\rho$. (Recall that for the torus discussed in Section \ref{subsec:torus}, also the imaginary part of the K\"ahler modulus $\rho$ was proportional to the volume $\propto\sqrt{\det g}$.) But at present we are in a warped background, therefore the damage done by supersymmetry breaking can be kept small if the contribution of the antibrane to the low-energy action is parametrically weak. At the bottom of a warped throat, this is the case: the anti-D3 prefers to sit at the IR end of the geometry where it can minimize its energy. The extra energy density provided by the anti-D3 then is
\beq
\delta V=2\,\frac{T_{3}}{h_{0}}\,\frac{1}{\sigma^{2}}\,,
\eeq
where he parameter $h_{0}$  can be tuned over an extremely wide range of values because it is given by the compactification as an exponential of the integers labeling the fluxes of the background. 
More generally, one can write the energy contribution of an undetermined number of anti-D3 branes in a warped throat with a given warp factor as $V\propto D/\sigma^{3}$, where $D$ depends on the value of the warp factor at the position(s) of the anti-branes.\\
If the parameter $D$ is finetuned, there now exists a de Sitter minimum for the scalar potential because after adding the uplifting contribution, $V$ reads (see Fig.~\ref{fig:KKLTpotentials})
\beq\label{eq:dSpotential}
V=\frac{aA e^{-a\sigma}}{2\sigma^{2}}\left(\frac{1}{3}\,\sigma a A e^{-a\sigma}+W_{0}+A e^{-a\sigma}\right)+\frac{D}{\sigma^{2}}\,.
\eeq
Note that the global minimum is still at $\sigma \rightarrow\infty$, \ie in the decompactification limit because all sources of energy vanish in this limit. The potential around the local de Sitter minimum is steep, the more so the closer one requires the value of energy density in the minimum to be to zero. Therefore, $\rho$ has become very massive. The position of the minimum is still at large $\sigma$, therefore the supergravity approximation holds. It can be shown that the de Sitter minimum obtained in this way is metastable, but stable enough to persist for a long time compared to cosmological time scales of $\order{10^{10}\mathrm{yrs}}$.

\bigskip
For a long time, progress in string cosmology was stalled because the potentials of scalar moduli in the four-dimensional theory were not very well understood: they were either identically zero, or, if lifted by non-perturbative effects, showed a runway behavior. In the KKLT setup, one can control a runaway direction like the overall volume modulus in the special background of warped compactifications, and moreover lift it to a de Sitter vacuum using anti-D3 branes. This is possible because the tensions of anti-D3-branes are screened while they sit at the IR end of a KS throat. Following this development, the number of string inflationary scenarios, in particular of the ``brane inflation'' type, grew almost without bounds. We present a small selection of them in the next Chapter.

\chapter{Exemplary Models of String Inflation}
\begin{quotation}
\emph{Following the development of stabilization techniques for both complex and K\"ahler structure moduli, a large variety of string inflationary scenarios was proposed. Different kinds of scalar fields were used for the inflaton, and in this Chapter we give but a taste of different classes. Our particular interest is in models of ``brane inflation''.}
\end{quotation}

A considerable number of review articles on string cosmology and string inflation exists, recent examples being \eg \cite{Kinney:2009vz,Baumann:2009ni, McAllister:2007bg,Burgess:2007pz,HenryTye:2006uv,Cline:2006hu}. An earlier overview (prior to the KKLT construction) is given in \cite{Quevedo:2002xw}. All types of scalar fields in string theory (of which there are, as we have seen, \emph{many}) have been used as inflaton candidates, among them moduli of the metric and the extra-dimensional gauge fields (among them axions), the distance between (anti-)branes, the dilaton itself \etc One cannot hope to do all of them justice, and we restrict ourselves to two short examples of ``moduli inflation'' before turning to the ``brane inflation'' class of scenarios. A subset of the latter class, brane--anti-brane inflation in a warped throat geometry, stands behind most of the original work presented in Part \ref{part:results}, and we return to it in detail in Chapter \ref{chapter:braneinflation}.

\section{The ``Racetrack'' Model}
\begin{figure}[t]
\begin{center}
\begin{minipage}{0.6\textwidth}
\includegraphics[width=\textwidth]{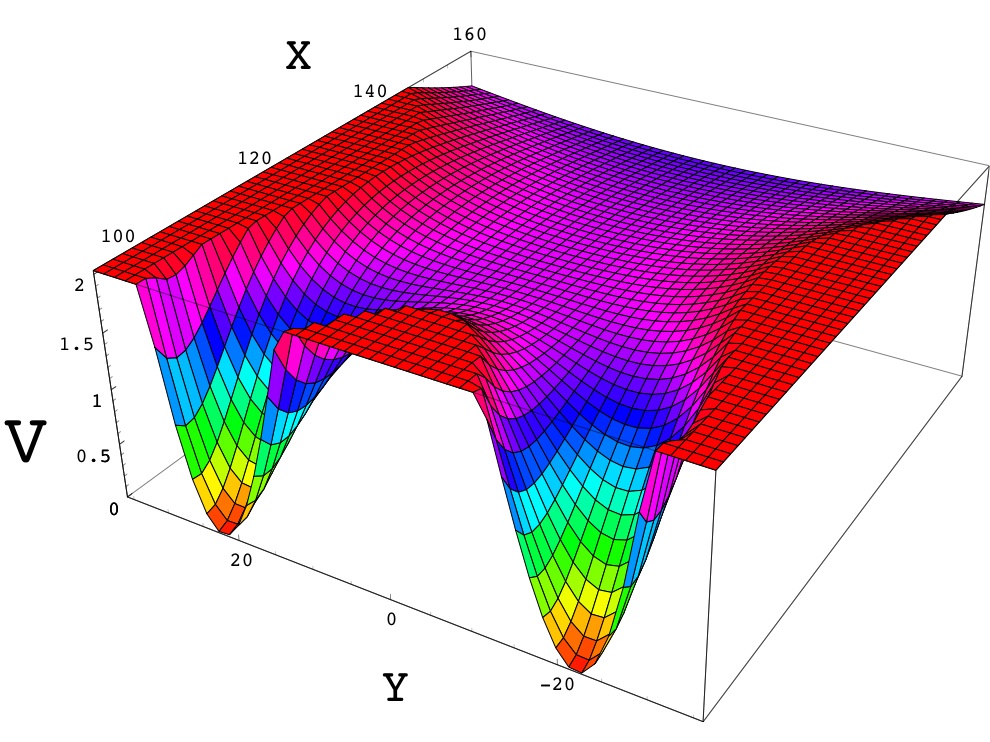}
\end{minipage}
\hfill
\begin{minipage}{0.3\textwidth}
\caption[The scalar inflaton potential in the racetrack model of inflation. (Source: \cite{BlancoPillado:2004ns})]{The scalar inflaton potential in the racetrack model of inflation, where the $Y$ direction is the imaginary part of the K\"ahler modulus $\phi$, and $X$ the real part. (Figure from \cite{BlancoPillado:2004ns})}
\end{minipage}
\label{fig:racetrack}
\end{center}
\end{figure}

Instead of stabilizing \emph{all} moduli fields, one can explore whether interesting inflationary scenarios arise by keeping a manageable number of them dynamic in four dimensions. For example, consider a slight generalization of the previously introduced non-perturbative superpotential for a K\"ahler modulus $\phi$,
\beq
W=W_{0}+A\,e^{-a\,\phi}+B\,e^{-b\,\phi}\,,
\eeq
which is historically called the ``racetrack'' superpotential. The inflating direction (see Fig.~\ref{fig:racetrack}) in this case is the axionic (real) part contained in the modulus $\rho$, and hence the ``racetrack'' model is an example of the axion inflation class of string cosmology \cite{BlancoPillado:2004ns}. In a later version, which can be derived rigorously from a specific Calabi Yau compactification on the manifold $\mathbb{P}^{4}_{[1,1,1,6,9]}$  (``better racetrack inflation'' \cite{BlancoPillado:2006he}), two complex moduli fields $\phi_{1}$ and $\phi_{2}$ are used, the K\"ahler potential for which is
\beq\label{eq:racetrackkaehler}
K(\phi,\bar{\phi})=-2\ln\left[\left(\tau^{1}\right)^{3/2}-\left(\tau^{2}\right)^{3/2}\right]\,,\qquad \tau^{i}=\Re(\phi_{i})\,,
\eeq
and the non-perturbative superpotential is computed to
\beq\label{eq:racetrack}
W=W_{0}+A\,e^{-a\,\phi_{1}}+B\,e^{-b\,\phi_{2}}\,,
\eeq
where $A,B,a,b$ are parametrically calculable constants. As in the KKLT construction, the resulting AdS vacuum can be uplifted to a dS one by adding anti-D3-branes. The resulting scalar potential $V$ then has a rather complicated form as a function of the four real fields $\tau^{1,2},\sigma^{1,2}$, where $\sigma^{i}=\Im(\phi_{i})$. This illustrates a general feature of many string inflation models: if the field space has several dimensions, the inflationary trajectory is often complicated, and one cannot hold other real fields fixed while making a single real or imaginary part of a scalar field dynamic. 
Inflation is not generic in the ``better racetrack'' model [it is very sensitive to the values of $A,B,a,b$ in the superpotential (\ref{eq:racetrack})], but can happen for specific parameter choices. However, it is not yet clear whether the required parameter values can be achieved from the particular underlying CY manifold \cite{Burgess:2007pz}.

\section{K\"ahler Moduli Inflation}
In the KKLT stabilization procedure, the value of the constant superpotential $W_{0}$ was very small. In ``K\"ahler moduli inflation'' \cite{Conlon:2005jm,Conlon:2005ki,Balasubramanian:2005zx,Balasubramanian:2004uy}, this constant is not so severely restricted, at the price of perturbative corrections (in $\alphap$) to the K\"ahler potential becoming important. This can produce new minima in the resulting scalar potential, but several moduli fields are required for this scenario. In the simplest case, three fields $\phi^{i}$ are involved: 
while the K\"ahler potential is of the same type as in Eq.~(\ref{eq:racetrackkaehler}), plus a perturbative correction, the superpotential for the three K\"ahler moduli reads
\beq
W=W_{0}+\sum_{i}A_{i}\,e^{-a_{i}\,\phi^{i}}\,.
\eeq
Again denoting the real parts of the scalar fields by $\tau^{i}$, this potential can lead to inflation in the regime where one of them, \eg $\tau^{3}$ is much larger than the other ones. Then $\tau^{3}$ is the most important direction in field space, and the potential has approximatively the form \cite{Burgess:2007pz}
\beq
V\simeq V_{0}-C\,(\tau_{\uc}^{3})^{4/3}\,\exp\left[-c(\tau_{\uc}^{3})^{4/3}\right]\,,
\eeq
where $\tau_{\uc}^{3}$ is the canonically normalized variable along the $\tau^{3}$ direction. The field values of $\tau_{c}^{3}$ must be sufficiently large to have slow roll. Compared to racetrack inflation, this potential is less dependent on the precise tuning of parameters.

\section{Brane Inflation}
In the two examples above, the inflaton was associated with a modulus of the entire compactification manifold. In ``brane inflation'', one instead considers world volume moduli associated with the positions and properties of embedded branes within the ten-dimensional string geometry. This class of models has given particularly rich offspring, and some scenarios have been investigated in very fine detail. (One of these is presented in the next Chapter.) Here, we assemble some historic developments and general features of brane inflation.

\subsection{Brane Inflation Prehistory}
\begin{figure}[h]
\begin{center}
\includegraphics[width=0.45\textwidth]{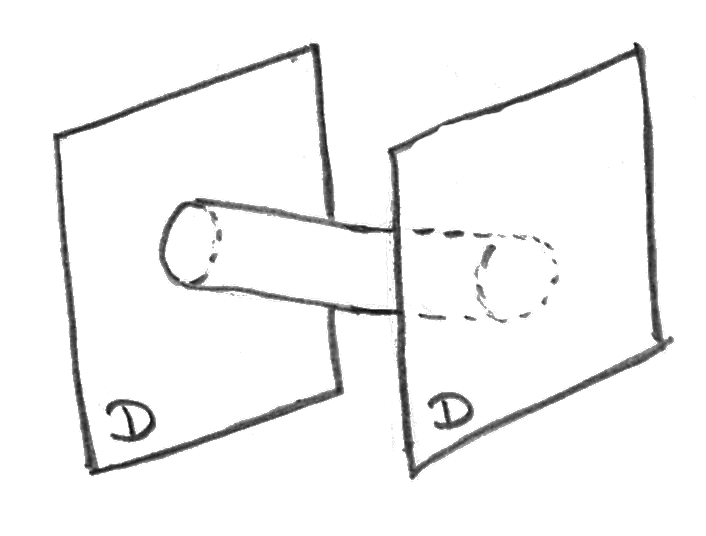}
\hfill
\includegraphics[width=0.45\textwidth]{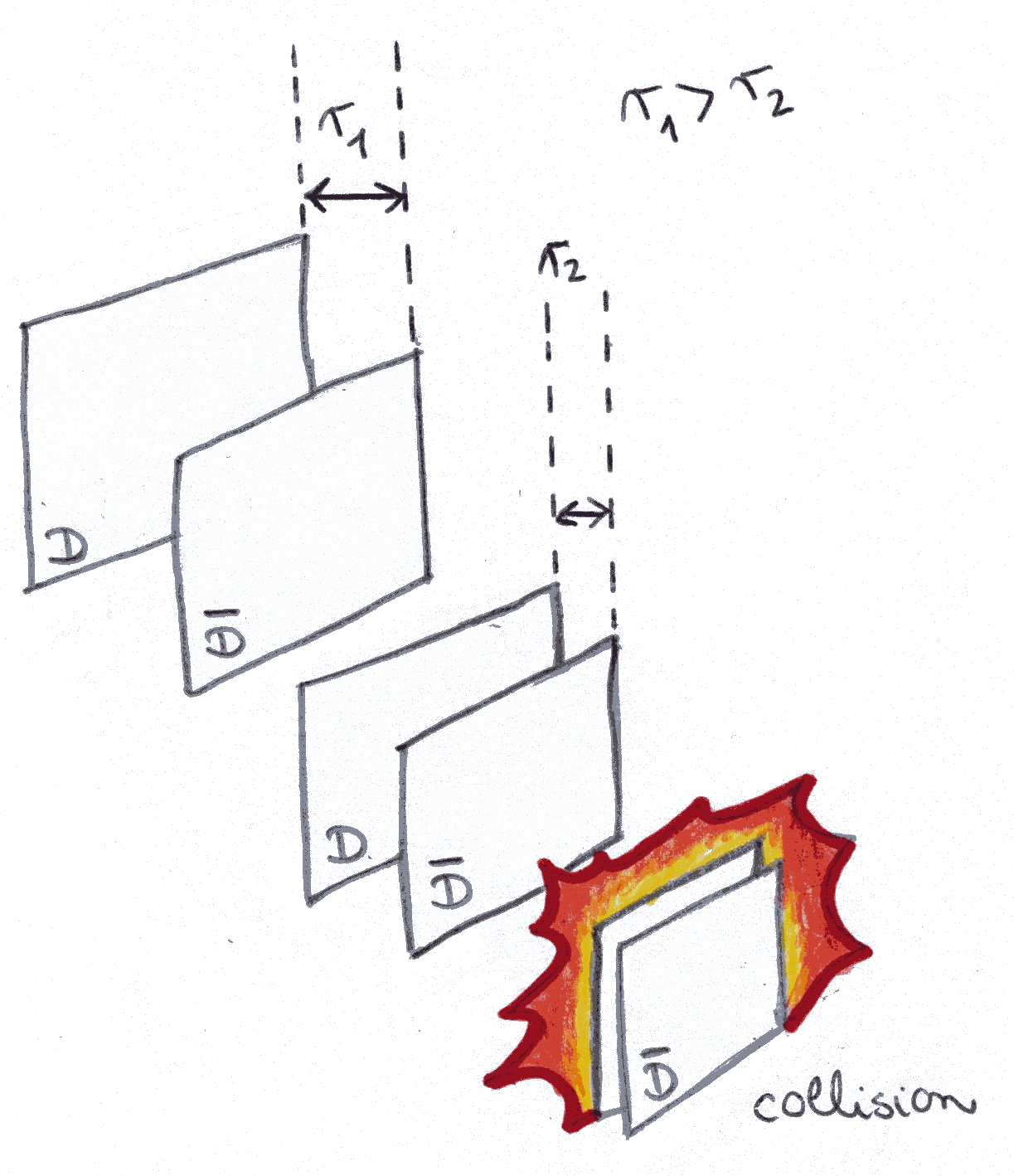}
\caption[Closed string exchange at tree level between D-branes. D-branes approaching and annihilating during brane--anti-brane inflation.]{\emph{Left:} Closed string exchange at tree level between D-branes. The cylinder diagram can be exactly calculated, and describes the exchange of graviton and R-R modes (as well as dilaton modes) between the branes. \emph{Right:} During brane--anti-brane inflation, the distance between the branes decreases because they attract each other both gravitationally and because of their opposite R-R charges. The brane distance $r$ plays the r\^ole af the inflaton. Inflation ends when the branes annihilate, releasing the energy stored in the system.}
\label{fig:closedexchange}
\end{center}
\end{figure}

The first string inflationary scenario using D-branes is due to Dvali and Tye in 1998 \cite{Dvali:1998pa}: two parallel BPS branes (\ie they preserve half of the original supersymmetry in the background) do not feel any force between them, due to the cancellation of their gravitational interaction and the R-R electrostatic repulsion. (The dilaton $\Phi$ also contributes an attractive force between the branes, but we will not mention it explicitly because in the scenarios we are eventually interested in, the dilaton is stabilized at a fixed value, \ie $\gs$ is a constant.) To show the vanishing of the net force in terms of explicit string amplitudes, one would have to calculate the ``cylinder'' diagram for exchange of closed strings between the branes, see Fig.~\ref{fig:closedexchange}. (Note that this is the same diagram as for the exchange of open strings at one loop order).\\
The idea presented in \cite{Dvali:1998pa} was that after supersymmetry is broken, only the graviton would stay massless, and with a massive R-R mode, subject to Yukawa suppression, the cancellation of forces would no longer be perfect. Instead, it would be replaced by an attractive force between the branes. The effective potential proposed by Dvali and Tye had the form \cite{Quevedo:2002xw}
\beq
V\approx 2T+\frac{a}{r^{d-2}}\left(1+\sum_{\textnormal{NS}}e^{-m_{_{\mathrm{NS}}}\,r}-2\sum_{\textnormal{R}}e^{-m_{_{\mathrm{R}}}\,r}\right)\,,
\eeq
where $r$ is the brane distance, and $d$ the number of compact spacetime dimensions. The first term in the brackets is the ``cosmological constant'' provided by the brane tensions, and the other two terms are due to the brane interactions, which cancel when the masses of the carrier modes are zero. If they are massive, the cancellation is no longer exact and the potential is very flat, possibly suitable for inflation with $r$ as the inflaton field. However, it is difficult to underpin this intuitive picture with actual calculations, and moreover, there is no natural explanation of the end of inflation and the transition to the reheating era.

\subsection{Brane--Anti-Brane Inflation}\label{sec:brane-antibrane}
Following the proposal of \cite{Dvali:1998pa}, the attraction between D-branes and anti-D-branes was considered, for which the R-R and graviton interactions do not cancel, but add up, since these branes have opposite charge \cite{Burgess:2001fx}.  Moreover, in this case the cylinder diagram can be computed explicitly. The corresponding calculation at the level of the supergravity action in type IIB theory is presented in the next Chapter, here let us state that the potential for the inter-brane distance $r$ is found to read \cite{Quevedo:2002xw}
\beq
V(r)\approx A-\frac{B}{r^{d_{o}-2}}\,,
\eeq
where $d_{o}$ is the number of dimensions orthogonal to the brane/anti-brane world volume. The constants $A$ and $B$ can be calculated explicitly in a given setup. In the case of the model considered in Chapter \ref{chapter:braneinflation}, they are determined by the higher-dimensional geometric background into which the branes are immersed.

\subsubsection{``D-celeration''}
In Sections \ref{subsec:k-inflation} and \ref{sec:kinflationpert} we discussed the generalization of the standard inflationary scenario (with a canonical scalar field rolling down a flat potential) to modified kinetic terms for the inflaton. These scenarios were dubbed ``$k$-inflation'' in Part \ref{part:cosmo}, and they can be used to capture an interesting features of an inter-brane distance inflaton: as we shall see in the next Chapter, from the DBI action of a D-brane (\ref{eq:DBIbraneaction}) one can show that the kinetic term of a D3-brane moving along one direction only is $\propto\left[1-\partial_{\mu}\phi\,\partial^{\mu}\phi/T(\phi)\right]^{-1/2}$. Here, $T(\phi)$ denotes the position-dependent brane tension, and $\phi$ is related to the radial coordinate $r$ by a renormalization.\\
For small velocities, the square root can be expanded and the inflaton kinetic term becomes canonical, but if $\phi$ starts to fast roll, its true DBI dynamics can lead to additional \efolds of inflation even if the potential is no longer flat. This has been termed ``acceleration from D-celeration'' in \cite{Silverstein:2003hf}, and a model in which (most of) the inflationary expansion is obtained in the fast roll regime is called ``DBI inflation'' \cite{Alishahiha:2004eh,Chen:2005ad,Cremades:2005ir}. In general, the DBI dynamics are present in any model where the inflation is an open string mode like an inter-brane distance, but it must be checked whether they in fact do play a r\^ole during inflation: for example, in the model investigated in Chapter \ref{chapter:WMAP3-paper}, the DBI regime is reached more or less at the same time as the end of inflation, therefore the expansion, up to small corrections, occurs entirely in the slow roll regime. In Chapters \ref{chapter:kinf-WMAP5-paper} and \ref{chapter:stochastic-project}, we study models with $k$-inflationary/DBI kinetic terms in detail with respect to their modified perturbation spectra and in the context of ``stochastic inflation'', which was introduced in Section \ref{subsec:eternalinf}. Because they make distinct predictions about non-Gaussianities and the tensor-to-scalar ratio (see Chapter \ref{chapter:kinf-WMAP5-paper}), DBI models are one way to distinguish string from usual field theory inflation.

\subsubsection{Remarks on reheating}
In our introductory remarks on cosmology in Part \ref{part:cosmo}, we emphasized the importance of the reheating era because it bridges the gap from the end of inflation to the onset of the Standard Big Bang Model evolution. In brane--anti-brane inflation, reheating has an intuitive interpretation as the annihilation between brane and anti-brane, which liberates their energy (\ie tension), see Fig.~\ref{fig:closedexchange} \cite{Burgess:2001fx}. Brane annihilation sets in when the system becomes unstable because the brane distance is of order of the string scale: a tachyon appears in the spectrum of a string stretching between the branes, making the instability manifest that leads to decay into closed string modes. The system's energy must then be channeled to the Standard Model degrees of freedom.\\
First, all energy stored in the brane--anti-brane system is transferred into closed string modes produced in the annihilation. These closed strings can penetrate the bulk and travel to the ``corner'' of the CY compactification where the Standard Model particles live (see Fig.~\ref{fig:compactifications}, \eg on another (stack of) D-branes in a second throat.  
Channeling the energy into the SM degrees of freedom can be difficult, and their coupling to the closed string modes not very efficient. 
Multiple throats attached to the CY bulk can have different length and energy scales: it seems that a long throat would be favorable for faster reheating \cite{Barnaby:2004gg,Frey:2005jk}, as it is also from the point of view of building a hierarchy.

\subsubsection{Cosmic strings}
At this point, let us briefly comment on the production of cosmic strings in the process of brane--anti-brane annihilation. A pair of D3-\Dbar that annihilates will produce D1- and $\overline{\mathrm{D}1}$-branes (D-strings) and F1-strings (fundamental strings), and these can stretch to cosmological scales, where they appear as cosmic strings (see \eg the reviews of \cite{Quevedo:2002xw,HenryTye:2006uv,Burgess:2007pz} and \cite{Sakellariadou:2009ev}). Note that in type IIB theory, there are no objects such as pointlike D0s or D2 domain walls, therefore cosmic strings are the only type of defect produced.\\
Once considered an alternative mechanism for seeding structure in the Universe, cosmic strings have lost this battle fair and square to the density perturbations of the inflaton, which correctly predict the observed coherent peak structure in the angular power spectrum. At present, the contribution of cosmic strings to the CMB anisotropies is bounded at the order of a few percent, see \eg \cite{Bouchet:2000hd,Pogosian:2003mz,Pogosian:2008am}. They are nevertheless interesting because they can form scaling networks, and would leave characteristic imprints on the CMB maps, see \eg \cite{Fraisse:2007nu}. One can also infer an upper bound on their tension $\mu$, and their number density, from observations. One advantage of the particular brane--anti-brane inflation setup presented in the following Chapter is that it ``redshifts'' the production rate and tension of cosmic strings \cite{Kachru:2003sx}. It is important to keep in mind, however, that cosmic (super-)strings are interesting as a second way to tell string-inspired inflationary models apart from purely field theoretic ones.

\subsubsection{Secrets of success}
Brane inflation models increasingly take over the field of string cosmology in the recent past. There are several reasons why this is rightfully so: brane inflation is a particularly robust and versatile subclass of string cosmological modes \cite{HenryTye:2006uv}. The attractive potential of Coulomb type between brane and anti-brane is intuitively easy to understand and technically not too difficult to calculate (see the following Chapter). We already mentioned that it has to be put into warped compactifications to be flat enough for inflation, as we show in detail in the next Chapter. Fortunately, this type of compactification is favored also from the point of view of  realistic string theory constructions. An additional mass term $\propto\phi^{2}$ (on top of the Coulombic interaction) in the inflaton's potential, while jeopardizing slow roll inflation, can nevertheless lead to additional \efolds in a DBI regime due to the non-canonical kinetic term of an open string mode inflaton like the brane distance. Hence, enough \efolds of inflation are easily achieved in brane inflation scenarios. The amount of parameter fine-tuning required will also be addressed later.

\section{String Inflation: User's Manual}\label{sec:usersmanual}
We end this Chapter with some guidelines for the handling of string cosmological scenarios. 
Note that, since this subject bridges the gap between two \apriori very different domains of physics, there are different vantage points: from a theoretical point of view, one may be interested in the question whether a model is in the closed or open string class. As a cosmologist, one may choose to rather focus on the regime in which inflation occurs (\ie standard slow roll or DBI, large or small range of field values).

The importance of a given scenario's string origin can be estimated by a comparison of the Hubble scale $H$ during inflation to both the string and compactification scales $\ms$ and $m_\uc$, respectively \cite{Burgess:2007pz}. If $H$ is $\order{\ms}$, the analysis requires the complete (and complex) machinery of string theory, but if the Hubble parameter is in the range $m_\uc<H<\ms$, one may reduce one's efforts to the study of a higher-dimensional field theory in which the physics of extra dimensions still is important. In particular, the full supergravity equations in ten dimensions must be consistently solved. However, if $H$ is below the compactification scale, a strictly four-dimensional description of the inflationary phase suffices because the available energy is too small to make the physics of the extra dimensions ``visible'' (\eg via excitation of their KK modes). The exemplary scenarios presented in this Chapter are in this last category, 
and it is not surprising that it can be difficult to tell them apart from their purely field theoretic cousins.

Ideally, the effective Lagrangian $\calL_{4d}$ of any string-derived scenario were completely fixed by a data set describing the underlying string compactification. This data set should tell us the precise geometry and topology of the compactification manifold as well as the locations of any embedded D-branes and orientifold planes (as well as other sources) \cite{McAllister:2007bg}. One would also like to know how many units of $n$-flux are threading which geometric $n$-cycle in the CY manifold. Up to corrections in $\alpha'$ and $\gs$ (and backreaction effects), this compactification data would encode all free parameters of the four-dimensional theory.\\
In today's string-inspired models, we are far from this idealized picture. Usually, a general stabilization framework is invoked for the moduli (where the precise form of the fluxes is unknown), and (anti-)branes or orientifold planes needed for tadpole cancellation are often ``outsourced'' to some corner of the overall Calabi Yau manifold, where they do not interfere with the dynamics of inflation. However, since the metric on the compact CY is not known, most models are restricted to events inside well defined areas such as the non-compact KS throat, where calculations are under control. 

\bigskip
Corrections beyond the leading supergravity order (\ie in $\alpha'$) as well as in $\gs$ are typically neglected. While not forbidden \emph{per se}, this approach demands special care: any model which seems promising at leading order must be investigated for its stability against higher order corrections. For example, in the famous case of the KKLMMT brane inflation model, whose details are the subject of the next Chapter, quantum corrections lead to significant modifications of the original proposal.

\chapter{Brane Inflation in a Warped Throat}\label{chapter:braneinflation}
\begin{quotation}
\emph{Using the background material of the previous Chapters, we now discuss in detail the so-called ``KKLMMT model'' of brane--anti-brane inflation, retracing each step from the ten-dimensional type IIB supergravity equations to the resulting four-dimensional inflationary cosmology. The comparison of this model's parameters to CMB measurements is the subject of a subsequent article (Chapter \ref{chapter:WMAP3-paper}). We also discuss more recent work on quantum corrections to the original KKLMMT scenario.}
\end{quotation}

An astute combination of the last Chapters' string model building tools was presented in 2003 by {\bf K}achru, {\bf K}allosh, {\bf L}inde, {\bf M}aldacena, {\bf M}cAllister and {\bf T}rivedi \cite{Kachru:2003sx}: in a type IIB flux compactification which stabilizes all complex structure moduli, they assumed that only the K\"ahler modulus $\rho$ corresponding to the total compactification volume remains unfixed. This last modulus is then stabilized by a non-peturbative superpotential as explained in Chapter \ref{chapter:kaehlerstabilized}, still preserving supersymmetry. By introducing an anti-D3-brane at the bottom of the warped geometry, supersymmetry is mildly broken and the minimum of the potential lifted from an anti-de Sitter to a de Sitter one. If a light test D3-brane then is launched far away from the throat's bottom, it feels only a small attractive force towards the anti-D3. The distance between D3 and \Dbar is interpreted as the inflaton field, and the resulting inflationary model, called the ``KKLMMT scenario'', has been abundantly studied in the literature. We now explain the separate steps of the KKLMMT construction.

\section{KKLMMT in a Nutshell}
We begin with a non-technical overview of this scenario: consider a D3- and an anti-D3-brane in a ten-dimensional type IIB supergravity background. Both branes are in the static gauge, \ie they are parallel and the 3+1 dimensions of their worldvolume are aligned with the \((x^{0},\,x^{i}),\,i=1\dots3\) of the background (which are the external, non-compact dimensions). In the six extra, compactified dimensions, the brane and the anti-brane are pointlike and are separated by certain distances \(y^{A},\,A=4\dots9\). We set all but one of these separations (the ``radial'' distance \(y^{4}=r\)) to zero. (The scenario can easily be turned into a multifield inflation model by making more than one of these distances dynamic.) 
It was then assumed in \cite{Kachru:2003sx} that the geometry of the compact spacetime is described by the Klebanov Strassler throat discussed in Section \ref{subsec:deformed}.\\
Into this warped but non-singular background, one then embeds one additional anti-D3 brane: seeking to minimize its energy, the anti-D3 will sink to the bottom of the throat (where it stays fixed at $r=r_{0}$). Since it is an additional source of flux and curvature, its effect on the background is calculated as a small perturbation to the deformed conifold. A test (\ie ``light'') D3-brane (inserted at position \(r_{1}\)) then probes this perturbed background, that is, it does not affect the geometry itself, but experiences forces due to gravity and Ramond-Ramond interaction with the anti-D3. Note that the \Dbar is sitting at the bottom of a KS throat with strong warping and therefore its tension and R-R charge (both of which attract the D3) appear redshifted when seen from a position up in the throat. To calculate the interaction potential between brane and anti-brane, it is in fact easier to consider the former as heavy, and have the anti-brane act as a probe, and we will do so below. This evidently gives the same result for their mutual interaction.\\
The distance \(r=r_{1}-r_{0}\) between the anti-D3-brane and the \((N+1)\)st test D3-brane corresponds (up to normalization) to the inflaton field \(\phi\), and its potential \(V(\phi)\)  can be calculated from the potential \(V(r)\) experienced by the test D3-brane in the limit where \(r\ll \ells\), \(\ells\) being the string length. Inflation therefore takes place while \(\phi\) ``rolls down'' the (extraordinary flat) interaction potential \(V(\phi)\), which corresponds to decreasing radial distance \(r\) between brane and anti-brane. When the branes become too close, \(r\approx l_{s}\), inflation ends due to the appearance of a tachyon, \ie the long distance potential \(V(r)\) ceases to be a good description of the brane interaction.

In this overview, we glossed over two essential features of the KKLMMT construction: following the introduction of fluxes, all of the complex structure moduli have been stabilized, but at least the overall volume modulus still remains unfixed. As mentioned above, one can use the non-perturbative KKLT stabilization procedure for the modulus $\rho$, which was discussed in Chapter \ref{chapter:kaehlerstabilized}. However, this turns out to be non-trivial in the presence of mobile branes. Moreover, one must make sure that the minimum of the potential has a positive non-zero cosmological constant, \ie the AdS minimum after moduli stabilization must be uplifted to a dS minimum. We first derive the naive interaction potential between the branes pretending stabilization of the total volume does not interfere, before turning to the challenges posed by these issues.

\section{Brane Interaction Potential}\label{sec:braneinteractionpot}

\subsection{Klebanov Strassler Throat and Fluxes}
In type IIB superstring theory with three-form fluxes $H_{(3)}$ and $F_{(3)}$ turned on, the complex structure moduli can be stabilized by choosing the (non-compact) Klebanov Strassler solution of Section \ref{subsec:deformed} for the six additional dimensions. At its tip $r_{0}$, the KS throat stays finite because the $S^{3}$ in its basis is stabilized by the $F_{(3)}$ flux threading it. Let us now consider this in somewhat more detail.\\
The definition of the deformed conifold in terms of four complex coordinates $w_{a}=w_{a}^{(1)}+w_{a}^{(2)}$ is given by Eq.~(\ref{eq:deformeddef}). ``Inside'' this definition, two (Poincar\'e dual) three-cycles $\mathscr{A},\mathscr{B}$ are described by
\beq
\mathscr{A}:\quad\sum_{i=1}^{4}\,\left[w^{(1)}_{i}\right]^{2}=\epsilon^{2}\,,\qquad\qquad\mathscr{B}:\quad\left[w^{(1)}_{4}\right]^{2}-\sum_{i=1}^{3}\,\left[w^{(2)}_{i}\right]^{2}=\epsilon^{2}\,.
\eeq
Note that $\mathscr{A}$ is just the $S^{3}$ remaining finite at the tip, and the dual three-cycle $\mathscr{B}$ can be understood as the $S^{2}$ of the basis plus the radial coordinate, see Fig.~11.1. (Since the KS throat is attached to a compact CY manifold in its UV limit, this cycle is also finite.) It is along $\mathscr{A},\mathscr{B}$ that the three-form fluxes are aligned, with
\beq\label{eq:fluxquantization}
\frac{1}{2\pi\alpha'}\int_{\mathscr{A}}F_{(3)}=2\pi M\,,\qquad\qquad\frac{1}{2\pi\alpha'}\int_{\mathscr{B}}H_{(3)}=-2\pi K\,,
\eeq
where $M,K$ are integers, illustrating the generalized Dirac quantization condition for non-localized sources. Moreover, we know from the Bianchi identities of the gauge invariant self-dual five-form flux $\tilde{F}_{(5)}$ that $\dd\tilde{F}_{(5)}=H_{(3)}\wedge F_{(3)}$, therefore the number $N$ of dissolved D3-branes that originally caused the warping is related to $K$ and $M$ by $N=M\cdot K$. 

\begin{figure}[t]
\begin{center}
\begin{minipage}{0.35\textwidth}
\includegraphics[width=\textwidth]{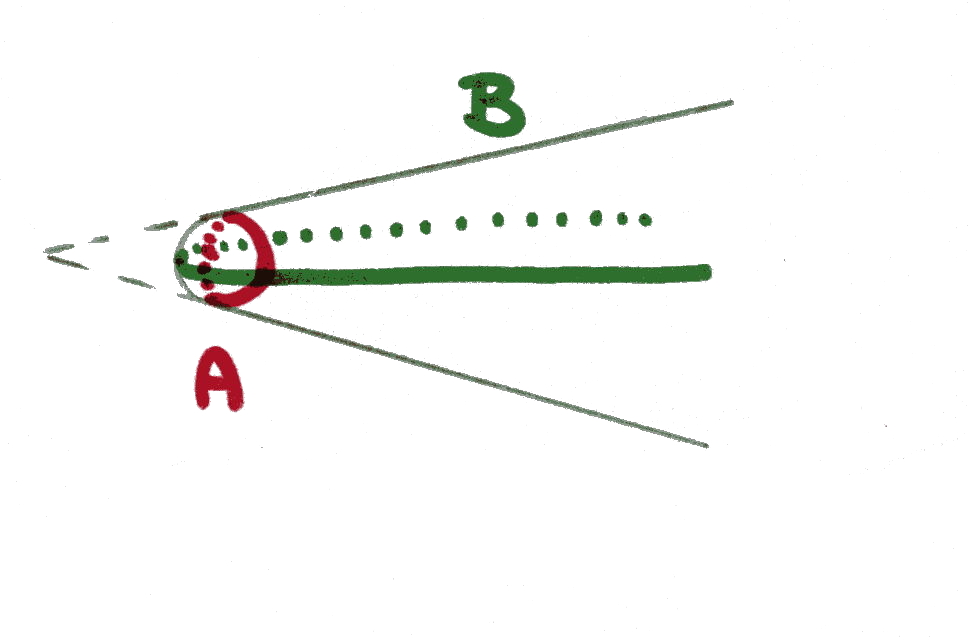}
\caption[Poincar\'e dual cycles on the conifold. D3- and anti-D3-branes embedded into a Klebanov Strassler throat with fluxes.]{\small \emph{Left:} Poincar\'e dual cyles $\mathscr{A}$ and $\mathscr{B}$ in the deformed conifold. The (green) $\mathscr{B}$ is closed once the compactified bulk is included. \emph{Right:} D3-brane and anti-D3-brane embedded in the Klebanov Strassler throat, \ie the deformed conifold. Fluxes deform the tip of the throat such that it is non-singular and ends at a finite value of the radial coordinate $r_{0}$. At its UV end $R$, the throat is joined to the a compact CY bulk manifold. The anti-D3-brane sits fixed at the bottom of the geometry, while the test D3 is launched close to the edge of the throat. Inflation occurs while the D3 moves downwards in the throat, attracted by the anti-D3-brane.}
\end{minipage}
\hfill
\begin{minipage}{0.55\textwidth}
\includegraphics[width=\textwidth]{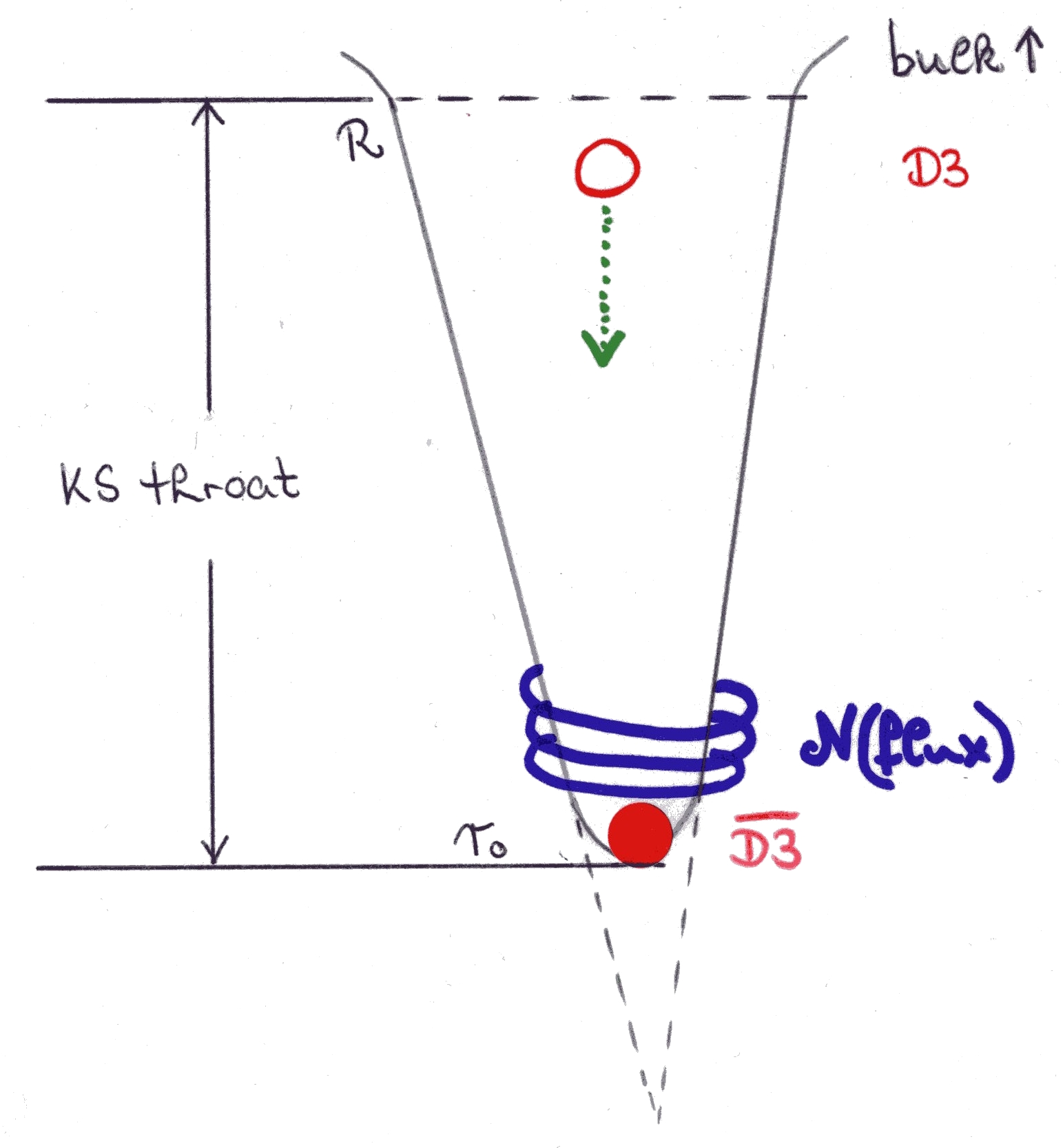}
\end{minipage}
\label{fig:dualcycles}
\end{center}
\end{figure}

In the UV, the KS throat ends at $r=R$ (where it is joined into the compact Calabi Yau bulk manifold). Let us state again the relation between $R$ and the background parameters given in Eq.~(\ref{eq:RonEinstein}),
\beq\label{eq:Ragain}
R^{4}=4\pi\,\gs\,\alphap^{2}\,\frac{N}{v}\,,.
\eeq
where $v=16/27$ for the particular case of the deformed conifold. We now want to find a similar expression for the IR end at $r_{0}$. To this end, recall that the superpotential $W$ on the deformed conifold has the form [compare Eq.~(\ref{eq:potW})]
\beq\label{eq:Wincycles}
W=\int_{\mathscr{M}_{6}}G_{(3)}\wedge\Omega=(2\pi)^{2}\,\alphap\left(M\int_{\mathscr{B}}\Omega-K\,\tau\int_{\mathscr{A}}\Omega\right)\,,
\eeq
where in the second equality, it has been used that $G_{(3)}=H_{(3)}-\tau\,F_{(3)}$ and that $\mathscr{A}$ and $\mathscr{B}$ are Poincar\'e dual on the internal manifold. In addition, the integrals over $H_{(3)}$ and $F_{(3)}$ have been performed, using Eq.~(\ref{eq:fluxquantization}). 
If we now define a complex coordinate $z$ from the period of the $\mathscr{A}$-cycle
\beq\label{eq:zA}
z=\int_{\mathscr{A}}\Omega\,,
\eeq
it can be shown using so-called special geometry (see \eg \cite{Becker:2007zj} for the application used here), that for the period on the $\mathscr{B}$-cycle it holds that
\beq
\int_{\mathscr{B}}\Omega=\frac{z}{2\pi\,i}\,\log z+\mathrm{holom.}\,,
\eeq
where the second piece is a holomorphic function on the internal manifold. Then, the K\"ahler covariant derivative of the superpotential (\ref{eq:Wincycles}) with respect to the coordinate $z$ defined by Eq.~(\ref{eq:zA}), in the limit where $z\ll1$ is
\beq
D_{z}W\simeq(2\pi)^{2}\,\alphap\left(\frac{M}{2\pi\,i}\,\log z-i\,\frac{K}{\gs}+\dots\right)\,.
\eeq
Setting $D_{z}W\equiv0$ then gives an expression for the period $z$ of $\mathscr{A}$ in terms of the flux quantum numbers,
\beq
z\simeq e^{-2\pi\,K/M\gs}\,.
\eeq
On the other hand, we also know that the deformed conifold is described by the equation (\ref{eq:deformeddef}) for the complex cooridinates $w_{a}$ with the deformation parameter $\epsilon^{2}$.
Since at the tip, the $S^{3}$ (here written as the Poincar\'e cycle $\mathscr{A}$) stays finite, we conclude that $\epsilon^{2}=z$. What does this tell us about the IR cutoff $r_{0}$? In Section \ref{subsec:deformed}, we showed that $r^{3}\propto \epsilon^{2}$ (strictly speaking, this is valid in the limit where $r$ is large enough such that the deformation is not felt yet). Then, we can conclude that $r\propto z^{1/3}$, and in exact terms we obtain for the ratio between $R$ and $r_{0}$ in terms of the fluxes that
\beq\label{eq:r0}
\frac{r_{0}}{R}=\exp\left(-\frac{2\pi K}{3\gs M}\right)\,.
\eeq
Note that this can be interpreted as the (simple) warp factor $h^{-1/4}$ evaluated at $r=r_{0}$ because $h(r)\propto R^{4}/r^{4}$. We therefore now have an expression of both $R$ and $r_{0}$ in terms of background parameters only (recall that $N=K\,M$). This is the underlying geometry of the KKLMMT model before any additional (anti-)branes (apart from $N$ branes and $M$ fractional branes ``dissolved'' into the background) are introduced.

\subsection{Additional D3-Brane}
We saw earlier that adding (unwrapped) branes to the supergravity background requires two additional terms per D$p$-brane in the action, namely the DBI part \cite{Polchinski:1998rr},
\begin{equation}\label{eq:DBIagain}
\action_{{\rm D}p}=-T_{p}\int_{\scrM_{p}}{\rm d}^{p+1}\xi\,\left[-\det\left(G_{ab}+B_{ab}+2\pi\alpha'F_{ab}\right)\right]^{1/2}\,,
\end{equation}
plus the Chern-Simons coupling to the $C_{(p+1)}$ gauge potential,
\begin{equation}\label{eq:ChernSimonsagain}
\action_{\rm CS}=\pm \mu_{p}\int_{\scrM_{p}}{\rm d}^{p+1}\xi\,C_{(p+1)}\,,
\end{equation}
where $\scrM_{p}$ is the brane's $(p+1)$-dimensional world volume, and $T_{p}$ and $\mu_{p}$ the tension and charge, respectively. [The upper sign in Eq.~(\ref{eq:ChernSimonsagain}) is for branes, the lower for anti-branes.] \(G_{ab}\) is the induced metric on the brane, and \(B_{ab}\) the pulled-back Kalb Ramond anti-symmetric tensor field. We set the gauge field $F_{ab}$ on the brane to zero. Since we work in type IIB string theory, the number of spatial brane dimensions $p$ is odd, and we concentrate on the case $p=3$. Then, the brane world volume is four-dimensional, and we align these dimensions with the external part $\scrM_{4}$ of spacetime. Replacing the indices $(a,b)\rightarrow(\alpha,\beta)$, the induced metric and field $B_{(2)}$ are given by
\beq
G_{\alpha\beta}=g_{mn}\frac{\partial x^{m}}{\partial \xi^{\alpha}}\frac{\partial x^{n}}{\partial \xi^{\beta}}\,,\qquad B_{\alpha\beta}=B_{mn}\frac{\partial x^{m}}{\partial \xi^{\alpha}}\frac{\partial x^{n}}{\partial \xi^{\beta}}\,.\label{eq:inducedfields}
\eeq
It was shown earlier [see Eq.~(\ref{eq:B2ansatz})] that the $B_{(2)}$ field only has components along the internal dimensions of spacetime, therefore there is no pullback onto the brane, $B_{\alpha\beta}=0$. To calculate the induced metric, let us make the assumption that the radial coordinate \(y_{1}^{4}=r_{1}\) of the brane depends on time \(x^{0}=t\) (but not on the \(x^{i}\)). (As a motivation, recall that \(r\) will eventually describe the distance between the D3 and anti-D3 brane, which is the time-dependent inflaton.) Then we have from Eq.~(\ref{eq:inducedfields}) that
\beq
G_{00}=g_{00}+g_{44}\left(\frac{\partial r_{1}}{\partial t}\right)^{2}\,,\qquad G_{ii}=g_{ii},\qquad i=1,2,3.
\eeq
With $g_{mn}$ the ten-dimensional warped metric with the KS throat as its six-dimensional piece, the determinant of this induced metric is
\begin{equation}\label{eq:detG}
-{\rm det}\,G_{\alpha\beta}=\frac{1}{h^{2}(r_{1})}\left[1-h(r_{1})\left(\frac{\partial r_{1}}{\partial t}\right)^{2}\right].
\end{equation}
The brane hence is dynamic and can move along the $r$ direction. Using Eq.~(\ref{eq:DBIagain}) and Eq.~(\ref{eq:ChernSimonsagain}) with the upper sign, we obtain for its Lagrangian
\beq\label{eq:D3Lagrangian}
\calL_{\mathrm{D}3}=-T_{3}\,h^{-1}(r_{1})\left[1-h(r_{1})\left(\frac{\partial r_{1}}{\partial t}\right)^{2}\right]^{1/2}+\mu_{3}\,h^{-1}(r_{1})\approx\frac{T_{3}}{2}\left(\frac{\partial r_{1}}{\partial t}\right)^{2}\,,
\eeq
where we have used the expansion of the square root for small $\partial r_{1}/\partial t$, and the fact that $T_{3}=\mu_{3}$ in our units. We see that up to a renormalization factor of $\sqrt{T_{3}}$, the radial brane coodinate $r_{1}$ behaves like a free scalar field without potential because the gravity and electrostatic (via the R-R charge) interactions precisely cancel. In a warped background, a D3-brane will therefore stay where it is put (in our case, at $r_{1}$), or can be moved around without energy cost, which means that its world volume scalars (here $r_{1}$, but potentially also the angular coordinates on the internal manifold) are massless moduli fields.

Note, however, that if the D3-brane is ``heavy'', it backreacts on the geometry and will manifest itself as a small perturbation to the function \(h(r)\) of the KS throat. We describe this using the ansatz $\tilde{h}=h+\delta h$, where $\delta h$ is the perturbation caused by the localized source. From the Einstein equations, we then obtain a differential equation for $\delta h$,
\begin{equation}
\left(\frac{\partial^{2}}{\partial r^{2}}\,\delta h\right)+\frac{5}{r}\left(\frac{\partial}{\partial r}\,\delta h\right)=C\,\delta(r-r_{1})\,,
\end{equation}
where $C$ is a constant. This is precisely the same differential equation as was found for $h(r)$ previously, and therefore has an analogous (localized) solution, $\delta h\propto1/r_{1}^{4}$. The full solution for the perturbed warp factor $\tilde{h}$ can be made plausible from physical intuition: recall that the unperturbed $h(r)$ was the consequence of a stack of $N$ D3-branes put into the background geometry, creating the throat. Now, we have added one more D3-brane at a fixed position $r_{1}$. Its effect should therefore be suppressed by a factor $\propto 1/N$. In terms of the notation in Eq.~(\ref{eq:hAdSsolution}), we may write
\begin{equation}\label{eq:fullh}
\tilde{h}(r)\approx\frac{C_{2}}{r^{4}}\left(1+\frac{1}{N}\,\frac{r^{4}}{r_{1}^{4}}\right)\,.
\end{equation}
The constant $C_{1}$ in Eq.~(\ref{eq:hAdSsolution}) responsible for the plateau has again be set to zero, and we already know that $C_{2}=R^{4}$. In this way, the warp factor is equal to 1 at the edge of the throat $R$ if the perturbation by the additional brane is neglected.

\subsection{Launching a Test Anti-Brane}
We are now ready to launch a test (\ie ``light'') anti-D3-brane. Consider the case of the unperturbed KS geometry first. Let us call the position where the anti-D3 is inserted \(y_{2}^{4}=\tilde{r}\) (for all other extra coordinates we take \(y_{2}^{A}=y_{1}^{A},\,A=5\dots9\), \ie the brane and anti-brane are coincident in these dimensions). As for the D3-brane above, $r_{0}$ depends on time. Then, since the anti-brane has the same tension but opposite charge, its Lagrangian is [compare Eq.~(\ref{eq:D3Lagrangian})]
\begin{equation}\label{eq:antiD3Lagrangian}
\calL_{\overline{\mathrm{D}3}}=-T_{3}\,h^{-1}(\tilde{r})\left[1-h(\tilde{r})\left(\frac{\partial \tilde{r}}{\partial t}\right)^{2}\right]^{1/2}-\mu_{3}\,h^{-1}(\tilde{r})\approx\frac{T_{3}}{2}\left(\frac{\partial \tilde{r}}{\partial t}\right)^{2}-2\,\frac{T_{3}}{h(\tilde{r})}\,,
\end{equation}
where we have performed the same expansion as before. For the anti-D3, the gravitational and R-R interactions add up, therefore there is a potential for the dynamical coordinate $\tilde{r}$, and the \Dbar will seek to minimize its energy. If we call $T(\tilde{r})=T_{3}\,h^{-1}(\tilde{r})$ the effective brane tension at the anti-branes position $\tilde{r}$, we see that for a KS warp factor $h=R^{4}/\tilde{r}^{4}$, the function $T(\tilde{r})$ is minimized at $\tilde{r}=r_{0}$. Therefore, unlike the D3 brane which stays fixed, the \Dbar will sink to the bottom of the throat under the influence of its DBI action and Cherm Simons term.

Now consider a background with both D3 and \Dbar embedded (see Fig.~11.1), and let us switch their r\^oles: the anti-D3, sitting at its minimal energy position $r_{0}$, is considered heavy and perturbs the background warp factor as in Eq.~(\ref{eq:fullh}). Because the \Dbar is fixed, we measure the position of the D3-brane relative to its location, \ie use the distance $r=r_{1}-r_{0}$ as the dynamic variable. The D3-brane is light and, when launched in the perturbed background, now feels the presence of the anti-brane attracting it. The potential for the combined D3-\Dbar system in the warped KS background can be written as
\begin{equation}\label{eq:Vofr}
V(r_{0},r)=\frac{2T_{3}}{\tilde{h}(r_{0},r)}=2T_{3}\,\frac{1}{\frac{R^{4}}{r_{0}^{4}}\left(1+\frac{1}{N}\,\frac{r^{4}_{0}}{r^{4}}\right)}\approx 2T_{3}\,\frac{r_{0}^{4}}{R^{4}}\left(1-\frac{1}{N}\,\frac{r^{4}_{0}}{r^{4}}\right),
\end{equation}
where we have used the expansion for $r\gg r_{0}$ in the last term. Using the canonic renormalization \(\phi=\sqrt{T_{3}}\,r_{1}\) [suggested by the kinetic term in Eq.~(\ref{eq:D3Lagrangian})], the brane--anti-brane system has the Lagrangian of a canonical scalar field,
\beq\label{eq:branedistanceL}
\calL_{\phi}=\frac{1}{2}\left(\frac{\partial\phi}{\partial t}\right)^{2}-V(\phi)\,,\qquad V(\phi)=\frac{2\,T_{3}\,r_{0}^{4}}{R^{4}}\left(1-\frac{r^{4}_{0}\,T_{3}^{2}}{N}\,\frac{1}{\phi^{4}}\right)\,.
\eeq
For large values of $\phi$, this potential is extremely flat and therefore makes $\phi$ a good candidate for the inflaton. Note that, because of the warping, $V(\phi)$ is flat \emph{inside} the throat (where $r$ can at most take the value $R$, \ie the UV end). A common problem in earlier models of brane--anti-brane inflation was that the brane interaction potential was only flat enough to inflate for brane distances exceeding the size of the compact manifold \cite{Kachru:2003sx}. In the KS throat, the situation is different: because it sits at the IR end of the geometry, the tension and charge and hence the attractive powers of the anti-D3 are only weakly felt up in the throat where the test D3 is launched.\\
The derivation of the above potential relied on the exchange of massless closed string modes only (gravitons and R-R modes), which means that once the branes get close and other (massive) interaction modes are no longer Yukawa suppressed, it becomes invalid. Moreover, there is a critical brane distance (the string length $\ells$) at which string with one end on each brane can appear. Its spectrum contains a tachyon, which signals that the brane--anti-branes system becomes unstable and mutual annihilation sets in. In summary, the Lagrangian (\ref{eq:branedistanceL}) may be used as an effective description while \emph{i)} the mobile D3-brane is inside the throat (since we do not know the metric of the bulk CY), $r<R$, and while \emph{ii)} the brane distance exceeds the string length, $r>\ells$.

\section{Issues of Volume Stabilization}
The simple Coulomb-like brane interaction that appears in the Lagrangian (\ref{eq:branedistanceL}) is not the full story, and we now revisit the assumptions that went into its derivation. After fluxes have been turned on, all the complex structure moduli are stabilized. Moreover, in concentrating on the brane motion, we implicitly assumed that all K\"ahler moduli including the overall volume were also fixed. (Otherwise, the volume modulus would have a  runaway potential, making it the direction of steepest descent in moduli space, and the flatness of the brane world volume modulus $r$ could not be exploited for inflation.) We saw in Chapter \ref{chapter:kaehlerstabilized} that the (complex) K\"ahler modulus $\rho$ can be stabilized by non-perturbative effects of the superpotential. But does the KKLT procedure carry through to a background that includes a mobile D3-brane? Note that the other ingredient of \cite{Kachru:2003aw}, \ie uplifting of the AdS minimum to a dS one by adding a small number of anti-branes is naturally incorporated in the above setup by the \Dbar sitting at the end of the throat.\\
We now adress the question whether the total compactification volume in the KKLMMT is indeed stabilized by the non-perturbative approach used to fix $\rho$ in \cite{Kachru:2003aw}. Recall that in Chapter \ref{chapter:kaehlerstabilized} the real part of $\rho$ was an axion (descending from the four-form potential), while its imaginary part $\sigma$ was the total compactification volume, and the goal was to stabilize $\sigma$. In the presence of branes, the definition of the compactification volume is more complicated.

\subsection{K\"ahler Potential}
After flux compactification, the modulus $\rho$ in a background without D-branes has the K\"ahler potential $K(\rho, \bar{\rho})=-3\log[-i(\rho-\bar{\rho})]$, compare Eq.~(\ref{eq:kaehlerrho}). However, when there are mobile D-branes embedded, the combined K\"ahler potential for $\rho$ and the world volume scalars of the D-branes (which we summarily denote by $\phi$, since in the KKLMMT scenario one of them is the inflaton) is\footnote{Note that there is a change in notation $\rho\rightarrow -i\rho$ between the papers \cite{Kachru:2003aw} and \cite{Kachru:2003sx}, which is why Eq.~(\ref{eq:KwithDbranes}) in the latter publication reads $K(\rho,\bar{\rho},\phi,\bar{\phi})=-3\log[\rho+\bar{\rho}-k(\phi,\bar{\phi})]$. In particular, this exchanges real and imaginary parts of $\rho$.}
\beq\label{eq:KwithDbranes}
K(\rho,\bar{\rho},\phi,\bar{\phi})=-3\log\left[-i(\rho-\bar{\rho})-k(\phi,\bar{\phi})\right]\,.
\eeq
Note that the K\"ahler potential $k(\phi,\bar{\phi})$ for the D3 world volume scalars should just be the K\"ahler potential of the Calabi Yau compactification, since the D3 are transverse to the the internal manifold, and the compact CY dimensions therefore represents their possible locations. 
The real part of $\rho$ is the axion arising from a four-form on the internal manifold. But we know that D3 branes couple to four-form potentials because their world volume is four-dimensional -- therefore the remaining K\"ahler moduli space after compactification [whose K\"ahler potential is (\ref{eq:KwithDbranes})] is \emph{not} a direct product between those of $\rho$ and $\phi$. Because of the non-trivial relation between the axion $\Re(\rho)$ and the brane moduli $\phi$, the good complex variable (in which the metric on moduli space takes a K\"ahler form) is no longer $\rho$, but a new complex quantity $\rho'$, where $\Re(\rho)=\Re(\rho')$ (\ie still the axion) and $\Im(\rho')$ is related to the proper volume modulus $r$ by
\beq\label{eq:define-r}
2r=-(\rho-\bar{\rho})-k(\phi,\bar{\phi})\,.
\eeq
It is the size of $r^{2}$ relative to the string scale $\alphap$ which controls the viability of the supergravity expansion for a background including D-branes. Give that the KKLT stabilization procedure addressed $\rho$, we have to consider what the fixation of $\rho$ means in terms of $r$.

\subsection{Superpotential}
From the flux compactification without D-branes, we found a constant superpotential $W\propto \int_{\scrM_{6}}G_{(3)}\wedge\Omega$, independent of $\rho$. For supersymmetry to be preserved, \ie for $D_{i}W=0$, the imaginary self-dual three-form flux $G_{(3)}$ must only contain a (2,1)-piece, which gives $W\equiv0$. If there is a contribution of the (3,0)-form, it will produce a non-zero constant, and we therefore write $W=W_{0}$ in general. Note that with this superpotential one still obtains the no scale cancellation in the scalar potential $V$, which leads to $V=0$, and $\rho$ is still massless at this level.

In \cite{Kachru:2003aw} it was suggested to fix $\rho$ by a non-perturbative correction to the constant flux superpotential as in Eq.~(\ref{eq:nonpert-correction}).  In the KKLMMT scenario, we are in a flux background that includes D-branes, 
and we want the $\phi$ (which represent the position of the D3s on the CY manifold, \ie in the above case of one D3-brane, there should be a triplet of complex $\phi$ fields, one of which is the radial coordinate we studied earlier) to move freely, but the volume stabilized. Note that fixing the six-dimensional volume modulus is obligatory if we want to build an inflationary scenario from the above Lagrangian in Eq.~(\ref{eq:branedistanceL}): for inflation, we must have a four-dimensional Einstein Hilbert action in which the Planck mass is constant. Very often, this is achieved by truncating the KS throat at its UV end $R$, because once the throat joins the bulk, calculational power is lost because the metric on the compact CY is unknown. (In Randall Sundrum models, the same effect is obtained from placing the second, ``Planck'' brane at a large distance.)

\section{The $\eta$ Problem}
To summarize, the true field which controls the size of the manifold and the viability of the $\alphap$ expansion is $r$ defined in Eq.~(\ref{eq:define-r}). What are the consequences for the potential of the D-brane world volume scalars $\phi$? The difficulty is that for any positive energy configuration (which we need for a de Sitter minimum) using known string theory sources (such as anti-D3-branes and fluxes), the scalar potential will fall off as a negative power of $r$, \ie schematically we can write \cite{Kachru:2003sx}
\beq
V(r,\rho)=\frac{X(\rho)}{r^{\alpha}}=\frac{2X(\rho)}{\left[-i\,(\rho-\bar{\rho})-\phi\bar{\phi}\right]^{\alpha}}\,.
\eeq
The form of the numerator depends on the chosen source of the energy. But the non-perturbative KKLT stabilization procedure is directed towards $\rho$ rather than $r$. Therefore, if the D3-brane with radial world volume modulus $\phi$ moves, the potential changes to
\beq
V=V_{0}\left(1+\alpha\frac{\phi\bar{\phi}}{2r}+\dots\right)\,.
\eeq
This generically yields a mass contribution of $\order{1}$ to the inflaton potential -- unless one can arrange for a cancellation. One option for this is to set $X(\rho,\phi)$ (\ie the superpotential $W$ must be a function not only of $\rho$, but also of $\phi$). Even if they are non-generic, there could then exist superpotentials for which the combined inflaton mass term vanishes. Note that this is not unlike the $\eta$ problem we encountered earlier in the context of inflationary models derived from supergravity (see Section \ref{subsec:susylagrangians}): while generically the second potential slow roll parameter $\eta_{V}$ is large, inhibiting prolonged slow roll inflation, there exist special combinations of K\"ahler and superpotentials for which $\eta_{V}$ is small due to cancellations.

Because this is a subtle effect, let us rephrase it again: while $W=W_{0}$ is a constant, a D3-brane put in the flux background does not feel a force, and we used this explicitly when calculating the brane--anti-brane interaction potential in Section \ref{sec:braneinteractionpot}. However, once the D3 starts to move (in the KKLMMT scenario, due to its attraction versus the anti-brane), the balancing mechanisms try to adjusts the force cancellation at the brane's new position \cite{Burgess:2007pz}. As long as the K\"ahler modulus $\rho$ is massless (\ie a modulus in the proper sense), it can be varied to this effect at no energy cost. But at present, $\rho$ has been fixed by a non-perturbative superpotential. Therefore, if $\rho$ adjusts to the new position of the D-brane, there will be an energy penalty imposed. (One possible result is that the brane might be localized at a specific position in the throat.)\\ 
There is yet another way to understand this effect. Once we cut off the throat [and therefore gain the right to add a four-dimensional Einstein Hilbert action to the Lagrangian (\ref{eq:branedistanceL})], there is in principle also a conformal coupling term for the brane position $r$, written as $(T_{3}/12)r^{2}R$ because it is a conformally coupled scalar \cite{Seiberg:1999xz}. Evidently, this gives a contribution to the inflaton mass, which is just the effect calculated above in the supergravity setup by considering the shift form $\rho$ to $r$ as the proper volume modulus.

This manifestation of the $\eta$ problem, \ie a large inflaton mass hindering slow roll inflation on the flat warped brane interaction potential, was already studied in the original paper \cite{Kachru:2003aw}. The main lesson learned was that inflationary model building in string theory cannot be considered as a separate issue form moduli stabilization (as it was often done in previous string cosmological scenarios). However, KKLMMT argued that a modest amount of fine-tuning [$\order{10^{-2}}$] could suffice to arrange for a cancellation of the dangerous mass term, given that there is a vast choice of flux compactifications. Moreover, the $\phi$-dependence of the superpotential was largely unknown, which inspired the hope that it could be just of the right form (\ie lead to a term quadratic in $\phi$ in the four-dimensional scalar potential) to cancel the contribution calculated above. Then, the only remaining term in the potential would be the very weak Coulomb attraction between the branes, and slow roll inflation could proceed successfully. In more recent work, the functional form of these superpotential corrections was obtained (see below), and unfortunately, this miracle does not occur. The Coulomb term is, however, unique among the contributions to the potential in the sense that its derivation is straightforwardly possible in the type IIB supergravity background studied above, and that its parameters are related in a transparent way to the basic string geometry of the KS throat, see Eqs.~(\ref{eq:Ragain}) and (\ref{eq:r0}).

\section{Reheating}
The above potential between the D3-brane and the anti-D3-brane was calculated in the limit of large distance: the anti-D3 sits fixed at the bottom of the throat, while the D3 probe is launched closer to the edge of the throat and starts moving downwards. 
However, the branes come closer and closer while inflation is under way, and eventually their distance will be of the order of the string scale $\ells$. It is at this point at the latest that our simplified calculation for the potential becomes invalid: once the branes are this close, a tachyon develops describing the lowest oscillatory mode of a string between the branes. This tachyon (and other degrees of freedom becoming massless) was is not taken into account by the Coulomb potential.

In the cosmological picture, the appearance of the tachyon triggers the phase of reheating, during which the brane and the anti-brane annihilate. In this process, the energy contained in the tensions is freed up and can be used for heating the Standard Model degrees of freedom, so that the SBBM evolution can set in. However, the reheating efficiency depends largely on where the SM particles are to be found: the D3 and \Dbar we considered so far (whose distance gave the inflaton $\phi$) cannot be used to harbor the Standard Model because they disappear once their annihilation is complete. However, there can be an additional D3-brane (or a stack of them) sitting in the same throat, which remains once the original D3--anti-D3 brane have decayed. However, the Standard Model brane could also live in a different corner of the overall Calabi Yau compactification.\\
The closed string loops produced by brane annihilation propagate to the Standard Model brane, where they couple to the gauge degrees of freedom living on it. Whether the process of reheating can be efficient depends largely on the nature of these couplings, and there is the danger that the brane tension energy could be channeled more efficiently into an unobservable sector, or into bulk degrees of freedom. It seems that strong warping can help to make reheating efficient, see Section \ref{sec:brane-antibrane} and the references given there.

\section{Quantum Corrections}
We saw that by embedding brane--anti-brane inflation into a warped compactification of type IIB string theory, one can at first sight make the brane interaction potential suitably flat for inflation. However, as was already discussed in the original publication \cite{Kachru:2003sx}, it would be inconsistent to regard brane motion on the KS geometric background while ignoring moduli stabilization (in particular the volume). These issues really are closely intertwined: while it is true that the fluxes of the warped background stabilize the complex structure moduli, the non-perturbative volume stabilization mechanism previously introduced in \cite{Kachru:2003aw} is compatible with inflation only under additional assumptions. If the superpotential $W$ depends on $\rho$ only, as was assumed in \cite{Kachru:2003aw}, it will generate mass terms in the inflaton potential because the presence of D3-branes changes the notion of the compactification volume: $\rho$ is no longer the good modulus to consider, but must be replaced by $r$.\\
Given that the non-perturbative stabilization procedure was directed towards $\rho$, the true volume modulus $r$ cannot be expected to stay fixed. It can be shown that the induced mass for the inflaton is generically of the same size as the Hubble parameter $H$, and therefore too large to support prolonged slow roll inflation. In \cite{Kachru:2003sx} it was nevertheless argued that the (largely unknown) dependence of the superpotential $W$ on $\phi$ (on top of its non-perturbative $\rho$ dependence) could allow to cancel this mass term. While such a cancellation would not be generic, modest fine-tuning might suffice to make it happen. Therefore, KKLMMT argued that inflationary scenarios dominated by the Coulomb-like attraction term between the branes can be feasible.

Since the original work of \cite{Kachru:2003sx}, the superpotential and its dependence on $\phi$ have been much better understood. The strategy employed is to cast all forces in the form of low energy supergravity. In the non-peturbative ansatz, this amounts to making $A$ a function of $\phi$ \cite{Baumann:2006th,Baumann:2007np,Baumann:2007ah,Krause:2007jk}. It was shown that the additional forces on the D3 can lead to a balance of forces for a range of field values towards the end of the throat, where inflation may then occur \cite{Pajer:2008uy}. The potential is of the ``inflection point'' type, and the flat stretch can be engineered to be flat enough to produce a sufficient number of \efolds. Unlike the hypothesis made in \cite{Kachru:2003sx}, the correction terms do not have the same functional form as the mass term generated by the conformal coupling. (Instead of being $\propto \phi^{2}$, they are rather proportional to $\phi^{3/2}$ \cite{Baumann:2007np,Baumann:2007ah}.) Therefore a cancellation can only occur for a limited range of $\phi$ values even when parameters are finetuned. 

\bigskip
The Coulomb term we calculated above is unique in the sense that it is always present, and in a simple and transparent way related to the background geometry. The other corrective terms are much more dependent on finer details such as wrapped D7-branes descending into the throat along four-cycles \etc In the spirit of Section \ref{sec:usersmanual}, one might say that it makes sense to use a less ambitious construction (\ie the ``pure'' KKLMMT  Coulomb term only) because complete compactification data (\ie all non-perturbative corrections) is not available in any case. Put a different way, we choose ``the devil we know'' over the unknown corrections. Note that there is also a cosmological argument to be made for restricting the potential in this way, which will be illustrated by the article presented in the next Chapter: the more parameters there are in a string cosmological model, the lesser the hope that the link to CMB observations may tell us something about the underlying geometry.

\newpage
\thispagestyle{empty}
\mbox{}
\newpage
\part{Results and Publications}\label{part:results}

\newpage
\thispagestyle{empty}
\mbox{}
\chapter{Contraints on Brane Inflation from WMAP3}\label{chapter:WMAP3-paper}
\begin{quotation}
\emph{The objective of the first article published during this thesis is a clear-cut comparison of the KKLMMT brane--anti-brane inflation model we studied in Chapter \ref{chapter:braneinflation} to the (at the time) most recent measurement of the cosmic microwave background temperature fluctuations, provided by the three-year data release of the WMAP satellite. The background of the KKLMMT model in type IIB string theory was studied thoroughly in this paper, and the string theoretic meaning of the ``visible'' cosmological parameters established. In particular, the string coupling $\gs$ and the string scale $\alphap$ were not fixed \apriori, but equally considered as free parameters to explore in the Monte Carlo Markov Chain analysis. Consistency relations derived from the underlying string model were imposed as priors on the parameters where applicable. A detailed description of the numerical methods as well as a careful interpretation of the obtained probability distributions is presented.}
\end{quotation}

\section{``Brane inflation and the WMAP data: a Bayesian analysis'' (article)}

\newpage
\thispagestyle{empty}
\mbox{}
\newpage


\chapter{Kinetically Modified Inflation and the WMAP5 Data}\label{chapter:kinf-WMAP5-paper}
\begin{quotation}
\emph{In the previous article on the original KKLMMT scenario, the inflaton potential was given by a pure Coulomb term due to the attraction between an anti-D3-brane (fixed at a ``redshifted'' position at the bottom of the Klebanov Strassler throat) and a mobile D3-brane. This potential is extremely flat and inflation takes place in the slow roll regime. However, additional terms generically appear in the potentials for string inflaton fields, which can render these potentials too steep for conventional slow roll. Accelerated expansion may still be possible because the inflaton has a non-canonical kinetic term.\\
Inflationary scenarios with modified dynamics are commonly called ``$k$-inflation'', and in this article we derive their scalar and tensor power spectra using the uniform approximation. The non-standard dynamics for the homogeneous background field lead to a non-trivial sound speed for the Fourier modes of the field perturbations. The standard solution to the perturbation equations in terms of Hankel functions can no longer be used, but if the $k$-inflationary slow roll conditions are satisfied (\ie when both the sound speed and the Hubble radius change only slowly), the equations can be solved with the so-called uniform approximation. In string theory, the kinetic term of the inflaton in brane inflation is of the Dirac Born Infeld (DBI) type, which is a special subclass of $k$-inflation.\\
In a second step, the resulting spectra are compared to the five-year release of the WMAP satellite. It is found that in $k$-inflation the notion of the parameters constrained by the data changes, and that therefore \eg there is no longer an upper limit on the first Hubble flow parameter $\epsilon_{1}$, as it is usually the case for the standard spectra. Instead, only the combination $\epsilon_{1}/\gamma$ is constrained. However, when restricted to the DBI subclass, this limit is recovered because in DBI scenarios the form of non-Gaussianities (an additional observable) is known and can be used to break the degeneracy between the sound speed and $\epsilon_{1}$.}
\end{quotation}

\section{``$k$-inflationary power spectra in the uniform approximation'' (article)}

\section{``Constraints on kinetically modified inflation from WMAP5'' (article)}

\newpage
\thispagestyle{empty}
\mbox{}
\newpage

\chapter{Tachyon Entropy Perturbations at the End of Brane Inflation}\label{chapter:tachyon-paper}
\begin{quotation}
\emph{A phase of brane--anti-brane inflation typically ends with the mutual annihilation of the branes, which corresponds to reheating of the Universe in the cosmological picture. The interest of the following article is to study the dynamics in the very early stages of reheating in more detail: a tachyon appears when the branes are sufficiently close to each other to start the annihilation process, and for a short period of time, there are two dynamical fields. Together, they can create perturbations of entropy type, and it is investigated under which conditions these can grow exponentially, resulting in an accumulated contribution to the comoving curvature (which, in one field models, is due to adiabatic perturbations only). It is shown that, in the absence of backreaction, there exist parameter values for which the part of the comoving curvature perturbation induced by entropy fluctuations is of the same size as the adiabatic contribution.}
\end{quotation}

\section{``Entropy Fluctuations in Brane Inflation Models'' (article)}

\newpage
\thispagestyle{empty}
\mbox{}
\newpage

\chapter{Brane Monodromy Inflation and Reheating}\label{chapter:monodromy-paper}
\begin{quotation}
\emph{In the previous articles, the underlying string picture behind the inflationary scenario under scrutiny was a mobile D3-brane in a type IIB superstring theory background enriched with fluxes. At present, we turn our attention to a model of brane inflation constructed in the dual type IIA background: a D4-brane is wrapped along a ``monodromic'' direction on the compactification manifold (which consists of two twisted tori). Initially, the world volume energy of the D4-brane is not minimized, and it will hence seek to reach its minimum by unwinding in the direction of the monodromy. This scenario is one of the rare realizations of large field inflation found in string theory.\\
The phenomenological picture of reheating is also significantly different in monodromy inflation: in type IIA theory, the Standard Model (SM) of particle physics can, for example, live on a D6-brane, to which the D4 can transfer energy through collisions. In this article we study the consequences of such a SM D6-brane localized at a fixed position in the monodromic coordinate (along which the D4 is unwinding). The D4 will therefore hit the D6-brane repeatedly while it travels towards its world volume energy minimum. At each collision, the branes align, reaching a state of enhanced symmetry. We use a simple Lagrangian to model the interaction between the inflaton and the SM particles,  and describe their coupling by strings between the branes that are created and then stretched at each brane encounter. It is, however, found that these strings are diluted to negligible density by the inflationary expansion, therefore no energy is transferred towards the SM when the branes meet during inflation. Reheating takes place instantaneously at the last brane collision after the D4 has completely unwound.}
\end{quotation}

\section{``Reheating in a brane monodromy inflation model'' (article)}

\newpage
\thispagestyle{empty}
\mbox{}
\newpage



\part{Conclusions}\label{part:conclusions}
\chapter{Conclusions}
\begin{quotation}
\emph{At the end of this thesis we look back to our starting point, the recent efforts to incorporate the cosmological scenario of inflation into string theory. The progress of the so-called brane inflation models is most striking, and several of them are able to match the most recent data with their predictions. We assemble our own contributions to the study of these models, which were presented in detail in Part \ref{part:results}, and we put them in the context of both experimental and theoretical advances expected in the near future.}
\end{quotation}

The goal of this thesis was to understand the origin and study the consequences of inflationary scenarios based on string theory. A natural motivation for this lies in the mutual completion promised by a combination of early Universe cosmology and high energy physics: phenomenologically successful, the scenario of inflation has yet to be put on the firm footing of an underlying theory. Ambitiously aimed at the unification of all physics, string theory is still in search of decisive signatures relevant at energy scales one can realistically hope to probe.\\
Today, we are in the fortunate situation that significant advances on both sides of the aisle enable us to bridge this gap: cosmological observations have become sufficiently precise to make quantitative statements about the very first moments of the Universe's history. String theoretic constructions, on the other hand, are now understood adequately enough to reduce the degrees of freedom during a stringy period of inflation to a manageable number and build well-defined scenarios.

Our first project, presented in Chapter \ref{chapter:WMAP3-paper}, was devoted to a detailed analysis of the KKLMMT brane--anti-brane inflationary scenario. Starting from its type IIB supergravity origin, we identified the effective cosmological parameters of this model and broke its evolution up into distinct regimes (quantum fluctuation dominated at very large field values, a slow roll phase producing nearly all observable \efolds, a negligibly short period of DBI dynamics, and finally brane annihilation/reheating). We then integrated the evolution of background and perturbations during inflation numerically and applied a phenomenological description to the complex reheating phase, before propagating the primordial spectra through the SBBM evolution using the \texttt{CAMB} code. The Monte Carlo Markov Chain comparison of the KKLMMT model to the WMAP3 data lead to probability distributions (and, in some cases, limits) for its cosmological and consequently its underlying stringy parameters.\\
While the DBI phase was found to be unimportant in the ``pure'' KKLMMT model with a Coulombic potential term (from brane--anti-brane attraction) only, the modified dynamics due to the the open string mode character of the inflaton can be crucial for more general potentials. Under the name of $k$-inflation, scenarios with a non-canonical kinetic term were even studied long before the current burst of activity in string-inspired model building. In Chapter \ref{chapter:kinf-WMAP5-paper}, we present the calculation of $k$-inflationary scalar and tensor perturbation spectra in the analogue of the slow roll limit for standard inflation: because of the modified kinetic term, the scalar perturbations have a sound speed $\cS\neq1$ which, to justify the slow roll limit, can only change slowly with $\dot{\cS}/\cS\ll1$ (in addition to the slow variation of the Hubble parameter required in the standard case). A comparison of both general $k$-inflation scenarios and their DBI type subclass to the WMAP5 data was provided in a separate publication (Chapter \ref{chapter:kinf-WMAP5-paper}).\\
Inflation in the KKLMMT scenario ends when the D3-brane alights onto the anti-D3 and they start annihilating into closed string modes. The tell-tale sign of this process is the appearance of a tachyon field. In Chapter \ref{chapter:tachyon-paper}, we studied the phenomenology of the temporary two-field model (inflaton and tachyon) at work during the early stages of reheating. As is always the case if more than one field is dynamic during inflation, entropy perturbations on top of the standard adiabatic ones can develop, and their growth was analyzed under simplifying assumptions in Chapter \ref{chapter:tachyon-paper}. It was found that there exist parameter choices where the induced curvature perturbation due to the entropy fluctuations is of the same size as the adiabatic contribution.\\
The KKLMMT scenario is derived from type IIB superstring theory. Recently, a new model of brane inflation based on the dual type IIA theory was proposed, in which the inflaton is identified with the recurring motion of a D4-brane around a ``monodromic'' coordinate of the compact background geometry (which, in the simplest case, is a product of twisted tori). The striking new feature of this construction is the large range of values the inflaton can cover in field space, which, in the brane--anti-brane models studied so far, was always limited by the maximum size of the compact dimensions. Reheating in ``monodromy inflation'' occurs when the D4-brane unwinding in the monodromic direction collides with \eg a (fixed) D6-brane on which the Standard Model particles are localized. It was shown in the work of Chapter \ref{chapter:monodromy-paper} that, even though the D4 and the D6 collide multiple times during inflation (the D6 is, figuratively speaking, ``stuck in the way'' of the D4), the energy stored in the inflaton potential and the brane dynamics is transfered to the Standard Model only at their last encounter after inflation has ended. The reheating temperature is found to be high.

As we have seen, the stringy point of view on various building blocks of the inflationary scenario requires the generalization of standard tools to account, for example, for the modified kinetic term of the inflaton. On the other hand, the fundamentally geometric interpretation string theory applies to physics can, in the context of inflation, be used to justify parameter ranges such as the distance covered by the inflaton in field space: for example, if the inflaton is the separation of branes along one of the compact dimensions of the string theoretic spacetime, it cannot take values exceeding the total size of this dimension.\\
In our work, we have tried to strengthen the link between string theory and cosmological scenarios by insisting on a consistent and, as far as possible, complete translation of the former's concepts into the latter's observational quantities. It was shown that, in principle, string cosmology makes it possible to ``read off'' string parameters (such as the volume of compactified dimensions) from precision measurements of cosmic microwave background anisotropies. In this undertaking we are, however, up against two intrinsic challenges: while a unique theory, it has been realized that string theory has a \emph{very large} number of possible vacuum states, and each of these vacua in turn can arise from many different combinations of its parameters, all of which are intertwined via their geometric interpretation. At the ``theory end'', fixing the quantities entering into a given string inflation model is therefore a  task that currently defies successful completion. Moreover, on the cosmological side, it is the built-in phenomenological robustness of inflation that may keep us from learning too much about its microphysical origin: inflationary predictions depend on few and generic observables, and the window of scales through which we may grasp a look at early Universe physics is restricted to a few \efolds. Therefore, distinguishing a string theoretic model from a purely field theoretic ansatz can be a delicate problem.\\
Nevertheless, we can today more than ever hope to rise to this challenge. The prospect of making contact with observations through cosmology has reinvigorated string theorists' efforts to deduce a full-scale description of our Universe from its earliest moments onwards from superstring theory. Cosmologists have every reason to share the excitement: the Planck satellite has just left its launch pad in Kourou and will start collecting several years' worth of data on both the CMB temperature fluctuations and their polarization. Thence, the decryption of intricate details about the era of inflation in the primordial Universe may be within our grasp before long.

\vspace{1cm}

\begin{figure}[h]
\begin{center}
\includegraphics[width=0.44\textwidth]{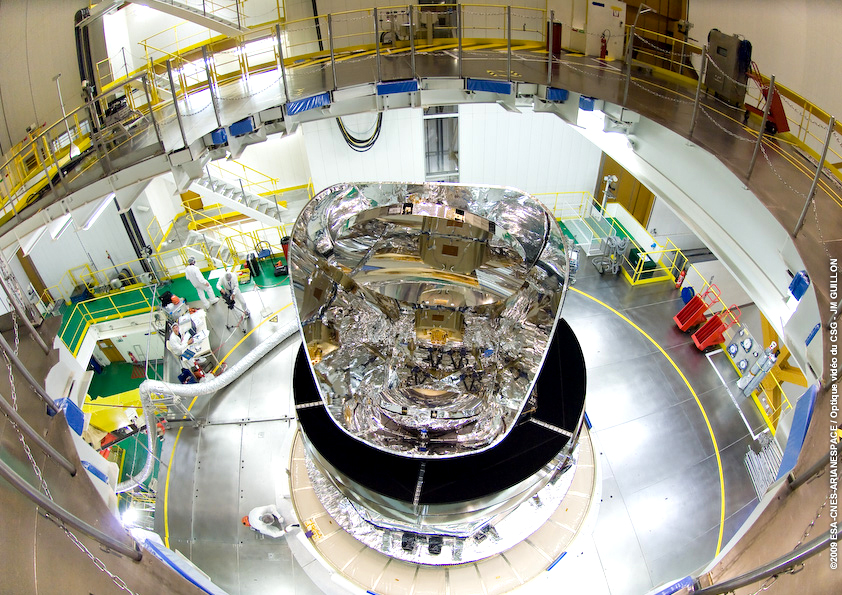}
\hfill
\includegraphics[width=0.46\textwidth]{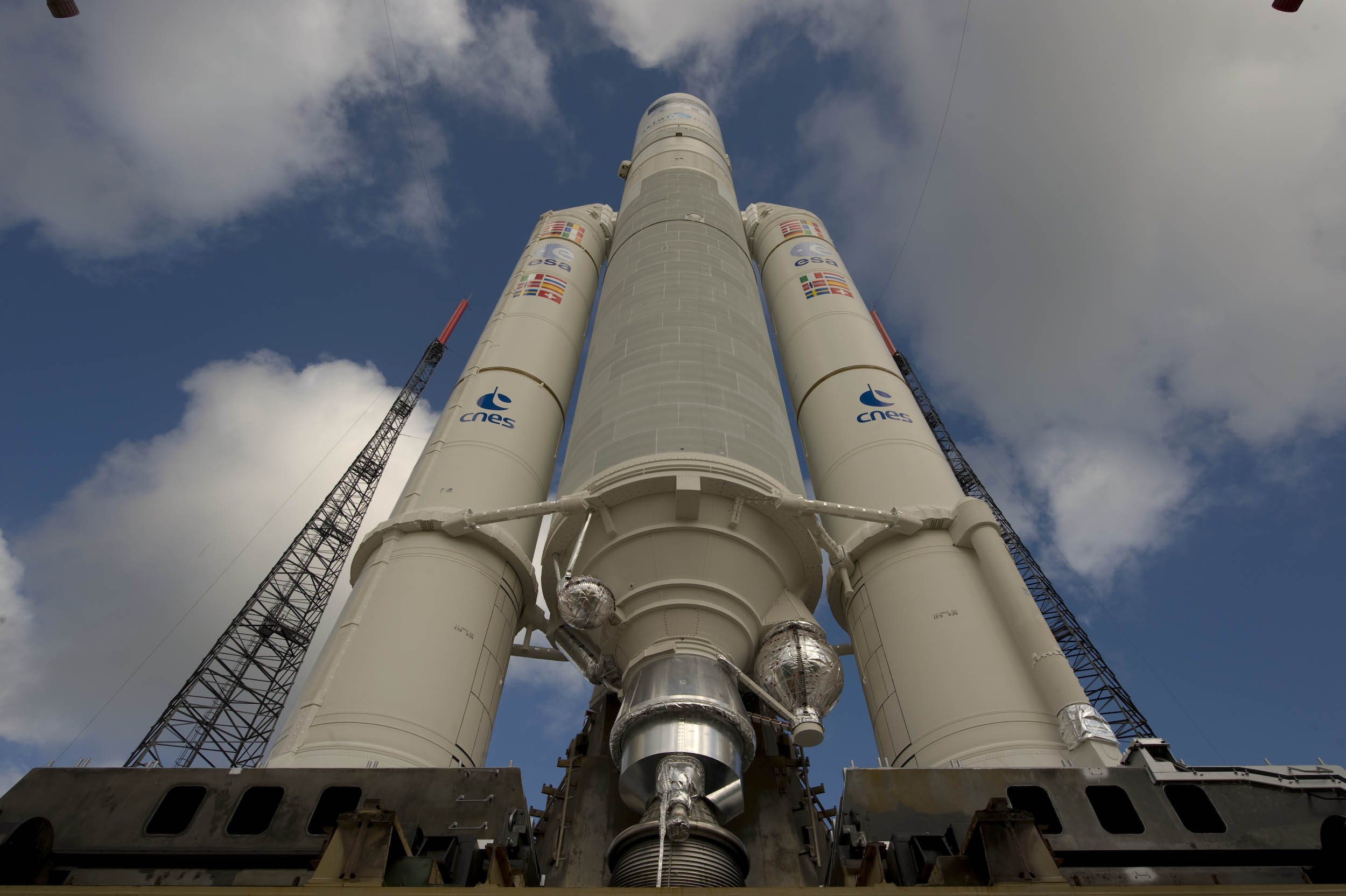}
\caption[Planck satellite before being integrated into the Ariane 5 launcher rocket, and the Herschel and Planck in their Ariane 5 launcher on the launch pad. (Source: website of the European Space Agency)]{\emph{Left:} The Planck satellite mirror before packaging. \emph{Right}: Herschel and Planck integrated in the Ariane 5 rocket, ready for take off on its launch pad. Lift off from the ESA basis in Kourou (French Guyana) took place as scheduled at 15:12 CEST on May 14th, 2009. [Pictures from the website of the European Space Agency (ESA)]}
\end{center}
\label{fig:Plancklaunch}
\end{figure}

\part{Appendices and Bibliography}\label{part:appendices}
\begin{appendix}

\chapter{Remarks on Geometry and Topology}\label{app:geotopo}
In this Appendix, we assemble some elements of geometry and topology used throughout Part \ref{part:string} of this thesis. An introduction to these topics accessible to physicists is found in \cite{Nakahara:1990th}, and short summaries for string theoretic applications are included in \cite{Becker:2007zj} and \cite{Polchinski:1998rq,Polchinski:1998rr}.

\section{Real Manifolds}
Let $\scrM$ be a compact real $d$-dimensional manifold with no boundaries. On $\scrM$, one may define $p$-forms $A_{(p)}$, with $p=0,1,\dots,d$, from
\beq
A_{(p)}=\frac{1}{p!}\,A_{\mu_{1}\cdots\mu_{p}}\,\dd x^{\mu_{1}}\wedge\cdots\wedge\dd x^{\mu_{p}}\,,
\eeq
where ``$\wedge$'' denotes the antisymmetrized (``wedge'') tensor product. The exterior derivative $\dd$, when applied to a $p$-form, turns it into a $(p+1)$-form as
\beq
\dd A_{(p)}=\frac{1}{p!}\,\partial_{\mu_{1}}A_{\mu_{2}\cdots\mu_{p+1}}\,\dd x^{\mu_{1}}\wedge\cdots\wedge\dd x^{\mu_{p+1}}\,.
\eeq
A $p$-form is called closed if $\dd A_{(p)}=0$, and it is exact if there exists a globally defined $(p-1)$-form such that $A_{(p)}=\dd A_{(p-1)}$. (From $\dd^{2}=0$, as can easily be shown from antisymmetry, it follows that an exact form is always closed.) 
The closed $p$-forms on $\scrM$ form the space $\mathscr{C}^{p}(\scrM)$, and the exact $p$-forms form the space $\mathscr{Z}^{p}(\scrM)$. The $p$th \emph{de Rham cohomology group} is defined as their quotient:
\beq
\mathscr{H}^{p}(\scrM)=\mathscr{C}^{p}(\scrM)/\mathscr{Z}^{p}(\scrM)
\eeq
Within $\mathscr{H}^{p}(\scrM)$, two closed $p$-forms therefore are equivalent if their difference is an exact form. The dimensionality of $\mathscr{H}^{p}(\scrM)$ gives the \emph{Betti number} $b_{p}$. The \emph{Euler characteristic} of the manifold $\chi(\scrM)$ then is obtained from an alternating sum of Betti numbers,
\beq
\chi(\scrM)=\sum_{i=0}^{d}(-1)^{i}\,b_{i}(\scrM)\,.
\eeq
Like the cohomology groups $\mathscr{H}^{p}(\scrM)$ of a manifold are defined from the action of the operator $\dd$, one can define the \emph{homology groups} from the boundary operator $\delta$ acting on submanifolds $\mathscr{N}$ of $\scrM$. (By $\delta\mathscr{N}$, we therefore mean the boundary of $\mathscr{N}$. The sign of $\delta\mathscr{N}$ accounts for its orientation.) Note that again $\delta^{2}=0$ because the boundary of a boundary is zero.

A $p$-chain $z_{p}$ is a linear combination of submanifolds of dimension $p$. (A $p$-chain that has no boundary is closed, and a $p$-chain that \emph{is} a boundary is called exact.) A closed chain is also called a cycle, and then $\delta z_{p}=0$. Two $p$-cycles are equivalent if and only if they differ by only a boundary. The \emph{simplical homology group} consists of equivalence classes of $p$-cycles and is denoted by $\mathscr{H}_{p}(\scrM)$.

\emph{Stokes' theorem} for a real manifold $\scrM$ of an arbitrary number of dimensions $d$ may be written as
\beq
\int_{\mathscr{N}}\dd A_{(p)}=\int_{\delta\mathscr{N}}A_{(p)}\,,
\eeq
where $\mathscr{N}$ is an arbitrary $(p+1)$-chain. This describes the so-called \emph{Poincar\'e duality}, which is an isomorphism between the groups $\mathscr{H}^{p}$ and $\mathscr{H}_{d-p}$ of $\mathscr{M}$. For every closed $p$-form $A_{(p)}$, and closed $(d-p)$-form $B_{(d-p)}$, there exists a relation between the manifold $\scrM$ and the $(d-p)$-cycle $\mathscr{N}$:
\beq
\int_{\scrM}A_{(p)}\wedge B_{(d-p)}=\int_{\mathscr{N}}B_{(d-p)}
\eeq
(Recall that $\scrM$ is closed and therefore has no boundary.) Using Poincare duality, one can determine the Betti number by counting those $p$-cycles of $\scrM$ which are not boundaries.

A manifold with a positive-definite metric is called a \emph{Riemannian manifold}, and with a metric of indefinite signature it is a \emph{pseudo-Riemannian manifold}. The metric allows one to calculate the (coordinate-independent) infinitesimal line element $\dd s^{2}=g_{mn}\,\dd x^{m}\,\dd x^{n}$. The Laplace operator acting on $p$-forms $\Delta_{p}$ in $d$ dimensions is given (in Euclidean signature) by
\beq\label{eq:Laplaceop-p}
\Delta_{p}=\dd^{\dagger}\dd+\dd\dd^{\dagger}=\left(\dd+\dd^{\dagger}\right)^{2}\,,\qquad\dd^{\dagger}=(-1)^{dp+d+1}*\dd*\,.
\eeq
(For Lorentzian signature, there is an additional minus sign.) The \emph{Hodge} or \emph{star operator} $*$ in Eq.~(\ref{eq:Laplaceop-p}) acts on $p$-forms and is given by
\beq\label{eq:Hodgestar}
*\left(\dd x^{\mu_{1}}\wedge\cdots\wedge\dd x^{\mu_{p}}\right)=\frac{\epsilon^{\mu_{1}\cdots\mu_{p}\mu_{p+1}\cdots\mu_{d}}}{(d-p)!\,|g|^{1/2}}\,g_{\mu_{p+1}\nu_{p+1}}\cdots g_{\mu_{d}\nu_{d}}\,\dd x^{\nu_{p+1}}\wedge\cdots\wedge\dd x^{\nu_{d}}\,,
\eeq
with $\epsilon^{\mu_{1}\cdots\mu_{p}\mu_{p+1}\cdots\mu_{d}}$ the Levi Civita symbol. A $p$-form is harmonic if and only if $\Delta_{p}\,A_{(p)}=0$, and consequently a harmonic $p$-form is closed and co-closed (meaning $\dd^{\dagger}A_{(p)}=0$). 
The Hodge dual turns a closed $p$-form into a co-closed $(d-p)$-form, therefore it can be understood as an isomorphism between the space of harmonic $p$-forms and the space of harmonic $(d-p)$-forms, and for the Betti numbers it follows that $b_{p}=b_{d-p}$.

\section{Complex Manifolds}\label{appsec:complexmanif}
A complex manifold $\mathscr{M}$ of complex dimension $n$ is defined using complex local coordinates $z^{a},\bar{z}^{\bar{a}}\,(a=1,\dots,n)$. A real manifold of even dimension $d=2n$ is a complex manifold if the following conditions are satisfied: firstly, determine whether a tensor called the almost complex structure ${J^{m}_{}}_{n}$, which satisfies ${J_{m}^{}}^{n}{J_{n}^{}}^{p}=-{\delta_{m}^{}}^{p}$, exists. Secondly, check whether ${J^{m}_{}}_{n}$ is an actual complex structure by calculating the so-called \emph{Nijenhuis tensor}. If this tensor vanishes, ${J^{m}_{}}_{n}$ is a complex structure. In that case, one can choose a local complex coordinate system $(z^{a},\bar{z}^{\bar{a}})$ in every open set on $\scrM$ such that ${J^{m}_{}}_{n}$ is given by
\beq
{J^{m}_{}}_{n}=i{\delta_{m}^{}}^{n},\qquad {J^{\bar{m}}_{}}_{\bar{n}}=i{\delta_{\bar{m}}^{}}^{\bar{n}},\qquad {J^{\bar{m}}_{}}_{n}={J^{m}_{}}_{\bar{n}}=0\,.
\eeq

Like $p$-forms on real manifolds, on a complex manifold one can define $(p,q)$-forms with $p$ holomorphic and $q$ anti-holomorphic indices:
\beq
A_{p,q}=\frac{1}{p!\,q!}\,A_{a_{1}\cdots a_{p}\bar{b}_{1}\cdots\bar{b}_{q}}\,\dd z^{a_{1}}\wedge\cdots\wedge\dd z^{a_{p}}\wedge\dd\bar{z}^{\bar{b}_{1}}\wedge\cdots\wedge\dd\bar{z}^{\bar{b}_{q}}
\eeq
The exterior derivative (which is real) can then be decomposed into a holomorphic and an anti-holomorphic piece, $\dd=\partial+\bar{\partial}$, where one uses the notation $\partial=\dd z^{a}\,(\partial/\partial z^{a})$ and $\bar{\partial}=\dd \bar{z}^{\bar{a}}\,(\partial/\partial \bar{z}^{\bar{a}})$. These operators take a $(p,q)$-form to a $(p+1,q)$- or a $(p,q+1)$-form, respectively.

On a complex Riemannian manifold, the line element can be written in terms of local complex coordinates as
\beq
\dd s^{2}=g_{ab}\,\dd z^{a}\,\dd z^{b}+g_{a\bar{b}}\,\dd z^{a}\,\dd \bar{z}^{\bar{b}}+g_{\bar{a}b}\,\dd \bar{z}^{\bar{a}}\,\dd z^{b}+g_{\bar{a}\bar{b}}\,\dd \bar{z}^{\bar{a}}\,\dd \bar{z}^{\bar{b}}\,.
\eeq
For the metric components, one has $g_{\bar{a}\bar{b}}=\left(g_{ab}\right)^{*}$ and $g_{a\bar{b}}=\left(g_{\bar{a}b}\right)^{*}$ from the requirement of reality. On a Hermitian manifold, $g_{\bar{a}\bar{b}}=g_{ab}=0$.

By the \emph{Dolbeault cohomology group} $\mathscr{H}^{p,q}_{\bar{\partial}}(\scrM)$ of a Hermitian manifold $\scrM$ we mean the equivalence classes of $\bar{\partial}$-closed $(p,q)$-forms, and the dimension of $\mathscr{H}^{p,q}_{\bar{\partial}}(\scrM)$ is called the \emph{Hodge number} $h^{p,q}$. The Laplacian operators for complex manifolds are
\beq
\Delta_{\partial}=\partial\partial^{\dagger}+\partial^{\dagger}\partial,\qquad \Delta_{\bar{\partial}}=\bar{\partial}\bar{\partial}^{\dagger}+\bar{\partial}^{\dagger}\bar{\partial}\,.
\eeq
A \emph{K\"ahler manifold} is defined to be a Hermitian manifold on which the \emph{K\"ahler form} $J$ is closed:
\beq
J=i\,g_{a\bar{b}}\,\dd z^{a}\wedge\dd\bar{z}^{\bar{b}},\qquad \dd J=0
\eeq
The metric on these manifolds therefore satisfies $\partial_{a}g_{b\bar{c}}=\partial_{b}g_{a\bar{c}}$ (and the complex conjugate), therefore one may write the metric as the derivative of a \emph{K\"ahler potential} $K$,
\beq
g_{a\bar{b}}=\frac{\partial}{\partial z^{a}}\,\frac{\partial}{\partial\bar{z}^{\bar{b}}}\,K(z,\bar{z})\,.
\eeq
Note that the K\"ahler form then is $J=i\partial\bar{\partial}K$, which means that the K\"ahler potential is only defined up to addition of arbitrary (anti-)holomorphic functions, $\tilde{K}(z,\bar{z})=K(z,\bar{z})+f(z)+\bar{f}(\bar{z})$. On K\"ahler manifolds, $\Delta_{\dd}=2\Delta_{\partial}=2\Delta_{\bar{\partial}}$, and for the cohomology groups defind with respect to the operators $\bar{\partial}$, $\partial$ and $\dd$, respectively, one has
\beq
\mathscr{H}^{p,q}_{\bar{\partial}}(\scrM)=\mathscr{H}^{p,q}_{\partial}(\scrM)=\mathscr{H}^{p,q}(\scrM)\,.
\eeq
This means that the Hodge and the Betti numbers are related by $b_{k}=\sum_{p=0}^{k}h^{p,k-p}$.

If $\omega$ is a $(p,q)$-form on a K\"ahler manifold with $n$ complex dimensions, then the complex conjugate form $\omega^{*}$ is a $(q,p)$-form. Therefore the Hodge numbers of a K\"ahler manifold are related by $h^{p,q}=h^{q,p}$. From the operation of the Hodge star given in Eq.~(\ref{eq:Hodgestar}), it follows that $*\omega$ is a $(n-p,n-q)$-form and $h^{n-p,n-q}=h^{p,q}$.

On a Hermitian manifold, only the mixed components of the Ricci tensor are different from zero, and one can define the $(1,1)$ \emph{Ricci form} from
\beq
\mathscr{R}=i\,R_{a\bar{b}}\,\dd z^{a}\wedge\dd\bar{z}^{\bar{b}}\,.
\eeq
The Ricci form is closed, $\dd\mathscr{R}=0$, and therefore belongs to the cohomology class of $\mathscr{H}^{1,1}(\mathscr{M})$, which is called the \emph{first Chern class} $c_{1}=(1/2\pi)\,[\mathscr{R}]$\,.

There is one more property of a Riemannian manifold $\scrM$ (with real dimension $d$) we refer to in Section \ref{subsec:compactify}: the holonomy group $\mathscr{H}(\scrM)$ describes the way various objects transform under parallel transport around closed curves. The most important examples (which have so-called special holonomy) are $\mathscr{H}\subseteq U(d/2)$ if $\scrM$ is a K\"ahler manifold, and $\mathscr{H}\subseteq SU(d/2)$ if $\scrM$ is a \emph{Calabi Yau manifold}.

\end{appendix}

\backmatter
 
\markboth{List of figures}{List of figures}
\listoffigures

\newpage
\markboth{Bibliography}{Bibliography}
\bibliographystyle{utphys}
\bibliography{thesis_references_lorenz}



\end{document}